\documentclass[11pt,xcolor=dvipsnames]{article}
\DeclareMathAlphabet{\scr}{U}{rsfs}{m}{n}
\pdfoutput=1
\usepackage{latexsym}
\usepackage{epsfig}
\usepackage[mathscr]{eucal}
\usepackage{amsfonts}
\usepackage{amscd}
\usepackage{cite}
\usepackage{array}
\usepackage{amssymb}
\usepackage{colordvi}
\usepackage[centertags]{amsmath}
\usepackage{enumerate}
\usepackage{graphicx}
\usepackage{booktabs}
\usepackage{theorem}
\usepackage[footnotesize]{caption}
\usepackage{soul}
\usepackage{mcite}
\usepackage{slashed}
\usepackage{xcolor}
\usepackage{bbm}
\usepackage[utf8]{inputenc}
\usepackage{fancyvrb}% provides improved Verbatim command
\usepackage{geometry}
\usepackage{epstopdf}
\setstcolor{red}
\usepackage{float}
\usepackage[caption = false]{subfig}

\newcommand{\newc}{\newcommand}
\newc{\be}{\begin{equation}}
\newc{\ee}{\end{equation}}
\newc{\bea}{\begin{eqnarray}}
\newc{\eea}{\end{eqnarray}}
\newc{\ol}{\overline}
\newc{\wt}{\widetilde}
\newc{\bs}{\boldsymbol}
\newc{\m}{\mathcal}
\newc{\la}{\langle}
\newc{\ra}{\rangle}

\newcommand{\beq}{\begin{eqnarray}}
\newcommand{\eeq}{\end{eqnarray}}
\newcommand{\bpmatrix}{\begin{pmatrix}}
\newcommand{\epmatrix}{\end{pmatrix}}
\newcommand{\ba}{\begin{array}}
\newcommand{\ea}{\end{array}}

\renewcommand{\ol}{\text{1l}}

\renewcommand{\eqref}[1]{Eq.~(\ref{#1})}

\newcommand{\bc}{\begin{center}}
\newcommand{\ec}{\end{center}}

\newcommand{\lsim}{\raisebox{-0.13cm}{~\shortstack{$<$ \\[-0.07cm]
      $\sim$}}~}
\newcommand{\s}{\newline \vspace*{-3.5mm}}

\begin{document}

\title{
\vspace*{-2.5cm}
%\phantom{h} \hfill\mbox{\small KA-TP-xx-2021}%\\[-1.1cm]
%\vspace*{1cm}
%\\[1cm]
%\vspace{13mm}
\textbf{Benchmarking Di-Higgs Production in Various Extended Higgs Sector Models\\[4mm]}}
%Di-higgs production in extended Higgs sector\\[4mm]}}

\date{}
\author{
Hamza Abouabid$^{1\,}$\footnote{E-mail:
\texttt{aabouabid@uae.ac.ma}} ,
Abdesslam Arhrib$^{1\,}$\footnote{E-mail:
\texttt{aarhrib@uae.ac.ma}} ,
Duarte Azevedo$^{3,4\,}$\footnote{E-mail:
\texttt{duarte.azevedo@kit.edu}} ,
Jaouad El Falaki$^{5\,}$\footnote{E-mail:
\texttt{jaouad.elfalaki@gmail.com}} , \\
Pedro. M. Ferreira $^{2,6\,}$\footnote{E-mail:
\texttt{pmmferreira@fc.ul.pt}} ,
Margarete M\"{u}hlleitner$^{3\,}$\footnote{E-mail:
\texttt{margarete.muehlleitner@kit.edu}} ,
Rui Santos$^{2,6\,}$\footnote{E-mail:
  \texttt{rasantos@fc.ul.pt}}
  \\[9mm]
\noindent
{\small\it
$^1$Abdelmalek Essaadi University, Faculty of Sciences 	and Techniques} \\
{\small\it Ancienne route de l’aéroport, Ziaten, B.P. 416, Tangier, Morocco
}\\[3mm]
{\small\it
$^2$ISEL -
 Instituto Superior de Engenharia de Lisboa,} \\
{\small \it Instituto Polit\'ecnico de Lisboa
 1959-007 Lisboa, Portugal} \\[3mm]
{\small\it
$^3$Institute for Theoretical Physics, Karlsruhe Institute of Technology,} \\
{\small\it  Wolfgang-Gaede Str. 1, 76128 Karlsruhe, Germany} \\[3mm]
{\small \it
$^4$Institute for Astroparticle Physics, Karlsruhe Institute of Technology,} \\
{\small \it Hermann-von-Helmholtz-Platz 1,
	76344 Eggenstein-Leopoldshafen, Germany} \\[3mm]
{\small\it
$^5$LPTHE, Physics Department, Faculty of Science, Ibn Zohr University,} \\
{\small\it P.O.B. 8106, Agadir, Morocco}\\[3mm]
{\small\it
$^6$Centro de F\'{\i}sica Te\'{o}rica e Computacional,
    Faculdade de Ci\^{e}ncias,} \\
{\small \it    Universidade de Lisboa, Campo Grande, Edif\'{\i}cio C8
  1749-016 Lisboa, Portugal}
}
\maketitle
\vspace*{-1cm}
\begin{abstract}
\noindent
We present a comprehensive study on Higgs pair production in various
archetypical extended Higgs sectors
such as the real and the complex 2-Higgs-Doublet Model, the
2-Higgs-Doublet Model augmented by a real singlet field and the
Next-to-Minimal Supersymmetric extension of the Standard Model. We take into
account all relevant theoretical and experimental constraints, in
particular the experimental limits on non-resonant and resonant
Higgs pair production. We present the allowed cross sections for
Standard Model (SM)-like Higgs pair production and the ranges of the
SM-like Yukawa and 
trilinear Higgs self-coupling that are still compatible with the applied
constraints. Furthermore, we give results for the pair production of a
SM-like with a non-SM-like Higgs boson and for the production of a
pair of non-SM-like Higgs bosons. We find that di-Higgs production in
the models under investigation can exceed the SM rate substantially,
not only in the non-resonance
region but also due to resonant enhancement. We give
several benchmarks with interesting features such as large cross
sections, the possibility to test CP violation, Higgs-to-Higgs cascade
decays or di-Higgs production beating single Higgs production. 
In all of our benchmark points, the next-to-leading order QCD corrections
are included in the large top-mass limit. For these points, we found
that, depending on the model and the Higgs pair final state, the corrections
increase the leading order cross section by a factor of 1.79 to 2.24. We also
discuss the relation between the description of Higgs pair production
in an effective field theory approach and in the specific models
investigated here.
\end{abstract}
\thispagestyle{empty}
\vfill
\newpage
\setcounter{page}{1}

%%%%%%%%%%%%%%%%%%%%%%%%%%%%%%%%%%%%%%%%%%%%%%%%%%%%
\vspace*{-2cm}
\tableofcontents
%%%%%%%%%%%%%%%%%%%%%%%%%%%%%%%%%%%%%%%%%%%%%%%%%%%%%%%
\section{Introduction} \label{sec:introduction}
The discovery of the Higgs boson in 2012 by the ATLAS
\cite{Aad:2012tfa } and CMS \cite{Chatrchyan:2012ufa}
collaborations at the Large Hadron Collider (LHC) structurally completes
the Standard Model (SM) of particle physics. The subsequent
investigation of its properties revealed the Higgs boson to be very
SM-like
\cite{Khachatryan:2014kca,Khachatryan:2014jba,Aad:2015mxa,Aad:2015gba}.
In order to verify that the Brout-Englert-Higgs mechanism
\cite{Higgs:1964pj,Englert:1964et,Guralnik:1964eu,Kibble:1967sv} is
indeed responsible for the generation of elementary particle masses we
not only need to measure the Higgs couplings to massive SM particles
at the highest precision but we also need to reconstruct the Higgs
potential itself to establish experimentally whether indeed its
Mexican hat shape is responsible for the Brout-Englert-Higgs mechanism
of spontaneous symmetry
breaking~\cite{Djouadi:1999gv,Djouadi:1999rca}. 
Moreover, despite the success of the SM, which has been tested very
accurately at the quantum level, some open problems remain that cannot be
solved by the SM, like {\it e.g.}~the nature of Dark Matter (DM) or
why there is more matter than antimatter in the universe.  Extensions of the
Higgs sector beyond the SM can provide a DM candidate
\cite{Hambye:2007vf, Ma:2006km,Deshpande:1977rw,Arcadi:2019lka,Drozd:2014yla,Chen:2013jvg,Cai:2013zga,Ko:2014uka,Arhrib:2015dez,Arhrib:2018eex}. 
Electroweak baryogenesis provides a mechanism to
dynamically generate the matter-anti-matter asymmetry if the three
Sakharov conditions \cite{Sakharov:1967dj} are fulfilled. This may be possible within
extended Higgs sectors with additional scalar degrees of freedom and
new sources of CP violation. In summary, the investigation
of the Higgs potential itself
will provide deep insights into the mechanism underlying electroweak
symmetry breaking and into the possible landscape of new physics
extensions of the Higgs sector. Understanding the Higgs potential will
help us to answer some of the open questions of contemporary
particle physics. \s

In the SM the trilinear and quartic Higgs self-couplings are given in
terms of the Higgs boson mass. This is not the case in extended Higgs sectors. Experimentally, the Higgs self-interactions are
accessible in multi-Higgs production, the trilinear Higgs
self-coupling in double and the quartic Higgs self-coupling in triple
Higgs production. At the LHC the dominant di-Higgs production process
is given by gluon fusion into Higgs pairs
\cite{Baglio:2012np,deFlorian:2016spz,DiMicco:2019ngk}.
At leading order the process is mediated by heavy quark triangle and
box diagrams \cite{Glover:1987nx,Dicus:1987ic,Plehn:1996wb}, implying
a small SM cross section of 31.05~fb
at FT$_{\text{approx}}$ \cite{Grazzini:2018bsd} and for a c.m.~energy
of $\sqrt{s}=13$~TeV.
At FT$_{\text{approx}}$, the cross section is computed at
next-to-next-to-leading order (NNLO) QCD in
the heavy-top limit with full leading order (LO) and next-to-leading
order (NLO) mass effects and full mass
dependence in the one-loop double real corrections at NNLO QCD.
Note that the NLO cross
  section in the heavy top limit was first presented in
  \cite{Dawson:1998py} supplemented by a large-top-mass expansion
  \cite{Grigo:2013rya,Grigo:2015dia} and the inclusion of the full
  real corrections \cite{Frederix:2014hta,Maltoni:2014eza}. The 
  NLO result including the full top-quark 
  mass dependence has been calculated by
  \cite{Borowka:2016ehy,Borowka:2016ypz,Baglio:2018lrj,Baglio:2020ini,Baglio:2020wgt}
  and confirmed in \cite{Grober:2017uho,Bonciani:2018omm} by applying
  suitable expansion methods. 
The NNLO corrections in the large $m_t$ limit have been calculated in
\cite{deFlorian:2013jea,Grigo:2014jma,Davies:2021kex}, the results at next-to-next-to-leading
logarithmic accuracy (NNLL) became available in
\cite{Shao:2013bz,deFlorian:2015moa}. And the corrections up to
next-to-next-to-next-to leading order (N$^3$LO) were presented in
\cite{Banerjee:2018lfq,Chen:2019lzz,Chen:2019fhs} for the heavy
top-mass limit. 
Note that these corrections apply to the SM
case and cannot necessarily be taken over to extended Higgs sectors
where {\it e.g.}~the bottom loops may play an important role or
additional diagrams are involved.
\s

In the SM the process suffers from a destructive interference between
the triangle and box diagrams which makes its observation very
challenging at the LHC. In extended Higgs sectors SM-like di-Higgs production
can be enhanced because Yukawa and trilinear Higgs
self-couplings are modified
 %of the involved Higgs bosons
relative to the SM. This can result in altering the interference
structure between the various diagrams. 
Additional Higgs bosons can enhance the cross section
resonantly and new colored particles that run in the loops can also 
lead to an increased cross section.
Therefore, extended Higgs sectors not only answer some of the open questions but
they can possibly also facilitate access to di-Higgs production. Furthermore, the
additional Higgs states lead to a large variety of di-Higgs final
states implying a plethora of multi-particle final state signatures.
To get a comprehensive picture of which final state signatures in Higgs
pair production processes are possible and to be able to give a meaningful guideline
to experimentalists in their searches for Higgs pair production, we
have to take into account all available experimental and
theoretical constraints on beyond-the-SM (BSM) extensions. Since
the discovered Higgs boson behaves very SM-like, new physics
extensions become increasingly constrained. The trilinear Higgs
self-coupling, however, is not as constrained as single Higgs
couplings yet \cite{ATLAS-CONF-2021-052,CMS:2020tkr} and we may still
expect some distinctive signatures from Higgs pair production. \s

The goal of this paper is to investigate Higgs pair production in some archetypical BSM
extensions. Imposing all available constraints - also from recent
di-Higgs searches - we derive the parameter space of these models
that is still allowed. Based on this data set we will derive limits
on the involved couplings in Higgs pair production, namely the
trilinear Higgs self-coupling and the Higgs-Yukawa coupling,
investigate the possible enhancement 
(or also suppression) of SM-like Higgs pair production and investigate
what kind of non-SM-like signatures might appear. We provide benchmark
points so that experimentalists can match their derived limits
on specific models for further interpretation. Since the experiments
derive limits from non-resonant and resonant Higgs pair production and
our extended Higgs sector models include both effects, we derive a
strategy on how we can apply the available di-Higgs limits on our
models. We included for all benchmark scenarios and
  parameter points fulfilling the applied constraints the information
  on the resonant part of the cross section, where applicable.
Our aim is to give a global and comprehensive overview on Higgs
pair production in BSM Higgs sectors. \s

The models that we consider are both non-supersymmetric and
supersymmetric ones. Supersymmetry (SUSY) \cite{Golfand:1971iw, Volkov:1973ix, Wess:1974tw, Fayet:1974pd,Fayet:1977yc, Fayet:1976cr, Nilles:1982dy,Nilles:1983ge, Frere:1983ag,Derendinger:1983bz,Haber:1984rc, Sohnius:1985qm,Gunion:1984yn, Gunion:1986nh} is able
to solve many of the open problems of the SM. The non-minimal
supersymmetric extension (NMSSM) \cite{Barbieri:1982eh,Dine:1981rt,Ellis:1988er,Drees:1988fc,Ellwanger:1993xa,Ellwanger:1995ru,Ellwanger:1996gw,Elliott:1994ht,King:1995vk,Franke:1995tc,Maniatis:2009re,Ellwanger:2009dp} solves the little hierarchy
problem and more easily complies with the discovered SM-like Higgs
mass after inclusion of the higher-order corrections
\cite{Slavich:2020zjv}. The Higgs sector consists of two Higgs
doublets to which a complex singlet superfield is added so that after
electroweak symmetry breaking (EWSB) we have three neutral CP-even,
two neutral CP-odd and two charged Higgs bosons in the spectrum.
Supersymmetric relations constrain the
Higgs potential parameters in a different way than non-SUSY models.
Therefore, we also investigate
non-SUSY Higgs sector extensions where the trilinear couplings are
less constrained from a theoretical point of view. This way we make sure
not to miss some possibly interesting di-Higgs signatures. We start
with one of the most popular extensions complying with $\rho=1$
at tree level, the CP-conserving 2-Higgs doublet model (R2HDM)
\cite{Lee:1973iz,Branco:1985aq} where a second Higgs doublet is added to
the SM sector. Incorporating a minimal set of BSM Higgs
bosons (five in total, three neutral and two charged ones) allows
for resonant di-Higgs enhancement. We
additionally take into account the possibility of CP violation (which
is required for electroweak baryogenesis) by
investigating the CP-violating 2HDM (C2HDM)
\cite{Branco:1985aq,Ginzburg:2002wt,Khater:2003wq,Fontes:2017zfn}
which consists of three CP-mixed and two
charged Higgs bosons. In this case the SM-like Higgs couplings can be
diluted by CP admixture, the same happens through
singlet admixture. Thus, light Higgs bosons may not be excluded yet
because they may have escaped discovery through small couplings to the SM
particles. Such a singlet admixture is realized in the next-to-2HDM
(N2HDM) \cite{Chen:2013jvg,Muhlleitner:2016mzt,Engeln:2018mbg}. By adding
a real singlet field to the 2HDM Higgs sector the Higgs spectrum then consists
of three neutral CP-even Higgs bosons, one neutral CP-odd and two
charged Higgs bosons, allowing for the possibility of Higgs-to-Higgs
cascade decays. This is also possible in the C2HDM and the NMSSM. For simplicity, we will focus on the type I and II versions of the
R2HDM, C2HDM and N2HDM.
With the models investigated in this
paper\footnote{For a selection of further works on Higgs pair production and the Higgs self-coupling in the framework of BSM Higgs sectors in non-SUSY models, see {\it e.g.}~\cite{Arhrib:2009hc,Arhrib:2015hoa,Gupta:2013zza,Hespel:2014sla,Baglio:2014nea,Chen:2014ask,Dawson:2015oha,Costa:2015llh,Dawson:2015haa,Bojarski:2015kra,Grober:2015cwa,Carvalho:2016rys,Bian:2016awe,Grober:2016wmf,Krause:2016xku,Dawson:2017vgm,DiLuzio:2017tfn,Lewis:2017dme,Basler:2017uxn,Grober:2017gut,Heinrich:2017kxx,Dawson:2017jja,Basler:2018dac,Babu:2018uik,Chen:2018wjl,Basler:2019nas,Alasfar:2019pmn,Chang:2017niy,Robens:2019kga,Papaefstathiou:2020lyp,Park:2020yps,Cheung:2020xij,Barman:2020ulr,Das:2020ujo,Heinrich:2020ckp,Arco:2020ucn,deFlorian:2021azd,Arco:2021bvf,Cao:2015oaa,Cao:2015oxx,Cao:2016zob,Li:2019uyy},
and in SUSY models, see {\it
  e.g.}~\cite{Cao:2013si,Brucherseifer:2013qva,Nhung:2013lpa,Ellwanger:2013ova,Cao:2014kya,Muhlleitner:2015dua,Costa:2015llh,Heng:2018kyd,Liebler:2018zul,Baum:2018zhf,Huang:2019bcs,Ellwanger:2017skc,Das:2020ujo,Chalons:2017wnz}.}
we
cover a broad range of interesting new physics and in particular a
large variety of possible new physics signatures in di-Higgs
production. In turn, we provide guidelines to the
experiment. We also take the occasion to confront our models with a
simple effective field theory (EFT) approach and investigate to which
extent this model-independent parametrisation of new physics,
becoming effective at high scales, can describe the effects in our
investigated specific UV-complete models. \s

The paper is organised as follows. In Sec.~\ref{sec:models}, we briefly
introduce our models. In Sec.~\ref{sec:scans}, we present the regions
which we scanned for each model and the theoretical and experimental
constraints that we take into account. After a brief re-capitulation
of the di-Higgs production process through gluon fusion, we explain in
detail in Sec.~\ref{sec:numerical} how we
apply the experimental limits from resonant and non-resonant di-Higgs
searches on our models. We then investigate the impact of the di-Higgs
constraints on our parameter sample. We present the distributions
of the Higgs mass spectra in the different models and give the ranges for
the SM-like Higgs top-Yukawa and Higgs self-couplings that are still allowed
after considering all constraints. Subsequently, we present scatter
plots for all models showing the cross section values for SM-like
Higgs pair production that are compatible with the constraints, and we list the
maximum values for resonant and non-resonant Higgs pair production possible
in each model. In Sec.~\ref{sec:benchmarks} we present the maximum values from resonant SM-like di-Higgs
production and present the corresponding
benchmark points along with their specific features. In
Sec.~\ref{sec:constraining} we investigate to which extent di-Higgs
production can constrain the parameter values of the
models. Section~\ref{sec:eft} is devoted to the comparison of the EFT
description of BSM Higgs pair production with the results in specific
UV-finite models. The last two sections, \ref{sec:mixed} and
\ref{sec:nonsm}, are devoted to the pair
production of a SM-like Higgs together with a non-SM-like Higgs, of a
pair of non-SM-like Higgs bosons and to the cascade decays leading to
multi-Higgs final states. We also present benchmark points where
di-Higgs production beats single Higgs production. We conclude in Sec.~\ref{sec:concl}. In
the Appendix we present cross sections for resonant
  and non-resonant production and discuss the conditions for alignment in the
C2HDM and the N2HDM.

%DM candidate \cite{He:2008qm , Grzadkowski:2009iz , Logan:2010nw ,Boucenna:2011hy  , He:2011gc , Bai:2012nv, He:2013suk,Cai:2013zga,Guo:2014bha,Wang:2014elb,Drozd:2014yla,Campbell:2015fra,Drozd:2015gda}.

\section{The Models \label{sec:models}}

In this section we provide a very brief description of the different
models we will be studying, highlighting the diverse scalar spectra of each
model as well as the different input parameter sets for each of them.

\subsection{The Real and Complex 2HDM}

The 2HDM is one of the simplest extensions of the SM, where instead of a single Higgs doublet we now have two,
carrying identical hypercharges. The model was first proposed by Lee in 1973~\cite{Lee:1973iz} to provide
an extra source of CP violation via spontaneous symmetry breaking, and has a rich phenomenology (for a review,
see~\cite{Branco:2011iw}). We considered the 2HDM version with a
softly broken discrete $\mathbb{Z}_2$ symmetry of the form $\Phi_1 \to \Phi_1\;$ and $\Phi_2 \to - \Phi_2\;$. In terms of the two $SU(2)_L$ Higgs
doublets $\Phi_{1,2}$ with hypercharge $Y=+1$, the most general scalar
potential which is $SU(2)_L\times U(1)_Y$ invariant and possesses a softly broken
$\mathbb{Z}_2$ symmetry is given by
\beq
V_{\text{(C)2HDM}} &=& m_{11}^2 |\Phi_1|^2 + m_{22}^2 |\Phi_2|^2 - m_{12}^2 (\Phi_1^\dagger
\Phi_2 + h.c.) + \frac{\lambda_1}{2} (\Phi_1^\dagger \Phi_1)^2 +
\frac{\lambda_2}{2} (\Phi_2^\dagger \Phi_2)^2 \nonumber \\
&& + \lambda_3
(\Phi_1^\dagger \Phi_1) (\Phi_2^\dagger \Phi_2) + \lambda_4
(\Phi_1^\dagger \Phi_2) (\Phi_2^\dagger \Phi_1) + \left[ \frac{\lambda_5}{2}
(\Phi_1^\dagger \Phi_2)^2 + h.c. \right] \;.
\label{eq:2hdmpot}
\eeq
The $\mathbb{Z}_2$ symmetry is introduced in the model in order to avoid dangerous flavour-changing
neutral currents (FCNCs) mediated by the neutral scalar. Since the $\mathbb{Z}_2$ symmetry is extended
to the fermion sector, it will force all families of same-charge fermions to couple to a single
doublet which eliminates tree-level
FCNCs~\cite{Glashow:1976nt,Branco:2011iw}. This implies
four different types of doublet couplings to the fermions listed in
Tab.~\ref{tab:yuycoup}. \s
\begin{table}
\begin{center}
\begin{tabular}{rccc|ccccc}
\hline
& $u$-type & $d$-type & leptons & $Q$ & $u_R$ & $d_R$ & $L$ & $l_R$ \\
  \hline
type I & $\Phi_2$ & $\Phi_2$ & $\Phi_2$ & + & $-$ & $-$ & + & $-$ \\
type II & $\Phi_2$ & $\Phi_1$ & $\Phi_1$ & + & $-$ & + & + & $-$ \\
flipped (FL) & $\Phi_2$ & $\Phi_1$ & $\Phi_2$ & + & $-$ & $-$ & + & + \\
 lepton-specific (LS) & $\Phi_2$ & $\Phi_2$ & $\Phi_1$ & + & $-$ & + & + & $-$
  \\ \hline
\end{tabular}
\caption{Four left rows: The four Yukawa types of the
  $\mathbb{Z}_2$-symmetric 2HDM, stating which Higgs
  doublet couples to the different fermion types. Five right columns:
  Corresponding $\mathbb{Z}_2$
  assignment for the quark doublet $Q$, the up-type
  quark singlet $u_R$, the down-type quark singlet $d_R$, the lepton
  doublet $L$, and the
  lepton singlet $l_R$. \label{tab:yuycoup}}
\end{center}
\end{table}

The dimension-2 coefficient $m_{12}^2$ which breaks
  the $\mathbb{Z}_2$ symmetry softly, is introduced to allow for the existence of a {\em
  decoupling limit}, in which all scalars other than the SM-like one
have very large masses, with 
suppressed couplings to fermions and gauge bosons. Furthermore, in the case where all parameters in Eq.~(\ref{eq:2hdmpot}) are real we are in the
CP-conserving 2HDM, which we will call R2HDM from now on. In the case of
complex $m_{12}^2$ and $\lambda_5$ parameters\footnote{The phases of these two parameters
must be such that $\mbox{arg}(m_{12}^2) \neq \mbox{arg}(\lambda_5)/2$, otherwise a trivial field rephasing would
render the potential real.} the model explicitly breaks the CP symmetry. This is the CP-violating
version of the 2HDM, called C2HDM.
The two complex doublet fields can be parametrised as
\beq
\Phi_i = \left( \begin{array}{c} \phi_i^+ \\ \frac{1}{\sqrt{2}} (v_i +
    \rho_i + i \eta_i) \end{array} \right) \;,  \ \ i=1,2 \;,
    \label{eq:phidef}
\eeq
with $v_{1,2}$ being the vacuum expectation values (VEVs) of the two
doublets $\Phi_{1,2}$. 
After EWSB three of the eight degrees of freedom initially present in
$\Phi_{1,2}$ are taken by the Goldstone bosons to give masses to the
gauge bosons $W^\pm$ and $Z$, and we are left with five
physical Higgs bosons. In the CP-conserving case, these are 
two neutral CP-even Higgs bosons, $h$ and $H$, where by convention $m_h < m_H$,
one neutral CP-odd $A$ and a pair of charged Higgs bosons $H^\pm$. In
the following we will denote $h$ and $H$ by $H_1$ and $H_2$,
respectively, in order to standardise the notation for all considered
models. The CP-even neutral Higgs mass matrix is diagonalised by a
mixing angle $\alpha$, whereas both the neutral CP-odd  and charged mass matrices
are diagonalised by a mixing angle $\beta$, such that
\beq
\tan\beta = \frac{v_2}{v_1} \;.
\eeq
In the C2HDM, the three neutral Higgs bosons mix, resulting in three
neutral Higgs mass eigenstates $H_i$ ($i=1,2,3$) with no definite CP
quantum number and which by 
convention are ordered as $m_{H_1} \le m_{H_2} \le m_{H_3}$. The
rotation matrix $R$ diagonalising the neutral Higgs sector can be
parametrised in terms of three mixing angles $\alpha_i$ ($i=1,2,3$) as
\beq
R = \left( \begin{array}{ccc}
c_{1} c_{2} & s_{1} c_{2} & s_{2}\\
-(c_{1} s_{2} s_{3} + s_{1} c_{3})
& c_{1} c_{3} - s_{1} s_{2} s_{3}
& c_{2} s_{3} \\
- c_{1} s_{2} c_{3} + s_{1} s_{3} &
-(c_{1} s_{3} + s_{1} s_{2} c_{3})
& c_{2}  c_{3}
\end{array} \right) \;, \label{eq:rmixmatrixc2hdm}
\eeq
where $s_i \equiv \sin \alpha_i$, $c_i \equiv \cos \alpha_i$, and,
without loss of generality, the angles vary in the range
\beq
- \frac{\pi}{2} \le \alpha_i \le \frac{\pi}{2} \;.
\eeq
The CP-conserving limit of the C2HDM is obtained by setting
$\alpha_2=\alpha_3=0$ and $\alpha_1 = \alpha + \pi/2$
\cite{Khater:2003wq}. The shift by $\pi/2$ in this limit is necessary
to match the usual 2HDM convention. We discuss the alignment limit,
where the C2HDM approaches the SM in
Appendix~\ref{sec:alignc2hdm}. \s

By identifying
\beq
v = \sqrt{v_1^2 + v_2^2} \;,
\eeq
where $v$ is the SM VEV, $v \approx 246$~GeV, and using the two
minimisation conditions, the scalar sector of the
2HDM can be described by eight independent input parameters. For
convenience in the parameter scans, {\it
  cf.}~Sec.~\ref{sec:scans}, we replace the mixing angle
$\alpha$ by the coupling of the $H_2$ state to massive gauge
bosons ($V=Z,W^\pm$), which we represent by $c_{H_2 VV}$. Thus we have
as input parameter set 
\beq
v  \;, \quad \tan\beta \;, \quad c_{H_2 VV} \;, \quad m_{H_1} \;, \quad m_{H_2} \;, \quad m_{A} \;, \quad m_{H^\pm} \quad
\mbox{and} \quad m_{12}^2 \;.
\label{eq:2hdminput}
\eeq
In the C2HDM, as stated above, the parameters
$\lambda_5$ and $m_{12}^2$ can be complex so that the C2HDM Higgs
sector at tree level is described by ten parameters. Notice that it is always possible to
perform a basis change to make one of these phases vanish so that we
end up with nine independent parameters.
In the C2HDM the three neutral Higgs boson masses are not independent. The third
neutral Higgs mass is a dependent quantity and is obtained from the
input parameters, {\it cf.}~\cite{ElKaffas:2007rq}. We hence choose two of the three
neutral Higgs boson masses as input values and calculate the third one. The
chosen input masses are called $m_{H_i}$ and $m_{H_j}$ with $H_i$ per
default denoting the lighter one, {\it i.e.}~$m_{H_i} <
m_{H_j}$. They denote any two of the three neutral Higgs
bosons among which we take one to be the 125 GeV SM-like scalar. 
We furthermore replace the three mixing
angles $\alpha_{1,2,3}$ by two coupling values of $H_i$ and by a matrix
element of our rotation matrix. These are the squared $H_i$ couplings to the
massive gauge bosons $V$ and to the top quarks $t$, $c^2_{H_i VV}$
and $c^2_{H_i tt}$, respectively, and the neutral mixing matrix entry
$R_{23}$. We furthermore fix the sign
of $R_{13}$, sg($R_{13}$), to either +1 or -1 in order to lift the
degeneracy that we introduce by specifying only the squared values of
the $H_i$ couplings. This choice of input parameters complies with
the input parameters of the program code {\tt ScannerS} that we will use
for our parameter scans as explained below. We hence have the 
input parameter set
\beq
v  \;, \quad \tan\beta \;, \quad c_{H_i VV}^2 \;, \quad c_{H_i tt}^2
\;, \quad R_{23} \;, \quad m_{H_i} \;, \quad m_{H_j} \;, \quad m_{H^\pm} \quad
\mbox{and} \quad \mbox{Re}(m_{12}^2) \;.
\label{eq:c2hdminput}
\eeq

%%%%%%%%%%%%%%%%%%%%%%%%%%%%%%%%%%%%%%%%%%%%%%%%%%%%%%%%%%%%
\subsection{The N2HDM}

In the following, we give a brief introduction to the N2HDM and refer
to \cite{Muhlleitner:2016mzt} for more details.
The scalar potential of the N2HDM is obtained from the 2HDM potential
by adding a real singlet field $\Phi_S$.
%\cite{He:2008qm,Grzadkowski:2009iz,Logan:2010nw,Boucenna:2011hy,He:2011gc,Bai:2012nv,He:2013suk,Cai:2013zga,Wang:2014elb,Drozd:2014yla,Campbell:2015fra,vonBuddenbrock:2016rmr}. In
%\cite{Chen:2013jvg}
In terms of the two $SU(2)_L$ Higgs doublets $\Phi_1$ and $\Phi_2$,
defined in  Eq.~(\ref{eq:phidef}), and the singlet field, defined as
\beq
\Phi_S=v_S+ \rho_S \;,
\eeq
the N2HDM potential is given by
\beq
V_{\text{N2HDM}} &=&  V_{\text{2HDM}}+ \frac{1}{2} m_S^2 \Phi_S^2 + \frac{\lambda_6}{8} \Phi_S^4 +
\frac{\lambda_7}{2} (\Phi_1^\dagger \Phi_1) \Phi_S^2 +
\frac{\lambda_8}{2} (\Phi_2^\dagger \Phi_2) \Phi_S^2 \;.
\label{eq:n2hdmpot}
\eeq
The above scalar potential is obtained by imposing two $\mathbb{Z}_2$ symmetries,
\begin{eqnarray}
 && \Phi_1 \to \Phi_1\;, \quad \Phi_2 \to - \Phi_2\;, \quad \Phi_S \to
    \Phi_S \quad \mbox{and} \nonumber\\
 && \Phi_1 \to \Phi_1\;, \quad \Phi_2 \to \Phi_2\;, \quad \Phi_S \to -\Phi_S \;.
\label{eq:Z2SYM}
\end{eqnarray}
The first (softly-broken) $\mathbb{Z}_2$ symmetry is the extension of the usual 2HDM
$\mathbb{Z}_2$ symmetry to the N2HDM which, once extended to the Yukawa sector,
will forbid FCNCs at tree level, implying four different N2HDM
versions just like in the 2HDM, {\it cf.}~Tab.~\ref{tab:yuycoup}.
The second $\mathbb{Z}_2$ symmetry is an exact symmetry which will be spontaneously
broken by the singlet VEV and as such does not allow the model to have a
DM candidate. Other versions of the model choose parameters such that $v_S = 0$
yielding very interesting DM phenomenology, but in the current work we
will not consider these possibilities. \s

After EWSB, we have three neutral CP-even Higgs bosons $H_{1,2,3}$
with masses ranked as 
$m_{H_1}<m_{H_2}<m_{H_3}$, one neutral CP-odd boson $A$ and a pair of charged
Higgs bosons $H^\pm$. The physical states $H_{1,2,3}$ are obtained
from the weak basis $(\rho_1, \rho_2, \rho_S)$ by an orthogonal
transformation $R$ which is defined by 3 mixing angles
$\alpha_{1,2,3}$ that are in the same range as in the C2HDM.
After exploiting the minimisation conditions, we are left with twelve
independent input parameters for the N2HDM. For the scan, we will again replace
the three mixing angles $\alpha_{1,2,3}$ by the squared $H_1$
couplings to massive gauge bosons $V$ and the top quarks $t$,
$c_{H_1VV}^2$ and $c_{H_1tt}^2$, respectively, and the 
neutral mixing matrix element $R_{23}$, so that our input parameters read
\beq
\tan\beta \ , \ c_{H_1 VV}^2 \ , \ c_{H_1 tt}^2 \ , \ R_{23} \ , \
m_{H_1}\ ,\  m_{H_2}\ ,\  m_{H_3}\ ,  \ m_A \ , \  m_{H^\pm}\ , \ v , \ v_s\ ,   \
\text{and} \ m_{12}^2 \;.
\label{eq:n2hdminput}
\eeq
Like in the 2HDM, we fix sg($R_{13}$) to either +1 or -1 in order to lift the
introduced degeneracy through the squared values of
the $H_1$ couplings.
The limit of the real 2HDM with an added decoupled singlet field is
obtained from the N2HDM spectrum by letting
$\alpha_{2,3} \to 0$ and $\alpha_1 \to \alpha + \pi/2$.
Again, the shift by $\pi/2$ in this limit is necessary to match the
usual 2HDM convention. The alignment limit for the N2HDM is discussed in
Appendix~\ref{sec:alignn2hdm}.

\subsection{The NMSSM}
As a supersymmetric benchmark model, we consider the Next-to Minimal Supersymmetric SM (NMSSM)
\cite{Ellis:1988er,Drees:1988fc,Ellwanger:1993xa,Ellwanger:1995ru,Ellwanger:1996gw,Elliott:1994ht,King:1995vk,Franke:1995tc,Maniatis:2009re,Ellwanger:2009dp}.
It extends the two doublet fields $\hat{H}_u$ and
$\hat{H}_d$ of the MSSM by a complex superfield $\hat{S}$.
When the singlet field acquires a non-vanishing VEV, this not
only solves the $\mu$ problem \cite{Kim:1983dt} but, compared to the
MSSM, it also relaxes the tension on the stop mass values that
need to be large for the SM-like Higgs boson mass value to be
compatible with the measured 125.09 GeV. Indeed in supersymmetry
the neutral Higgs masses are given in terms of the gauge parameters at
tree level so that there is an upper mass bound on the lightest
neutral scalar which, in the MSSM, is given by the $Z$ boson
mass. Substantial higher-order corrections to the Higgs boson mass are
therefore required to obtain phenomenologically valid mass
values for the SM-like Higgs boson. The additional singlet contribution to the
tree-level mass of the lightest neutral Higgs boson shifts its mass to
larger values compared to the MSSM prediction, thus no longer requiring large radiative
corrections. The scale-invariant NMSSM superpotential that is added to
the MSSM superpotential $W^{\text{MSSM}}$ reads
\beq
W^{\text{NMSSM}}&=&- \lambda \hat{S} \hat{H}_u \cdot\hat{H}_d + \frac{\kappa}{3}
\hat{S}^3+ W^{\text{MSSM}}\,, \quad \mbox{with} \nonumber\\
W^{\text{MSSM}}&=&-  y_t
\widehat{Q}_3\widehat{H}_u\widehat{t}_R^c + h_b \widehat{Q}_3
\widehat{H}_d\widehat{b}_R^c  + y_\tau \widehat{L}_3 \widehat{H}_d
\widehat{\tau}_R^c \; ,
\label{eq:nmssmsuperpot}
\eeq
where for simplicity we only included the third generation fermion
superfields, given by the left-handed doublet quark ($\widehat{Q}_3$)
and lepton ($\widehat{L}_3$) superfields, and the right-handed singlet
quark ($\widehat{t}_R^c,\widehat{b}_R^c$) and lepton
($\widehat{\tau}_R^c$) superfields. The NMSSM-type couplings $\lambda$
and $\kappa$ are dimensionless and taken real since we consider the
CP-conserving NMSSM. The Yukawa couplings $y_t, y_b, y_\tau$ can
always be taken real. The scalar part of $\hat{S}$ will develop a VEV
$v_S/\sqrt{2}$, which dynamically generates the effective $\mu$
parameter $\mu_{\text{eff}}=\lambda v_S/\sqrt{2}$ through the first term in
the superpotential. The second term, cubic in $\hat{S}$, breaks the
Peccei-Quinn symmetry and thus avoids a massless axion, and
$W^{\text{MSSM}}$ contains the Yukawa interactions. The symplectic
product $x \cdot y = \epsilon_{ij} x^i y^j$ ($i,j=1,2$) is built by
the antisymmetric tensor $\epsilon_{12}= \epsilon^{12} = 1$.
The soft SUSY breaking Lagrangian reads
\beq
\label{eq:Lagmass}
 {\cal L}_{\text{soft,NMSSM}} &=&
 - m_{H_u}^2 | H_u |^2 -  m_{H_d}^2 | H_d|^2 - m_{{\widetilde
     Q}_3}^2|{\widetilde Q}_3^2|-  m_{\widetilde t_R}^2 |{\widetilde t}_R^2|
 -  m_{\widetilde b_R}^2|{\widetilde b}_R^2| -
 m_{{\widetilde L}_3}^2|{\widetilde L}_3^2| \nonumber \\
&& - m_{\widetilde  \tau_R}^2|{\widetilde \tau}_R^2|
+ (y_t A_t H_u \cdot \widetilde Q_3 \widetilde t_R^c - y_b A_b H_d \cdot
\widetilde Q_3  \widetilde  b_R^c - y_\tau A_\tau H_d \cdot \widetilde L_3 \widetilde \tau_R^c + \mathrm{h.c.})
\nonumber \\
&& - \frac{1}{2} \bigg( M_1 \widetilde{B}
\widetilde{B} + M_2 \sum_{a=1}^3 \widetilde{W}^a \widetilde{W}_a +
M_3 \sum_{a=1}^8 \widetilde{G}^a \widetilde{G}_a  \ + \ {\rm h.c.}
\bigg) \nonumber \\
&& - m_S^2 |S|^2 + (\lambda A_\lambda S H_d \cdot H_u - \frac{1}{3}
\kappa  A_\kappa S^3 + \mathrm{h.c.}) \;,
\eeq
where again only the third generation of fermions and sfermions have been taken into
account. The tilde over the fields denotes the complex scalar component
of the corresponding superfields. The soft SUSY
breaking gaugino parameters $M_k$ ($k=1,2,3$) of the bino, wino and
gluino fields $\widetilde{B},$ $\widetilde{W}$ and $\widetilde{G}$, as
well as the soft SUSY breaking trilinear couplings $A_x$ ($x=\lambda,
\kappa, t, b, \tau$) are in general complex, whereas the soft SUSY
breaking mass parameters of the scalar fields, $m_X^2$
($X=S,H_d,H_u,\widetilde{Q}, \widetilde{u}_R, \widetilde{b}_R,
\widetilde{L}, \widetilde{\tau}_R$) are real. Since we consider
the CP-conserving NMSSM, they are all taken real.
In what follows, we will use conventions such that $\lambda$ and $\tan\beta$
are positive, whereas $\kappa, A_\lambda, A_\kappa$ and
$\mu_{\text{eff}}$ are allowed to have both signs. \s

After EWSB, we expand the Higgs fields around their VEVs $v_u$,
$v_d$, and $v_S$, respectively, which are chosen to be real and
positive
\beq
H_d = \left( \begin{array}{c} (v_d + h_d + i a_d)/\sqrt{2} \\
   h_d^- \end{array} \right) \,, \;
H_u = \left( \begin{array}{c} h_u^+ \\ (v_u + h_u + i a_u)/\sqrt{2}
 \end{array} \right) \,, \;
S= \frac{v_s+h_s+ia_s}{\sqrt{2}}.
\label{eq:Higgs-para}
\eeq
This leads to the mass matrices of the three scalars $h_d, h_u,
h_s$, the three pseudoscalars $a_d, a_u, a_s$, and the charged Higgs
states $h_u^\pm,h_d^\mp$, obtained from the second derivatives of the
scalar potential. The mass matrix is diagonalised with orthogonal
rotation matrices, mapping the gauge eigenstates to the mass
eigenstates. These are the three neutral CP-even Higgs bosons $H_1, H_2,
H_3$ that are ordered by ascending mass with $m_{H_1} \le m_{H_2} \le
m_{H_3}$, the two CP-odd mass eigenstates $A_1$ and $A_2$ with
$m_{A_1} \le m_{A_2}$, and a pair of charged Higgs bosons $H^\pm$. \s

After applying the minimisation conditions, we choose as independent
input parameters for the tree-level NMSSM Higgs sector the following,
\beq
\lambda\ , \ \kappa\ , \ A_{\lambda} \ , \ A_{\kappa}, \
\tan \beta =v_u/ v_d \quad \mathrm{and}
\quad \mu_\text{eff} = \lambda v_s/\sqrt{2}\; .
\eeq
The sign conventions are chosen such that $\lambda$ and $\tan\beta$
are positive, while $\kappa, A_\lambda, A_\kappa$ and
$\mu_{\text{eff}}$ are allowed to have both signs. Further parameters
will become relevant upon inclusion of the higher-order corrections to
the Higgs boson mass that are crucial to shift the SM-like Higgs boson
mass to the measured value.

%%%%%%%%%%%%%%%%%%%%%%%%%%%%%%%%%%%%%%%%%%%%
\section{Scans and Theoretical and Experimental Constraints}
\label{sec:scans}
Our goal is to investigate the landscape of Higgs pair production in extended Higgs
sectors by taking into account the relevant theoretical and
experimental constraints. In order to do so, we performed a scan in the
various parameter spaces of the models and checked each parameter
point for compatibility with our applied constraints. In this section,
we briefly describe the constraints used in our study. For
details, we refer to our previous papers on the different models
\cite{Arhrib:2000is,Costa:2015llh,Muhlleitner:2016mzt,Basler:2016obg,Muhlleitner:2017dkd,Arhrib:2018qmw,Azevedo:2018llq}. \s

We performed the scans with the help of the program {\tt ScannerS}
\cite{Coimbra:2013qq,ScannerS,Muhlleitner:2020wwk} for all models
except for the NMSSM.
In Tables \ref{tab:r2hdmranges}, \ref{tab:c2hdmranges}, and
\ref{tab:n2hdmranges}, we list the scan ranges of the R2HDM, the
C2HDM, and the N2HDM, respectively. We give them for the various
set-ups with respect to which neutral Higgs boson takes the role of
the SM-like Higgs which we will denote $H_{\text{SM}}$ from now
on. We distinguish the cases ``light'' where the lightest of the
neutral Higgs bosons is SM-like ($H_1 \equiv H_{\text{SM}}$),
``medium'' with $H_2 \equiv H_{\text{SM}}$, and ``heavy'' with the
heaviest being SM-like ($H_3 \equiv H_{\text{SM}}$). In the R2HDM, with
only two neutral Higgs bosons, we have the cases ``light'' ($H_1 \equiv
H_{\text{SM}}$) and ``heavy'' ($H_2 \equiv H_{\text{SM}}$) only.
Note that in the C2HDM, only two of the
three neutral Higgs boson masses are independent input quantities
whereas the third one is dependent and computed from the input
parameters. Therefore, in the generation of the data points two cases are
considered, namely $H_1/H_2 \equiv H_{\text{SM}}$. The third one is
calculated and all three masses are subsequently ordered by ascending
mass so that all three set-ups, with either of the $H_i$ ($i=1,2,3$)
respresenting the $H_{\text{SM}}$, are covered by the two scans described in
Tab.~\ref{tab:c2hdmranges}. Note also that we restrict ourselves to the type
I and II models. For all these models, the R2HDM, the C2HDM and the N2HDM, we apply
the same theoretical constraints, which have different expressions for each model,
requiring that all potentials are bounded from below,
that perturbative unitarity holds and that the electroweak vacuum is
the global minimum. 
In the R2HDM we use for the latter the
discriminant from \cite{Barroso:2013awa} and 
for the C2HDM the one from \cite{Ivanov:2015nea}. \s

\renewcommand{\arraystretch}{1.2}
\begin{table}[ht!]
\begin{center}
\begin{tabular}{|c|c|c|c|c|c|c|}
\hline
$m_{H_1}$ [GeV] & $m_{H_2}$ [GeV] & $m_A$ [GeV] & $m_{H^\pm}$ [GeV] & $\tan\beta$ &
$c_{H_2 VV}$ & $m_{12}^2$ [GeV$^2$] \\ \hline \hline
\multicolumn{7}{|c|}{R2HDM I/II (light)} \\ \hline
125.09 & 130...3000 & 30...3000 & 85/800...3000 & 0.8...30 & -0.3...0.3 &
 $10^{-3}$...10$^7$ \\ \hline \hline
\multicolumn{7}{|c|}{R2HDM I/II (heavy)} \\ \hline
30...120 & 125.09 & 30...3000 & 85/800...3000 & 0.8...30 & 0.8...1 &
 $10^{-3}$...10$^7$ \\ \hline
\end{tabular}
\end{center}
\caption{Scan ranges of the R2HDM input parameters, {\it
    cf.}~Eq.~(\ref{eq:2hdminput}), where light (heavy) refers
to the set-up where the lightest (heaviest) of the two CP-even neutral Higgs
bosons is the SM-like Higgs $H_{\text{SM}}$, {\it i.e.}~$H_1 (H_2) \equiv
H_{\text{SM}}$.  \label{tab:r2hdmranges} }
\end{table}
\renewcommand{\arraystretch}{1}
\renewcommand{\arraystretch}{1.2}
\begin{table}[h!]
\begin{center}
\begin{tabular}{|c|c|c|c|c|c|c|c|c|}
\hline
$m_{H_i}$& $m_{H_j}$ & $m_{H^\pm}$ & $\tan\beta$ &
$c_{H_i VV}^2$ & $c_{H_i tt}^2$ & sg($R_{13}$) & $R_{23}$ &
                                                                    Re($m_{12}^2$) \\
$\mbox{[GeV]}$ & [GeV] & [TeV] & & & & & & [GeV$^2$] \\\hline \hline
\multicolumn{9}{|c|}{C2HDM I/II (light)} \\ \hline
125.09 & 130...3000 & 0.08/0.8...3 & 0.8...30 & 0.8...1 & 0.7...1.3 & -1/1 & -1...1 &
 $10^{-3}$...10$^7$ \\ \hline \hline
\multicolumn{9}{|c|}{C2HDM I/II (medium/heavy)} \\ \hline
30...120 & 125.09 & 0.08/0.8...3 & 0.8...30 & 0...0.1 & 0...1.2 & -1/1 & -1...1 &
 $10^{-3}$...$5\! \cdot \!\!  10^5$ \\ \hline
\end{tabular}
\end{center}
\caption{Scan ranges of the C2HDM input parameters, {\it
    cf.}~Eq.~(\ref{eq:c2hdminput}), where light/medium/heavy refers
to the set-up where the lightest/medium/heaviest of
the three CP-mixed neutral Higgs
bosons is the SM-like Higgs $H_{\text{SM}}$, {\it i.e.}~$H_1/H_2/H_3
\equiv H_{\text{SM}}$.  \label{tab:c2hdmranges} }
\end{table}

\renewcommand{\arraystretch}{1}
\renewcommand{\arraystretch}{1.2}
\begin{table}[h!]
\begin{center}
\begin{tabular}{|c|c|c|c|c|c|}
\hline
$m_{H_1}$& $m_{H_2}$ & $m_{H_3}$ & $m_A$ & $m_{H^\pm}$ & $m_{12}^2$ \\
$\mbox{[GeV]}$ & [GeV] & [GeV] & [GeV] & [GeV] & [GeV$^2$] \\\hline
$\tan\beta$ & $c_{H_1 VV}^2$ & $c_{H_1 tt}^2$ & sg($R_{13}$) &
$R_{23}$ & $v_S$ [GeV] \\ \hline \hline
\multicolumn{6}{|c|}{N2HDM I/II (light)} \\ \hline
125.09 & 130...2995 & 130...3000 & 30...3000 & 85/800...3000 & 10$^{-3}$...10$^7$
  \\ \hline
0.8...30 & 0.9...1 & 0.8...1.2 & -1/1 & -1...1 & 0...10000\\ \hline \hline
\multicolumn{6}{|c|}{N2HDM I/II (medium)} \\ \hline
30...120 & 125.09 & 135...3000 & 30...3000 & 85/800...3000 & 10$^{-3}$...$5
\, \cdot \,\, 10^5$
  \\ \hline
0.8...30 & 0...0.1 & 0...1.2 & -1/1 & -1...1 & 0...5000 \\ \hline \hline
\multicolumn{6}{|c|}{N2HDM I/II (heavy)} \\ \hline
30...115 & 35...120 & 125.09 & 30...3000 & 85/800...3000 & 10$^{-3}$...$5
\, \cdot \,\, 10^5$
  \\ \hline
0.8...30 & 0...0.1 & 0...1.2 & -1/1 & -1...1 & 0...5000 \\ \hline
\end{tabular}
\end{center}
\caption{Scan ranges of the N2HDM input parameters, {\it
    cf.}~Eq.~(\ref{eq:n2hdminput}), where light/medium/heavy refers
to the set-up where the lightest/medium/heaviest of the three neutral Higgs
bosons is the SM-like Higgs $H_{\text{SM}}$, {\it i.e.}~$H_1/H_2/H_3 \equiv
H_{\text{SM}}$.  \label{tab:n2hdmranges} }
\end{table}
\renewcommand{\arraystretch}{1}
As for experimental constraints, we impose compatibility with the
electroweak precision data by demanding the computed $S$,
$T$ and $U$ values to be within $2\sigma$ of the SM fit \cite{Baak:2014ora}, taking into account the full
correlation among the three parameters.
We require one of the Higgs bosons to have a mass of \cite{Aad:2015zhl}
\beq
m_{H_{\text{SM}}} = 125.09 \, \mbox{GeV} \,,
\eeq
and to behave SM-like. Compatibility with the Higgs signal data is checked
through {\tt HiggsSignals} version 2.6.1 \cite{Bechtle:2013xfa} which is linked to
{\tt ScannerS}. We furthermore suppress interfering Higgs signals by
forcing any other neutral scalar mass to deviate by more than $\pm 2.5$ GeV
from $m_{H_{\text{SM}}}$. 
Scenarios with neutral Higgs bosons that are close in mass are
particularly interesting for non-resonant di-Higgs production as they
may have discriminating power with respect to the SM case. The
appearance of non-trivial interference effects requires, however,  a
dedicated thorough study that is beyond the focus of this study and
is left for future work.
We require 95\% C.L. exclusion limits on non-observed scalar states by using {\tt
HiggsBounds} version 5.9.0 \cite{Bechtle:2008jh,Bechtle:2011sb,Bechtle:2013wla}.
Additionally, we checked our sample with respect to the recent ATLAS
  analyses in the $ZZ$ \cite{ATLAS:2020tlo} and $\gamma\gamma$
  \cite{ATL20-37} final 
  states that were not yet included in {\tt HiggsBounds}.
Consistency with recent flavour constraints is ensured by testing for
the compatibility 
with $\mathcal{R}_b$ \cite{Haber:1999zh,Deschamps:2009rh} and
$B\rightarrow X_s \gamma $
\cite{Deschamps:2009rh,Mahmoudi:2009zx,Hermann:2012fc,Misiak:2015xwa,Misiak:2017bgg, Misiak:2020vlo} in the
$m_{H^{\pm}}-\tan\beta$ plane. For the
  non-supersymmetric type II models, we imposed the
latest bound on the charged Higgs mass given in \cite{Misiak:2020vlo},
$m_{H^\pm} \ge 800$~GeV for essentially all values of $\tan\beta$, whereas
in the type I models this bound is much weaker and is strongly correlated
with $\tan\beta$.
%More recent results for the type II $B\rightarrow X_s \gamma$
%calculation~\cite{Misiak:2020vlo} have yielded an even stricter bound
%on the charged mass, $m_{H^{\pm}} > 800$ GeV, so our lower bound for
%this parameter is a more conservative one. We chose this lower bound
%to account for the possibility of higher mass fields not considered
%in this study (extra fermions, for instance) being able, conceivably,
%to significantly relax the $B\rightarrow X_s \gamma$ constraints all
%the while yielding the same lower energy scalar sector.
Lower values for $m_{H^{\pm}}$ allow, via electroweak precision
constraints, different ranges for the masses of the neutral Higgs bosons, which will
therefore affect our predictions for di-Higgs production. \s

In the C2HDM, we additionally have to take into account
constraints on CP violation  in the Higgs sector arising from
electric dipole moment (EDM) measurements. Among these, the data from the EDM
of the electron imposes the strongest constraints~\cite{Inoue:2014nva}, with the
current best experimental limit given by the ACME
collaboration~\cite{Baron:2013eja}. We demand compatibility with the
values given in~\cite{Baron:2013eja} at 90\% C.L. \s

In the NMSSM, we use the program {\tt NMSSMCALC} \cite{Baglio:2013iia,King:2015oxa} and compute
the Higgs mass corrections up to ${\cal
  O}((\alpha_t+\alpha_\lambda+\alpha_\kappa)^2 + \alpha_t \alpha_s)$
\cite{Muhlleitner:2014vsa,Dao:2019qaz,Dao:2021khm}
with on-shell renormalisation in the top/stop sector. We demand the
computed SM-like Higgs boson mass to
lie in the range $122 \mbox{ GeV}... 128 \mbox{GeV}$ which accounts for
the present typically applied theoretical error of 3~GeV
\cite{Slavich:2020zjv}. We use {\tt HiggsBounds} and {\tt HiggSignals}
to check for compatibility with the Higgs constraints. Furthermore, we
omit parameter points with the
following mass configurations for the lightest chargino
$\tilde{\chi}_1^\pm$ and the lightest stop $\tilde{t}_1$,
\beq
m_{\tilde{\chi}_1^\pm} < 94 \mbox{ GeV} \,, \; m_{\tilde{t}_1} <
1\mbox{ TeV} \;,
\eeq
to take into account lower limits on the
lightest chargino and the lightest stop mass.
The experimental limits given by the LHC experiments ATLAS and CMS
rely on assumptions on the mass
spectra and are often based on simplified models. The quotation of a
lower limit therefore necessarily requires a scenario that matches the
assumptions made by the experiments. For our parameter scan we therefore
chose a conservative approach to apply limits that roughly
comply with the recent limits given by ATLAS and CMS \cite{atlaslim,cmslim}.
For further details of the Higgs mass
computation and of the input parameters as well as their scan
ranges, we refer to \cite{Dao:2021khm}.

%%%%%%%%%%%%%%%%%%%%%%%%%%%%%%%%%%%%%%%%%%%%%%%%%%%%%%%%%%%%%
\section{Numerical Results for SM-Like Higgs Pair Production}
\label{sec:numerical}
In this section, we will present our numerical results for Higgs pair
production at the LHC with the dominant process given by gluon fusion
\cite{Baglio:2012np}. After recapitulating the details of the process, we
will outline how we applied the experimental limits on resonant
and non-resonant di-Higgs production. We will subsequently present the
mass values allowed in the various models as well as the ranges of the
SM-like top-Yukawa and trilinear Higgs self-coupling that are still
compatible with the data. Finally we will show our results for SM-like Higgs pair
production in all discussed models that are in accordance with the applied
constraints.

\subsection{Gluon Fusion into Higgs Pairs}
\begin{figure}[t]
\includegraphics[width=\textwidth]{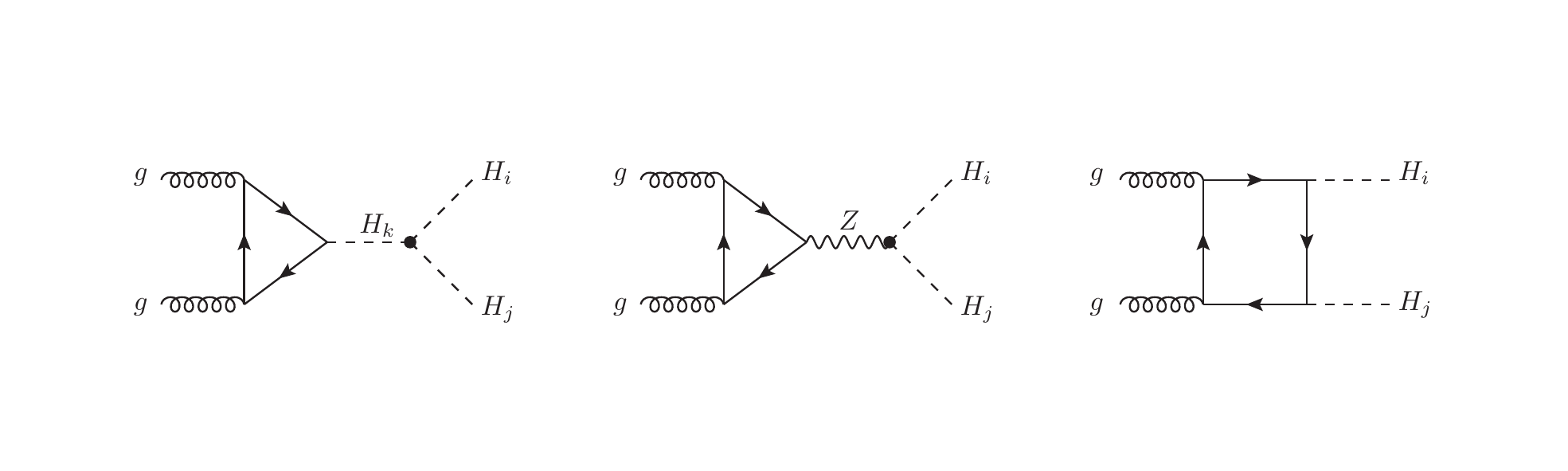}
\vspace*{-1.5cm}
\caption{Generic diagrams contributing to leading-order C2HDM Higgs
  $H_i H_j$ ($i,j,k=1,2,3$) pair production in
  gluon fusion. \label{fig:c2hdmdiags}}
\end{figure}

All our di-Higgs production cross sections through gluon fusion
have been computed by adapting the public code {\tt HPAIR}
\cite{hpair} to the R2HDM, the C2HDM
\cite{Grober:2017gut,Basler:2018dac}, the N2HDM and the NMSSM
\cite{Nhung:2013lpa,Basler:2018dac}. The process is mediated through
heavy quark loops already at leading order (LO). Generic diagrams are
shown for the example of $H_i H_j$ ($i,j=1,2,3$) production in the
C2HDM in Fig.~\ref{fig:c2hdmdiags}. The diagrams that contribute are
triangle diagrams and box diagrams. The first triangle diagram
contains a Higgs boson $H_k$ ($k=1,2,3$) in
the $s$-channel that couples to the final state Higgs bosons $H_i$ and $H_j$ through the
trilinear coupling $\lambda_{H_iH_jH_k}$. The box diagrams (third diagram) are
proportional solely to the Yukawa couplings of $H_i$ and $H_j$ to the top and
bottom quarks. In the C2HDM, which is CP-violating, we additionally have
a $Z$ boson exchange in the $s$-channel (second diagram) which couples
to the CP-mixed Higgs boson final states. This diagram also has to be
taken into account in the CP-conserving models when the production of a mixed pair of
one CP-even and one CP-odd Higgs boson in the final state is
considered. 
Note that the contribution of the $Z$ boson exchange
  diagram to the overall cross section is small. Furthermore, the QCD
  corrections from the SM cannot be taken 
over here. Our implementation of the BSM models in {\tt HPAIR} allows us
to take the QCD corrections (in the heavy top limit) correctly into
account also for this diagram. \s

\begin{figure}[t!]
  \centering
  \includegraphics[width=0.42\linewidth]{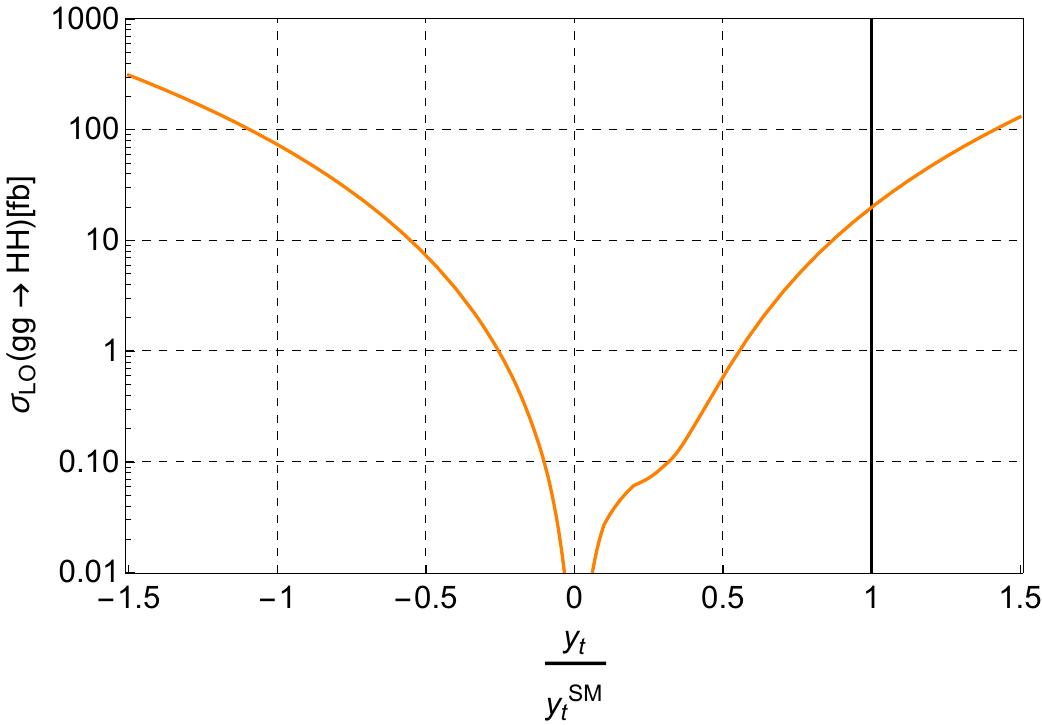}
  \includegraphics[width=0.42\linewidth]{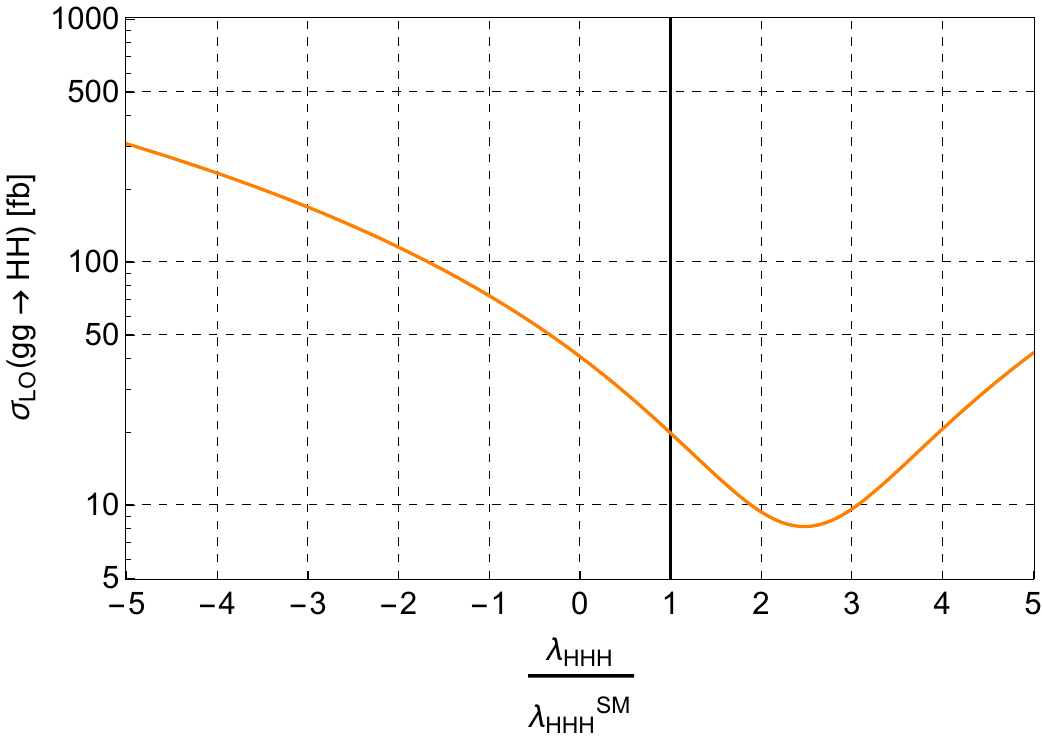} \\[0.2cm]
  \includegraphics[width=0.42\linewidth]{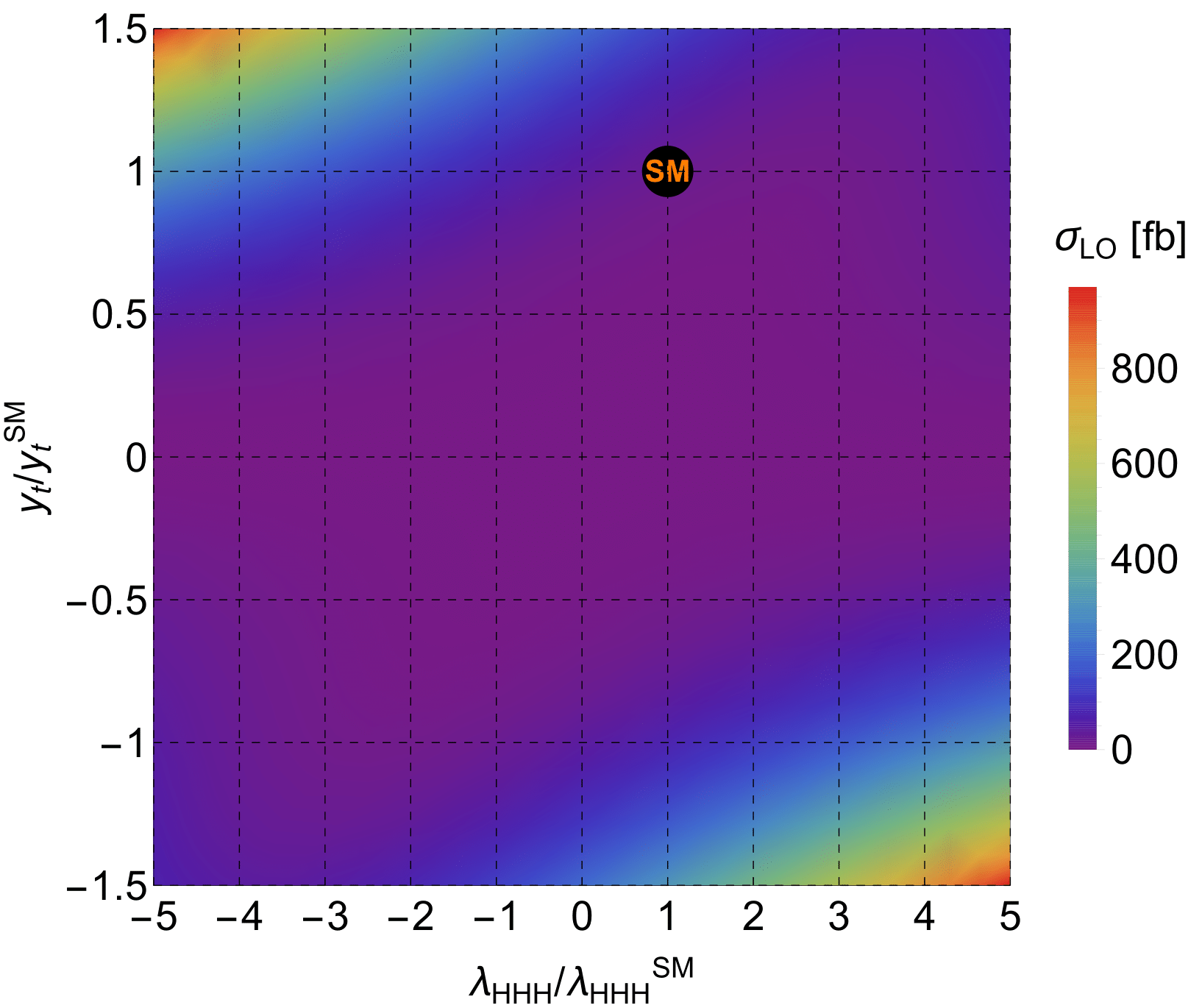}
  \caption{Leading-order SM Higgs pair production in gluon fusion at
    $\sqrt{s}=14$~TeV as function of the Higgs top-Yukawa coupling $y_t$ normalized to the
    SM value $y_t^{\text{SM}}$ for the trilinear coupling fixed to the
    SM value (upper left); as a function of the trilinear coupling
    $\lambda_{HHH}$ normalized to the SM value
    $\lambda_{HHH}^{\text{SM}}$ for the top Yukawa coupling fixed to
    the SM value (upper right); as a function of the trilinear and the
    top-Yukawa couplings in terms of the respective SM values
    (lower). The color code indicates the size of the LO cross section
  in fb. The SM point is marked in orange.}
    \label{fig:smvar}
\end{figure}

In extended Higgs sectors we have several modifications
compared to the SM. The additional Higgs bosons $H_k$ can lead to
resonant enhancement of the di-Higgs cross section compared to the SM
in case $m_{H_k} > m_{H_i} + m_{H_j}$. In Higgs pair production we
will call parameter configurations where the resonant
rates makes up for a significant part of the cross section
``resonant production''. For mediator masses of $m_{H_k}
< m_{H_i} + m_{H_j}$ resonant enhancement is kinematically not
possible. This is a clear case of ``non-resonant''
production. However, note that, for parameter configurations with
$m_{H_k} > m_{H_i} + m_{H_j}$, the resonance contribution may be very
suppressed if the 
involved couplings are small, the mediator mass is very
heavy, its total width is large, or if there are destructive interferences between
different diagrams. From an experimental point of
view, the cross section would not
be distinguishable from ``non-resonant'' production then. The transition
between ``resonant'' and ``non-resonant'' is of course fluid. We will
address this in detail in the discussion of our application of the
experimental limits from resonant and non-resonant di-Higgs
production.
%We will call this the {\it resonant} part
%of the cross section in the following. The {\it continuum} contribution is
%built up by all remaining triangle diagrams with $H_k$ exchange and
%$m_{H_k} < m_{H_i} + m_{H_j}$ as well as the box diagrams plus the
%$Z$-boson exchange diagrams where applicable (in C2HDM Higgs pair
%production and in final states consisting of a CP-even plus a CP-odd%
%Higgs boson).
Further differences from the SM case arise from Higgs-Yukawa
and trilinear Higgs couplings deviating from those of the SM Higgs
boson and from additional particles running in the loop. The latter
is the case for the NMSSM where supersymmetric partners of the top and
bottom quark contribute to the loop. An interesting feature is that in
the SM we have a destructive
interference between the triangle and the box diagrams, implying possible
enhancements in extended Higgs sectors where the couplings
differ from the SM case. This can be inferred from Fig.~\ref{fig:smvar},
where we show the LO Higgs pair production cross section when we vary the
SM Higgs top-Yukawa coupling (upper left), the trilinear Higgs
self-coupling (upper right) and both couplings (lower) while keeping
all other couplings fixed to the SM values. Note,
that for the sake of illustration we varied the top-Yukawa coupling in ranges
beyond the experimental exclusion limits.\footnote{In the subsequently
presented analyses, the experimental limits on the couplings are taken
into account.}
We see the destructive interference which becomes
largest for $\lambda_{HHH}/\lambda_{HHH}^{\text{SM}}=2.48$. The cross
section drops to zero
(modulo the small bottom quark contribution) for the top-Yukawa
coupling $y_t=0$ as the Higgs does not couple to the top quarks any more.
Note finally that the di-Higgs cross section values through the
$s$-channel exchange triangle diagrams are sensitive to the total
widths of the exchanged Higgs bosons as well, that have to be provided
for the computation of the cross section. \s

For the calculation of the gluon fusion di-Higgs cross section shown
here and in the following, we use the following input values for
the c.m.~energy $\sqrt{s}$, the top and bottom quark mass values $m_t$
and $m_b$, the $Z$ boson mass $m_Z$, the strong coupling constant
$\alpha_s$ at $m_Z$, the renormalisation and factorisation scale
$\mu$ at which the cross section is evaluated, and the pdf set,
\beq
&& \sqrt{s} = 14 \mbox{ TeV}, \; m_t = 173.2 \mbox{ GeV}, \; m_b= 4.75
\mbox{ GeV}, \\
&& m_Z = 91.187 \mbox{ GeV}, \; \alpha_s (m_Z) = 0.118,
\; \mu = 0.5 \, m_{HH}, \; \mbox{pdfset: CT14lo/nlo} ,  \nonumber
\eeq
where $m_{HH}$ denotes the invariant Higgs pair mass and where we take
the corresponding CT14 pdf set \cite{Dulat:2015mca} for the LO and the
next-to-leading order (NLO) computation.
The NLO QCD corrections which are of two-loop order are
computed in the limit of heavy loop particle masses. When we
explicitly present NLO results we consistently set the bottom-quark
mass to zero both at LO and at NLO. For presented $K$-factors, {\it
  i.e.}~the ratio of the NLO
to the LO cross section, we consistently set the pdf sets to LO and NLO
for the LO and the NLO calculation, respectively. \s

Note that, in order to keep our scans economic in time, we compute the Higgs pair
production cross sections for the results covering the whole scanned
parameter space of the models at leading order. The NLO QCD
corrections are roughly taken into account by applying a factor of two. This
rough approximation works reasonably well for SM-like Higgs pair
production
\cite{Dawson:1998py,Nhung:2013lpa,Grober:2017gut,Borowka:2016ehy,Borowka:2016ypz,Baglio:2018lrj,Baglio:2020ini,Baglio:2020wgt}. When
we present specific benchmark scenarios, however, we compute the cross sections
explicitly at NLO QCD in the heavy loop particle limit for the
considered model by using our adapted {\tt HPAIR} codes.
Be aware, however, that the QCD corrections should be
  taken with caution for scenarios with large values of
  $\tan\beta$. In this case the bottom 
loop contributions become more important so that the limit of heavy
loop particle masses cannot be applied any more.

\subsection{The Impact of Di-Higgs Constraints \label{sec:sep}}
We first want to discuss how we take into account the already
available results from the experimental SM-like di-Higgs searches, 
resonant and non-resonant. Since the
experimental limits in the two cases rely on different topologies, the
question arises how these limits can be applied in our cases where
both resonant and non-resonant production are included in the
evaluated cross sections. Moreover, in some of the models we
can have more than one resonance decaying into a SM-like Higgs
pair. In the following, we will outline our applied strategy. \s

For the generation of the data we turned off in {\tt
  HiggsBounds} the experimental limits from resonant
di-Higgs production while applying all other constraints described
above. We then computed, for each data point that fulfils the kinematic constraint $m_{H_k} > 2
m_{H_{\text{SM}}}$, the production cross section $\sigma (H_k)$, for all possible intermediate resonances $H_k$.
The latter has been obtained with the code {\tt SusHi v1.6.1}
\cite{Harlander:2012pb,Liebler:2015bka,Harlander:2016hcx} at NNLO QCD
and it includes both production in gluon fusion and in association with a $b$-quark
pair. Note, however, that associated production with a $b$-quark pair
does not play an important role for our scenarios. This production
cross section is subsequently multiplied with the branching
ratio of the $H_k$ decay into $H_{\text{SM}} H_{\text{SM}}$. The
branching ratio has been obtained for the R2HDM, C2HDM, N2HDM, and
NMSSM by using the public codes {\tt HDECAY}
\cite{Djouadi:1997yw,Djouadi:2018xqq,Djouadi:2006bz},
{\tt C2HDM\_HDECAY} \cite{Fontes:2017zfn}, {\tt N2HDECAY}
\cite{Muhlleitner:2016mzt,Engeln:2018mbg}, and {\tt NMSSMCALC}
\cite{Baglio:2013iia}, respectively. We then compare the cross section
$\sigma (H_k) \times \mbox{BR} (H_k \to H_{\text{SM}} H_{\text{SM}})$ for each
possible intermediate resonance\footnote{In the case of the N2HDM {\it
  e.g.} with three neutral Higgs bosons $H_{1,2,3}$ for scenarios with
$H_1 \equiv H_{\text{SM}}$ both $H_2$ and $H_3$ could be resonantly produced
and decay into an $H_{\text{SM}}$ pair provided they fulfill the kinematic
constraint.} with the experimental limits on resonant di-Higgs
production. Since these limits are obtained from the LHC run at
$\sqrt{s}=13$~TeV, we also compute the {\tt SusHi} cross sections at
this c.m.~energy (in contrast to the Higgs pair production cross
sections that are all evaluated at $\sqrt{s}=14$~TeV).
We took into account the experimental limits that were the
stringest ones at the time of the production of our plots. They are
given in Refs.~\cite{ATLAS:2018rnh} for the $4b$,
\cite{ATLAS:2018uni,ATLAS:2020azv} for the $(2b)(2\tau)$,
\cite{ATLAS:2018dpp} for the $(2b)(2\gamma)$, \cite{ATLAS:2018fpd} for
the $(2b)(2W)$, \cite{CMS:2020jeo} for the $(2b)(ZZ)$, \cite{ATLAS:2018hqk} for the
$(2W)(2\gamma)$ and \cite{ATLAS:2018ili} for the $4W$ final
state.\footnote{Combinations of all searches have been published in
\cite{ATLAS-CONF-2021-052} and \cite{CMS:2018ipl}.
We, however, only place limits on the individual final states as the
correct combination of all limits would require a sophisticated
treatment that goes beyond the rough estimate of the experimental
results  on our channels applied in this paper.}
We furthermore  included the recently published results on
$4b$ by ATLAS~\cite{ATLAS-CONF-2021-035} and CMS~\cite{CMS-PAS-B2G-20-004},
$(2b)(2\tau)$ \cite{ATLAS-CONF-2021-030} and
$(2b)(2\gamma)$ \cite{ATLAS-CONF-2021-016}.
Parameter points where at least for one possible resonance $H_k$
the experimental limit for any of the final state signatures is
exceeded, are rejected. Nevertheless, there is one exception. Since the
experimental limits are given assuming narrow resonances, we do not
reject points where the ratio of the total width $\Gamma_{\text{tot}}
(H_k)$ of $H_k$ and its mass $m_{H_k}$ exceeds the
value $(\Gamma_{\text{tot}}
(H_k)/m_{H_k})_{\text{limit}}=5$\%.\footnote{Note that by default,
  {\it i.e.}~independently of the applied experimental constraints, we
  rejected all parameter points with $\Gamma_{\text{tot}}
  (H_i)/m_{H_i}>50$\%.} \s
%After applying the limits from the
%resonant production all remaining parameter points induce di-Higgs
%cross sections that are below the sensitivity of the limits from
%non-resonant production (which are all implemented in {\tt
 % HiggsBounds}). \s

%Actually, in the non-minimal Higgs
%sectors of the models investigated here, the SM Higgs boson pair
%production can be largely enhanced compared to the SM case through the
%resonant production of a heavy Higgs boson present in the model that
%subsequently decays into a pair of SM-like Higgs bosons. Also the
%continuum non-resonant SM-like Higgs pair production can exceed the SM
%value because the trilinear Higgs self-coupling of the SM-like Higgs
%boson stil can deviate from the SM value. The top-Yukawa coupling on
%the other hand is very much to be constrained SM-like due to the Higgs
%signal data. Nevertheless, it turns out that the continuum of our
%models is not affected by the experimental results on continuum SM
%Higgs pair production. This is not the case for resonant searches,
%however. In the following, we show the effect of these
%searches exemplary for the real 2HDM. \s

\begin{figure}[h!]
\centering
%\subfloat{\includegraphics[width = 2.5in]{MH2_T1_HsmHsm_bbbb_R2HDM.eps}}
%\subfloat{\includegraphics[width =
%2.5in]{MH2_T1_HsmHsm_bbbb_R2HDM_WITHCONST.eps}} \\
\hspace*{-0.5cm}
\subfloat{\includegraphics[width = 3.3in]{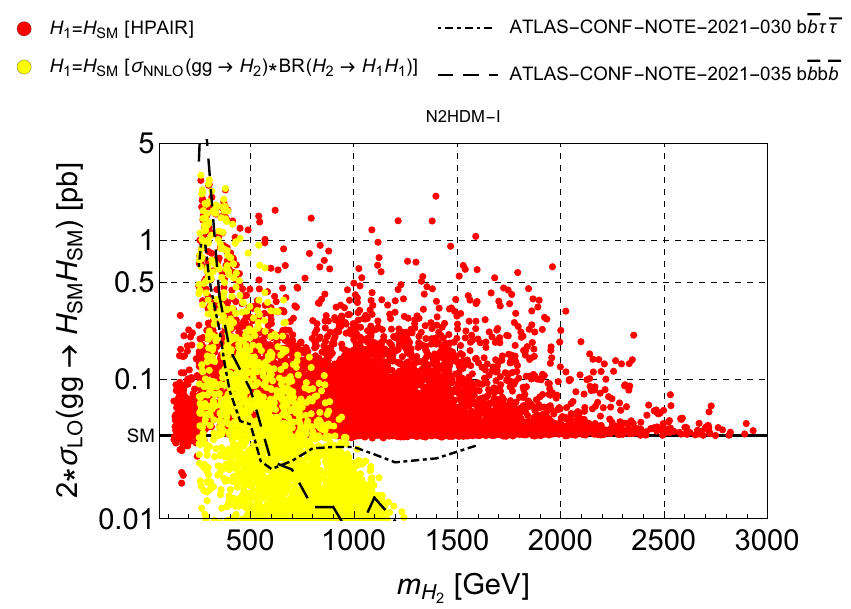}}
\subfloat{\includegraphics[width = 3.3in]{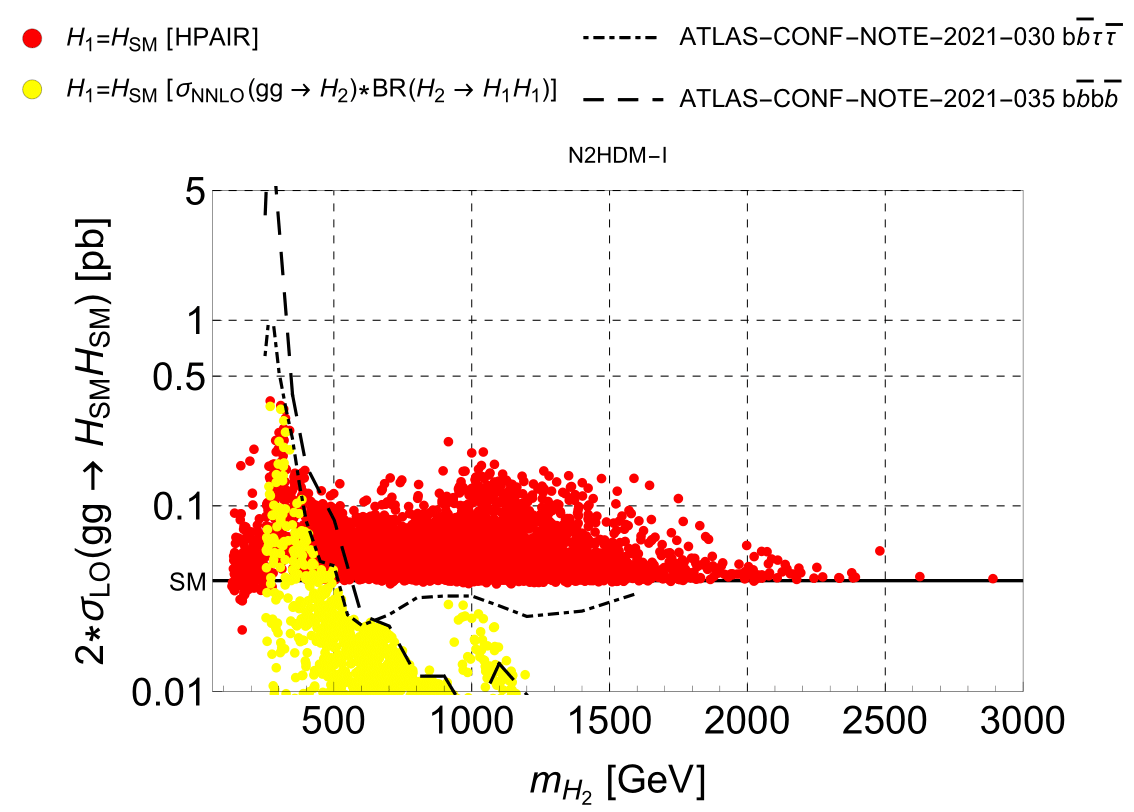}} \\
\caption{N2HDM-I: Yellow points: $\sigma(pp\to
  H_2)_{\text{NNLO}}^{\text{SusHi}}\times \mbox{BR}(H_2\to H_1 H_1)$.
  Red points: $2\times
  \sigma_{\text{LO}}^{\text{HPAIR}}(gg \to H_{\text{SM}}
  H_{\text{SM}})$, with $H_{\text{SM}}
  \equiv H_1$, as function of $m_{H_2}$. 
  Left (right) panel without (with) the constraints from experimental
  resonant di-Higgs searches, {\it cf.}~text for details. The dashed (dot-dashed)
  line denotes the ATLAS limit from the $(b\bar{b}) (b\bar{b})$
  ($(b\bar{b}) (\tau \tau)$) final state. Horizontal line: Higgs pair
  production cross section in the 
  SM.}
\label{fig:hh1}
\end{figure}

For illustration, we show the impact of the limits from resonant
searches for the N2HDM type I (N2HDM-I) where the lightest CP-even
scalar $H_1$ is the SM-like Higgs boson $H_{{\text{SM}}}$. 
The yellow points in Fig.~\ref{fig:hh1} (left) show for the
  points of our scan in the N2HDM-I parameter space that pass
the constraints described in Sec.~\ref{sec:scans}, the single
production cross sections of the heavy Higgs
boson $H_2$, computed with {\tt SusHi}, and multiplied with the branching ratio
into a pair of two SM-like Higgs bosons $H_1$. In other words, they
represent one of the resonant production modes of a SM-like Higgs
pair. The dashed line and the dot-dashed line are the experimental
limits obtained from resonant di-Higgs production searches in the 4$b$
final state \cite{ATLAS-CONF-2021-035} and the ($2b$)($2\tau$) final
state \cite{ATLAS-CONF-2021-030}, respectively. These limits are
now applied on all yellow points. Note, however, that we not only
apply them on resonant $H_2$ but also on resonant $H_3$ production. 
The right plot in Fig.~\ref{fig:hh1} shows the situation after
applying the aforementioned experimental constraints plus the bounds
from $(2b)(2\gamma)$~\cite{ATLAS-CONF-2021-016} and the CMS bounds
from $4b$~\cite{CMS-PAS-B2G-20-004}, which only affect the 
very low and heavy mass region, respectively, and, due to better
visibility, were not added in the 
plot. All previous experimental results are weaker in the whole heavy
resonance mass range and thus automatically satisfied. We
see that some of the yellow points above the experimental limits are
left over. Here we do not fulfil our criteria of $(\Gamma_{\text{tot}}
(H_k)/m_{H_k})_{\text{limit}}<5$\% so that the experimental limits
cannot be applied and, thus, no statement about the validity 
of these points can be made. \s

The red points in Fig.~\ref{fig:hh1} (left) show for all parameter
scenarios passing the constraints of Sec.~\ref{sec:scans}, the cross
sections for SM-like Higgs pair production as a function of the mass
of the non-SM-like Higgs boson $H_2$. As
described above, the cross section is calculated at LO and multiplied
by a factor two to approximately take into account NLO QCD
corrections. The constraints from resonant di-Higgs searches are
taken into account by referring to the yellow points. Only those
scenarios where the yellow points passed the resonant search limits are
retained for the di-Higgs cross sections and result in the allowed red points
presented in Fig.~\ref{fig:hh1} (right). The comparison of the left and right plot in
Fig.~\ref{fig:hh1} clearly shows that the present di-Higgs searches
are already sensitive to the N2HDM parameter space and exclude parts
of it beyond single Higgs data constraints. \s 
%We explicitly compared the results of the corresponding
%yellow points with those obtained from Higgs pair production where we
%turned off in {\tt HPAIR} the contribution from all non-resonant
%diagrams, {\it i.e.}~$s$-channel Higgs $H_k$ exchange with $m_{H_k}
%\le 2 m_{H_{\text{SM}}}$ and box diagrams. The results deviate
%slightly (not more than 20\%) which can be explained by the
%fact that the {\tt SusHi} gluon fusion results are computed at NNLO
%QCD and $\sqrt{s}=13$~TeV whereas
%{\tt HPAIR} is run at $\sqrt{s}=14$~TeV for LO QCD and mutliplies the
%result afterwards by a
%factor of two, and different pdfs are used in the
%computation. Furthermore, {\tt HPAIR} includes all $s$-channel Higgs
%exchange and the box diagrams in the computation. Only where the resonant
%production dominates the two results approach each other (see also
%Fig.~\ref{fig:continuum} below). Note
%that the SM Higgs pair production cross sections
%at FT$_{\text{approx}}$ \cite{Grazzini:2018bsd}\footnote{At
%  FT$_{\text{approx}}$, the cross section is computed at NNLO QCD in
%  the heavy-top limit with full LO and NLO mass effects and full mass
%  dependence in the one-loop double real corrections at NNLO QCD.}
%differ by 18\% at $\sqrt{s}=13$~TeV and 14 TeV, respectively.\s
%
%The horizontal line shows the result obtained for the SM, also
%computed with \texttt{HPAIR}, with a cross-section of $2\times
%\sigma_{HH}^{\text{SM,HPAIR}}@\mbox{LO}=39$ fb, from which it becomes
%clear that resonant production can by far enhance the
%cross section compared to the SM
%result. 

From the right plot, we infer that there are many points left
after application of the resonant search limits. In many of them, the contribution
from resonant diagrams is suppressed or kinematically forbidden. 
%From
%an experimental point of view, they correspond to non-resonant di-Higgs
%searches so that the corresponding non-resonant limits have to be
%applied. 

Looking only at the
total cross section values, as we do it here, and not at distributions,
the sizes of the resonant Higgs pair production cross
sections in the suppressed cases are similar to the non-resonant ones
or even smaller, so that they cannot be distinguished based on the
total rates.
Also, invariant mass distributions barely change as we explicitely verified,
since the contributions from the resonances are largely suppressed
w.r.t.~to those from the non-resonant parts. We therefore applied on
these points the corresponding non-resonant limits.
We come back to this point below. 
One can also see from this plot that for $m_{H_2}\lsim 2 m_{H_1}$ the
cross section can be suppressed relative to the SM value. This is to
be attributed to destructive interferences between the various diagrams
contributing to the di-Higgs cross section. \s

\begin{figure}[h!]
\includegraphics[width=0.45\textwidth]{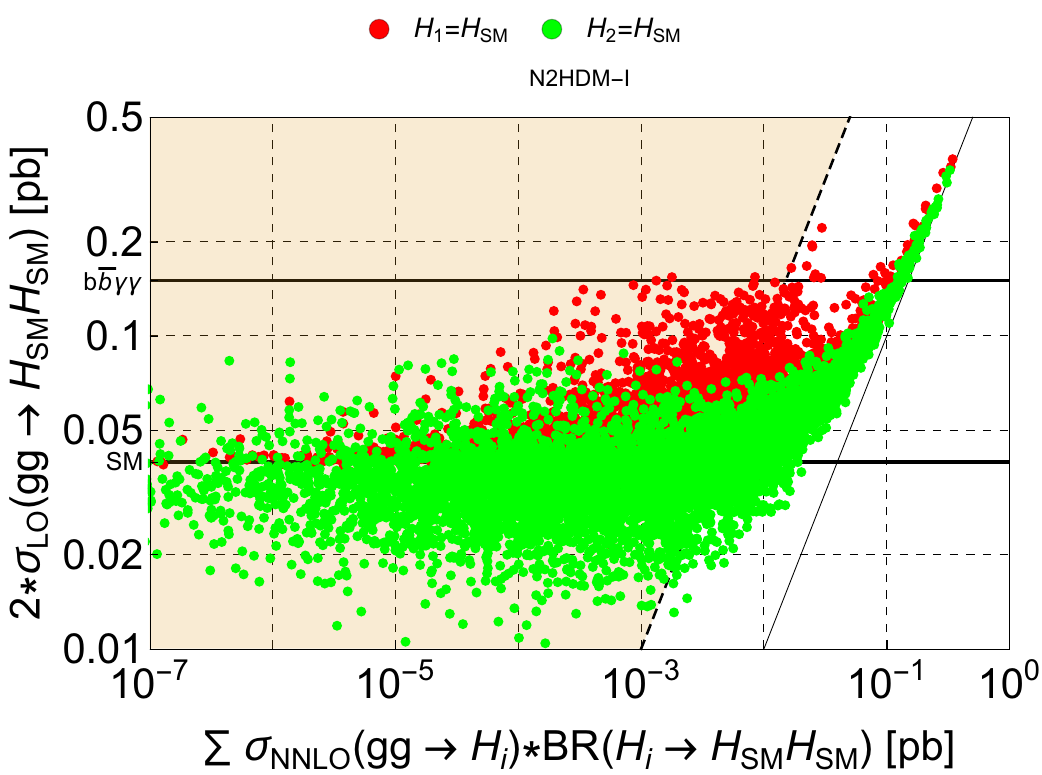}
\hspace*{0.5cm}
\includegraphics[width=0.45\textwidth]{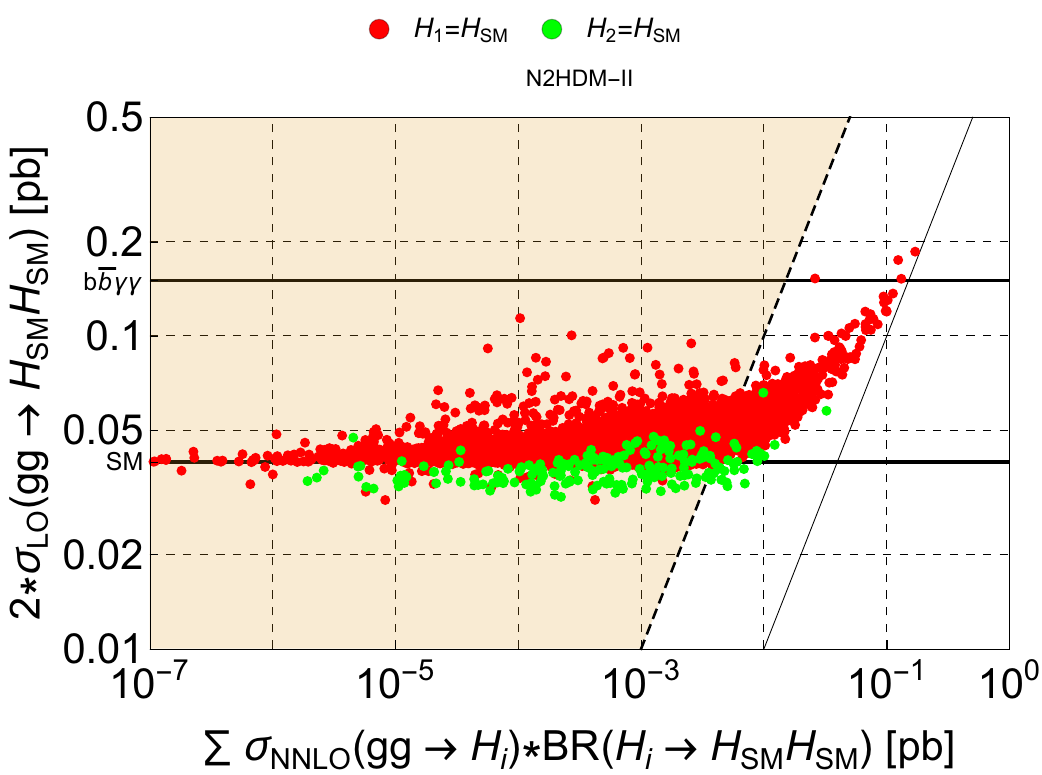}
\caption{N2HDM: LO gluon fusion cross sections, multiplied by a factor
  2 to approximate the NLO QCD corrections, into a SM-like Higgs
  pair with $H_1 \equiv H_{\text{SM}}$ (red) and $H_2 \equiv
  H_{\text{SM}}$ (green) for all points passing our constraints, as a
  function of the NNLO cross section for resonant heavy Higgs $H_i$
  production with subsequent decay into a SM-like Higgs pair. For $H_1
  \equiv H_{\text{SM}}$ (red points) we sum up the single Higgs
  production cross sections of $H_2$ and $H_3$.
  Left: type I, right: type II. The horizontal line denoted by SM corresponds to the SM
  Higgs pair production value, and the one denoted by
  $b\bar{b}\gamma\gamma$ to the limit from non-resonant di-Higgs
  searches in the $2b 2\gamma$ final state. For the shaded region,
  see text. \label{fig:continuum}}
\end{figure}

As for the experimental limits from non-resonant searches, they mostly do not constrain
our models. The latest results in $(b\bar{b})(\gamma\gamma)$
\cite{ATLAS-CONF-2021-016}, however,
start cutting on the N2HDM-I (with $H_1\equiv H_{\text{SM}}$) parameter space as
is illustrated in Fig.~\ref{fig:continuum} for type I (left) and type
II (right). For all points passing our constraints, we
plot the NLO QCD (approximated by a factor 2) gluon fusion SM-like Higgs pair
production cross sections for the N2HDM-I (left) and II (right),
versus the NNLO QCD gluon fusion production cross section of a heavy
non-SM-like Higgs $H_i$ that subsequently decays into the SM-like Higgs
pair. For $H_2 \equiv H_{\text{SM}}$, we have $H_i = H_3$, for $H_1
\equiv H_{\text{SM}}$ we sum over the two possibilities $H_i=H_2,
H_3$. From the plot, we can infer that for parameter  points
where $H_{\text{SM}}H_{\text{SM}}$ production from resonant heavy
Higgs production dominates the di-Higgs process, both cross sections,
di-Higgs and single Higgs times Higgs-to-Higgs decay, approach each
other (see diagonal line in the plot).\footnote{Note that the di-Higgs
and single Higgs cross sections are not exactly the same for several
reasons. The {\tt SusHi} single Higgs gluon fusion results are computed at NNLO
QCD and $\sqrt{s}=13$~TeV whereas
{\tt HPAIR} for di-Higgs production is run at $\sqrt{s}=14$~TeV for LO
QCD and multiplies the result afterwards by a factor of
two. (The SM Higgs pair production cross sections
at FT$_{\text{approx}}$ \cite{Grazzini:2018bsd} differ by 18\% at
$\sqrt{s}=13$~TeV and 14 TeV, respectively.)
Furthermore, different pdfs were used in the computation. Also,
{\tt HPAIR} includes all $s$-channel Higgs exchange and the box
diagrams in the computation of the cross section. The impact of the
difference between the
cross sections w.r.t. to the application of the experimental limits
is negligible, however, as we explicitly verified.} For the smaller cross sections, resonant production
stops playing a significant role and the experimental limits from
non-resonant di-Higgs searches can be applied. The most stringent one
among the various final states is presently given by the $(2b) (2\gamma)$
final state \cite{ATLAS-CONF-2021-016}\footnote{Apart from the
  combined limit which we do not apply.}, which is visualized in the
plots by the horizontal lines.
We see that in the N2HDM-I this already cuts into the parameter space so that
    non-resonant Higgs search constraints start to play a role for certain
    models.\footnote{We remind the reader that our di-Higgs cross sections are
 computed at $\sqrt{s}=14$~TeV while the limits are obtained at 13
 TeV. With future experimental results at 14 TeV, even more points will
be excluded. For our rough analysis, however, this is a good enough approach.}
The transition between non-resonant and resonant production is fluid
of course. In order to be able to apply the experimental limits we are
forced to define a separation between the two cases which is
arbitrary. We define a cross section to be resonantly dominated when
the single non-SM Higgs production with subsequent decay into
SM-like Higgs bosons makes up for more than 10\% of the di-Higgs
result and accordingly apply the resonant limits. This region separation is
shown by the diagonal dashed line in each plot. The shaded region
is hence the region where we apply the non-resonant search limits. Apart from the
N2HDM-I case, we found that non-resonant searches do not constrain the
investigated models at present.  The previous definition is arbitrary and a sophisticated
experimental analysis taking into account distributions would be
required. This is beyond the scope of this paper. Since at present the
non-resonant searches are not very sensitive, this approach is good
enough for our purpose of drawing an overall phenomenological
picture.

%It is interesting to see that the continuum cross section exhibits the
%decoupling behavior of the 2HDM. When $m_{H_2}$ becomes large, one can
%see that the total cross section is rather close to the SM value. This
%is because for large $m_{H_2}$, $\cos(\beta-\alpha)\to 0$ and
%therefore the contribution of $pp\to H_2\to H_1 H_1$ becomes
%suppressed as the involved trilinear coupling $\lambda_{H_2 H_1 H_1}$
%is proportional to $\cos(\beta-\alpha)\to 0$. \s

\subsection{Parameter Dependences}
The size of the cross section for SM-like Higgs pair production
depends on the SM-like Higgs values of the trilinear Higgs
self-couplings and of the top and bottom Yukawa couplings. The influence of the
latter is only relevant for BSM models in the case of large values of
$\tan \beta$. The cross section value furthermore depends on the
masses of all additionally involved
non-SM-like Higgs bosons $H_k$, their total widths, their Yukawa
couplings and the trilinear Higgs 
self-couplings $\lambda_{H_k H_\text{SM} H_\text{SM}}$. In this
subsection we present the distributions of the mass spectra of 
the non-SM-like neutral Higgs bosons and discuss the allowed
sizes of the trilinear Higgs
self-coupling and top-Yukawa coupling of the SM-like Higgs boson in
our investigated BSM models.

\subsubsection{Distributions of the Mass Spectra}
\begin{figure}[h!]
\centering
\includegraphics[width=0.45\textwidth]{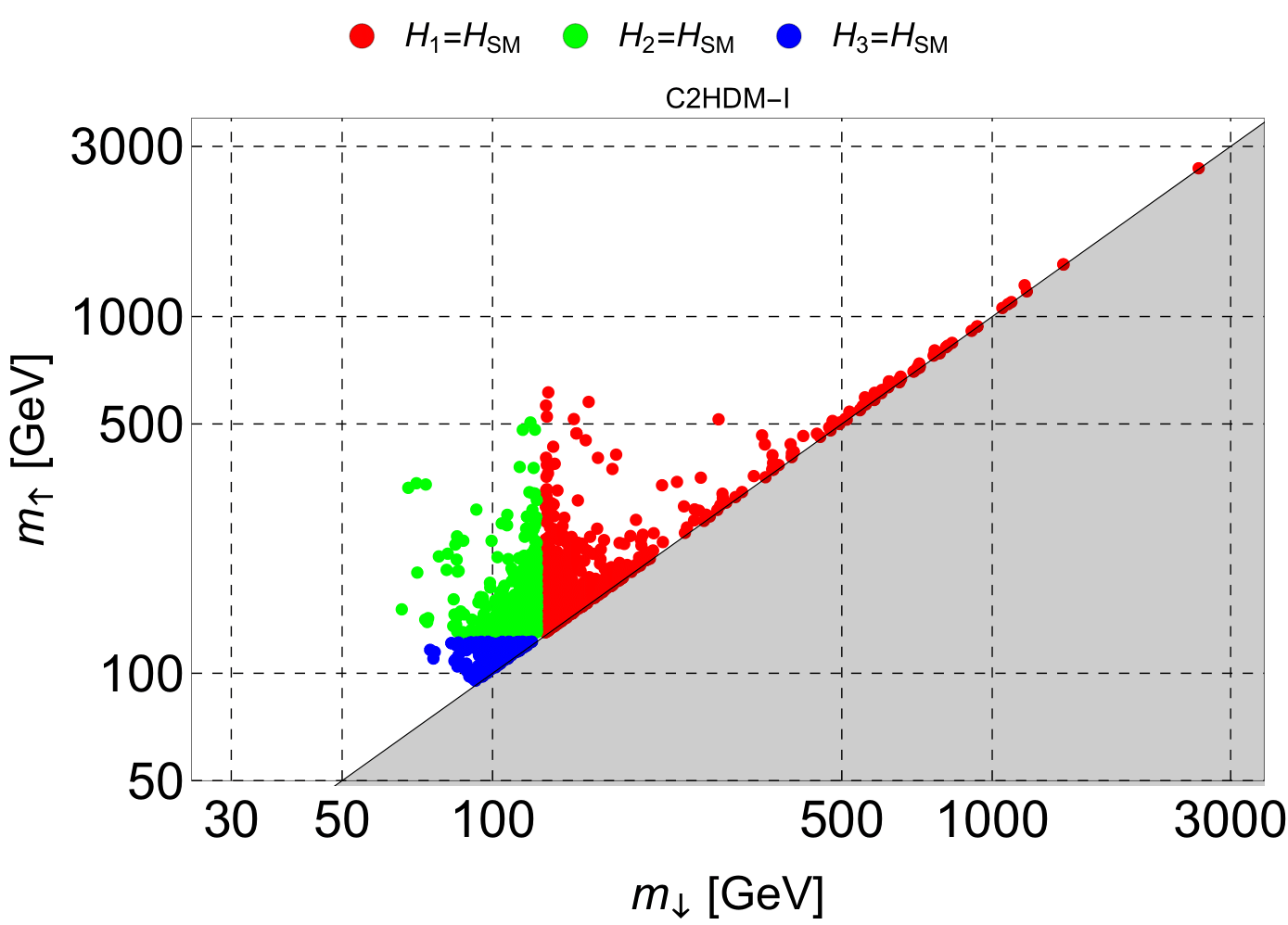}
\hspace*{0.5cm}
\includegraphics[width=0.45\textwidth]{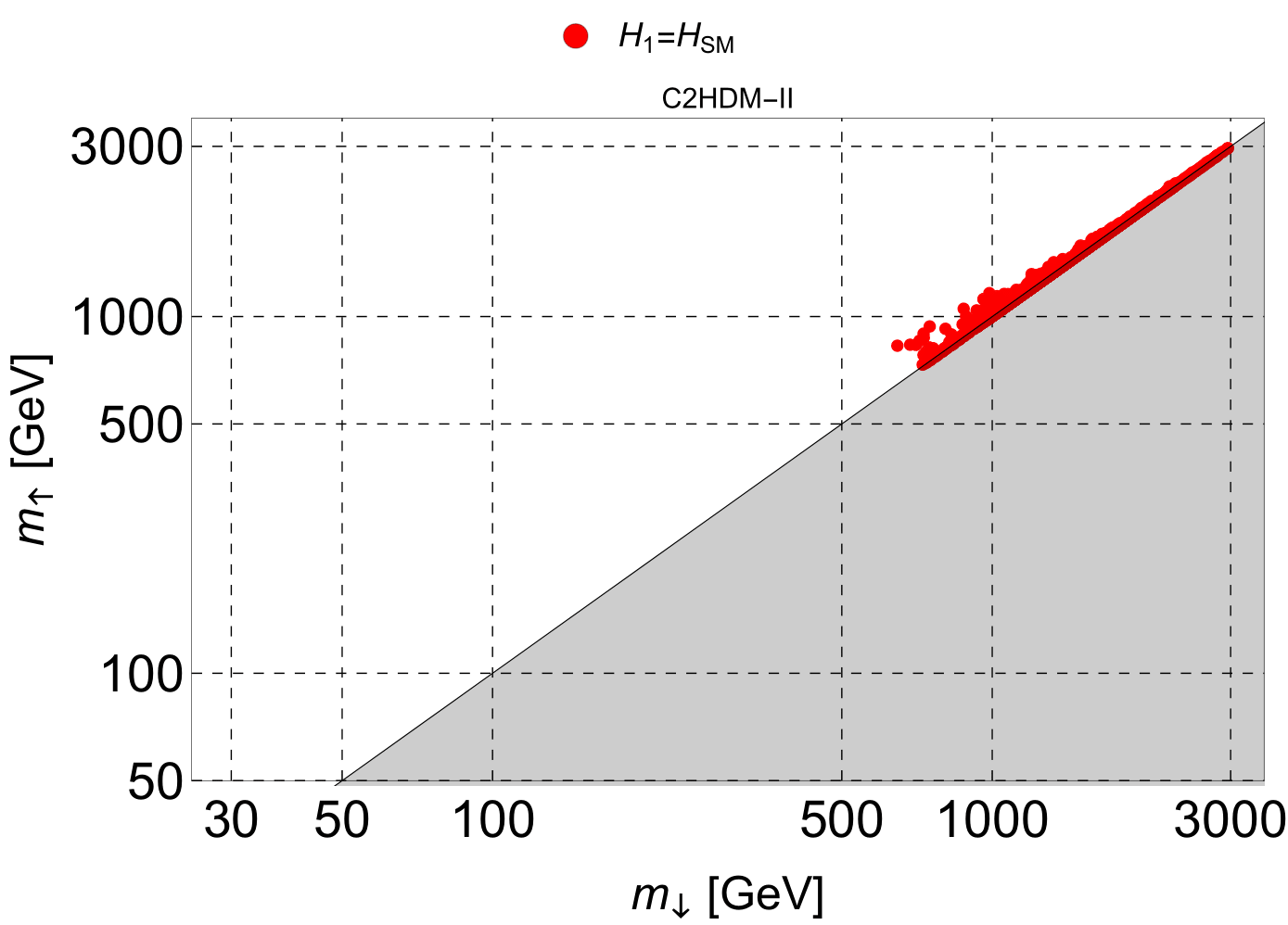}\\
\includegraphics[width=0.45\textwidth]{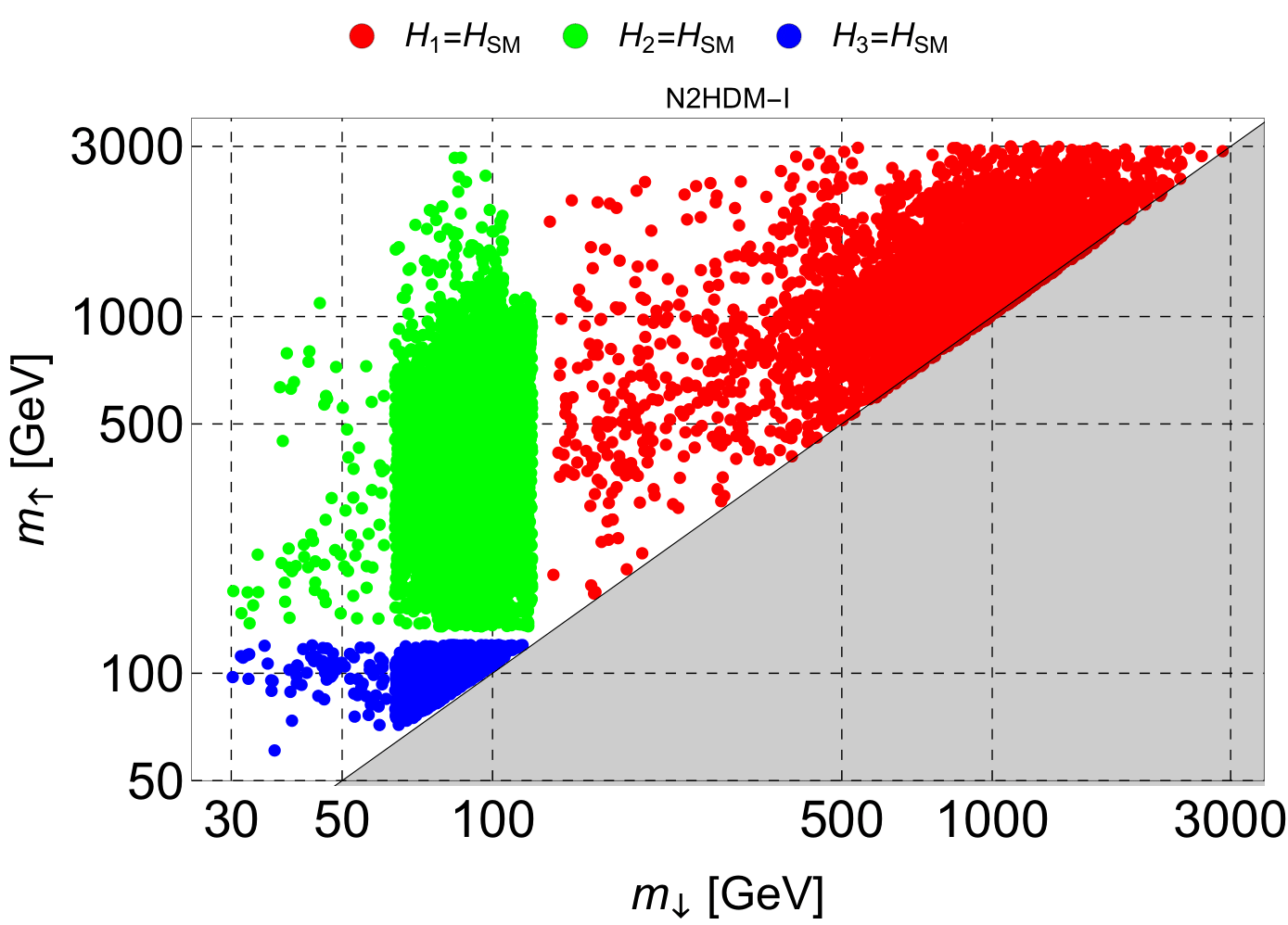}
\hspace*{0.5cm}
\includegraphics[width=0.45\textwidth]{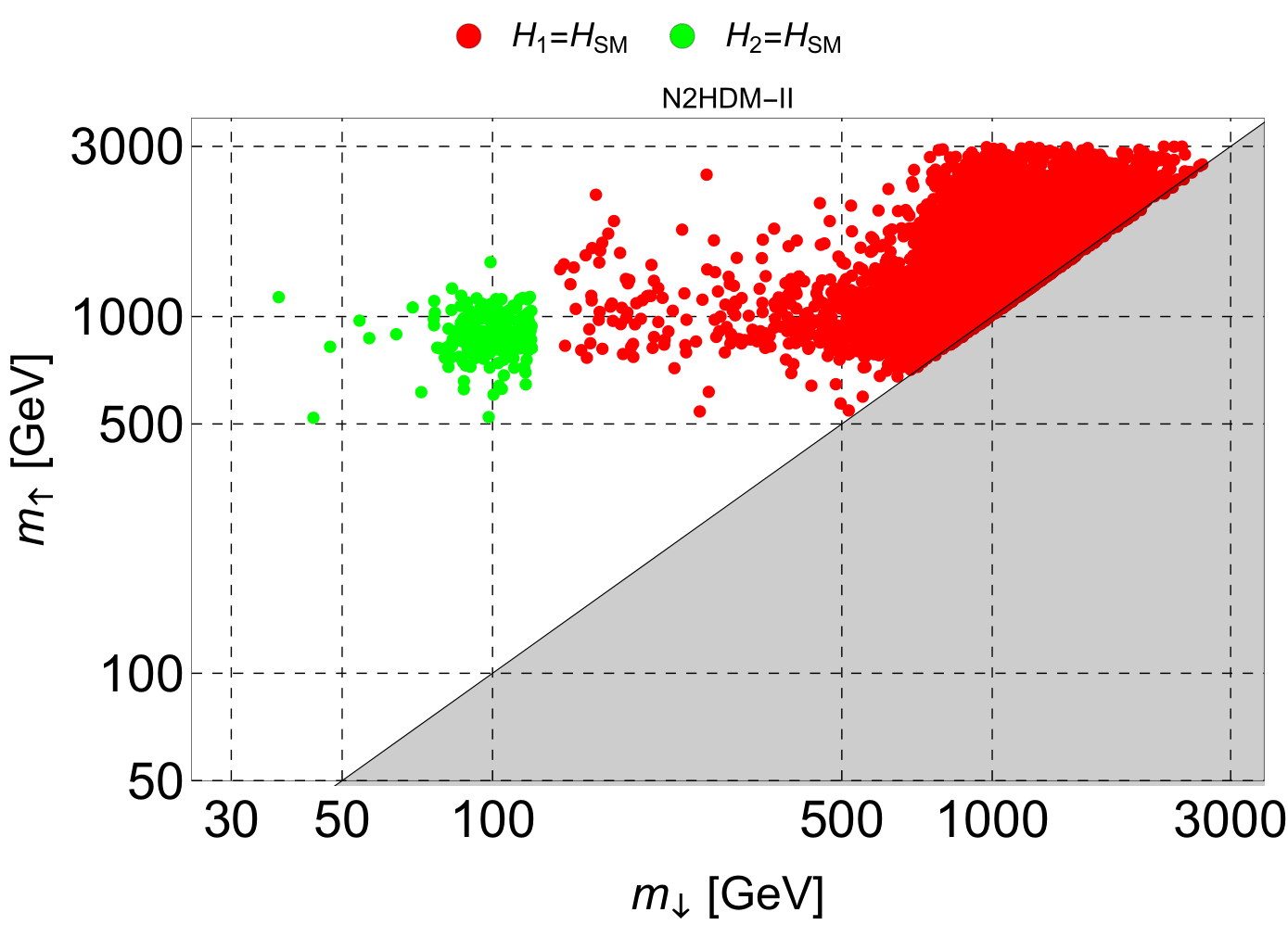}
\\
\includegraphics[width=0.45\textwidth]{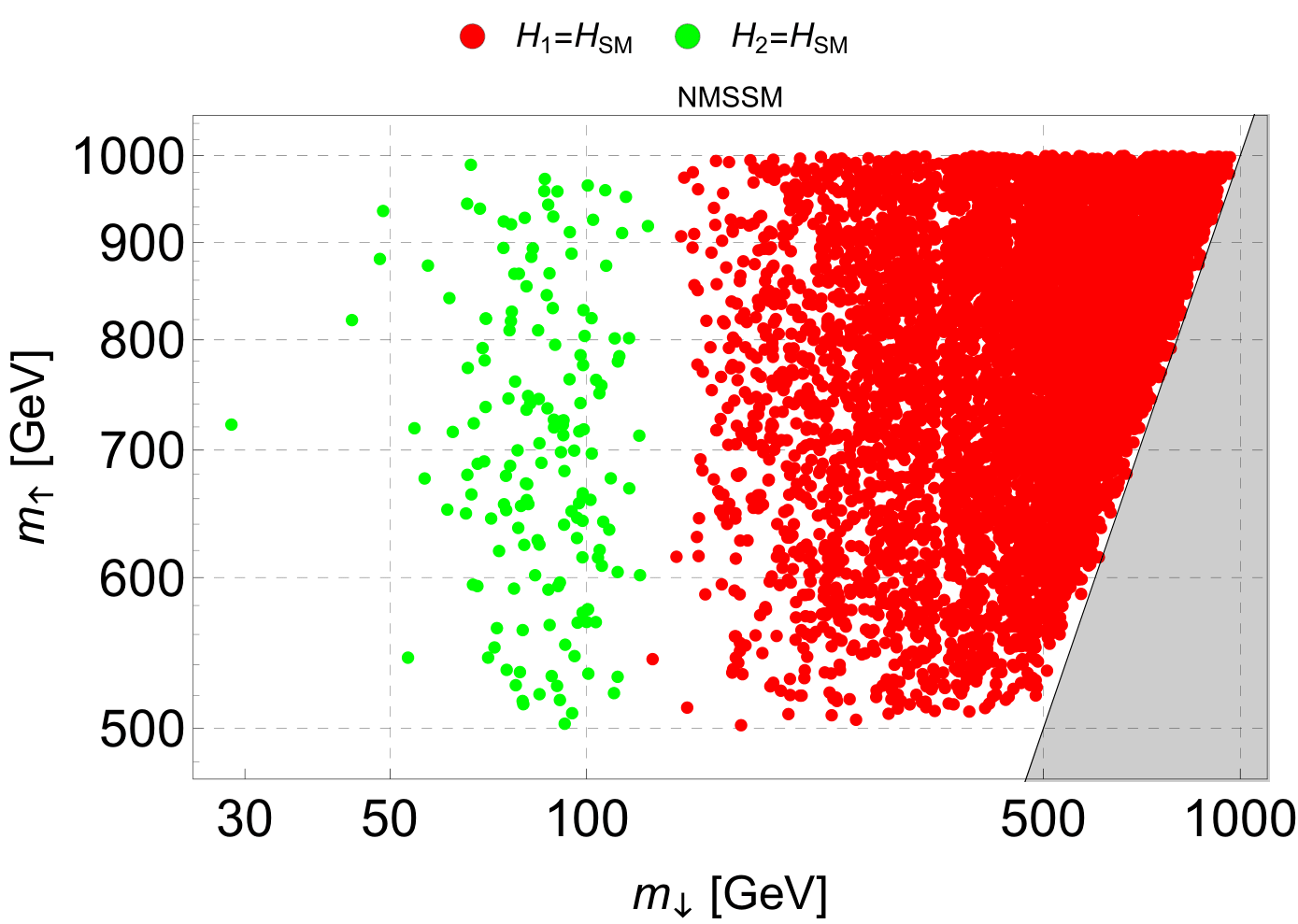}

\caption{Mass distributions of the non-SM-like neutral Higgs
  bosons in the C2HDM-I and II (upper left and right), the N2HDM-I and
II (middle left and right), and the NMSSM (lower) for the parameter
points respecting all applied constraints. The colour of the points
denotes which Higgs boson is SM-like. In the shaded region
$m_{\uparrow} \le m_{\downarrow}$.  \label{fig:massdistr}}
\end{figure}

Figure~\ref{fig:massdistr} displays the mass distributions of the two neutral
non-SM-like Higgs bosons contributing to the cross section for
SM-like Higgs pair production in the C2HDM (upper), the N2HDM (middle)
and NMSSM (lower) for type I (left) and type II (right) in the former
two models.\footnote{For a previous phenomenological investigation of
  the models, {\it cf.}~\cite{Muhlleitner:2017dkd}. Due to less strict
  constraints at that time, the mass spectra were less constrained.}
We denote the masses of the heavier one by $m_{\uparrow}$ and the
one of the lighter Higgs by $m_{\downarrow}$, respectively. In the N2HDM and
the NMSSM these Higgs bosons are CP-even, in the C2HDM they have no definite CP quantum number. We found in our scans that the
largest freedom in the distributions, after applying all considered
constraints, is found in the N2HDM (middle row) where in type I all
three neutral Higgs bosons $H_{1,2,3}$ can be SM-like (red, green,
blue points). The overall lightest Higgs mass spectrum is realized for $H_3
\equiv H_{\text{SM}}$ and becomes increasingly heavier if instead $H_2$ or
$H_1$ are SM-like, respectively. For the latter two cases large mass gaps can occur
between $m_\uparrow$ and $m_\downarrow$. In the
NMSSM large mass gaps are also possible. In the C2HDM-I, where all three SM-like cases can be realized, this is not the case any
more. In particular for $H_1 = H_{\text{SM}}$ the masses
$m_{\uparrow}$ and $m_{\downarrow}$ quickly become nearly
degenerate with increasing values. This is even more pronounced for
the C2HDM-II. Large differences in the 
mass hierarchies allow for Higgs-to-Higgs cascade decays which is
particularly interesting for non-SM-like single Higgs searches or Higgs pair
production and is a clear sign of non-minimally extended Higgs sectors
in contrast to the R2HDM or MSSM {\it e.g.}~which feature only two
CP-even neutral Higgs bosons. We can hence expect such decays in the
N2HDM-I, II, and the NMSSM, and to a less extent in the C2HDM-I because of
the more compressed mass spectrum.
We further found that in the C2HDM-II only the case $H_1\equiv
H_{\text{SM}}$ is realized in our scan after including the
constraints. In the N2HDM-II and the NMSSM,
$H_1$ and $H_2$ being SM-like is still possible but not $H_3$.
In the R2HDM, not shown in the plots, in type I, for $H_1 \equiv
H_{\text{SM}}$, the heavier $H_2$ mass ranges between 130~GeV and, the
upper scan limit of 3~TeV. For $H_2 \equiv H_{\text{SM}}$, the lighter
$H_1$ mass varies between 30 and 122.5~GeV.\footnote{Note that we
  applied a gap of $\pm 2.5$~GeV around 125.09 GeV.} In type II, $m_{H_2}$ ranges
between 800 GeV and the upper scan limit. Also for all other models,
the lightest Higgs masses found to be allowed are betweeen around 30 and
122.5~GeV, where the lower bound is due to our scan limit.
In the type-II models the overall mass spectrum is pushed to higher
values because of the lower limit on the charged Higgs mass. \s

\subsubsection{Higgs Couplings}
In the SM, the triangle and box diagrams
contributing to Higgs pair production interfere
destructively. Deviations of the trilinear Higgs self-coupling
$\lambda_{3H_{\text{SM}}}$ and Yukawa coupling to top quarks
$y_{t,H_\text{SM}}$ of the SM-like Higgs $H_{\text{SM}}$ with respect
to the SM values $\lambda_{3H}$ and  $y_{t,H}$, respectively, will mitigate this destructive
interference. In turn, it will allow for larger cross sections in (the non-resonant) SM-like di-Higgs
production. Experimental limits on the trilinear Higgs self-coupling
derived from Higgs pair production results often assume the top-Yukawa
coupling to be SM-like.\footnote{Results where this assumption has been
dropped are given in \cite{CMS:2016cma,CMS:2017rpp,CMS-PAS-HIG-17-008,
  ATLAS-CONF-2021-016, ATLAS-CONF-2021-052}.} We
therefore want to answer the following questions:
\begin{itemize}
\item After applying constraints from single Higgs data, what are the allowed
  ranges for $\lambda_{3H_{\text{SM}}}$ and $y_{t,H_\text{SM}}$ in our
  models? What is the impact of di-Higgs constraints?
\item How do di-Higgs cross sections behave as a function of
  $\lambda_{3H_{\text{SM}}}$ and $y_{t,{H_\text{SM}}}$?
\item What is the relation of the $\lambda_{3H_{\text{SM}}}$ and
  $y_{t,H_\text{SM}}$ limits to the effective coupling parameters in
  SM effective field theory (SMEFT)?
\end{itemize}

\subsubsection{Allowed Coupling Ranges and Impact of Di-Higgs Constraints}
\begin{figure}[t!]
\centering
%\subfloat{\includegraphics[width = 2.5in]{MH2_T1_HsmHsm_bbbb_R2HDM.eps}}
%\subfloat{\includegraphics[width = 2.5in]{MH2_T1_HsmHsm_bbbb_R2HDM_WITHCONST.eps}} \\
\subfloat{\includegraphics[width = 3in]{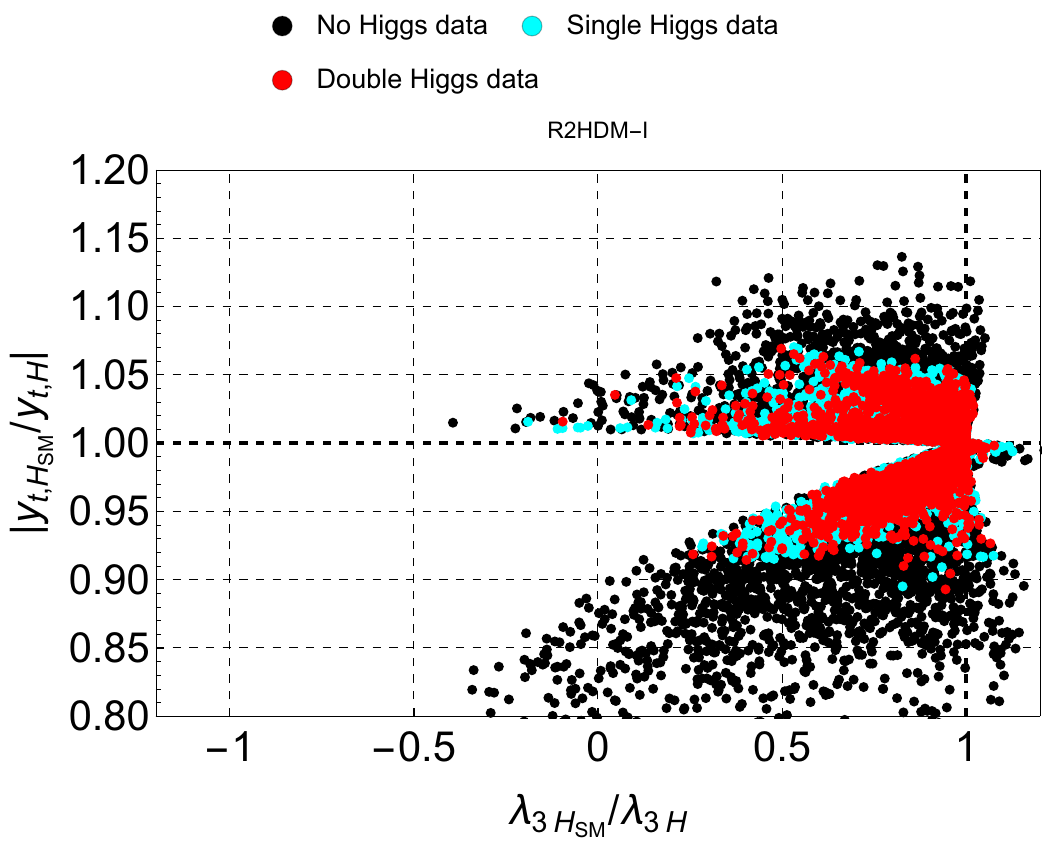}}
\subfloat{\includegraphics[width = 3in]{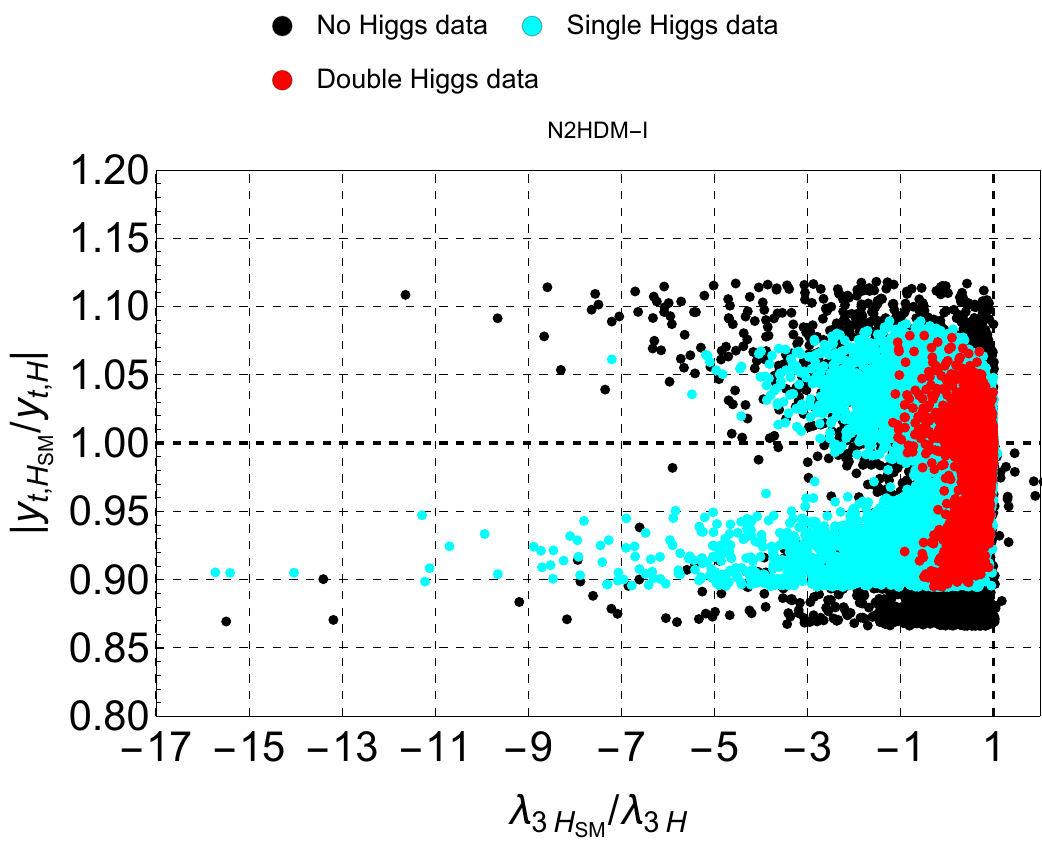}} \\
\caption{Absolute value of the Higgs top Yukawa coupling
  $y_{t,H_{\text{SM}}}$ versus the trilinear Higgs
  self-coupling $\lambda_{3H_{\text{SM}}}$ of the SM-like Higgs boson
  $H_{\text{SM}}$ given by $H_1$, $H_1\equiv H_{\text{SM}}$,
  normalised each to the SM values $y_{t,H}$ and $\lambda_{3H}$ for
  the R2HDM-I (left) and the N2HDM-I (right) for the parameter points passing all
  constraints without single and double Higgs constraints (black),
  including single Higgs constraints (blue) and
  also including di-Higgs search constraints
  (red). Dashed lines correspond to the SM-case of each coupling
  ratio. }
\label{fig:triltopyuk}
\end{figure}

Figure~\ref{fig:triltopyuk} depicts for the R2HDM-I (left) and the
N2HDM-I (right), with $H_1 \equiv H_{\text{SM}}$, the
absolute top-Yukawa versus the trilinear self-coupling values of the
SM-like Higgs normalized each to the corresponding SM-values. Black
points are allowed parameter points after applying all but the single Higgs
constraints, blue points additionally include the single Higgs
constraints (blue), and red points finally also comply with the
di-Higgs constraints. The comparison of the 
red with the blue points hence shows the impact of the di-Higgs
measurements on the coupling values. \s

As we can see from the left plot, in the R2HDM-I, after applying single Higgs constraints, the Yukawa coupling is restricted to
values close to the SM, $0.87 \le (y_{t,H_{\text{SM}}}^{\text{R2HDM}}/y_{t,H}) \le
1.07$ (blue points). The trilinear Higgs self-coupling is less restricted with
values in the range $-0.2 \le (\lambda_{3H_{\text{SM}}}^{\text{R2HDM}}/\lambda_{3H})\le
1.15$ (blue points). Note that the restriction of the trilinear Higgs
self-coupling at this stage basically stems from the constraints on the
Yukawa coupling from the single Higgs data.
Di-Higgs constraints (red points) slightly reduce these values,
to about $-0.1 \le (\lambda_{3H_{\text{SM}}}^{\text{R2HDM}}/\lambda_{3H})\le
1.1$.
Departures from the SM
value in the trilinear couplings come along with non-SM-like Yukawa
couplings ({\it cf.}~wedge region). Turning to
the N2HDM-I (right plot) we see that the usual plus single Higgs
constraints reduce the Yukawa
coupling to about the same values as in the R2HDM-I (blue points). The
trilinear coupling on the other hand can vary
in a larger negative range with values between -16 and 1 times the SM
value. After including the di-Higgs constraints the trilinear coupling
range is reduced to about -1 to 1 times the SM value (red points). This is
primarily caused by the unitarity constraint. In addition to the
perturbative unitarity check performed by {\tt
  ScannerS} we applied the following approximate limit on all trilinear Higgs
self-couplings $\lambda_{H_i H_j H_k}$. We required $|\lambda_{H_i H_j
  H_k}/\lambda_{3H}| \le 30$ for all $i,j,k$ combinations in the
respective model.\footnote{The 
value is derived by assuming a rough perturbative limit on the Higgs
mass of $M_H= 700$~GeV, implying a limit on the trilinear coupling of
$\lambda_{3H}^{\text{perturb}}=3 M_H^2/v=5975$~GeV compared to the value
of $\lambda_{3H}\approx 190$~GeV for the SM-like Higgs mass
$M_H=125.09$~GeV.} We found that this additional constraint only
affects the N2HDM. In all other models the
inclusion of the constraints through {\tt ScannerS} left over only
scenarios that already respect this unitarity constraint. Besides the latter, we also found that the di-Higgs searches cut on the allowed trilinear Higgs
self-coupling values, though to a lesser extent. \s

\renewcommand{\arraystretch}{1.2}
\begin{table}[t!]
\begin{center}
\begin{tabular}{|c|c|c||c|c|}
\hline
& \multicolumn{2}{c||}{R2HDM} & \multicolumn{2}{c|}{C2HDM} \\
\hline
& $y_{t,H_{\text{SM}}}^{\text{R2HDM}}/y_{t,H}$ &
$\lambda_{3H_{\text{SM}}}^{\text{R2HDM}}/\lambda_{3H}$
& $y_{t,H_{\text{SM}}}^{\text{C2HDM}}/y_{t,H}$ &
$\lambda_{3H_{\text{SM}}}^{\text{C2HDM}}/\lambda_{3H}$ \\
\hline
light I & 0.893...1.069 & -0.096...1.076 & 0.898...1.035 &
-0.035...1.227 \\
medium I & n.a. & n.a. & 0.889...1.028 & 0.251...1.172 \\
heavy I & 0.946...1.054 & 0.481...1.026 & 0.893...1.019 &
0.671...1.229 \\ \hline
light II & 0.951...1.040& 0.692...0.999 & 0.956...1.040 &
0.096...0.999 \\
medium II & n.a. & n.a. & -- & -- \\
heavy II & -- & -- & -- & -- \\
 \hline \hline
& \multicolumn{2}{c||}{N2HDM} & \multicolumn{2}{c|}{NMSSM} \\
\hline
& $y_{t,H_{\text{SM}}}^{\text{N2HDM}}/y_{t,H}$ &
$\lambda_{3H_{\text{SM}}}^{\text{N2HDM}}/\lambda_{3H}$
& $y_{t,H_{\text{SM}}}^{\text{NMSSM}}/y_{t,H}$ &
$\lambda_{3H_{\text{SM}}}^{\text{NMSSM}}/\lambda_{3H}$ \\
\hline
light I & 0.895...1.079 & -1.160...1.004 & n.a. & n.a.\\
medium I & 0.874...1.049 & -1.247...1.168 & n.a. & n.a. \\
heavy I & 0.893...1.030& 0.770...1.112 & n.a. & n.a. \\ \hline
light II & 0.942...1.038 & -0.608...0.999 & 0.826...1.003 &
                                                            0.024...0.747 \\
medium II & 0.942...1.029 & 0.613...0.994 &
%0.410...1.000
0.916...1.000 & -0.502...0.666\\
heavy II& -- & -- & -- & -- \\ \hline
\end{tabular}
\caption{Allowed ranges for the top-Yukawa and trilinear Higgs self-coupling
  of the SM-like Higgs boson after application of all constraints, normalized to the
  corresponding SM value,
  for the R2HDM, C2HDM, N2HDM, and NMSSM, respectively, for type 1 (I) and
  type 2 (II). Light/medium/heavy correspond to $H_{1/2/3}$ being the SM-like
  Higgs boson. The medium case is not applicable (n.a.) to the R2HDM,
  type 1 is not applicable to the NMSSM. In our scans, for type 2 some
  of the cases were found not
  to be compatible with the constraints any more (marked by a  dash in the table).  \label{tab:trilyuklimits}}
\end{center}
\vspace*{-0.8cm}
\end{table}
\renewcommand{\arraystretch}{1.0}

The unitarity constraints are responsible for the wedge regions in the
plots. Comparing the shape of the wedge regions
in the R2HDM-I and the N2HDM-I we see that an increased
precision in the Yukawa coupling will affect the allowed deviation in the trilinear
coupling in the N2HDM-I more than in the R2HDM-I.
Overall, we find that the trilinear
coupling gets more and more restricted but significant deviations are
still possible and that they come along with a
non-SM-like Yukawa coupling. The present (observed) limits on the
trilinear Higgs self-coupling assuming a SM top-Yukawa coupling are
-1.0 to 6.6 times the SM trilinear Higgs self-coupling at 95\% CL
as derived by ATLAS \cite{ATLAS-CONF-2021-052} and -3.3 to 8.5 as
given by CMS \cite{CMS:2020tkr}.
These experimental sensitivities to the trilinear
  Higgs self-coupling of the SM start to constrain the parameter space
  of our models, namely the N2HDM.\footnote{This is
  only true, however, if we assume a SM-like Yukawa coupling which is
  not appropriate in all models. We will come back to this point later.}
This can be inferred from Tab.~\ref{tab:trilyuklimits} where we list
the allowed ranges for the
top-Yukawa and the trilinear Higgs self-couplings in our investigated
models after applying all described constraints. For all models, due to the single
Higgs constraints, the top-Yukawa
coupling is bounded to a range of at most $\pm 0.1$ around the
SM case, with the exception of the NMSSM where it can deviate by up to
17\%.\footnote{Note that we excluded all scenarios where the mass
  gap between the SM-like and one non-SM-like Higgs boson is less than
2.5~GeV. Would we allow for these scenarios as well then the top-Yukawa
coupling could substantially deviate from the SM case, as the Higgs
signal is now built up by two Higgs bosons close in mass.} The
trilinear couplings are less constrained. For the N2HDM-I with $H_1$
or $H_2$ being SM-like
they are outside the lower ATLAS limit; however, only assuming SM-like
Yukawa couplings which is not the case as can be inferred from
Fig.~\ref{fig:triltopyuk}. Note also that a vanishing trilinear
SM-like Higgs self-coupling is also still allowed in some of the 
models.\s

There is one caveat to be made on the values given in
Tab.~\ref{tab:trilyuklimits}. These limits have been obtained from the
scans in the chosen parameter space with application of all constraints. Hence, they depend on the constraints that we apply, and they also depend on
our scanning procedure and sampling. More
extended scan ranges and scans adapted to specific parameter regions
could possibly find more points and extend these allowed coupling values
somewhat. With the given coupling values, however, we are on the
conservative side. Furthermore, also note that the C2HDM contains per
definition the limit of the R2HDM. This is not reflected, however, in
the coupling ranges (and will not be in the plots shown below
either). The reason is, that the scan in the C2HDM is performed in
different input parameters than in the R2HDM and for finite scan
ranges necessarily leads to differences. We explicitly checked that
larger R2HDM ranges than in the C2HDM indeed coincide with the CP-conserving limit in
the C2HDM and that larger C2HDM ranges compared to the R2HDM are due
to truly CP-violating points. We chose not to merge the C2HDM sample
with the R2HDM as it allows us to investigate CP-violating effects. As
a side remark we add that for the values of our scan the SM-like Higgs
boson in the C2HDM-I can
still have a CP-violating admixture\footnote{It is defined by the
  rotation matrix element squared $|R_{i3}|^2$, the index $i$ denotes the
  SM-like Higgs boson in the mass basis, the index 3 the
  CP-violating degree of freedom in the interaction basis.}
of up to 16\%, 20\% and 10\% for $H_1$, $H_2$ and $H_3$ being SM-like,
respectively, and of up to 2\% in the C2HDM-II with $H_1 \equiv
H_{\text{SM}}$.

\vspace*{-0.2cm}
\subsection{SM-Like Higgs Pair Production}
\begin{figure}[h!]
%\vspace*{-2cm}
  \centering
  \includegraphics[width=0.47\linewidth]{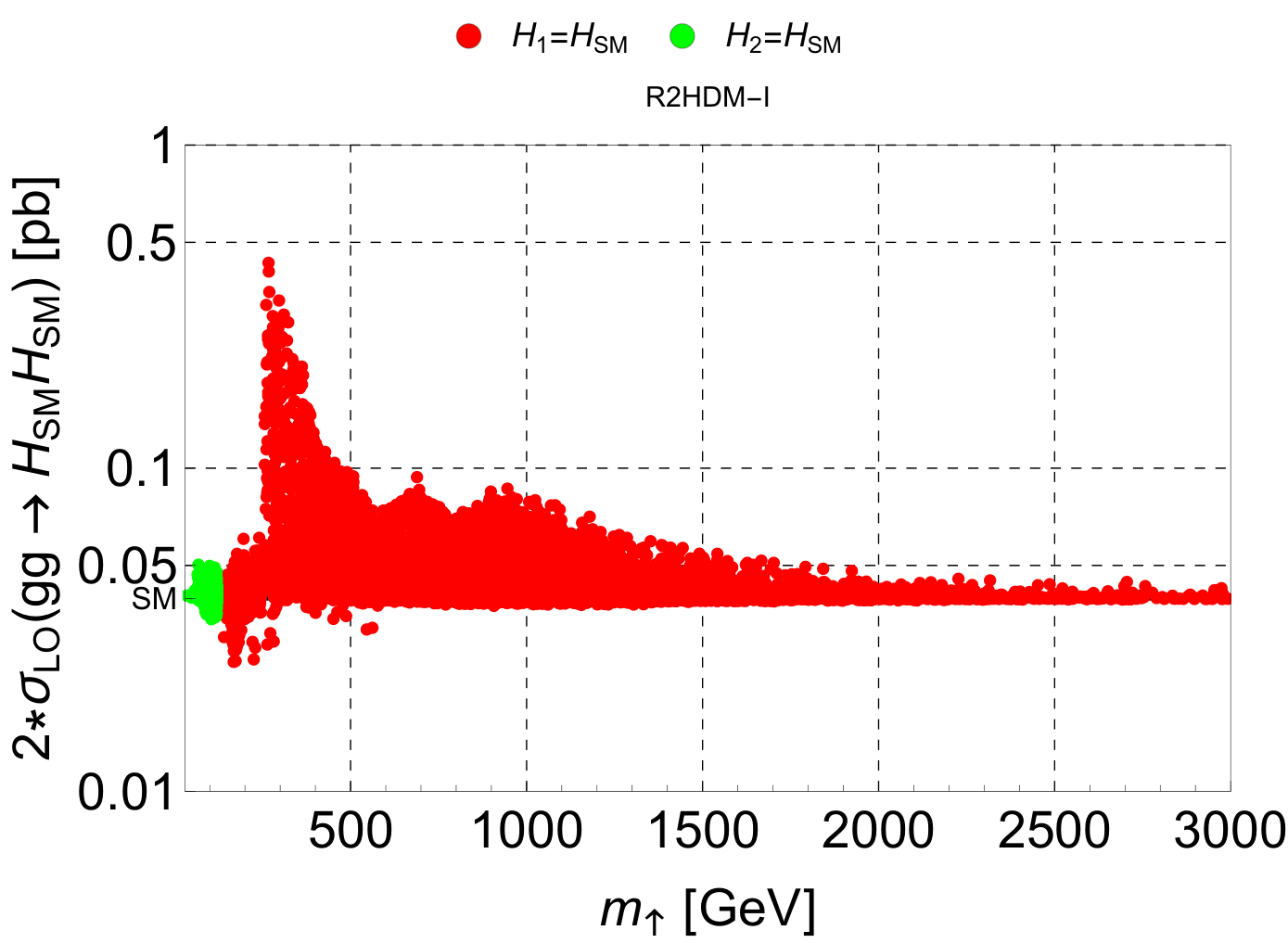}
  \includegraphics[width=0.47\linewidth]{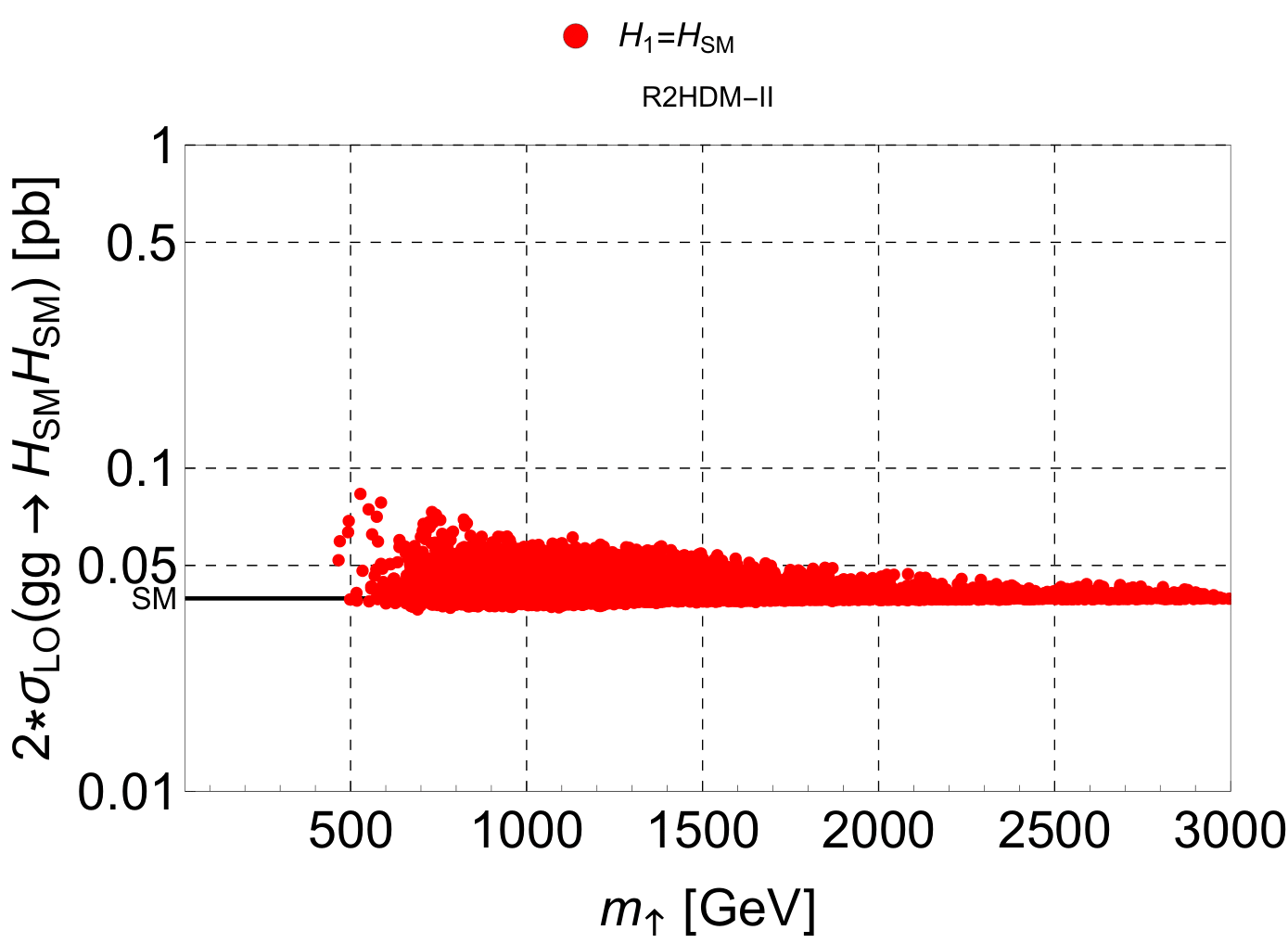}
  \\
 \includegraphics[width=0.47\linewidth]{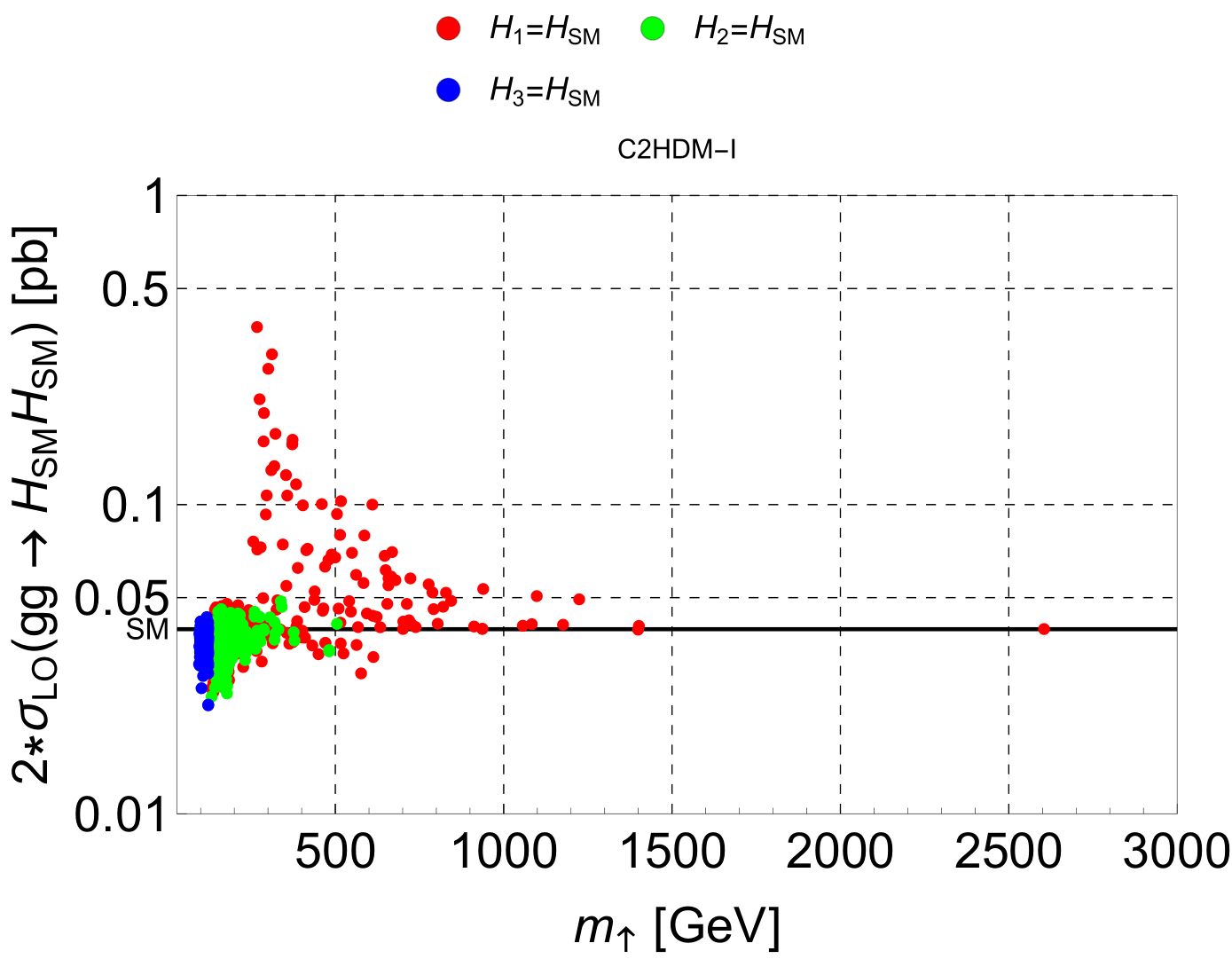}
 \includegraphics[width=0.47\linewidth]{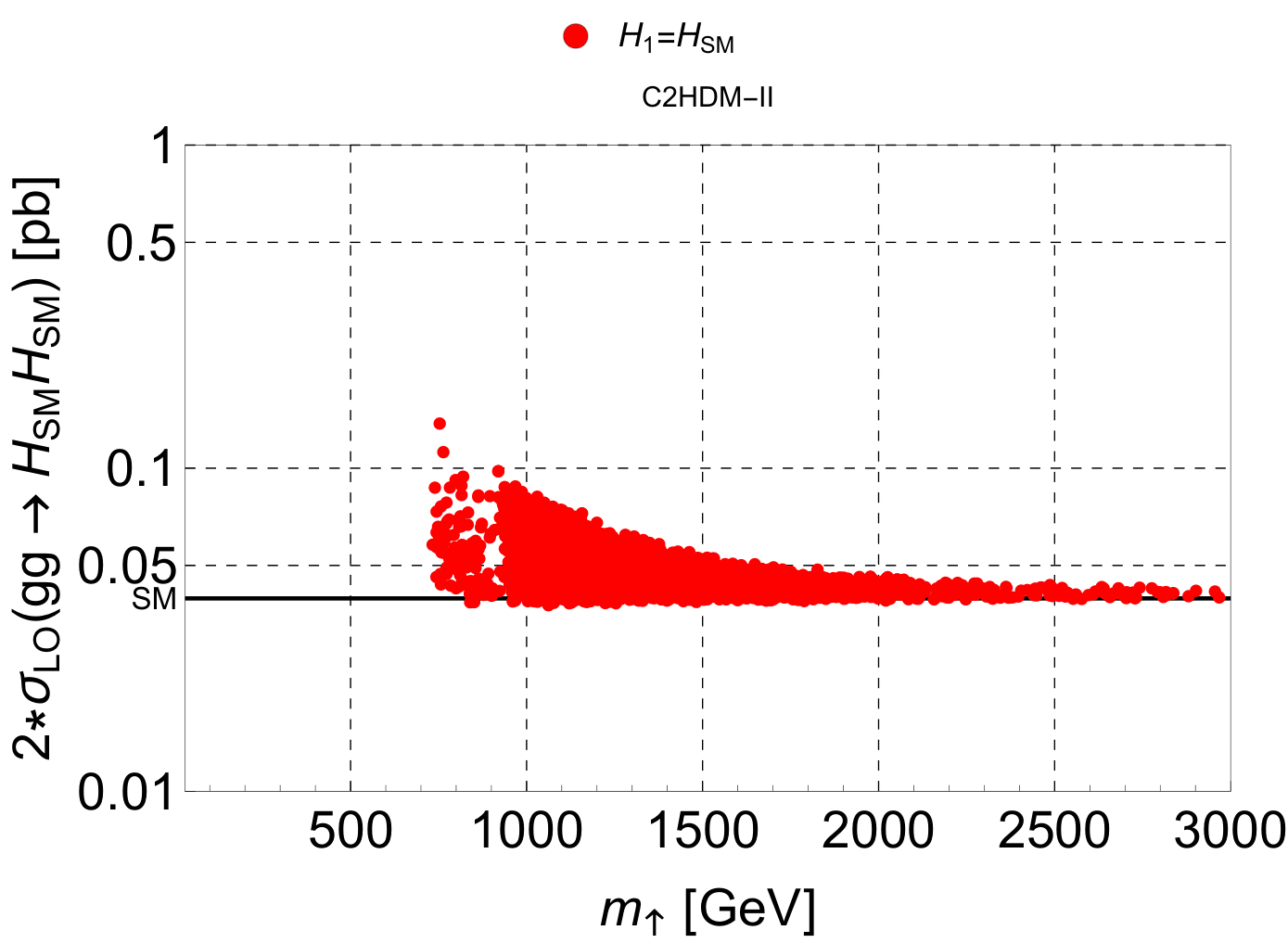}
 \\
 \includegraphics[width=0.47\linewidth]{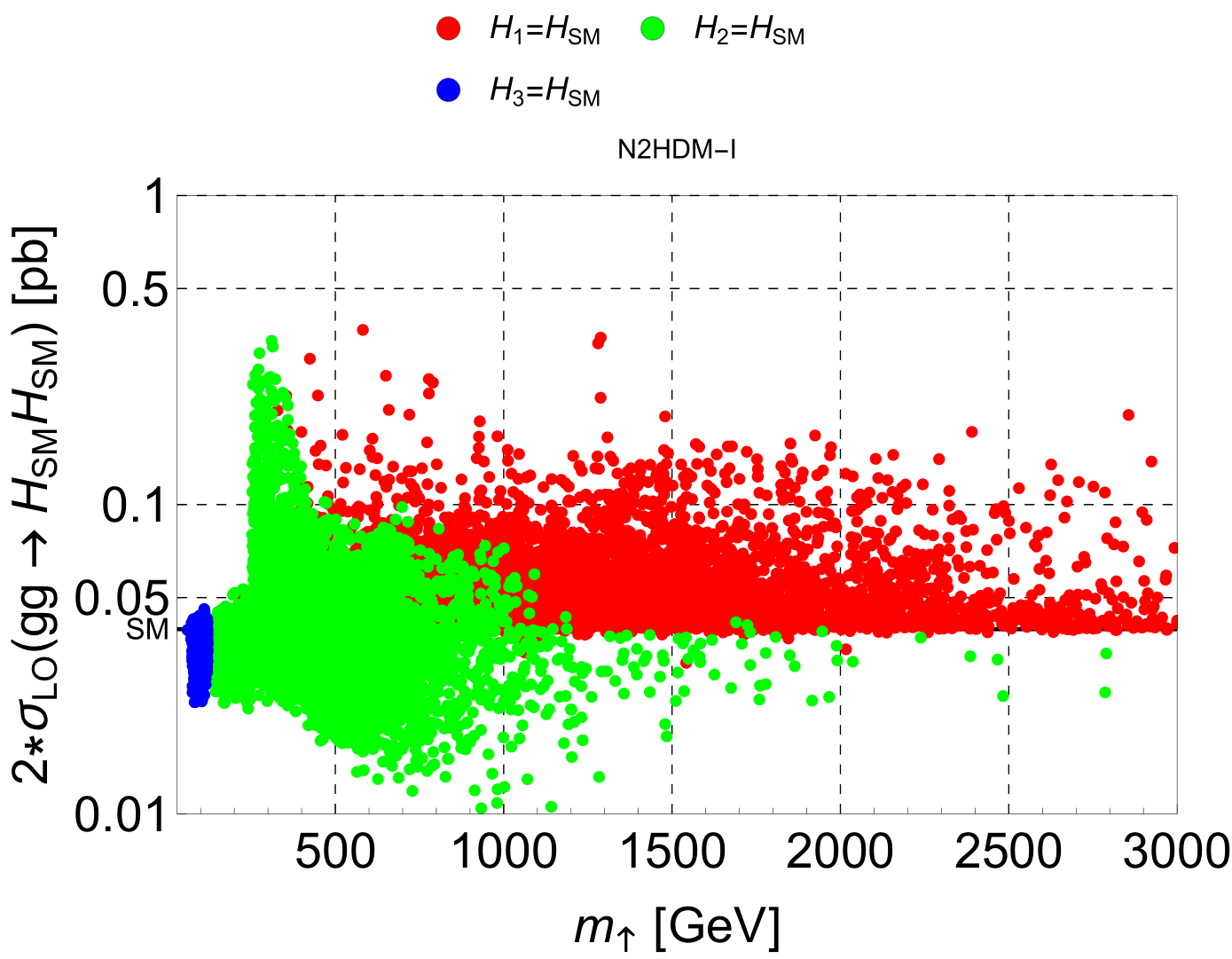}
 \includegraphics[width=0.47\linewidth]{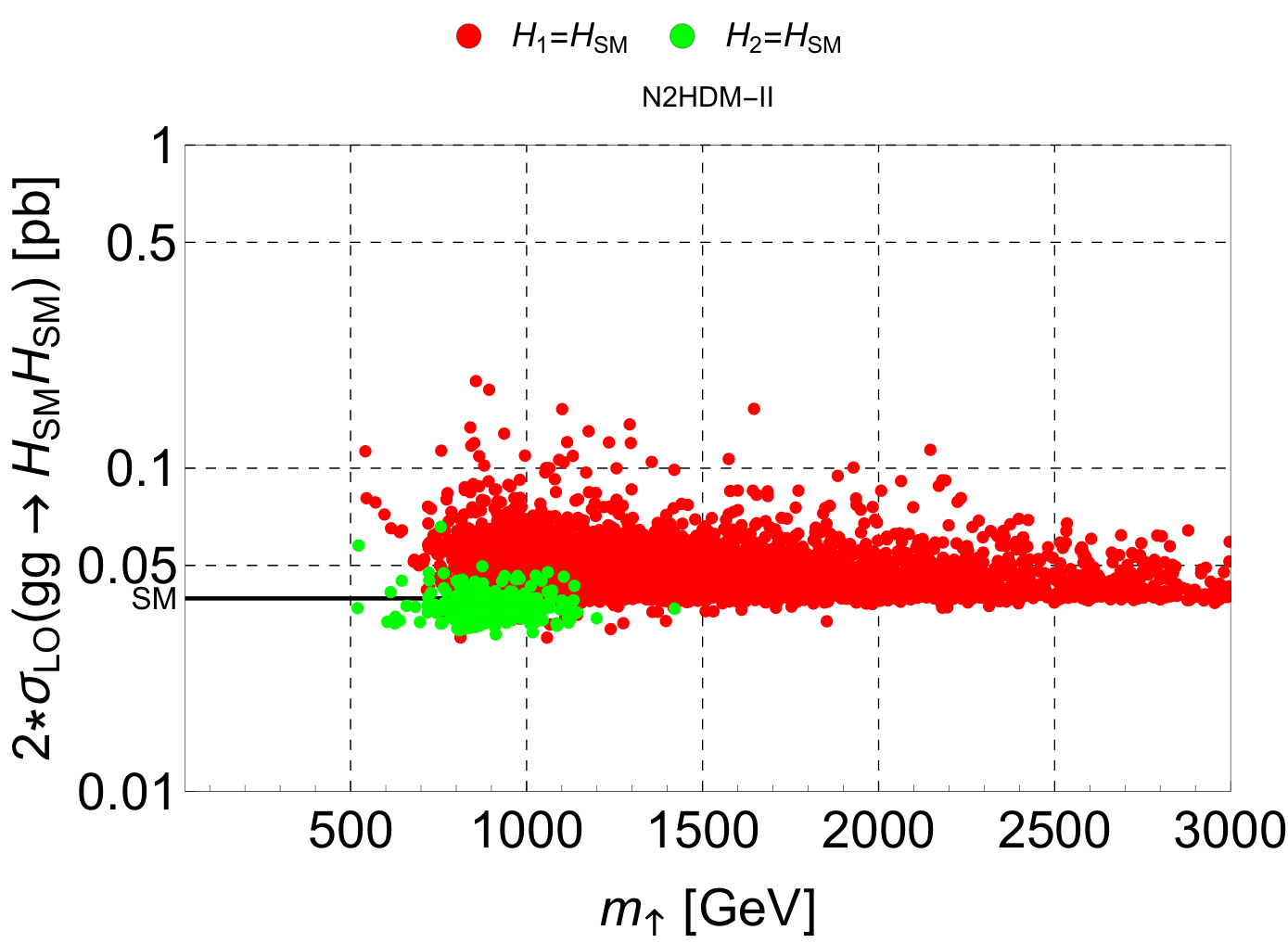}
 \caption{Scatter plots for SM-like Higgs pair production including a factor of 2 for
    estimating the NLO QCD corrections, for all points passing the
    constraints, as a function of the mass of the heavier of the
    non-SM-like Higgs bosons $m_{\uparrow}$ for the R2HDM (upper), the C2HDM
      (middle), and the  N2HDM  (lower line), for type 1 (left column)
      and type 2 (right column). Red points for
    scenarios with $H_1 \equiv H_{\text{SM}}$, green ones for those
    where $H_2 \equiv H_{\text{SM}}$, and blue points for those where
    $H_3 \equiv H_{\text{SM}}$. 
%Dotted and dashed lines:  experimental  limits as indicated in the legend;
    Horizontal line: SM result.}
    \label{fig:scattersm1}
\end{figure}

\begin{figure}[h!]
\vspace*{-2cm}
  \centering
  \includegraphics[width=0.47\linewidth]{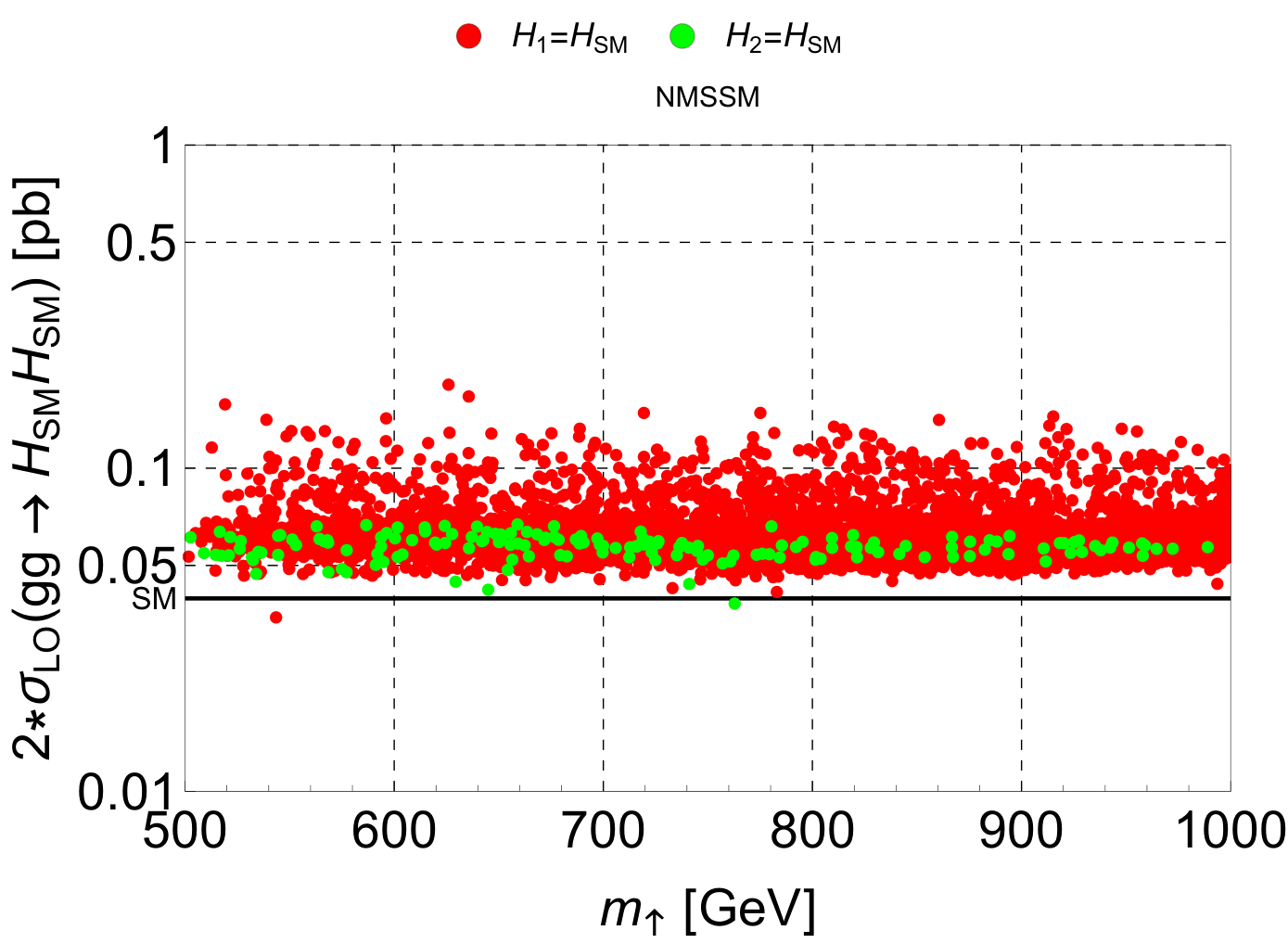}
  \caption{Same as Fig.~\ref{fig:scattersm1}, but for the NMSSM.}
    \label{fig:scattersm2}
\end{figure}

In this section we discuss the production of two SM-like Higgs bosons
$H_{\text{SM}}$ in our four considered models R2HDM, C2HDM, N2HDM, and
NMSSM. We display, for all valid parameter points
passing the applied constraints (including those from di-Higgs
searches), the cross sections for SM-like Higgs pair production in the
R2HDM, C2HDM, and N2HDM in Fig.~\ref{fig:scattersm1} and for the NMSSM
in Fig.~\ref{fig:scattersm2}.\footnote{The
values of the signal rates for the various final states after the
$H_{\text{SM}}$ decays can be
obtained by multiplying the cross section by the SM branching ratio into
the final state of interest. To a very good approximation the values of
the branching ratios of the SM Higgs boson can be applied here as the
present LHC results push any SM extension strongly towards the SM
limit.} The plots are shown as a function of the heavier of the
neutral non-SM-like Higgs bosons, denoted
by $m_{\uparrow}$. Red points depict the results where the SM-like Higgs boson is
given by the lightest Higgs boson $H_1$, green points are those for
$H_{\text{SM}} \equiv H_2$, and blue ones correspond to $H_{\text{SM}}
\equiv H_3$. In the R2HDM, we of course only have red and green
points. The overall lower density of the
points in the C2HDM compared to the R2HDM is an artifact of the scan,
as in the C2HDM one of the neutral masses is not an input parameter but computed
from the other two neutral input masses in contrast to the R2HDM so
that the same coverage of the parameter space would require a
dedicated scan in specific regions, {\it cf.}~also the remark made
above on the comparison of the coupling ranges of the R2HDM and
C2HDM. Qualitatively, a larger scan would not change the overall
picture, however. \s

From the plots we can infer from the steep rise of the cross
section values, once $m_{\uparrow}\ge 2 m_{H_\text{SM}}$, the resonant enhancement of
the cross sections. In the case of the R2HDM this is due to
resonant $H_2$ production and for the C2HDM, N2HDM or NMSSM this can
be resonant $H_2$ and/or $H_3$ production. For the latter models, in the case of
$H_{\text{SM}} \equiv H_3$, and for the R2HDM in the case of $
H_{\text{SM}} \equiv H_2$ we can only have non-resonant
di-Higgs production. We can also see that the cross section values can be
suppressed compared to the SM case which is given by the horizontal
line in the plots.
The R2HDM and C2HDM cross sections approach the SM-like cross section
for large mass
values (red points) whereas this is not the case for the N2HDM with
$H_1 \equiv H_{\text{SM}}$. In the R2HDM, we approach the decoupling limit
when $m_{H_2}$ becomes large. For large $H_2$ mass the
production cross section decreases, and furthermore with increasing
$m_{H_2}$ the trilinear coupling $\lambda_{H_2 H_1 H_1}$ that is relevant for
resonant $H_1 H_1$ production goes to zero as it is proportional to
$\cos(\beta-\alpha)$ which approaches zero in the decoupling limit. In the N2HDM,
for $H_1 \equiv H_{\text{SM}}$, we can have resonant enhancement both
from $H_2$ and $H_3$ production so that although $m_{H_3} = m_{\uparrow}$
grows in Fig.~\ref{fig:scattersm1} (middle) large cross sections are
still possible due to resonant $H_2$ production. For $H_2
\equiv H_{\text{SM}}$ (green), on the other hand, we also see the decoupling
behavior. Note that for the C2HDM-I we 
observe the decoupling behaviour for $H_1 \equiv H_{\text{SM}}$ in
contrast to the N2HDM. In the C2HDM $m_{\downarrow}$ and
$m_{\uparrow}$ become nearly degenerate for large
non-SM-like masses so that both the $H_2$ and the $H_3$ resonant
contributions are small. Also, for lower masses $m_{\uparrow}$ the
masses of $H_{\downarrow}$ and $H_{\uparrow}$ do not differ much so that
for $H_2 \equiv H_{\text{SM}}$  (green points) the
resonant enhancement is not efficient in increasing the cross
section. \s

%%%%%%%%%%%%%%%%%%%%%%%%%%%%%%%%%%%%%%%%%%%%%%%%%%%%%%%

In Tab.~\ref{tab:compnonres} (left), we compare the maximum cross section values in
the different models for all parameter scenarios where $m_{H_i\ne
  H_{\text{SM}}} < 2 m_{H_{\text{SM}}}$, in case the SM-like
Higgs boson is not given by the heaviest neutral one. For these
scenarios resonance enhancement is kinematically forbidden. We have
included the NLO QCD correction in the large top-mass limit. In the
R2HDM-II and C2HDM-II for $H_{\text{SM}} \equiv H_1$, in the N2HDM-II
and NMSSM for $H_{\text{SM}} \equiv H_{1,2}$ we did not
find any points where $m_{H_i\ne H_{\text{SM}}} < 2
m_{H_{\text{SM}}}$, because of the constraints. Furthermore, in the
C2HDM-II the application of the experimental constraints excludes scenarios with
$H_{\text{SM}} \equiv H_2$ or $H_3$, and in the N2HDM-II and the NMSSM for
$H_{\text{SM}} \equiv H_3$. In the table, we put dashes for all these
cases.
We observe that the non-resonant cross sections for all models (where
they are available) are above the SM value which is given by
$\sigma_{\text{SM}} = 38$~fb.\footnote{This value has been obtained with {\tt
  HPAIR} in the heavy top-quark limit which we apply in the scans. In
this limit the $K$-factor, {\it i.e.}~the ratio between NLO and LO
cross section, amounts to 1.93.}
The largest value is obtained for the R2HDM-I with $H_1 \equiv
H_{\text{SM}}$. The enhancements are due to a combination of Yukawa
and self-coupling values deviating from the SM.
Altogether the cross sections deviate by not more 
than a factor 1.5 from the SM value in the defined non-resonant regions. \s

%%%%%%%%%%%%%%%%%%%%%%%%%%%%%%%%%%%%%%%%%%%%%
%
%   NOT UPDATED FROM HERE REVIEWD
%
%%%%%%%%%%%%%%%%%%%%%%%%%%%%%%%%%%%%%%%%%%%%%

\begin{table}
\begin{center}
\begin{tabular}{l|ccc}
& $H_1$ & $H_2$ & $H_3$ \\ \hline
R2HDM-I & 59 & 49 &  \\
R2HDM-II & -- & -- \\ \hline
C2HDM-I &46 & 44 & 42\\
C2HDM-II & -- & -- & -- \\ \hline
N2HDM-I & 50 & 52 & 44 \\
N2HDM-II & -- &-- &-- \\ \hline
NMSSM & -- & -- & --
\end{tabular}
\hspace*{2cm}
\begin{tabular}{l|cc}
& $H_1$ & $H_2$ \\ \hline
R2HDM-I & 92 & --  \\
R2HDM-II &  59 & -- \\ \hline
C2HDM-I & 98 & 42 \\
C2HDM-II & 75 & -- \\ \hline
N2HDM-I & 151 & 96 \\
N2HDM-II & 112 & 48 \\ \hline
NMSSM & 73 & 65
\end{tabular}
%
%\begin{tabular}{l|ccc}
%& $H_1$ & $H_2$ & $H_3$ \\ \hline
%R2HDM-I & 93 & 50 & n.a.  \\
%R2HDM-II &  61 & -- & n.a.\\ \hline
%C2HDM-I & 99 & 43 & 35 \\
%C2HDM-II & 78 & -- & -- \\ \hline
%N2HDM-I & 154 & 98 & 46 \\
%N2HDM-II & 114 & 50 & -- \\ \hline
%NMSSM & 72 & 67 & --
%\end{tabular}
\caption{Maximum NLO QCD gluon
  fusion cross section values in fb for
  the case where $H_{\text{SM}} \equiv H_1$,
  $H_2$, and $H_3$, respectively, in the R2HDM-I/II, C2HDM-I/II,
  N2HDM-I and NMSSM. Left:
  scenarios with $m_{H_i\ne H_{\text{SM}}} < 2 m_{H_{\text{SM}}}$;
  right: scenarios with suppressed resonance contributions ({\it cf.}~text).
\label{tab:compnonres}}
\end{center}
% R2HDM-I 3696 (H1) kappa_lambda=0.83, sigma(Ah)=58 fb / 930 (H2) kappa_lambda=1.00 ;
% mH1=68 GeV, sigma(hH)=68 fb
% R2HDM-II 1 (H2) mH1=78 GeV, kappa_lambda=0.999288
% C2HDM-I 52 (H1) mH2=168 GeV, kappa_lambda=0.94 (did not change)
% C2HDM-I 143 (H2) mH1=119, mH3=161 kappa_lambda=0.976 (did not change)
% C2HDM-I 2001 (H3) mH1=85, mH2=104, tgbeta=16.1, sigma(H1H1)=27 fb,
% kappa_lambda=1.04378
% N2HDM-I 9629 (H1) mH2=199, mH3=217, kappa_lambda=0.917 (did not change)
% N2HDM-I 469 (H2) mH1=69.74, mH3=244.11, kappa_lambda=0.874,
% sigma(H1H1)=82 fb (did not change)
% N2HDM-I 131 (3) mH1=98.56, mH2=110.67, kappa_lambda=0.983 (did not
% change)
% N2HDM-II 8998 (H1) mH2=177, mH3=190, kappa_lambda=0.856 (did not
% change)
% N2HDM-II 53 (H2) mH1=79, mH3=176, kappa_lambda=0.868,
% sigma(H1H1)=330 fb (did not change)
% N2HDM-II 16 (H3) mH1=48.79, mH3=102.14, BR(H2->H1H1)=0.95, BR(H3->H1H1)=0.09,
% sigma(H1H1)=26 pb, sigma(H2)=28 pb, sigma(H3)=48 pb,
% sigma(H1H2)=80fb, sigma(H1H3)=207 fb, sigma(AH3)=56fb,
% sigma(H2H3)=56 fb, kappa_lambda=0.916 (did not change)
\end{table}
%
%

% R2HDM-I (H1) 8736 kappa_lambda=-0.0956, mH=689 GeV, BR(H->hh)=0.14,
% sigma(H)=0.8fb Why so small?
% R2HDM-II (H1) 8742 kappa_lambda=0.992, mH=1131, BR(H->hh)=0.028,
% sigma(H)=88 fb
% C2HDM-I (H1) 76 mH2=621, mH3=658, BR(H2->H1H1)=0.17,
% BR(H3->H1H1)=0.02, sigma(H2)=4.6fb, sigma(H3)=25 fb,
% kappa_lambda=0.595
% C2HDM-I 6 mH1=121.68, mH3=260, kappa_lambda=0.9777
% C2HDM-II (H1) mH2=1184, mH3=1200, tgbeta=0.9, BR(H2->H1H1)=0.005,
% BR(H3->H1H1)=0.048, sigma(H2)=53 fb, sigma(H3)=58 fb,
% kappa_lambda=0.513
% N2HDM-I (H1) 1802 mH2=1261, mH3=1665, kappa_lambda=-1.16
% N2HDM-I (H2) 1578 mH1=111.74, mH3=698, kappa_lambda=0.138,
% sigma(H1H1)=148 fb
% N2HDM-II (H1) 1147 mH2=1655, mH3=2147, kappa_lambda=-0.608
% N2HDM-II (H2) 44 mH1=85.97, mH3=638, kappa_lambda=0.882
% NMSSM 6717 (H1) mH2=653, mH3=992, kappa_lambda=0.29
% NMSSM 11642 (H2) mH1=80.97, mH3=658.96, kappa_lambda=0.333

It can also be that, although kinematically possible, the di-Higgs
production cross section is not much enhanced compared to the SM value
through resonance production. This is the case for suppressed
Yukawa and/or trilinear coupling values of the $s$-channel
exchanged heavy Higgs boson or because it is very heavy or
  its total width is large. Also
destructive interferences can lead to a suppression. 
Therefore there is no clear
  correlation between the computed resonantly enhanced cross sections
  and the Higgs mass values of the resonantly produced Higgs boson as
  can also be inferred from Figs.~\ref{fig:scattersm1} and
  \ref{fig:scattersm2}. Only the obtained maximum values 
  show the correlation with the mass value, namely that these maximum
  possible values decrease with increasing mass of the non-SM-like heavy Higgs as
  it is expected in models with a decoupling limit.
In the suppressed
case, from an experimental point of view, these scenarios would be interpreted
as non-resonant di-Higgs production.
As discussed above, the transition between resonant and non-resonant
cases is not trivial. In accordance with our choice of separation given
above, we also give the maximum cross section values for those scenarios
where the resonance contribution makes up for less than 10\% of the total
di-Higgs cross section.\footnote{The values are obtained from
  the plateau region, {\it cf.}~the horizontal branches in
  Figs.~\ref{fig:continuum}.} The NLO QCD values are summarised in
Tab.~\ref{tab:compnonres} (right) for all models with $H_{\text{SM}}=H_{1,2}$
(only $H_{1}$ for the R2HDM).\footnote{For $H_3\equiv H_{\text{SM}}$
  ($H_2$ in the R2HDM) resonant production is not possible per definition.} As
expected, they exceed the values given in
Tab.~\ref{tab:compnonres} (left) but by at most a factor 3. \s

\begin{figure}[t!]
  \centering
  \includegraphics[width=0.47\linewidth]{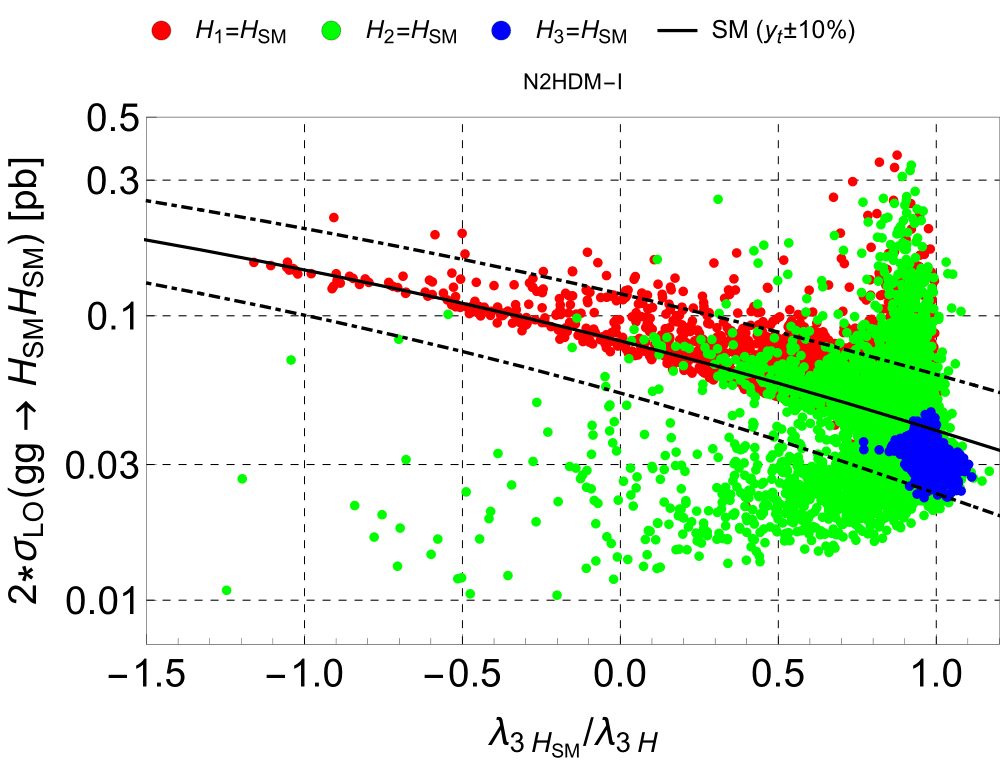}
  \caption{N2HDM-I: SM-like Higgs pair production including a factor 2 to
    roughly account for the NLO QCD corrections, for $H_{1,2,3} \equiv
    H_{\text{SM}}$ (red, green, blue), as a
    function of the trilinear SM-like Higgs self-coupling
    $\lambda_{3H_{\text{SM}}}$
    normalized to the SM value $\lambda_{3H}$. The full line shows
    SM di-Higgs production as a function of
    $\lambda_{3H_{\text{SM}}}/\lambda_{3H}$. The dot-dashed lines show
    the additional change of this cross section for a variation of
    $y_{t,H_{\text{SM}}}/y_{t,H}=1\pm 0.1$.
    }\label{fig:resenhance}
\end{figure}

If not suppressed, the cross sections can be
much more enhanced compared to the SM case in the resonance case.
This can nicely be inferred from
  Fig.~\ref{fig:resenhance} which depicts the cross section values for
SM-like Higgs pair production in the N2HDM-I as a function of the
SM-like trilinear Higgs self-coupling ratio for all possible
$H_{\text{SM}}$ scenarios. The full line in the plot shows the SM
di-Higgs production cross section as a function of the trilinear Higgs
self-coupling variation, the dot-dashed lines additionally depict the
change if we allow for a $\pm 10$\% variation of the top Yukawa
coupling as is still compatible with the data. The comparison with
these lines allows us to estimate the resonant enhancement effects beyond
the cross section change due to modified trilinear Higgs self- and
top-Yukawa-couplings. Close to the SM case
$\lambda_{3H_\text{SM}}/\lambda_{3H}=1$ and $y_{t,H_{\text{SM}}}/y_{t,H}=1$,
we observe enhancements of up to 10 (9) times the
SM expectation for
$H_1 (H_2) \equiv H_{\text{SM}}$.The enhancement
stems from resonant heavy Higgs production which can be $H_2$ and/or
$H_3$ (only $H_3$) for $H_1 \equiv H_{\text{SM}}$ ($H_2 \equiv
H_{\text{SM}}$). For $H_3 \equiv H_{\text{SM}}$, no such
enhancement is possible and the (small) deviations from the SM case
(in the vinicity of
$\lambda_{3H_\text{SM}}^{\text{N2HDM}}/\lambda_{3H}=1$)
are due to the small deviations of the involved couplings from
the SM values or due to interference effects with nearby Higgs
bosons. In Tab.~\ref{tab:compres}, we summarize
the maximum cross section values in each model for the different
SM-like Higgs configurations and for the parameter sets where
resonant production is kinematically possible. The values in the tables are to be
compared to the corresponding NLO QCD SM value (in the heavy top-limit) of 38~fb.
\begin{table}
\begin{center}
\begin{tabular}{l|cc}
& $H_1$ & $H_2$   \\ \hline
R2HDM-I & 444 & n.a.   \\
R2HDM-II &  81 & n.a. \\ \hline
C2HDM-I & 387 & 47 \\
C2HDM-II & 130 & --   \\ \hline
N2HDM-I & 376 & 344\\
N2HDM-II & 188 & 63 \\ \hline
NMSSM & 183 &  65
\end{tabular}
\caption{Maximum NLO QCD gluon fusion cross section values for
  resonant SM-like Higgs pair 
  production in fb for the case where $H_{\text{SM}} \equiv H_1$,
  and $H_2$, respectively, in the various considered models. In the R2HDM the case
$H_{\text{SM}}= H_2$ is not applicable (n.a.) as we cannot have
resonant $H_2H_2$ production. \label{tab:compres}}
\end{center}
\end{table}
The largest cross section with about 444~fb is obtained in the R2HDM-I,
followed by the C2HDM-I with a value
of 387~fb and the N2HDM-I with 376~fb, all for $H_1 \equiv
H_{\text{SM}}$. The resonant BSM cross sections
largely exceed the SM value
by up to a factor 12. The clear hierarchy in the
cross sections gives us a handle to at least partially distinguish the
models by investigating solely Higgs pair production. Clearly cross
section values above 400 fb would exclude all
investigated models but the R2HDM-I with the lightest neutral Higgs
boson being the SM-like one. A value above 100~fb on the other
hand would exclude the R2HDM-II with $H_1 \equiv H_{\text{SM}}$, the
C2HDM-I with $H_2 \equiv H_{\text{SM}}$, the 
N2HDM-II with $H_2 \equiv H_{\text{SM}}$ and
the NMSSM with $H_2 \equiv H_{\text{SM}}$.
The numbers have to be taken
with some care, however, as a more complete scan could find additional
values. They should rather be taken as a rough guideline.
Note furthermore, that in the NMSSM with $H_2 \equiv
H_{\text{SM}}$ the resonant production makes up only a small contribution of
the full cross section meaning that it is the
same point as the one quoted in Tab.~\ref{tab:compnonres} (right).
The enhancement of the cross section is rather due to
coupling deviations of the SM-like trilinear Higgs coupling, while the
$H_2$ top-Yukawa coupling is close to the SM value, and from an
experimental point of view this would look like non-resonant di-Higgs production.  \s

For completion, we include in Appendix \ref{sec:prodsec} separately
  the information on resonant and 
non-resonant cross section values corresponding to
Figs.~\ref{fig:scattersm1} and \ref{fig:scattersm2}. In the transition
region, where resonant and non-resonant cross sections are similar in
size, they should be taken, however, with a grain of salt.

%%%%%%%%%%%%%%%%%%%%%%%%%%%%%%%%%%%%%%%%%%%%%%%%%%%%%%%%%%%%
\section{Benchmarks for SM-Like Higgs Pair Production \label{sec:benchmarks}}
In the following, we give the input parameters and basic
features of all scenarios listed in
Tab.~\ref{tab:compres}. The various models and mass hierarchies
have different interesting features: large di-Higgs
cross sections, large rates for di-Higgs associated production
with gauge bosons but also large triple Higgs production rates from
Higgs-to-Higgs cascade decays that are only possible in non-minimal
Higgs sectors. Additionally, in the C2HDM, we have the possibility to test CP
violation through the measurement of Higgs-to-Higgs or Higgs-to-gauge
plus Higgs boson decays. These features will be specified
for the various benchmarks. We will also give the
dominant decay channels of the non-SM-like Higgs bosons. Due to space
limitations, we cannot give all decay channels.
They can be obtained, however, from the publicly
available programs {\tt HDECAY}
\cite{Djouadi:1997yw,Djouadi:2018xqq,Djouadi:2006bz}, {\tt C2HDM\_HDECAY}
\cite{Fontes:2017zfn}, {\tt N2HDECAY}
\cite{Muhlleitner:2016mzt,Engeln:2018mbg}, and {\tt NMSSMCALC}
\cite{Baglio:2013iia} by using the input values that we give for each
benchmark point. 
The benchmark points comply with the present and past Higgs searches
as explicitly checked by applying the code {\tt HiggsBounds} where we
made sure to take into account the latest search limits. Note in this context that
light Higgs bosons still escape detection due to suppressed couplings.
Moreover, for some light mass regions there are no experimental analyses
available yet. This emphasizes once again the necessity to investigate the
low BSM Higgs mass region to constrain extended Higgs sectors. Also
the heavy Higgs bosons given in the following benchmarks are escaping
all present exclusion limits with rates lying below the current
experimental sensitivity.
\s

All Higgs pair cross sections have been obtained with our codes based on
{\tt HPAIR} \cite{hpair}, that has been extended to include the various models, at
NLO QCD in the heavy loop particle limit. The single Higgs cross
sections are calculated at  NNLO with the
program {\tt SusHi  v1.6.1}
\cite{Harlander:2012pb,Liebler:2015bka,Harlander:2016hcx}. 
The single and double Higgs production cross sections are given for
$\sqrt{s}=14$~TeV. Further information
on all benchmarks can be provided on request. Please write an email
to the authors of the paper in this case.

\subsection{Resonant SM-like $H_1 H_1$ Production in the R2HDM-I}
This benchmark scenario provides maximum SM-like $H_1 H_1$ production
through resonant enhancement in the R2HDM-I, and is defined by the
input parameters given in Tab.~\ref{tab:bp1r2hdm1h1} (upper). Further
information is provided in
Tab.~\ref{tab:bp1r2hdm1h1} (lower) where we give the NLO QCD $H_1 H_1$ cross
section value computed with {\tt HPAIR} at $\sqrt{s}=14$~TeV along
with the total widths of the Higgs bosons. We also give the values of
the $K$-factor, {\it i.e.}~the ratio between NLO and LO cross section,
of the SM-like Yukawa and trilinear
Higgs self-couplings normalized to their SM values and we list the single Higgs
cross sections at NNLO QCD obtained from {\tt SusHi} at the same
c.m.~energy.\footnote{All single Higgs production cross sections quoted in this section are computed at
  NNLO QCD for $\sqrt{s}=14$~TeV. }
\begin{table}[h!]
\begin{center}
\begin{tabular}{|c|c|c|c|c|c|c|c|}
\hline
$m_{H_1}$ [GeV] & $m_{H_2}$ [GeV] & $m_A$ [GeV] & $m_{H^\pm}$ [GeV] & $\alpha$ & $\tan\beta$ & $m_{12}^2$ [GeV$^2$] \\ \hline
125.09 & 267 & 512 & 516 & -0.259 & 4.276 & 15020\\ \hline
\end{tabular}

\vspace*{0.3cm}
\begin{tabular}{|c|c|c|c|c|c|}
\hline
$\sigma_{H_1 H_1}^{\text{NLO}}$ [fb] & $K$-factor & $\Gamma^{\text{tot}}_{H_1}$ [GeV] &
$\Gamma^{\text{tot}}_{H_2}$ [GeV] & $\Gamma^{\text{tot}}_{A}$ [GeV]
& $\Gamma^{\text{tot}}_{H^\pm}$ [GeV] \\ \hline
 444 & 2.06 & 4.029
  $\times\,10^{-3}$ & 0.011 & 14.57
   & 16.07
  \\ \hline \hline
$\lambda_{3H_1}/\lambda_{3H}$ & $y_{t,H_1}/y_{t,H}$ & $\sigma_{H_1}^{\text{NNLO}}$ [pb] & $\sigma_{H_2}^{\text{NNLO}}$ [pb] & $\sigma_{A}^{\text{NNLO}}$ [pb] & \\ \hline
0.993 & 0.993 & 48.56 & 0.916 & 0.489 & \\
\hline
\end{tabular}
\caption{\underline{{\tt BP1}} Upper: R2HDM-1 input parameters. Lower:
  NLO QCD $H_1H_1$ pair production cross section
  in gluon fusion in the heavy loop particle limit, $K$-factor, the total
  widths of the Higgs bosons, the SM-like ($H_1$) Yukawa and trilinear Higgs couplings
  normalized to their SM values and the single Higgs production cross
 sections at NNLO QCD for $H_{1,2}$ and $A$.
% Point number 3285
\label{tab:bp1r2hdm1h1}}
\end{center}
\vspace*{-0.6cm}
\end{table}
Besides a large SM-like Higgs pair production cross section, the
benchmark scenario features a large rate for SM-like Higgs pair
production in association with a $Z$ boson.
The dominant branching ratios of $H_2$, $A$, and $H^\pm$ are given by
\beq
\begin{array}{lcllcllcl}
\mbox{BR} (H_2 \to H_1 H_1) &=& 0.544 \;, & \;
\mbox{BR} (H_2 \to W W) &=& 0.280 \;, & \;
\mbox{BR} (H_2 \to Z Z) &=& 0.121 \;, \\[0.1cm]
\mbox{BR} (A \to ZH_2) &=& 0.912 \;, & \;
\mbox{BR} (A \to t\bar{t}) &=& 0.086 \;, & \\[0.1cm]
\mbox{BR} (H^\pm \to W^\pm H_2) &=& 0.922
\;, & \; \mbox{BR} (H^+ \to t\bar{b}) &=& 0.076\;.
\end{array}
\eeq
With the values given we see that indeed $H_2$ resonant production is
responsible for the enhancement of the cross section as
\beq
\sigma_{H_2}^{\text{NNLO}} \times \mbox{BR} (H_2 \to H_1 H_1) =
498~\mbox{fb} \;.
\eeq
The fact that the {\tt HPAIR} result for $H_1 H_1$
  production is somewhat lower is due to the finite width of $H_2$
  that is taken into account in the {\tt HPAIR} calculation when we
  integrate across the $H_2$ resonance in the $s$-channel.
We find that for associated $H_1 H_1$ production with a $Z$ boson we have
\beq
\sigma_A^{\text{NNLO}} \times \mbox{ BR}(A \to ZH_2) \times \mbox{ BR}(H_2 \to H_1
H_1) = 0.489 \mbox{ pb} \times 0.912 \times 0.544 = 247 \mbox{ fb} \;.
\eeq
This leads to a $Z+(4b)$ final state with a signal rate of 84~fb so that this discovery
channel for $A$ exceeds $A$ production in the $t\bar{t}$
final state which amounts to 43~fb.

\subsection{Resonant SM-like $H_1 H_1$ Production in the R2HDM-II}
This is the benchmark scenario for maximum SM-like $H_1 H_1$ production
through resonant enhancement in the R2HDM-II, with the
input parameters given in Tab.~\ref{tab:bp2r2hdm2h1} (upper) and Higgs
pair and single Higgs production information in
Tab.~\ref{tab:bp2r2hdm2h1} (lower).  As we
are in the type 2 model, the overall Higgs spectrum is heavier than {\tt BP1}.
\begin{table}[h!]
\begin{center}
\begin{tabular}{|c|c|c|c|c|c|c|c|}
\hline
$m_{H_1}$ [GeV] & $m_{H_2}$ [GeV] & $m_A$ [GeV] & $m_{H^\pm}$ [GeV] & $\alpha$ & $\tan\beta$ & $m_{12}^2$ [GeV$^2$] \\ \hline
125.09 & 528 & 798 & 809 & -0.695 & 1.268 & 130388\\ \hline
\end{tabular}

\vspace*{0.3cm}
\begin{tabular}{|c|c|c|c|c|c|}
\hline
$\sigma_{H_1 H_1}^{\text{NLO}}$ [fb] & $K$-factor & $\Gamma^{\text{tot}}_{H_1}$ [GeV] &
$\Gamma^{\text{tot}}_{H_2}$ [GeV] & $\Gamma^{\text{tot}}_{A}$ [GeV]
& $\Gamma^{\text{tot}}_{H^\pm}$ [GeV]  \\ \hline
 81 & 1.94 & 4.239 $\times\,10^{-3}$ & 10.15 & 47.82 & 51.56
\\ \hline \hline
$\lambda_{3H_1}/\lambda_{3H}$ & $y_{t,H_1}/y_{t,H}$ &
$\sigma_{H_1}^{\text{NNLO}}$ [pb] & $\sigma_{H_2}^{\text{NNLO}}$ [pb]
& $\sigma_{A}^{\text{NNLO}}$ [pb] &\\ \hline
0.974 & 0.978 & 47.02 & 2.84 &
0.47 &\\ \hline
\end{tabular}
\caption{\underline{{\tt BP2}} Upper: R2HDM-II input
  parameters. Lower: Additional double and single Higgs production
  related information.
% Point number 9791
\label{tab:bp2r2hdm2h1}}
\end{center}
\vspace*{-0.8cm}
\end{table}
Here we also have, apart from a large SM-like Higgs pair production
cross section, SM-like Higgs pair production in
association with a $Z$ boson, though at a much lower rate. The dominant
$H_2$, $A$, and $H^\pm$ branching ratios are given by
\beq
\begin{array}{lcllcllcl}
\mbox{BR} (H_2 \to H_1 H_1) &=& 0.012 \;, & \;
\mbox{BR} (H_2 \to t\bar{t}) &=& 0.979 \;, \\[0.1cm]
\mbox{BR} (A \to ZH_2) &=& 0.514 \;, & \;
\mbox{BR} (A \to t\bar{t}) &=& 0.482 \;, \\[0.1cm]
\mbox{BR} (H^\pm \to W^\pm H_2) &=& 0.560 \;, & \;
\mbox{BR} (H^+ \to t\bar{b}) &=& 0.437 \;.
\end{array}
\eeq
From the values given we calculate the di-Higgs $H_1 H_1$ cross
section from resonant $H_2$ production,
\beq
\sigma_{H_2}^{\text{NNLO}} \times \mbox{BR} (H_2 \to H_1 H_1) =
34~\mbox{fb} \;.
\eeq
Comparing this with the $H_1 H_1$ Higgs pair production cross section
we see that resonant $H_2$ production is responsible for the enhanced
cross section (as can be seen when taking into account also the
SM-like contribution of 38~fb included in {\tt HPAIR}).\footnote{Note
  that deviations between the full 
  {\tt HPAIR} result and the {\tt SusHi} result are to be expected as
  {\tt HPAIR} takes into account all $s$-channel (including the total
  widths of the $s$-channel particles) and box contributions
and their interferences.} Also the deviations of the $H_1$ trilinear and
Yukawa couplings from the SM values cause a slight enhancement.
For $H_1 H_1$ associated production with a $Z$ boson we find
\beq
\sigma_{\text{prod}} (A) \times \mbox{ BR}(A \to ZH_2) \times \mbox{ BR}(H_2 \to H_1
H_1) = 0.47 \mbox{ pb} \times 0.514 \times 0.012 =
2.9 \mbox{ fb} \;,
\eeq
leading to a $Z+(4b)$ rate of 1.04 fb.

\subsection{Resonant SM-like $H_1 H_1$ Production in the C2HDM-I}
In Tab.~\ref{tab:bpc2hdm1h1} (upper), the input parameters for the scenario
with maximum resonant SM-like Higgs production with $H_1 \equiv
H_{\text{SM}}$ in the C2HDM-I are given. Further Higgs pair and single
Higgs production information is summarized in
Tab.~\ref{tab:bpc2hdm1h1} (lower). Apart
from its large SM di-Higgs cross section this scenario is special as
it allows for the test of CP violation through Higgs decays when decays are combined~\cite{Fontes:2015xva}.
\begin{table}[h!]
\begin{center}
\begin{tabular}{|c|c|c|c|c|c|c|c|}
\hline
$m_{H_1}$ [GeV] & $m_{H_2}$ [GeV] & $m_{H^\pm}$ [GeV] & $\alpha_1$ & $\alpha_2$
  & $\alpha_3$ & $\tan\beta$ & $\mbox{Re}(m_{12}^2)$
 [GeV$^2$] \\ \hline
125.09 & 265& 236 & 1.419 &0.004 & -0.731 & 5.474 &
9929 \\ \hline
\end{tabular}

\vspace*{0.3cm}
\begin{tabular}{|c|c|c|c|c|c|}
\hline
$\sigma_{H_1 H_1}^{\text{NLO}}$ [fb] & $K$-factor & $\Gamma^{\text{tot}}_{H_1}$ [GeV] &
$\Gamma^{\text{tot}}_{H_2}$ [GeV] & $\Gamma^{\text{tot}}_{H_3}$ [GeV]
& $\Gamma^{\text{tot}}_{H^\pm}$ [GeV] \\ \hline
 387 & 2.06 & 4.106
  $\times\,10^{-3}$ & 3.625 $\times\,10^{-3}$ & 4.880 $\times\,10^{-3}$
   & 0.127
\\ \hline \hline
$\lambda_{3H_1}/\lambda_{3H}$ & $y_{t,H_1}^e/y_{t,H}$ &
$\sigma_{H_1}^{\text{NNLO}}$ [pb] & $\sigma_{H_2}^{\text{NNLO}}$ [pb]
& $\sigma_{H_3}^{\text{NNLO}}$ [pb] & \\ \hline
0.995& 1.005 & 49.75 & 0.76 & 0.84 &\\ \hline
\end{tabular}
\caption{\underline{{\tt BP3}} Upper: C2HDM-I input parameters. The
  third Higgs boson mass calculated from the input parameters is given by
  $m_{H_3}=267$~GeV. Lower: Additional double and single Higgs production
  related information. The value for $y_{t,H_1}^e$ is the CP-even part
  of the Yukawa coupling. The CP-odd part for the SM-like Higgs is
  tiny.
\label{tab:bpc2hdm1h1}}
% This is point 88.
\end{center}
\vspace*{-0.2cm}
\end{table}
The dominant branching ratios of the non-SM-like Higgs bosons are
\beq
% \begin{tabular}{ll}
% \st{BR($H_2 \to H_1 H_1$) = 0.685}\textcolor{blue}{BR($H_2 \to H_1 H_1$) = 0.703} \;,&  BR ($H_2 \to WW$) = 0.174 \;, \\
% BR ($H_2 \to ZZ$) = 0.075 \;, BR ($H_2 \to ZH_1$) = 0.046 \;, \\
% \st{BR(H$H_3 \to H_1 H_1$)=0.703}\textcolor{blue}{BR($H_3 \to H_1 H_1$)=0.684} \;,& BR ($H_3 \to WW$) = 0.166 \;, \\
% BR ($H_3 \to ZZ$) = 0.072 \;, BR ($H_3 \to ZH_1$) = 0.075 \;, \\
% BR($H^\pm \to tb$) = 0.644 \;, \st{BR($H^\pm \to W^\pm H_1$)=0.359}\textcolor{blue}{BR($H^\pm \to W^\pm H_1$)=0.354} \;.
% \end{tabular}
\begin{tabular}{ll}
BR($H_2 \to H_1 H_1$) = 0.252 \;,&  BR ($H_2 \to WW$) = 0.335 \;, \\
BR ($H_2 \to ZZ$) = 0.144 \;,& BR ($H_2 \to ZH_1$) = 0.161 \;, \\
BR($H_3 \to H_1 H_1$) = 0.280 \;,& BR ($H_3 \to WW$) = 0.376 \;, \\
BR ($H_3 \to ZZ$) = 0.162 \;,& BR ($H_3 \to ZH_1$) = 0.090 \;, \\
BR($H^\pm \to tb$) = 0.995 \;.&
\end{tabular}
\eeq
With the single Higgs production cross sections given in
Tab.~\ref{tab:bpc2hdm1h1} (lower) the large cross section is due to
resonant enhancement from both $H_2$ and $H_3$,
\beq
\sigma_{H_2}^{\text{NNLO}} \times \mbox{BR} (H_2 \to H_1 H_1) &=&
 191~\mbox{fb} \nonumber \\
\sigma_{H_3}^{\text{NNLO}} \times \mbox{BR} (H_3 \to H_1 H_1) &=&
 235~\mbox{fb} \;,
\eeq
and we arrive at the following Higgs-to-Higgs, Higgs-to-gauge bosons and
Higgs-to-gauge+Higgs rates
\beq
\begin{array}{lcllcl}
\sigma(H_2) \times \mbox{BR}(H_2 \to H_1H_1) &=& 191 \mbox{ fb} \;,
  & \;
\sigma(H_2) \times \mbox{BR}(H_2 \to WW) &=& 254 \mbox{ fb} \;, \\
\sigma(H_2) \times \mbox{BR}(H_2 \to ZZ) &=& 109 \mbox{ fb} \;, & \;
\sigma(H_2) \times \mbox{BR}(H_2 \to ZH_1) &=& 122 \mbox{ fb} \;,  \\[0.1cm]
\sigma(H_3) \times \mbox{BR}(H_3 \to H_1H_1) &=& 235 \mbox{ fb} \;,
  & \;
\sigma(H_3) \times \mbox{BR}(H_3 \to WW) &=& 315 \mbox{ fb} \;, \\
\sigma(H_3) \times \mbox{BR}(H_3 \to ZZ) &=& 136 \mbox{ fb} \;, & \;
\sigma(H_3) \times \mbox{BR}(H_3 \to ZH_1) &=& 76 \mbox{ fb} \;.
\end{array}
\eeq
These large rates allow for the test of CP violation through Higgs
decays. The decays of $H_{2/3}$ each into $WW/ZZ$ and the SM-like Higgs
boson pair $H_1 H_1$ assuming $H_1$ is CP-even\footnote{Note that although it is by now clear that
the SM-like Higgs cannot be a pure CP-odd state, we are far from excluding large CP-odd components in its Yukawa couplings.
In fact, there are so far only direct measurements of the $\bar tt
H_{\text{SM}}$ and $\tau^+ \tau^- H_{\text{SM}}$ couplings.
Both ATLAS and CMS~\cite{ATLAS:2020ior, CMS:2020cga} were able to exclude the purely CP-odd
hypothesis in the process  $pp \to \bar t t (H_{\text{SM}} \to \gamma \gamma)$ with 3.9 standard deviations and to establish a 95\% CL observed (expected)
exclusion upper limit for the mixing angle of $43^o$ ($63^o$). Recently CMS~\cite{CMS:2021sdq} has performed the
first measurement of the CP mixing angle of the tau lepton Yukawa coupling, using 13 TeV data, and an integrated luminosity of 137 fb$^{-1}$
The CP mixing angle was found to be $4^o \pm 17^o$, allowing to set an observed (expected) exclusion upper limit for the
mixing angle of $36^o$ ($55^o$). This angle is defined as $\arctan(b/a)$, if the generic Yukawa coupling is written as $a+ib\gamma_5$.}
attribute the CP quantum number $+1$ to $H_{2/3}$. On the other hand,
the decays into $ZH_1$ of $H_{2/3}$ identify them to be CP-odd. The
simultaneous measurement of all these decays with substantial rates
would clearly identify $H_{2/3}$ to be states with mixed CP quantum
numbers. Note that BR($H_1 \to b\bar{b}$) = 0.592.

\subsection{Resonant SM-like $H_2 H_2$ Production in the C2HDM-I}
Information on this benchmark point is gathered in
Tab.~\ref{tab:bpc2hdm1h2}. It
features an overall light Higgs mass spectrum. In contrast to the
previous C2HDM scenario, the corresponding rates are too small to allow
for the test of CP violation through Higgs decays. This is also the
case for the other C2HDM scenario presented in the following.
\begin{table}[h!]
\begin{center}
\begin{tabular}{|c|c|c|c|c|c|c|c|}
\hline
$m_{H_1}$ [GeV] & $m_{H_2}$ [GeV] & $m_{H^\pm}$ [GeV] & $\alpha_1$ & $\alpha_2$
  & $\alpha_3$ & $\tan\beta$ & $\mbox{Re}(m_{12}^2)$
 [GeV$^2$] \\ \hline
74 & 125.09 & 347 & -0.308 & -1.328 & -0.434 & 10.69 & 9758 \\ \hline
\end{tabular}

\vspace*{0.3cm}
\begin{tabular}{|c|c|c|c|c|c|}
\hline
$\sigma_{H_2 H_2}^{\text{NLO}}$ [fb] & $K$-factor& $\Gamma^{\text{tot}}_{H_1}$ [GeV] &
$\Gamma^{\text{tot}}_{H_2}$ [GeV] & $\Gamma^{\text{tot}}_{H_3}$ [GeV]
& $\Gamma^{\text{tot}}_{H^\pm}$ [GeV] \\ \hline
 47 & 1.95 & 2.662 $\times\,10^{-5}$ & 3.990 $\times\,10^{-3}$ & 9.22 & 10.33
\\ \hline \hline
$\lambda_{3H_2}/\lambda_{3H}$ & $y_{t,H_2}^e/y_{t,H}$ &
$\sigma_{H_1}^{\text{NNLO}}$ [pb] & $\sigma_{H_2}^{\text{NNLO}}$ [pb]
& $\sigma_{H_3}^{\text{NNLO}}$ [pb] & \\ \hline
1.127 & 0.993 & 3.37 & 48.56 &
0.22 &\\ \hline
\end{tabular}
\caption{\underline{{\tt BP4}} Upper: C2HDM-I input parameters. The
  third Higgs boson mass calculated from the input parameters is given by
  $m_{H_3}=338$~GeV. Lower: Additional double and single Higgs production
  related information. The value for $y_{t,H_2}^e$ is the CP-even part
  of the Yukawa coupling. The CP-odd part for the SM-like Higgs is
  tiny.
\label{tab:bpc2hdm1h2}}
% This is point 91.
\end{center}
\vspace*{-0.6cm}
\end{table}
The dominant branching ratios of the non-SM-like Higgs bosons are
given by

\vspace*{0.2cm}
\begin{tabular}{lll}
BR($H_1 \to b\bar{b}$)=0.797\,,& BR ($H_3 \to ZH_1$) = 0.865\,, & BR($H_3\to H_2 H_2$) = 0.047\,,\\
BR($H_3 \to WW$) = 0.048\,, & 
BR ($H^\pm \to W^\pm H_1$) = 0.954\,.
\end{tabular}

\vspace*{0.2cm}
\noindent
In this scenario, the enhancement of the cross section is
due to resonant $H_3$ production. We have
\beq
\sigma_{H_3}^{\text{NNLO}}  \times \mbox{BR} (H_3 \to H_2 H_2) =
10.34 \mbox{ fb} \;.
\eeq

\subsection{Resonant SM-like $H_1 H_1$ Production in the C2HDM-II}
For the C2HDM-II the maximum resonant production of a SM-like
Higgs pair with $H_{\text{SM}} \equiv H_1$ is given by {\tt BP5} with
the input parameters defined in Tab.~\ref{tab:bpc2hdm2h1} (upper) and
additional information related to double and single Higgs production
in Tab.~\ref{tab:bpc2hdm2h1} (lower).
\begin{table}[h!]
\begin{center}
\begin{tabular}{|c|c|c|c|c|c|c|c|}
\hline
$m_{H_1}$ [GeV] & $m_{H_2}$ [GeV] & $m_{H^\pm}$ [GeV] & $\alpha_1$ & $\alpha_2$
  & $\alpha_3$ & $\tan\beta$ & $\mbox{Re}(m_{12}^2)$
 [GeV$^2$] \\ \hline
125.09 & 743 & 820 & 0.717 & 0.096 & 1.151 & 0.924 & 206750
\\ \hline
\end{tabular}

\vspace*{0.3cm}
\begin{tabular}{|c|c|c|c|c|c|}
\hline
$\sigma_{H_1 H_1}^{\text{NLO}}$ [fb] & $K$-factor & $\Gamma^{\text{tot}}_{H_1}$ [GeV] &
$\Gamma^{\text{tot}}_{H_2}$ [GeV] & $\Gamma^{\text{tot}}_{H_3}$ [GeV]
& $\Gamma^{\text{tot}}_{H^\pm}$ [GeV] \\ \hline
 130 & 1.88 & 4.179
  $\times\,10^{-3}$ & 40.26  & 41.54 & 44.66
\\ \hline \hline
$\lambda_{3H_1}/\lambda_{3H}$ & $y_{t,H_1}^e/y_{t,H}$ &
$\sigma_{H_1}^{\text{NNLO}}$ [pb] & $\sigma_{H_2}^{\text{NNLO}}$ [pb]
& $\sigma_{H_3}^{\text{NNLO}}$ [pb] & \\ \hline
0.494 & 0.964 & 46.92 & 1.20 & 1.03 &\\ \hline
\end{tabular}
\caption{\underline{{\tt BP5}} Upper: C2HDM-II input parameters. The
  third Higgs boson mass calculated from the input parameters is given by
  $m_{H_3}=753$~GeV. Lower: Additional double and single Higgs production
  related information. The value for $y_{t,H_1}^e$ is the CP-even part
  of the Yukawa coupling. The CP-odd part for the SM-like Higgs is
  very small.}
\label{tab:bpc2hdm2h1}
% This is point 134.
\end{center}
\vspace*{-0.6cm}
\end{table}
Overall, the Higgs spectrum is rather heavy as expected in type-2
models. The dominant branching ratios of the non-SM-like Higgs bosons are
given by
\beq
\mbox{BR} (H_2 \to t\bar{t})&=& 0.928\,, \quad
\mbox{BR} (H_3 \to t\bar{t}) = 0.939\,, \,, \nonumber \\
\mbox{BR} (H^\pm \to tb) &=& 0.961\,.
\eeq
The dominant contribution to resonant production stems from
$H_2$. More specifically, we have
\beq
\sigma (H_2) \times \mbox{ BR} (H_2 \to H_1 H_1) &=&
1.20 \mbox{ pb}
\times 0.031 = 37 \mbox{ fb} \\
\sigma (H_3) \times \mbox{ BR} (H_3 \to H_1 H_1) &=&
1.03 \mbox{ pb}
\times 0.022= 23 \mbox{ fb} \;.
\eeq
The sum of the two resonant contributions makes up for about half of
the total cross section. The remaining part is given by non-resonant
production which is enhanced compared to the SM case because the
trilinear coupling between three SM-like Higgs bosons deviates from
the SM case so that the destructive interference between box and
triangle diagrams is not effective.

\subsection{Resonant SM-like $H_1 H_1$ Production  in
    the N2HDM-I \label{sec:bp6}}
The input parameters for the scenario {\tt BP6} with maximum resonant SM-like
Higgs production with $H^{\text{SM}} \equiv H_1$ in the N2HDM-I are
given in Tab.~\ref{tab:bpn2hdm1h1} (upper) with additional information
related to double and single Higgs production in
Tab.~\ref{tab:bpn2hdm1h1} (lower). For this N2HDM point it is
basically the resonant contribution of both $H_2$ that leads to
the enhanced $H_1H_1$ production with $H_2$ being rather light, namely
$m_{H_2}=269$~GeV. Additionally we have a large rate for Higgs pair
production in association with a $Z$ boson, and we can
produce three SM-like Higgs bosons at a rate that might be accessible at a
high-luminosity collider.
\begin{table}[t!]
\begin{center}
\begin{tabular}{|c|c|c|c|c|c|}
\hline
$m_{H_1}$ [GeV] & $m_{H_2}$ [GeV] & $m_{H_3}$ [GeV] & $m_{A}$ [GeV]
  & $m_{H^\pm}$ [GeV] & $\tan\beta$ \\ \hline
125.09 & 269 & 582 & 390 & 380 & 4.190 \\ \hline \hline
$\alpha_1$ & $\alpha_2$ & $\alpha_3$ & $v_s$ [GeV] &
$\mbox{Re}(m_{12}^2)$ [GeV$^2$] & \\ \hline
1.432 & -0.109 & 0.535 & 1250 & 28112 & \\ \hline
\end{tabular}

\vspace*{0.3cm}
\begin{tabular}{|c|c|c|c|c|c|c|}
\hline
$\sigma_{H_1 H_1}^{\text{NLO}}$ [fb] & $K$-factor & $\Gamma^{\text{tot}}_{H_1}$ [GeV] &
$\Gamma^{\text{tot}}_{H_2}$ [GeV] & $\Gamma^{\text{tot}}_{H_3}$ [GeV] & $\Gamma^{\text{tot}}_{A}$ [GeV]
& $\Gamma^{\text{tot}}_{H^\pm}$ [GeV] \\ \hline
 376 & 2.05 & 4.130 $\times\,10^{-3}$ & 0.0752 & 15.279 & 1.483 & 1.477
\\ \hline \hline
$\lambda_{3H_1}/\lambda_{3H}$ & $y_{t,H_1}/y_{t,H}$ &
$\sigma_{H_1}^{\text{NNLO}}$ [pb] & $\sigma_{H_2}^{\text{NNLO}}$ [pb]
& $\sigma_{H_3}^{\text{NNLO}}$ [pb] & $\sigma_{A}^{\text{NNLO}}$ &\\ \hline
0.876 & 1.012 & 50.47 & 0.42 & 0.002 & 2.16 &\\ \hline
\end{tabular}
\caption{\underline{{\tt BP6}} Upper: N2HDM-I input parameters. Lower:
  Additional double and single Higgs production
  related information.}
\label{tab:bpn2hdm1h1}
% This is point 543.
\end{center}
\vspace*{-0.8cm}
\end{table}
The dominant branching ratios of the non-SM-like Higgs bosons into
observable final states are
\beq
\mbox{BR} (H_2\to H_1 H_1) &=& 0.946 \;, \;
\mbox{BR} (H_2\to WW) = 0.035 \;, \;
\mbox{BR} (H_2\to ZZ) = 0.015 \;, \nonumber \\
\mbox{BR} (H_3\to H_2 H_2) &=& 0.314 \;, \;
\mbox{BR} (H_3\to W^+ H^-) = 0.299 \;, \;
\mbox{BR} (H_3\to Z A) = 0.117 \;, \nonumber \\
\mbox{BR} (A\to tt) &=& 0.533 \;, \; 
\mbox{BR} (A\to ZH_2) = 0.396 \;, \;
\mbox{BR} (H^\pm \to tb) = 0.560 \;.
\eeq
%sigma_A = 2.472 pb.
The resonant contribution stems here basically from $H_2$ resonant
production where we have
\beq
\sigma (H_2) \times \mbox{BR} (H_2 \to H_1 H_1) =
397\mbox{ fb}  \label{eq:cxnh2h31}
\eeq
The resonant contribution from $H_3$ amounts only to 0.63~fb.
We find that the $H_1 H_1$ production value computed with
{\tt HPAIR} is lower than the value given in
Eq.~(\ref{eq:cxnh2h31}). The comparatively lower value of 
the {\tt HPAIR} result is caused by the finite width of $H_2$ as we
explicitly checked. \s

We note that in this scenario SM-like plus additional Higgs boson pair
production can amount to 
\beq
\sigma (H_1 H_2) =5.17\mbox{ fb} \;,
\eeq
which leads to a triple SM-like $H_1$ rate of 4.89 fb and a $6b$-quark
final state at 1.04~fb as BR($H_1 \to b\bar{b}$)=0.597. Higgs pair
production in association with a $Z$ boson can be large, 
\beq
\sigma_{\text{prod}} (A) \times \mbox{ BR}(A \to ZH_2) \times \mbox{ BR}(H_2 \to H_1
H_1) = 2.16 \mbox{ pb} \times 0.396 \times 0.946 = 809 \mbox{fb} \;,
\eeq
leading to a $Z+(4b)$ rate of 288~fb.

\subsection{Resonant SM-like $H_2 H_2$ Production in the N2HDM-I}
Here we have the scenario {\tt BP7} where $H_2 \equiv
H_{\text{SM}}$. The point is interesting not only because of its large
$H_2 H_2$ cross section but also because it allows for significant
production of a Higgs pair with a $Z$ boson in the final state and it
leads to a large pair production rate for the SM-like Higgs together
with a non-SM-like lighter one. All $ZH_1 H_2$, $Z H_1 H_1$ and $H_1
H_2$ production channels have significant rates in the $4b$ (plus $Z$) final
states. The input parameters are listed in Tab.~\ref{tab:bpn2hdm1h2}
(upper) together with double and single Higgs production related
information in Tab.~\ref{tab:bpn2hdm1h2} (lower). The CP-even Higgs
bosons are rather light. \s

\begin{table}[h!]
\begin{center}
\begin{tabular}{|c|c|c|c|c|c|}
\hline
$m_{H_1}$ [GeV] & $m_{H_2}$ [GeV] & $m_{H_3}$ [GeV] & $m_{A}$ [GeV]
  & $m_{H^\pm}$ [GeV] & $\tan\beta$ \\ \hline
75 & 125.09 & 311 & 646 & 659 & 1.619 \\ \hline \hline
$\alpha_1$ & $\alpha_2$ & $\alpha_3$ & $v_s$ [GeV] & $\mbox{Re}(m_{12}^2)$
 [GeV$^2$] & \\ \hline
-0.936 & -1.020 & -0.341 & 1432 & 20022 &
 \\ \hline
\end{tabular}

\vspace*{0.3cm}
\begin{tabular}{|c|c|c|c|c|c|c|}
\hline
$\sigma_{H_2 H_2}^{\text{NLO}}$ [fb]& $K$-factor & $\Gamma^{\text{tot}}_{H_1}$ [GeV] &
$\Gamma^{\text{tot}}_{H_2}$ [GeV] & $\Gamma^{\text{tot}}_{H_3}$ [GeV]
& $\Gamma^{\text{tot}}_{A}$ [GeV]
& $\Gamma^{\text{tot}}_{H^\pm}$ [GeV]  \\ \hline
344 & 2.04& 4.666 $\times\,10^{-4}$ & 3.605 $\times\,10^{-3}$ & 0.137 & 57.43 & 62.72
\\ \hline \hline
$\lambda_{3H_2}/\lambda_{3H}$ & $y_{t,H_2}/y_{t,H}$ &
$\sigma_{H_1}^{\text{NNLO}}$ [pb] & $\sigma_{H_2}^{\text{NNLO}}$ [pb]
& $\sigma_{H_3}^{\text{NNLO}}$ [pb] & $\sigma_{A}^{\text{NNLO}}$ &\\ \hline
0.921 & 0.928 & 29.98 & 42.39 & 3.08 & 0.95 & \\ \hline
\end{tabular}
\caption{\underline{{\tt BP7}} Upper: N2HDM-I input parameters. Lower:
  Additional double and single Higgs production
  related information.
\label{tab:bpn2hdm1h2}}
\end{center}
\vspace*{-0.8cm}
\end{table}
% This is point 710.
%
The dominant branching ratios of the non-SM-like Higgs bosons are
\beq
\mbox{BR} (H_1 \to b\bar{b}) &=& 0.838\,, \;
\mbox{BR} (H_3 \to H_1 H_2) = 0.831\,, \;
\mbox{BR} (H_3 \to H_2 H_2) = 0.123\,, \nonumber \\
\mbox{BR} (A\to Z H_1) &=& 0.327\,, \;
\mbox{BR} (A \to ZH_3) = 0.464\,, \;
\mbox{BR} (A\to t\bar{t})=0.199\,, \\
\mbox{BR} (H^\pm \to tb) &=& 0.179\,,\; 
\mbox{BR} (H^\pm \to W^\pm H_1) = 0.324\,, \;
\mbox{BR} (H^\pm \to W^\pm H_3)=0.487\,. \nonumber
\eeq
This leads to substantial
production rates for Higgs pair plus gauge boson final states, namely
\beq
\sigma_{\text{prod}} (A) \times \mbox{BR} (A \to Z H_3) \times \mbox{BR} (H_3
\to H_1 H_2) = 0.95 \mbox{ pb }  \times 0.464 \times
0.831 =366 \mbox{ fb} \;, \nonumber \\
\sigma_{\text{prod}} (A) \times \mbox{BR} (A \to Z H_3) \times \mbox{BR} (H_3
\to H_2 H_2) =0.95 \mbox{ pb } \times 0.464 \times 0.123 = 54
\mbox{ fb}  \;.
\eeq
With {\it e.g.}  $\mbox{BR} (H_2 \to b\bar{b}) = 0.575$ and $\mbox{BR}(H_1
\to b\bar{b}) = 0.838$, this leads to $Z+4b$ final states with rates of
176 fb in $Z H_1 H_2$ production and 18 fb in $ZH_2 H_2$
production. Since the branching ratio of 
$H_3$ into $H_1H_2$ is rather large and also the $H_3$ production
cross section is significant with $\sigma (H_3) = 3.08$~pb, we can
expect $H_1 H_2$ production to be large due to resonant
enhancement. And indeed we find at NLO QCD a large cross section of 
\beq
\sigma (H_1 H_2) = 2.15 \mbox{ pb} \;,
\eeq
leading to 1.04~pb in the $4b$ final state.

\subsection{Resonant SM-like $H_1 H_1$ Production
  in the N2HDM-II}
For this scenario, the information is given in
Tab.~\ref{tab:bpn2hdm2h1}.
\begin{table}[h!]
\begin{center}
\begin{tabular}{|c|c|c|c|c|c|}
\hline
$m_{H_1}$ [GeV] & $m_{H_2}$ [GeV] & $m_{H_3}$ [GeV] & $m_{A}$ [GeV]
  & $m_{H^\pm}$ [GeV] & $\tan\beta$ \\ \hline
125.09 & 302 & 856 & 959 & 946 & 1.650
\\ \hline \hline
$\alpha_1$ & $\alpha_2$ & $\alpha_3$ & $v_s$ [GeV] &
$\mbox{Re}(m_{12}^2)$ [GeV$^2$] & \\ \hline
1.077 & -0.258 & -1.444 & 4548 & 277693
& \\ \hline
\end{tabular}

\vspace*{0.3cm}
\begin{tabular}{|c|c|c|c|c|c|c|}
\hline
$\sigma_{H_1 H_1}^{\text{NLO}}$ [fb] & $K$-factor & $\Gamma^{\text{tot}}_{H_1}$ [GeV] &
$\Gamma^{\text{tot}}_{H_2}$ [GeV] & $\Gamma^{\text{tot}}_{H_3}$ [GeV] & $\Gamma^{\text{tot}}_{A}$ [GeV]
& $\Gamma^{\text{tot}}_{H^\pm}$ [GeV]  \\ \hline
 188 &2.02& 3.426 $\times\,10^{-3}$ & 1.076& 12.97
  & 18.84
& 23.36
\\ \hline \hline
$\lambda_{3H_1}/\lambda_{3H}$ & $y_{t,H_1}/y_{t,H}$ &
$\sigma_{H_1}^{\text{NNLO}}$ [pb] & $\sigma_{H_2}^{\text{NNLO}}$ [pb]
& $\sigma_{H_3}^{\text{NNLO}}$ [pb] & $\sigma_{A}^{\text{NNLO}}$ [pb] &\\ \hline
0.719077 & 0.996 & 49.11 & 0.40 &  0.12 & 0.09 & \\ \hline
\end{tabular}
\caption{\underline{{\tt BP8}} Upper: N2HDM-II input parameters. Lower:
  Additional double and single Higgs production
  related information.
\label{tab:bpn2hdm2h1}}
% This is point 4489
\end{center}
\vspace*{-0.8cm}
\end{table}
The dominant branching ratios of the non-SM-like Higgs bosons are
\beq
\begin{array}{lcllcl}
\mbox{BR} (H_2 \to H_1 H_1) &=& 0.478 \;, & \; \mbox{BR} (H_2 \to WW)
  &=& 0.361\;, \\
\mbox{BR} (H_2 \to ZZ) &=& 0.160 \;, \\[0.1cm]
\mbox{BR} (H_3 \to  t\bar{t}) &=& 0.907\;, \\[0.1cm]
\mbox{BR} (A \to Z H_2) &=& 0.133 \;, & \; \mbox{BR} (A \to t\bar{t})
  &=& 0.818 \;, \\[0.1cm]
\mbox{BR} (H^\pm \to W^\pm H_2) &=& 0.134 \;, & \; \mbox{BR} (H^\pm
                                                \to tb) &=& 0.822 \;.
\end{array}
\eeq
With BR($H_3 \to H_1 H_1$)=$0.0458$, we have the
resonant $H_1 H_1$ production rates
\beq
\sigma (H_2) \times \mbox{BR} (H_2 \to H_1 H_1) &=&  191 \mbox{ fb} \;,\\
\sigma (H_3) \times \mbox{BR} (H_3 \to H_1 H_1) &=&
 5.50\mbox{ fb} \;.
\eeq
The di-Higgs cross section enhancement hence basically stems from the $H_2$
exchange in the triangle diagram. Higgs pair
production in association with a $Z$ boson has a rather low rate of
\beq
\sigma_{\text{prod}} (A) \times \mbox{ BR}(A \to ZH_2) \times \mbox{ BR}(H_2 \to H_1
H_1) = 90 \mbox{ fb} \times 0.133 \times 0.478=  5.72 \mbox{ fb} \;,
\eeq
leading to a $Z+(4b)$ rate of 1.72~fb.

\subsection{Resonant SM-like $H_2 H_2$ Production
  in the N2HDM-II}
In Tab.~\ref{tab:bpn2hdm2h2}, we
summarize information on this scenario. 
\begin{table}[h!]
\begin{center}
\begin{tabular}{|c|c|c|c|c|c|}
\hline
$m_{H_1}$ [GeV] & $m_{H_2}$ [GeV] & $m_{H_3}$ [GeV] & $m_{A}$ [GeV]
  & $m_{H^\pm}$ [GeV] & $\tan\beta$ \\ \hline
117 & 125.09 & 756 & 792 & 836 & 1.040
\\ \hline \hline
$\alpha_1$ & $\alpha_2$ & $\alpha_3$ & $v_s$ [GeV] &
$\mbox{Re}(m_{12}^2)$ [GeV$^2$] & \\ \hline
0.392 & 1.484 & -1.193 & 865 & 252856
& \\ \hline
\end{tabular}

\vspace*{0.3cm}
\begin{tabular}{|c|c|c|c|c|c|c|}
\hline
$\sigma_{H_2 H_2}^{\text{NLO}}$ [fb] & $K$-factor & $\Gamma^{\text{tot}}_{H_1}$ [GeV] &
$\Gamma^{\text{tot}}_{H_2}$ [GeV] & $\Gamma^{\text{tot}}_{H_3}$ [GeV] & $\Gamma^{\text{tot}}_A$ [GeV]
& $\Gamma^{\text{tot}}_{H^\pm}$ [GeV] \\ \hline
 63 & 1.91 & $3.644 \,\times\, 10^{-5}$ & 4.204 $\times\,10^{-3}$ &
                                                                      30.14 & 34.49 & 35.03
\\ \hline \hline
$\lambda_{3H_2}/\lambda_{3H}$ & $y_{t,H_2}/y_{t,H}$ &
$\sigma_{H_1}^{\text{NNLO}}$ [pb] & $\sigma_{H_2}^{\text{NNLO}}$ [pb]
& $\sigma_{H_3}^{\text{NNLO}}$ [pb] & $\sigma_{A}^{\text{NNLO}}$ [pb] &\\ \hline
0.917 & 0.964 & 0.11 & 45.61 &
0.73 & 0.72 & \\ \hline
\end{tabular}
\caption{\underline{{\tt BP9}} Upper: N2HDM-II input
  parameters. Lower: Additional double and single Higgs production
  related information.
\label{tab:bpn2hdm2h2}}
\end{center}
% This is point 32
\vspace*{-0.8cm}
\end{table}
For the non-SM-like Higgs bosons, the dominant branching ratios into
detectable final states are
\beq
\mbox{BR} (H_1 \to b\bar{b}) &=& 0.826 \;, \quad
\mbox{BR} (H_3 \to t\bar{t}) = 0.968 \;, \nonumber \\
\mbox{BR} (A \to t\bar{t}) &=& 0.987 \;, \quad
\mbox{BR} (H^\pm \to tb) = 0.986 \;.
\eeq
The resonant contribution to the cross section amounts to
\beq
\sigma_{H_3}^{\text{NNLO}} \times \mbox{BR} (H_3 \to H_2 H_2) =
720 \times 0.0167 \mbox{ fb } = 12.02 \mbox{ fb} \;,
\eeq
so that the larger SM-like $H_2 H_2$ production compared to
the SM value is partly caused by resonant enhancement from $H_3$ production. 

%%%%%%%%%%%%%%%%%%%%%%%%%%%%%%%%%%%%%%%%%%%%%%%%%%%%%%%%%%%%

\subsection{Resonant SM-like $H_1 H_1$ Production
  in the NMSSM}
This NMSSM benchmark point features, besides a large $H_1 H_1$
production cross section,
large rates for $Z H_1 H_1$ and triple $H_1$
production.
As stated above, in the NMSSM, the Higgs
boson masses are computed from the input parameters of the model and
higher-order corrections are important to shift the SM-like Higgs mass
to the observed 125~GeV. We have computed these masses using the new
version of {\tt NMSSMCALC} which includes the two-loop Higgs mass
corrections at ${\cal O}((\alpha_\lambda+ \alpha_\kappa + \alpha_t)^2+
\alpha_t \alpha_s)$ \cite{Dao:2021khm}.
In Tab.~\ref{bench:h1nmssmres_1}, we list all input parameters for this
benchmark point\footnote{In accordance
  with the SUSY Les Houches Accord (SLHA)
\cite{Skands:2003cj,Allanach:2008qq} the soft SUSY breaking masses and
trilinear couplings are understood as $\overline{\mbox{DR}}$
parameters at the scale $M_{\text{SUSY}}=\sqrt{m_{\tilde{Q}_3}
      m_{\tilde{t}_R}}$ which is also the renormalisation scale used
  in the computation of the higher-order corrections to the Higgs
  masses. The soft SUSY breaking parameters of the first two
  generations are not listed as their influence is negligible. The
  remaining SM input parameters are given in
  Ref.~\cite{Dao:2021khm}. The Higgs mass corrections are computed
  with on-shell renormalisation in the top/stop sector and on-shell
  renormalised charged Higgs mass, {\it cf.}~\cite{Dao:2021khm} for
  details. For completeness, we also list the corresponding value of
  $A_\lambda$.} that are required
by {\tt NMSSMCALC} to compute the Higgs spectrum. Higgs masses, mixing angles
and the total widths of the Higgs bosons,
are given in Tab.~\ref{bench:h1nmssmres_2}. The table contains
additional information related to double and single Higgs
production. The given CP-even
mixing elements $h_{ij}$ ($i,j=1,2,3$) comply with the SLHA
\cite{Skands:2003cj,Allanach:2008qq}, while the elements $a_{ij}$
relate to the SLHA definition $a_{ij}^{\text{SLHA}}$ through
\beq
a_{11}^{\text{SLHA}} &=& \frac{a_{11}}{\sin\beta} \;, \quad
a_{21}^{\text{SLHA}} = \frac{a_{21}}{\sin\beta} \;, \nonumber\\
a_{12}^{\text{SLHA}} &=& a_{13}  \;, \quad
a_{22}^{\text{SLHA}} = a_{23}  \;.
\eeq
\begin{table}[h!]
\begin{center}
\begin{tabular}{|c|c|c|c|c|c|}
\hline
$\lambda$ & $\kappa$ & $A_\lambda$ [GeV] &  $A_\kappa$ [GeV] &
$\mu_{\text{eff}}$ [GeV] & $\tan\beta$
\\ \hline
0.650 & 0.645 & 359.27 & -432.19 & 224.95 & 2.622
\\ \hline \hline
$m_{H^\pm}$ [GeV] & $M_1$ [GeV] & $M_2$ [GeV] & $M_3$ [TeV] & $A_t$
[GeV] & $A_b$ [GeV] \\ \hline
610.64 & 810 & 642 & 2 & -46 & -1790  \\ \hline \hline
$m_{\tilde{Q}_3}$ [GeV] & $m_{\tilde{t}_R}$ [GeV] & $m_{\tilde{b}_R}$ [GeV] & $A_\tau$
[GeV] & $m_{\tilde{L}_3}$  [GeV] & $m_{\tilde{\tau}_R}$ [GeV] \\ \hline
1304 & 3000 & 3000 & -93 & 3000 & 3000 \\ \hline
\end{tabular}
\caption{\underline{{\tt BP10}} NMSSM input parameters required
  by {\tt NMSSMCALC} for the computation of the NMSSM
  spectrum. Lower: Additional double and single Higgs production
  related information. \label{bench:h1nmssmres_1}}
\end{center}
% This is the point 1863.
\vspace*{-0.5cm}
\end{table}
\begin{table}[h!]
\begin{center}
\begin{tabular}{|c|c|c|c|c|c|}
\hline
$m_{H_1}$ [GeV] & $m_{H_2}$ [GeV] & $m_{H_3}$ [GeV] & $m_{A_1}$ [GeV] &
$m_{A_2}$ [GeV] & $m_{H^\pm}$ \\ \hline
122.39 & 300 & 626 & 543 & 616 & 611
\\ \hline \hline
$\Gamma^{\text{tot}}_{H_1}$ [GeV] & $\Gamma^{\text{tot}}_{H_2}$ [GeV] &
$\Gamma^{\text{tot}}_{H_3}$ [GeV] & $\Gamma^{\text{tot}}_{A_1}$ [GeV] &
$\Gamma^{\text{tot}}_{A_2}$ [GeV] & $\Gamma^{\text{tot}}_{H^\pm}$ [GeV]
\\ \hline
3.947 $\times\,10^{-3}$ & 0.127 & 4.81 & 5.52 & 5.82 & 5.61
\\ \hline \hline
$h_{11}$ & $h_{12}$ & $h_{13}$ & $h_{21}$ & $h_{22}$ & $h_{23}$ \\
\hline
0.372 & 0.924 & 0.092 & 0.170 & -0.165 & 0.971
\\ \hline \hline
$h_{31}$ & $h_{32}$ & $h_{33}$ & $a_{11}$  & $a_{21}$ & $a_{13}$\\ \hline
0.912 & -0.346 & -0.219 & 0.133 & 0.925 & 0.990
\\ \hline \hline
$a_{23}$ & $\lambda_{3H_1}/\lambda_{3H}$ & $y_{t,H_1}/y_{t,H}$ & & & \\
\hline
-0.143 & 0.594 & 0.988 & & & \\ \hline
  \hline
$\sigma_{H_1 H_1}^{\text{NLO}}$ [fb]
& $\sigma_{H_1}^{\text{NNLO}}$ [pb] & $\sigma_{H_2}^{\text{NNLO}}$ [pb]
& $\sigma_{H_3}^{\text{NNLO}}$ [pb] & $\sigma_{A_1}^{\text{NNLO}}$ [pb]
& $\sigma_{A_2}^{\text{NNLO}}$ [pb] \\ \hline
183 & 49.93 & 0.38 & 0.26 & 0.017 & 0.42
  \\ \hline \hline
  $K$-factor&&&&&\\ \hline
  2.01 & &&&&\\ \hline
\end{tabular}
\caption{\underline{{\tt BP10}} Additional information related to
  double and single Higgs production. \label{bench:h1nmssmres_2}}
\end{center}
\vspace*{-0.5cm}
\end{table}
The dominant branching ratios of the non-SM-like Higgs bosons are
given by
\beq
\begin{array}{lcllcllcllcl}
\mbox{BR}(H_2 \to WW) &=& 0.404 \,, & \; \mbox{BR}(H_2 \to
                                            ZZ) &=& 0.179 \,, & \;
\mbox{BR}(H_2 \to H_1 H_1) &=& 0.404 \;, \\
\mbox{BR}(H_3 \to t\bar{t}) &=& 0.626 \,, & \;
\mbox{BR}(H_3 \to H_1 H_2) &=& 0.298 \,, & \;
\mbox{BR}(A_1 \to \tilde{\chi}_1^+ \tilde{\chi}_1^-) &=& 0.506\;, \\
\mbox{BR}(A_1 \to \tilde{\chi}_1^0 \tilde{\chi}_1^0) &=& 0.300 \,, &
\; \mbox{BR}(H^+ \to t\bar{b}) & = & 0.731\;, &
\; \mbox{BR}(A_2 \to tt) &=& 0.714 \,,  \\
\mbox{BR}(A_2 \to Z H_2) &=& 0.235 \,, 
% mail by Duarte on Aug, 27, 2021.
\nonumber \\
\end{array}
\eeq
where $\tilde{\chi}^{\pm,0}$ denote the charginos and neutralinos, respectively.
We can infer that the dominant contribution to $H_1 H_1$ production
stems from the $H_2$ triangle resonance diagram as
\beq
\sigma_{H_2}^{\text{NNLO}} \times \mbox{BR} (H_2 \to H_1 H_1) =
154 \mbox{ fb}.
\eeq
Interesting signatures are the $H_1$ pair production in
association with a $Z$ boson and triple $H_1$ production from Higgs
cascades
\beq
&& \sigma_{\text{prod}} (A_2) \times \mbox{BR}(A_2 \to ZH_2) \times
\mbox{BR} (H_2 \to H_1 H_1) = 420 \mbox{ fb} \times 0.235 \times
0.404 = 39.87 \mbox{ fb}  \;, \nonumber\\
&& \sigma_{\text{prod}} (H_3) \times \mbox{BR}(H_3 \to H_1 H_2) \times
\mbox{BR} (H_2 \to H_1 H_1) = 260 \mbox{ fb} \times 0.298 \times
0.404 = 31.30 \mbox{ fb} \;. \nonumber \\
\eeq
With a branching ratio of BR($H_1 \to b\bar{b}$) = 0.652, this results
in a $Zb\bar{b}b\bar{b}$ final state with a rate of 16.95 fb
for $ZH_1 H_1$ production
and in $6b$-quark final state with a rate of 8.68 fb
for triple $H_1$ production.

%%%%%%%%%%%%%%%%%%%%%%%%%%%%%%%%%%%%%%%%%%%%%%%%%%%%%%%%%%%%

\subsection{Resonant SM-like $H_2 H_2$ Production
  in the NMSSM}
Although for this point resonant production is kinematically possible,
it is very suppressed. The (small) enhancement compared to the SM is
caused by the deviation of the trilinear coupling from the SM
value. The relevant information on this scenario is given in
Tabs.~\ref{bench:h2nmssmres_1} and \ref{bench:h2nmssmres_2}. \s
\begin{table}[h!]
\begin{center}
\begin{tabular}{|c|c|c|c|c|c|}
\hline
$\lambda$ & $\kappa$ & $A_\lambda$ [GeV] &  $A_\kappa$ [GeV] &
$\mu_{\text{eff}}$ [GeV] & $\tan\beta$
\\ \hline
0.0749 & 0.0646 & 425.77 & -705.59 & 215.06 & 3.007
\\ \hline \hline
$m_{H^\pm}$ [GeV] & $M_1$ [GeV] & $M_2$ [GeV] & $M_3$ [TeV] & $A_t$ [GeV] & $A_b$ [GeV] \\ \hline
660.53 & 728 & 430
& 2 & 3226.33 & 584.37  \\ \hline \hline
$m_{\tilde{Q}_3}$ [GeV] & $m_{\tilde{t}_R}$ [GeV] & $m_{\tilde{b}_R}$ [GeV] & $A_\tau$
[GeV] & $m_{\tilde{L}_3}$  [GeV] & $m_{\tilde{\tau}_R}$ [GeV] \\ \hline
1196.41 & 1440.66 & 3000 &1796 & 3000 & 3000 \\ \hline
\end{tabular}
\caption{\underline{{\tt BP11:}} NMSSM input parameters required by
  {\tt NMSSMCALC} for the  computation of the NMSSM spectrum.
\label{bench:h2nmssmres_1}}
\end{center}
% This is the point 11642.
\end{table}
\begin{table}[h!]
\begin{center}
\begin{tabular}{|c|c|c|c|c|c|}
\hline
$m_{H_1}$ [GeV] & $m_{H_2}$ [GeV] & $m_{H_3}$ [GeV] & $m_{A_1}$ [GeV] &
$m_{A_2}$ [GeV] & $m_{H^\pm}$\\ \hline
81 & 125.05 & 659 & 626 & 656 & 661
\\ \hline \hline
$\Gamma^{\text{tot}}_{H_1}$ [GeV] & $\Gamma^{\text{tot}}_{H_2}$ [GeV] &
$\Gamma^{\text{tot}}_{H_3}$ [GeV] & $\Gamma^{\text{tot}}_{A_1}$ [GeV] &
$\Gamma^{\text{tot}}_{A_2}$ [GeV] & $\Gamma^{\text{tot}}_{H^\pm}$ [GeV]
\\ \hline
2.888 $\times\,10^{-5}$ & 4.281 $\times\,10^{-3}$ & 4.17 & 0.088 & 5.23 & 4.40
\\ \hline \hline
$h_{11}$ & $h_{12}$ & $h_{13}$ & $h_{21}$ & $h_{22}$ & $h_{23}$ \\
0.039 & 0.055 & 0.998 & 0.331 & 0.941 & -0.065
\\ \hline \hline
$h_{31}$ & $h_{32}$ & $h_{33}$ & $a_{11}$ & $a_{21}$ & $a_{13}$  \\ \hline
0.943& -0.333 & -0.018 & -0.012 & 0.949 & 0.999
\\ \hline \hline
 $a_{23}$ & $\lambda_{3H_2}/\lambda_{3H}$ & $y_{t,H_2}/y_{t,H}$ & & & \\
\hline
 0.013 & 0.333 & 0.992 & & & \\ \hline \hline
$\sigma_{H_2 H_2}^{\text{NLO}}$ [fb]
& $\sigma_{H_1}^{\text{NNLO}}$ [pb] & $\sigma_{H_2}^{\text{NNLO}}$ [pb]
& $\sigma_{H_3}^{\text{NNLO}}$ [pb] & $\sigma_{A_1}^{\text{NNLO}}$ [pb]
& $\sigma_{A_2}^{\text{NNLO}}$ [pb] \\ \hline
65 & 0.32 & 42.76 & 0.16 &
$5.10^{-5}$ & 0.25
  \\ \hline \hline
$K$-factor&&&&&\\ \hline
1.95 & &&&&\\ \hline
\end{tabular}
\caption{\underline{{\tt BP11:}} Additional information related to
  double and single Higgs production. \label{bench:h2nmssmres_2}}
\end{center}
\end{table}

The dominant branching ratios of the non-SM-like Higgs bosons are
given by
\beq
\begin{array}{lcllcllcllcl}
\mbox{BR}(H_1 \to b\bar{b}) &=&0.89 \,, & \;
\mbox{BR}(H_3 \to t\bar{t}) &=&0.709 \,, & \;
\mbox{BR}(A_1 \to \tilde{\chi}_1^+\tilde{\chi}_1^-) &=& 0.47 \;, \nonumber \\
\mbox{BR}(A_2 \to t\bar{t}) &=&0.66 \;, & \;
\mbox{BR}(A_2 \to \tilde{\chi}_1^+ \tilde{\chi}_1^-) &=& 0.32 \,, &
\mbox{BR}(A_2 \to \tilde{\chi}_1^0 \tilde{\chi}_1^0) &=& 0.152 \,, \nonumber \\
\mbox{BR}(A_2 \to \tilde{\chi}_2^0 \tilde{\chi}_2^0) &=& 0.202 \,, & \;
\mbox{BR}(H^+ \to  t\bar{b}) & = & 0.778 \;.
% mail by Duarte on Aug, 27, 2021.
\end{array}
\eeq
Since the branching ratio BR($H_3\to H_2 H_2$)= 6.5 $\times\,10^{-5}$ is tiny,
it is not resonance $H_3$ production that enhances the cross section
but rather the SM-like trilinear coupling deviation from
the SM as stated above. 
%\subsection{4W Final State}

%%%%%%%%%%%%%%%%%%%%%%%%%%%%%%%%%%%%%%%%%%%%%%%%%%%%%%%%%%%%
\section{Constraining Model Parameters \label{sec:constraining}}
With the SM-like Higgs pair production cross section in the
non-resonant case being
three orders of magnitude smaller than single Higgs production,
deriving constraints on the parameter spaces of the models from
di-Higgs production may not be very efficient. This picture changes of
course in case of resonance enhancements. To get a rough picture of
what can be learnt from di-Higgs production, we present a few selected
heat plots for the SM-like Higgs pair production cross sections as a
function of relevant model parameters. \s

\begin{figure}[t!]
\centering
\includegraphics[width=0.5\linewidth]{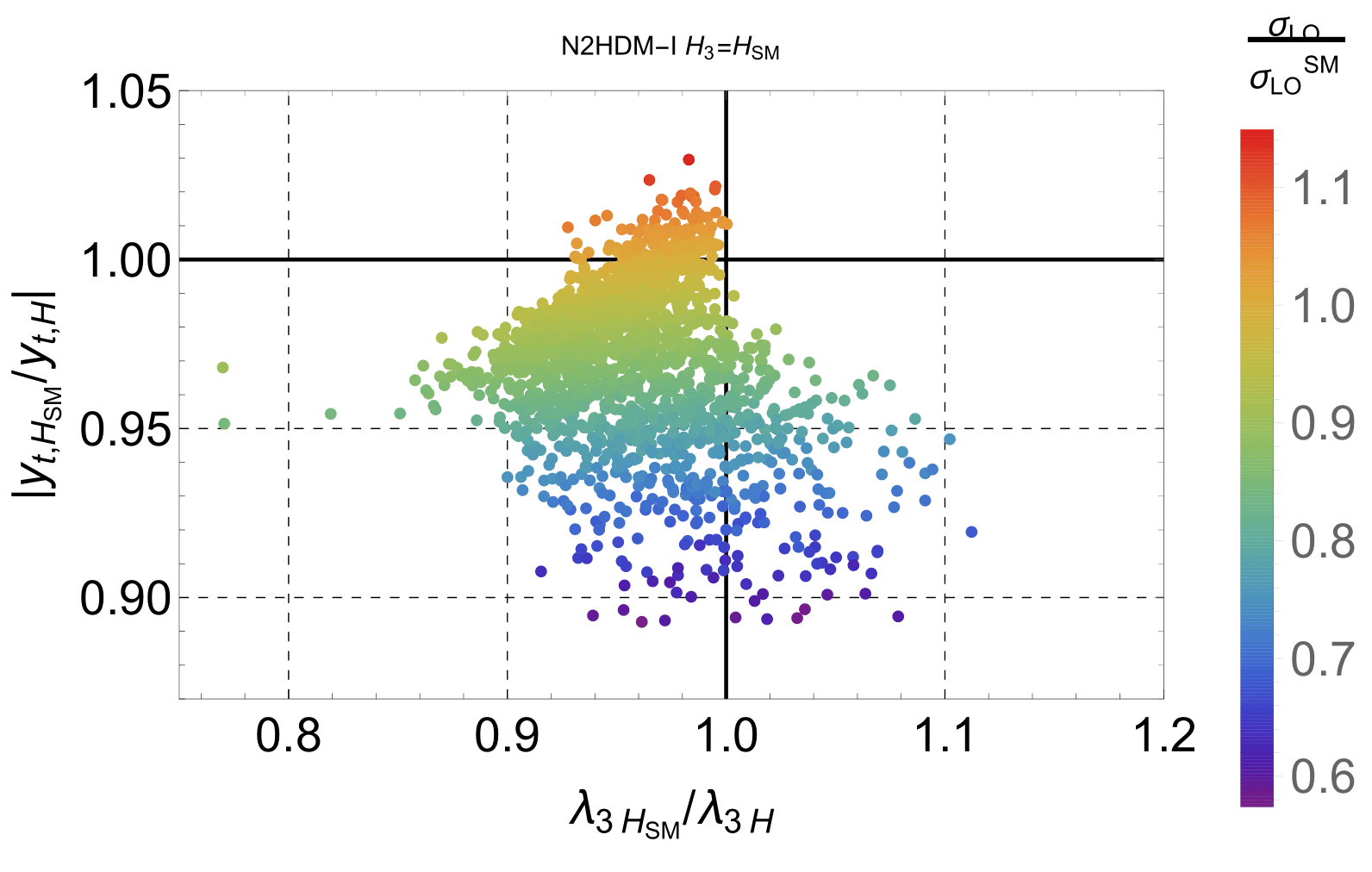}
\caption{N2DHM-I with $H_3 \equiv H_{\text{SM}}$: Cross section
  values normalized to the SM value (color code) as function of the
  top-Yukawa and trilinear Higgs self-couplings of
  $H_3$ normalized to the respective SM values. A factor of two is
  applied to the LO cross section to roughly account for the QCD
  corrections. \label{fig:cxnchange}}
\end{figure}

In the non-resonant case, it is the top Yukawa and trilinear Higgs
self-couplings of the SM-like Higgs boson that determine the size of
the SM-like Higgs pair production cross section. In Fig.~\ref{fig:cxnchange}, we show
as colour code the size  of the cross section for SM-like Higgs pair
production normalized to the SM value
in the N2HDM-I for the case where $H_3 \equiv H_{\text{SM}}$, as a
function of its top-Yukawa and trilinear Higgs self-coupling
normalized to their SM values, respectively. 
By choosing $H_3 \equiv H_{\text{SM}}$, we make sure that the
contributions from lighter $s$-channel Higgs boson exchanges to the
cross section are subdominant and that it is the SM-like couplings
that determine its size. As can be inferred from the
plot, the size of the cross section is mostly insensitive to the
value of the trilinear Higgs self-coupling in the limited range that is still allowed for it,
while it shows a strong dependence on the top-Yukawa coupling.
Taking off the resonance contribution for the case where $H_1 \equiv
H_{\text{SM}}$ and where the allowed range for the trilinear Higgs self-coupling is
larger we find a significant dependence of the cross section on
the trilinear coupling. \s

%\footnote{In
%  case the lighter Higgs bosons are SM-like the total cross section is
%superseeded by the resonant contribution from heavy Higgs bosons so
%that the dependence on the couplings is not visible in the plot.}
\begin{figure}[t!]
\centering
\includegraphics[width = 0.49\textwidth]{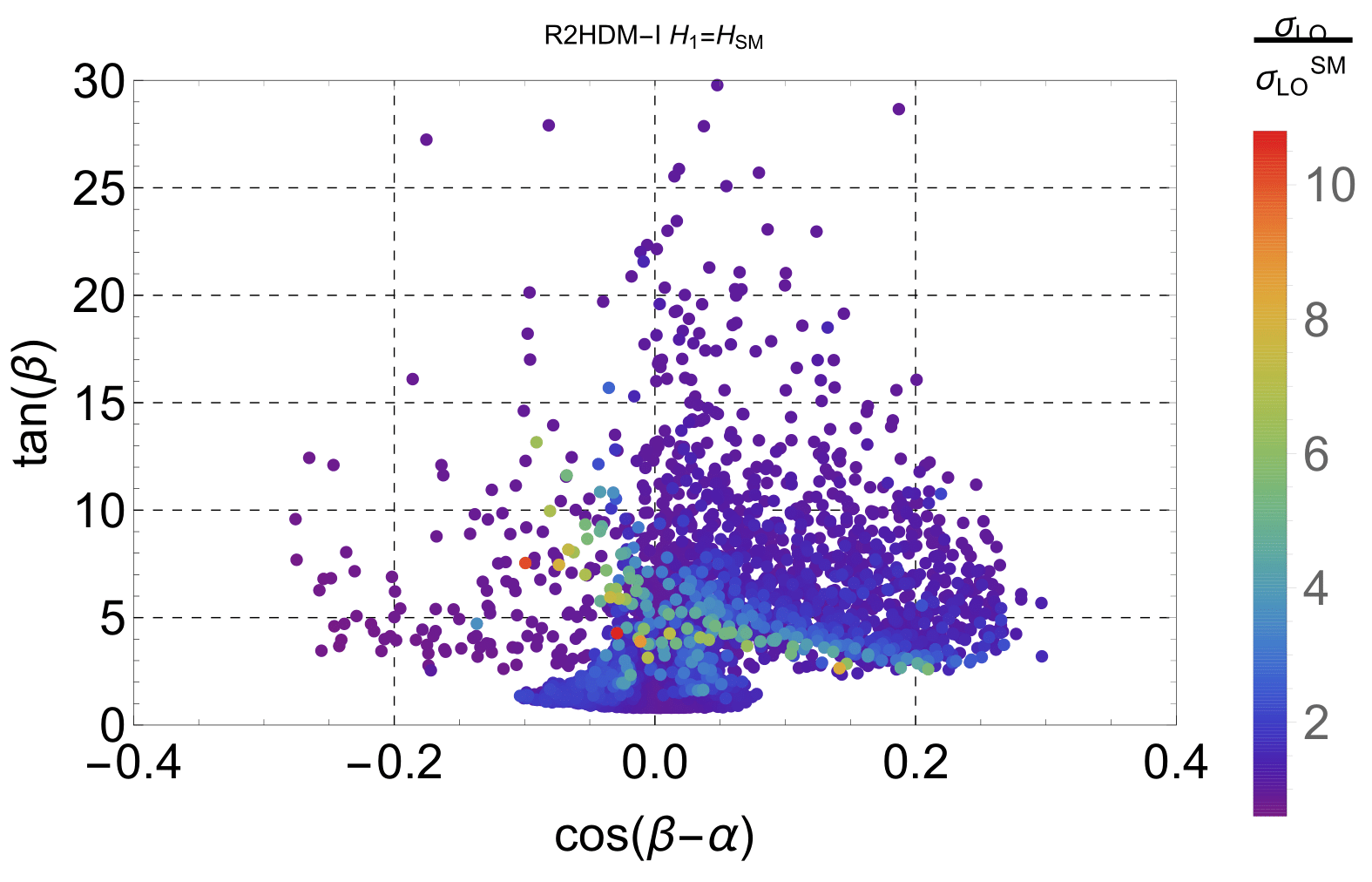}
\includegraphics[width = 0.49\textwidth]{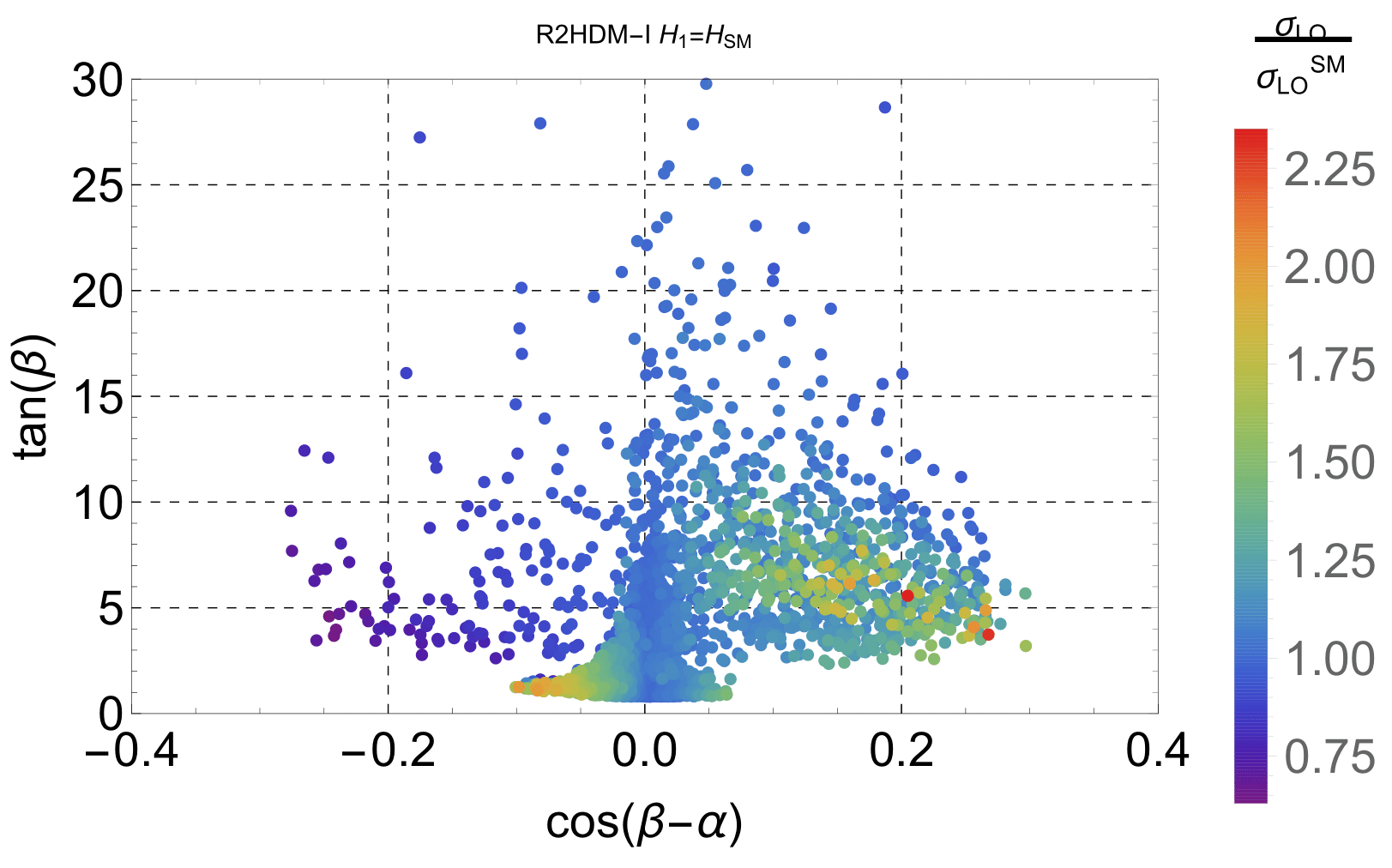}
\caption{R2HDM-I with $H_1 \equiv H_{\text{SM}}$: NLO (through a
  factor of 2) QCD cross section
  values normalized to the SM value (color code) as function of $\tan\beta$ and $\cos
  (\beta-\alpha)$. Left: complete cross section, right: only points
  with $\sigma_{H_2}^{\text{NNLO}} (H_2) \times
  \mbox{BR} (H_2 \to H_1 H_1) \le 0.1 \sigma^{\text{NLO}} (H_1
  H_1)$ \label{fig:t1lightr2hdm}}
\end{figure}
Let us investigate if di-Higgs production can possibly
contribute to constraining the parameter space of the model. For this, we
resort to the simpler R2HDM whose tree-level Higgs couplings are described
by only two mixing angles. In Fig.~\ref{fig:t1lightr2hdm}, we depict in
the $\tan\beta$--$\cos(\beta-\alpha)$ plane, through the colour code,
the NLO QCD-corrected (by including a factor 2) di-Higgs cross section
normalized to the SM value
for the R2HDM-I, where the lighter Higgs $H_1
\equiv H_{\text{SM}}$. On the
left, the complete cross section is plotted. On the right plot, we consider only points where the
resonant contribution from $H_2\to H_1 H_1$ makes up for less than 10\% of the total cross section. More
specifically, we only include points where $\sigma_{H_2}^{\text{NNLO}}
\times \mbox{BR} (H_2 \to H_1 H_1) \le 0.1 \times \sigma^{\text{NLO}} (H_1
H_1)$. From an experimental point of view this would correspond to non-resonant
$H_1 H_1$ production (according to our definition). From the right
plot, we could infer that cross section values deviating from the SM
value allow us to constrain $\cos (\beta - \alpha)$. The true picture
is more complex though, as shown by the left plot: the possible resonance
enhancements in BSM di-Higgs production also allow for larger cross
sections very close to the SM Higgs alignment limit $\cos
(\beta-\alpha)=0$. Indeed we found by comparing the
  constraints on $\cos(\beta-\alpha)$ before and after applying the
  di-Higgs constraints that the impact of di-Higgs constraints is vanishingly small. \s

With increasing complexity of the Higgs sectors, the superposition of
the various Higgs contributions to the cross sections make it more and
more complicated to derive conclusive statements on the various
trilinear and Yukawa couplings and require the combination of different
cross sections to extract all involved couplings ({\it
  cf.}~\cite{Djouadi:1999gv} for a discussion in the MSSM).

%%%%%%%%%%%%%%%%%%%%%%%%%%%%%%%%%%%%%%%%%%%%%%%%%%%%%%%%%%%%%%
\section{Effective Field Theory versus Specific Models \label{sec:eft}}
In this paper so far, we discussed mainly Higgs pair production in specific models. In this section, we switch gears towards  a different approach to describe new physics effects. This is given by the effective field theory (EFT) framework where BSM physics is
expected to appear at some high new physics scale $\Lambda$. In the linear approach called SMEFT
\cite{Berthier:2015oma,Ghezzi:2015vva,Brivio:2017bnu,Ellis:2018gqa}
new physics is formulated as a power series in the dimensionful
parameter $1/\Lambda$. The non-linearly realized EFT, on the other hand,
can be viewed as organised by chiral dimension
\cite{Feruglio:1992wf, Bagger:1993zf, Koulovassilopoulos:1993pw, Burgess:1999ha, Wang:2006im,
Grinstein:2007iv, Contino:2010mh,Contino:2010rs,Alonso:2012px,Buchalla:2013rka,Delgado:2013hxa,Buchalla:2013eza,
Buchalla:2015qju,deBlas:2018tjm}.
If we choose to describe Higgs pair production in the EFT approach
this means that effects from additional
non-SM-like light Higgs bosons cannot be
described.\footnote{For an extension of the EFT
    approach to include an extended particle content, an EFT for the
    2HDM, {\it cf.}~\cite{Crivellin:2016ihg}. Also for composite Higgs
  models a concrete model with two Higgs doublets has been proposed, {\it cf.}~\cite{DeCurtis:2018zvh,DeCurtis:2021uqx}.} A discussion of the
higher-dimensional operators relevant for Higgs pair production can be
found in
\cite{Azatov:2015oxa,Contino:2012xk,Chen:2014xra,Goertz:2014qta,Edelhauser:2015msp}. The
QCD corrections in the infinite top mass limit, $m_t \to \infty$ have
been provided at NLO QCD in \cite{Grober:2015cwa} and also extended to the
CP-violating case in \cite{Grober:2017gut}. At NNLO QCD they have been
calculated in \cite{deFlorian:2017qfk}. The authors of \cite{Buchalla:2018yce}
presented the NLO QCD corrections including the full top quark mass
effects in a non-linearly realized EFT. An interface with {\tt POWHEG}
\cite{Nason:2004rx,Frixione:2007vw,Alioli:2010xd} has been provided in
\cite{Heinrich:2020ckp}. \s

\begin{figure}[h!]
\begin{center}
\vspace*{-4cm}
\includegraphics[width=0.7 \textwidth, angle=-90]{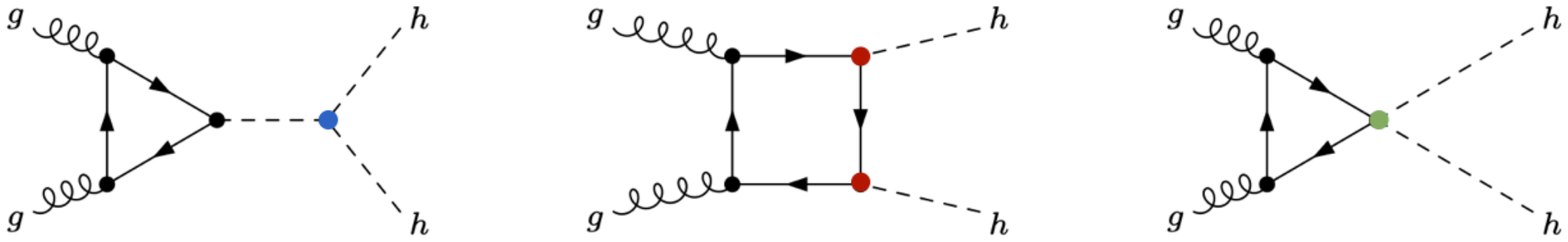}
\vspace*{-4cm}
\caption{Diagrams contributing to Higgs pair production in the EFT
  approach (with $c_g = c_{gg}=0$ and neglecting the chromomagnetic
  operator). The blue, red and green blobs
  denote the modified Higgs trilinear, Higgs top-Yukawa and the new
  two-Higgs-two-top-quark couplings,
  respectively. \label{fig:eftdiags}}
\end{center}
\vspace*{-0.6cm}
\end{figure}

For the models that we considered, only the new physics operators that modify the
trilinear Higgs self-coupling and the Yukawa coupling are
relevant. The induced effective couplings of one or two Higgs bosons
to two gluons only appear in the NMSSM, where integrating out
heavy stops and sbottoms in the Higgs-to-gluon loop couplings would
induce such couplings. We neglect that effect in the
present discussion for simplicity, by setting the associated couplings
($c_{g}$ and $c_{gg}$ in the notation of the non-linear Lagrangian of
Ref.~\cite{Grober:2015cwa}) to zero. In the R2HDM, C2HDM and N2HDM, these couplings do not appear as long as we do not
  include additional heavy coloured particles beyond the
  SM. Furthermore, we do not consider 
effects from the chromomagnetic operator as they are
  of different order in the chiral expansion.\footnote{For
  a discussion on the chromomagnetic operator, see
  \cite{Buchalla:2018yce}.} Finally, integrating out a 
possible heavy Higgs boson exchange in the $s$-channel leads to an effective
two-Higgs-two-fermion coupling. Denoting by $c_3$ the trilinear
coupling modification and by $c_t$ the top-Yukawa coupling
modification with respect to the SM and by $c_{tt}$ the effective
two-Higgs-two-fermion coupling coefficient, {\it i.e.}~adapting the
notation of \cite{Grober:2015cwa}, our considered correction $\Delta {\cal
  L}_{\text{non-lin}}$ to the SM Lagrangian reads,
\beq
\Delta {\cal L}_{\text{non-lin}} \supset -m_t t\bar{t} \left( c_t
\frac{h}{v} + c_{tt} \frac{h^2}{2 v^2} \right) - c_3 \frac{1}{6}
\left( \frac{3M_h^2}{v} \right) h^3 \;,
\eeq
where $h$ denotes the physical Higgs boson.
Since the single and double-Higgs coefficients $c_t$ and $c_{tt}$ to
the top-quark pair are taken to be independent, we adapt the
non-linear effective Lagrangian approach here. In SMEFT, they are
correlated (as well as $c_g$ and $c_{gg}$). In Fig.~\ref{fig:eftdiags},
we show the generic diagrams that contribute to our EFT approach to
Higgs pair production and indicate the EFT coupling modifiers.
In the notation of \cite{Plehn:1996wb} and \cite{Grober:2015cwa}, we
have the following matching relations of our specific models to the
EFT Lagrangian,
\beq
\begin{array}{llccc}
\mbox{Higgs-top Yukawa coupling} &:& g^{\scriptsize H_{\text{SM}}}_t(\alpha_i,\beta) & \to & c_t \\
\mbox{trilinear Higgs coupling} &:& \frac{g_3^{\scriptsize H_{\text{SM}} H_{\text{SM}} H_{\text{SM}}} (p_i)}{3M_{H_{\text{SM}}}^2/v} & \to & c_3 \\
\mbox{two-Higgs-two-top quark coupling} &:&
\sum_{k=1}^{k_{\text{max}}} \left(\frac{-v}{m_{H_k}^2}\right)
g_3^{\scriptsize H_k H_{\text{SM}} H_{\text{SM}}}(p_i) \, g^{\scriptsize H_k}_t (\alpha_i, \beta) & \to & c_{tt} \\
\end{array}
\eeq
Here $g^{\phi}_t(\alpha_i,\beta)$ denotes
the dimensionless function of the mixing angles
$\alpha_{i}$ and $\beta$ that specifies for each model under
consideration the modification of the Yukawa coupling of a Higgs boson
$\phi$ of the model with respect to
the SM Yukawa coupling. The function $g_3^{\scriptsize H_{\text{SM}}
  H_{\text{SM}}H_{\text{SM}}}(p_i)$ denotes
the dimensionful trilinear coupling of three SM-like Higgs bosons
$H_{\text{SM}}$ in our BSM model that in the SM case would approach
$3M_{H_{\text{SM}}}^2/v$. We denote by $p_i$ the various parameters on
which the trilinear coupling depends in the respective model. The third
matching relation to $c_{tt}$ is obtained by assuming a possible heavy Higgs
$H_k$ $s$-channel exchange ({\it cf.}~the first diagram in
Fig.~\ref{fig:c2hdmdiags} for $H_k \ne H_{\text{SM}}$) where the mass of the
exchanged Higgs boson, denoted by $m_{H_k}$, is very large. By
$g_3^{\scriptsize H_k H_{\text{SM}} H_{\text{SM}}}$ we denote the
corresponding $H_k$ trilinear coupling to two SM-like Higgs bosons.
Note that, in non-minimal models, we would have two such contributions. Hence
$k_{\text{max}}=1$ in the R2HDM and 2 in the C2HDM, N2HDM and
NMSSM.\footnote{We only take into account in the cross section linear EFT
  contributions and no squared ones. For details on the LO partonic
  cross section, we refer to Eqs.~(2.5)-(2.13) of
  Ref.~\cite{Grober:2015cwa}.} Table~\ref{tab:trilyuklimits} gives us an overview of the
$c_t$ and $c_3$ values that are allowed by the bulk of the parameter
points. \s

We have chosen a few benchmark points from our samples in order to
investigate the validity of the EFT approach. In
Tab.~\ref{tab:smeftbp1}, we present the benchmark point {\tt
  SMEFTBP1} for the R2HDM-II
with the heavy scalar Higgs $M_{H_2}$ mass above 1~TeV so that the EFT
approach should be justified. The corresponding SMEFT coupling
coefficients are
\beq
\mbox{{\tt SMEFTBP1:}} \; c_3 = 0.782\,, \; c_t = 0.951 \,, \;
c_{tt} = -0.122 \;.
\eeq
\begin{table}[h!]
\begin{center}
\begin{tabular}{|c|c|c|c|c|c|c|c|}
\hline
$m_{H_1}$ [GeV] & $m_{H_2}$ [GeV] & $m_A$ [GeV] & $m_{H^\pm}$ [GeV] & $\alpha$ & $\tan\beta$ & $m_{12}^2$ [GeV$^2$] \\ \hline
125.09 & 1131 & 1082 & 1067 & -0.924 & 0.820 & 552749\\ \hline
\end{tabular}
\caption{{\tt SMEFTBP1}: R2HDM-II input parameters.
\label{tab:smeftbp1}}
\end{center}
% This is point 8742
\end{table}
\begin{table}[h!]
\begin{center}
\begin{tabular}{|c|c|c|c|c|c|c|}
\hline
$m_{H_2}$ [GeV] & $\Gamma_{H_2}$ [GeV] & $c_{tt}$ & $g_3^{H_2  H_1 H_1}$ [GeV] & $\sigma_{\text{R2HDM}}^{\text{w/ res}}$ [fb] & $\sigma_{\text{SMEFT}}^{c_{tt}\ne 0}$ [fb] & ratio  \\ \hline
1131 & 78.80 & -0.1222 & -504.52 & 30.5 & 26.1 & 86\% \\ \hline
1200 & 89.74 & -0.1031 & -479.29 & 27.7 & 24.8 & 90\% \\ \hline
1500 & 470.2 & -4.853$\, 10^{-2}$ & -352.42 & 21.8 & 21.4 & 98\% \\ \hline
\end{tabular}
\caption{{\tt SMEFTBP1}: Value of $m_{H_2}$ and corresponding
  $\Gamma_{H_2}$, $c_{tt}$ and $g_3^{H_2  H_1 H_1}$
  values together with the R2HDM-II and the SMEFT results for LO $H_2 H_2$
  production including the resonance contribution.
\label{tab:smeftbp1_add}}
\end{center}
\vspace*{-0.5cm}
\end{table}
In Tab.~\ref{tab:smeftbp1_add} we give, for the listed $m_{H_2}$,
$c_{tt}$, the corresponding R2HDM cross
section values and the results in the SMEFT approach, as well as the
ratios of these two cross sections. Note that $g_3^{H_2  H_1 H_1}$ also
changes when we change $m_{H_2}$, whereas $g_t^{H_2}=-1.126$ remains
the same.  Thus, we list $g_3^{H_2  H_1 H_1}$ in
Tab.~\ref{tab:smeftbp1_add}.
We also give the value of the total width $\Gamma_{H_2}$ which changes
as well. 
Note that in this subsection all gluon fusion cross sections are given
at LO. From the second line, we read off
that in our scenario the SMEFT approach
approximates the cross section in the full model by only 86\% for
a Higgs mass $m_{H_2}$ of the order of 1~TeV. When we turn off the
$H_2$ resonance and compare the results with the one in the SMEFT
approach where we accordingly set $c_{tt}=0$ we get
\beq
\sigma_{\text{R2HDM}}^{\text{w/o res}}=18.6 \mbox{ fb} \quad
\mbox{and} \quad
\sigma_{\text{SMEFT}}^{c_{tt}= 0} = 18.6 \mbox{ fb} \;.
\eeq
Both cross sections agree as expected in contrast to the case with the
resonance included.
Since in the di-Higgs cross section we integrate $\sqrt{s}$ in the
$s$-channel exchange across
the resonance, the SMEFT approach is not a good approximation. We want
to investigate the minimum mass values from which the SMEFT rate is close to the full R2HDM result. For this, we gradually increase
$m_{H_2}$ and calculate for the corresponding $c_{tt}$ and trilinear
Higgs coupling values the SMEFT
cross section and also the full R2HDM cross section.\footnote{The
  resulting scenarios do not then necessarily fulfil all applied constraints
  any more. We still take them for illustrative purposes.} The values are
given in the third and fourth line in Tab.~\ref{tab:smeftbp1_add}. We
clearly see that with increasing $m_{H_2}$, and hence decreasing
contribution of the resonance to the cross section, the SMEFT and the
full R2HDM results approach each other. Starting from about $m_{H_2}=1200$~GeV
the deviation is less than 10\%, continuously decreasing with
increasing $m_{H_2}$. \s

We perform the same investigation but now for the N2HDM-I with $H_1 =
H_{\text{SM}}$ where we have two resonance contributions. As benchmark
point {\tt SMEFTBP2} we take the N2HDM benchmark point {\tt BP6} given
in Tab.~\ref{tab:bpn2hdm1h1}. For convenience, we repeat the input
parameters in Tab.~\ref{tab:smeftbp2}. The SMEFT coupling coefficients read
\beq
\mbox{{\tt SMEFTBP2:}} \; c_3 = 0.877 \,, \; c_t = 1.012 \,, \;
c_{tt} = 4.127 \,\times\, 10^{-2} \;.
\eeq
And we have for this scenario
\beq
g_t ^{H_2} = 0.179 \quad \mbox{and} \quad g_t^{H_3} = 2.337 \,\times\,
10^{-2}\;.
\eeq
\begin{table}
\begin{center}
\begin{tabular}{|c|c|c|c|c|c|}
\hline
$m_{H_1}$ [GeV] & $m_{H_2}$ [GeV] & $m_{H_3}$ [GeV] & $m_{A}$ [GeV]
  & $m_{H^\pm}$ [GeV] & $\tan\beta$ \\ \hline
125.09 & 269 & 582 & 390 & 380 & 4.190 \\ \hline \hline
$\alpha_1$ & $\alpha_2$ & $\alpha_3$ & $v_s$ [GeV] &
$\mbox{Re}(m_{12}^2)$ [GeV$^2$] & \\ \hline
1.432 & -0.109 & 0.535 & 1250 & 28112 & \\ \hline
\end{tabular}
\caption{{\tt SMEFTBP2=BP6}: N2HDM-I input parameters
\label{tab:smeftbp2}}
\end{center}
\vspace*{-0.5cm}
% This is point 5094
\end{table}
\begin{table}[h!]
\begin{center}
\begin{tabular}{|c|c|c|c|c|c|c|c|}
\hline
$m_{H_2}$ & $\Gamma_{H_2}$ & $c_{tt}^{H_2}$ & $c_{tt}$ & $g_3^{H_2 H_1 H_1}$ 
  & $\sigma_{\text{N2HDM}}^{\text{w/ res}}$ [fb] & $\sigma_{\text{SMEFT}}^{c_{tt}\ne 0}$ [fb] & ratio \\ \hline
269 & 0.075 & $4.410 \,\times\, 10^{-2}$ & $4.127 \,\times\, 10^{-2}$ & -72.42 & 183.70 & 20.56 &  11\%
  \\ \hline
300 & 0.083 & $3.170\,\times\,10^{-2}$ & $2.877\,\times\,10^{-2}$ &
-64.80 &  162.80 & 21.28 & 13\% \\ \hline
400 & 0.177 & $9.544\,\times\,10^{-3}$ & $6.721\,\times\,10^{-3}$ & -34.68 & 43.33 & 22.60 & 52\% \\ \hline
420 & 0.229 & $6.895\,\times\,10^{-3}$ & $4.063\,\times\,10^{-3}$ &
-27.62 &  31.70  & 22.76 & 72\% \\ \hline
440 & 0.284 & $4.600\,\times\,10^{-3}$ & $1.767\,\times\,10^{-3}$ &
-20.22 &  26.26  & 22.90 & 87\% \\ \hline
450 & 0.315 & $3.564\,\times\,10^{-3}$ & $7.323\,\times\,10^{-4}$ &
-16.39 &  24.84  & 22.96 & 92\% \\ \hline
500 & 2.567 & $-7.132\,\times\,10^{-4}$ & $-3.545\,\times\,10^{-3}$ &
4.05 &  23.56  & 23.22 & 99\% \\ \hline
\end{tabular}
\caption{{\tt SMEFTBP2}: Values of $m_{H_2}$ in GeV, $\Gamma_{H_2}$ in
  GeV, $c_{tt}^{H_2}$, $c_{tt}$, and $g_3^{H_2 H_1 H_1}$ in GeV together with the
  N2HDM-I and the SMEFT result for $H_1 H_1$ production at LO
  including the resonance contribution. 
\label{tab:smeftbp2_add}}
\end{center}
\vspace*{-0.6cm}
\end{table}
The cross section values for the N2HDM and the SMEFT calculation are
given in the second line of Tab.~\ref{tab:smeftbp2_add}. We vary
$m_{H_2}$ together with the
corresponding total width $\Gamma_{H_2}$, accordingly. The
mass $m_{H_3}$ is kept at its original value. Its total width is given
by $\Gamma_{H_3}=15.28$~GeV. We also
list the corresponding $c_{tt}$ value for $H_2$ only, named
$c_{tt}^{H_2}$, as well as the sum $c_{tt}$ of the $H_2$
and $H_3$ contributions, {\it i.e.}~$c_{tt}= c_{tt}^{H_2}+
c_{tt}^{H_3}$ with $c_{tt}^{H_3}=-2.832 \, \times\,10^{-3}$.
We furthermore give the corresponding trilinear coupling $g_3^{H_2 H_1
  H_1}$. For $g_3^{H_3 H_1 H_1}$ which does not change as we keep
$m_{H_3}$ at its original value, we have $g_3^{H_3 H_1 H_1}=167$~GeV. With a rather
light $H_2$ mass and a medium-valued $H_3$ mass we expect significant
resonance contributions. This was already confirmed by the
investigation of this parameter point in Subsec.~\ref{sec:bp6} where
we found that the resonance contribution is given by the $s$-channel
$H_2$ exchange whereas the $H_3$ resonance contribution is negligible
which can be explained by the tiny $H_3$ top-Yukawa coupling. Due to
the large resonance contribution of the rather light $H_2$ the result
in the SMEFT approach is completely off. When we turn off the
$H_2$ and $H_3$ resonances (where $H_3$ has only a tiny effect) and
compare the result with the one in the SMEFT 
approach where we accordingly set $c_{tt}=0$ we obtain
\beq
\sigma_{\text{N2HDM}}^{\text{w/o res}}= 23.05 \mbox{ fb} \quad
\mbox{and} \quad
\sigma_{\text{SMEFT}}^{c_{tt}= 0} = 23.01 \mbox{ fb} \;.
\eeq
The cross sections in the two approaches agree as expected.
Starting from our original N2HDM-I scenario we then gradually increase the
$m_{H_2}$ mass which hence changes $c_{tt}^{H_2}$ and thereby $c_{tt}$
in order to investigate when the SMEFT result starts to reproduce the full
result. The corresponding values are given in
Tab.~\ref{tab:smeftbp2_add} from the third line onwards. The
SMEFT and the N2HDM results start to deviate by less than 10\% for
$H_2$ masses above about 440~GeV. This agreement also depends on the
total width $\Gamma_{H_2}$ of the $s$-channel resonance. Keeping {\it
  e.g.}~the total width at the value $\Gamma_{H_2}=0.075$~GeV
(corresponding to the mass $m_{H_2}=269$~GeV) 
the agreement between full and EFT approach within 10\% would be
reached around $M_{H_2}=465$~GeV.\footnote{In our investigation of
  different benchmark scenarios we also found cases where the total
  width has a much more dramatic effect, moving agreement {\it
    e.g.}~from a resonance mass of 410~GeV to 2.1~TeV if the width is
  kept at its original value.} We hence find in this scenario with two
possible heavy resonances that the Higgs mass limit, from which the SMEFT
approach starts to approximate the full result, ranges at lower values. \s

The investigation of these two benchmarks with additional Higgs bosons
has shown that first, the results calculated in the full theory and in
the EFT approach can differ severly. Second, the agreement between the
full theory and the EFT approach depends on the masses of the additionally present Higgs bosons and their
total widths. The total width plays an important role when we integrate
across the resonance in the $s$-channel within {\tt HPAIR}. A priori, one
cannot predict to which extent the full theory and the EFT
approach agree, as this depends on the parameters of the
full model.

%%%%%%%%%%%%%%%%%%%%%%%%%%%%%%%%%%%%%%%%%%%%%%%%%%%%%%%%%%%%%%
\section{Mixed Higgs Pair Final States - $H_{\text{SM}}+\Phi$ \label{sec:mixed}}
In all presented models it is possible to produce Higgs pair final
states that consist of a SM-like Higgs boson $H_{\text{SM}}$ plus a non-SM-like
one $\Phi$. In the R2HDM, N2HDM and
the NMSSM the non-SM-like Higgs boson can be a scalar or a pseudoscalar. In
the C2HDM it would be a CP-mixed state. There is a plethora of
final states with substantial production rates possible depending on
the major decay modes of the non-SM-like Higgs boson. They range
from pure multi-fermion, multi-photon, mixed fermion-photon to
multi-Higgs boson, multi-gauge boson, mixed Higgs-plus-gauge boson or
mixed fermion-plus-Higgs or gauge boson final states. In the NMSSM, we
can additionally have supersymmetric particles in the final state
which we will not discuss in this paper. In the following, we
present some selected benchmark points. All of them have the common
feature that their rates exceed 10~fb at NLO. We have many more final state
signatures beyond the presented ones that we can provide on request.

\subsection{The $(b\bar{b})(b\bar{b})$ Final State}
The largest $4b$ final state from mixed Higgs pair production is found
in the N2HDM-I. We present a benchmark point for
\beq
\mbox{\underline{N2HDM-I:} } \quad \sigma(pp \to H_1 H_2(\equiv H_{\text{SM}}) \to
(b\bar{b})(b\bar{b}) ) = 2.11 \mbox{ pb}\;.
\eeq
The related branching ratios are BR$(H_1 \to b\bar{b})=0.857$ and
BR$(H_2 \to b\bar{b})=0.559$.
In Tab.~\ref{tab:bpn2hdm4b} (upper), we give the input parameters for this
point and in Tab.~\ref{tab:bpn2hdm4b} (lower) the $H_1 H_2$ Higgs pair cross
section value at NLO QCD, the total widths of the particles and other interesting
final states rates. The SM Higgs pair ($H_2 H_2$) cross section in this case
is very SM-like with 18~fb at LO, and $H_1 H_1$ production reaches
778~fb at LO and is hence rather large. \s

\begin{table}[h!]
\begin{center}
\begin{tabular}{|c|c|c|c|c|c|}
\hline
$m_{H_1}$ [GeV] & $m_{H_2}$ [GeV] & $m_{H_3}$ [GeV] & $m_{A}$ [GeV]
  & $m_{H^\pm}$ [GeV] & $\tan\beta$ \\ \hline
54 & 125.09 & 229 & 664 & 676 & 1.442 \\ \hline \hline
$\alpha_1$ & $\alpha_2$ & $\alpha_3$ & $v_s$ [GeV] & $m_{12}^2$
 [GeV$^2$] & \\ \hline
-0.989 & 1.010 & 0.381 & 1420 & 5760 &
 \\ \hline
\end{tabular}

\vspace*{0.4cm}
\begin{tabular}{|c|c|c|c|c|c|}
\hline
$\sigma_{H_1 H_2(\equiv H_{\text{SM}})}^{\text{NLO}}$ [pb] & $\Gamma^{\text{tot}}_{H_1}$ [GeV] &
$\Gamma^{\text{tot}}_{H_2}$ [GeV] & $\Gamma^{\text{tot}}_{H_3}$ [GeV] &
$\Gamma^{\text{tot}}_{A}$ [GeV]  & $\Gamma^{\text{tot}}_{H^\pm}$ [GeV] \\ \hline
 4.403 & 3.062 $\times\,10^{-4}$ & 3.947 $\times\,10^{-3}$ &
0.248 & 80.67 & 85.89
  \\ \hline \hline
$(b\bar{b})(\tau\bar{\tau})$ [fb] & $(\tau\bar{\tau})(b\bar{b})$ [fb]
&  $(b\bar{b})(\gamma\gamma)$ [fb] & $(\gamma\gamma)(b\bar{b})$ [fb] & $(b\bar{b})(WW)$ [fb] & $(WW)(b\bar{b})$ [fb]
\\ \hline
227 & 190 & 8& 0.03 & 790 & 0.01\\ \hline%\hline 

%K-factor & & & & & \\ \hline
%2.09 & & & & & \\
%\hline  
\end{tabular}
\caption{\underline{{\tt BP12}} Upper: N2HDM-I input
    parameters. Lower: Further information on this
    point.
\label{tab:bpn2hdm4b}}
\end{center}
% This is point 73; BR(H1tautau)=0.07698, BR(H1gamgam)=1.1766e-05,
% BR(H1WW)=4.383e06; BR(H2tautau)=0.0602, BR(H2gamgam)=0.002128,
% BR(H2WW)=0.209.
\vspace*{-0.6cm}
\end{table}
\begin{table}
\begin{center}
\begin{tabular}{|c|c|c|c|c|c|c|}
\hline
Model & Mixed Higgs State & $m_\text{res.}$ [GeV] & res. rate [fb]  & $m_\Phi$ [GeV] & Rate [fb] & $K$-factor \\ \hline \hline
R2HDM-I & $A H_1(\equiv H_{\text{SM}})$ & --- & --- & 82 & 46 & 2.02 \\ % 3794 x
& $H_1 H_2(\equiv H_{\text{SM}})$ & --- & ---&68 & 35 & 1.97 \\ \hline  % 930 x
C2HDM-I & $H_2 H_1(\equiv H_{\text{SM}})$&266& 9 & 128 & 19 & 2.02\\ % 173 x
& $H_1 H_2(\equiv H_{\text{SM}})$ &---&---&122 & 14 & 2.01\\ % 79 x
& $H_1 H_3(\equiv H_{\text{SM}})$ &---&---& 99 & 11 & 1.96 \\ \hline % 1895 x
N2HDM-I
& $H_2 H_1(\equiv H_{\text{SM}})$ & 360& 109& 146 & 105 & 2.01 \\ % 3611 x
& $A H_1(\equiv H_{\text{SM}})$ &---&---& 75 & 830 & 2.06 \\ % 100 x
& $H_1 H_2(\equiv H_{\text{SM}})$ &229&2260& 54 & 2110 & 2.09 \\ % 73 x
& $A H_2(\equiv H_{\text{SM}})$ &---&---& 101 & 277 & 2.04\\ % 257 x
& $H_1 H_3(\equiv H_{\text{SM}})$ &---&---& 73 & 44 & 1.97 \\ % 64 x
 & $H_2 H_3(\equiv H_{\text{SM}})$ &---&---& 83 & 30 & 1.97 \\ % 85 x
& $A H_3(\equiv H_{\text{SM}})$ &---&---& 69 & 19  & 2.01\\ \hline % 246 x
N2HDM-II &
$H_1 H_2(\equiv H_{\text{SM}})$ &640 &18&103 & 18 & 1.86\\ \hline% 25 x
NMSSM & $A_1 H_1(\equiv H_{\text{SM}})$ &553&210& 113  & 201 & 1.92 \\ % 5109 x
& $H_2 H_1(\equiv H_{\text{SM}})$ & 535 & 42 & 167 & 43 & 1.91\\ % 1143 x
& $A_1 H_2(\equiv H_{\text{SM}})$ &511&42& 87 & 40 & 1.94 \\ % 9934 x
& $H_1 H_2(\equiv H_{\text{SM}})$ &714&58& 80 & 59 & 1.90 \\ \hline % 9512 x
\end{tabular}
\caption{Maximum rates at NLO QCD in the $4b$ final
  state for different mixed Higgs pair final states in the
  investigated models; the corresponding $K$-factor is given in the
  last column. In case of resonantly enhanced production, we
  give in the third and fourth column, respectively, the mass of the resonant Higgs boson and
  the resonant cross section as defined in the text. The fifth
  column contains the mass of the non-SM-like final state Higgs
  boson. More details on these points can be provided on request.} \label{tab:mixed4b}
\end{center}
\vspace*{-0.5cm}
\end{table}
The same channel in the N2HDM-II reaches 18~fb. We present in
Tab.~\ref{tab:mixed4b} for the $4b$ final state the maximum NLO QCD
values in the heavy loop particle limit together with the respective $K$-factors, for these channels and other mixed 
Higgs pair combinations in the models under
investigation. In the case of resonantly enhanced di-Higgs 
  cross sections, we give the mass of the ``resonant'' Higgs boson and the corresponding
resonant production cross section. It is obtained by calculating the
production cross section of the ``resonant'' particle with {\tt SusHi} at
NNLO QCD and subsequently multiplying it with the branching ratio into
the investigated Higgs pair final state.

\subsection{The $(b\bar{b})(WW)$ Final State}
If the SM-like Higgs boson decays into $WW$ then the rates  are
easily obtained from those of the previous subsection in the $4b$ final state
by multiplying them with $\mbox{BR} (H_{\text{SM}} \to WW)/\mbox{BR}
(H_{\text{SM}} \to b\bar{b}) \approx 1/3$. However, we can also have the case
that the non-SM-like Higgs boson decays into $WW$, which are
the benchmark points that we list here. The maximum rate (at NLO) is obtained
for
\beq
\mbox{\underline{N2HDM-I:} } \quad \sigma(pp \to H_1 H_2(\equiv H_{\text{SM}}) \to
(WW)(b\bar{b}) ) = 590 \mbox{ fb} \;. % 1268
\eeq
The related branching ratios are given by BR($H_1 \to WW$)
 = 0.402 and BR($H_2 \to bb$) = 0.598.
The input parameters of the corresponding benchmark point and
additional relevant information together with the rates in other final
states are given in
Tabs.~\ref{tab: bpn2hdm2b2W} (upper) and (lower).
The maximum rates at NLO QCD for all investigated models in the various final
state Higgs pair combinations, where the non-SM-like Higgs decays into
$WW$, are summarized in Tab.~\ref{tab:other2b2W} (provided they exceed
10~fb at NLO). The approximate rates for the $4W$ final state are obtained from those
given in the table by multiplying them with a factor $1/3$.

\begin{table}[h!]
\begin{center}
\begin{tabular}{|c|c|c|c|c|c|}
\hline
$m_{H_1}$ [GeV] & $m_{H_2}$ [GeV] & $m_{H_3}$ [GeV] & $m_{A}$ [GeV]
  & $m_{H^\pm}$ [GeV] & $\tan\beta$ \\ \hline
113 & 125.09 & 304 & 581 & 581 & 1.804 \\ \hline \hline
$\alpha_1$ & $\alpha_2$ & $\alpha_3$ & $v_s$ [GeV] & $m_{12}^2$
 [GeV$^2$] & \\ \hline
0.173 & 1.276 & -0.651 & 414 & 999 &
 \\ \hline
\end{tabular}

\vspace*{0.4cm}
\begin{tabular}{|c|c|c|c|c|c|}
\hline
$\sigma_{H_1 H_2(\equiv H_{\text{SM}})}^{\text{NLO}}$ [pb] & $\Gamma^{\text{tot}}_{H_1}$ [GeV] &
$\Gamma^{\text{tot}}_{H_2}$ [GeV] & $\Gamma^{\text{tot}}_{H_3}$ [GeV] & $\Gamma^{\text{tot}}_{A}$ [GeV]  & $\Gamma^{\text{tot}}_{H^\pm}$ [GeV] \\ \hline
2.453 & 1.691$\times\,10^{-5}$  & 4.103$\times\,10^{-3}$ & 0.477  &
30.41 & 32.10 \\ \hline  \hline
$(b\bar{b})(\tau\bar{\tau})$ [fb] & $(\tau\bar{\tau})(b\bar{b})$ [fb]
&  $(b\bar{b})(\gamma\gamma)$ [fb] & $(\gamma\gamma)(b\bar{b})$ [fb] & $(b\bar{b})(WW)$ [fb] & $(WW)(b\bar{b})$ [fb] \\ \hline
67 & 66 & 2 & 23 & 210 & 590\\ \hline
\end{tabular}
\caption{\underline{{\tt BP13}} Upper: N2HDM-I input
    parameters. Lower: Further information on this
    point.
\label{tab: bpn2hdm2b2W}}
% This is point 440
\end{center}
\vspace*{-0.6cm}
\end{table}
\begin{table}[h!]
\begin{center}
\begin{tabular}{|c|c|c|c|c|c|c|}
\hline
Model & Mixed Higgs State  & $m_\text{res.}$ [GeV] & res. rate [fb]  & $m_\Phi$ [GeV] & Rate [fb] & $K$-factor \\ \hline \hline
N2HDM-I
& $H_2 H_1(\equiv H_{\text{SM}})$&406& 497& 179 & 498 &1.98 \\ % 1093 x
& $H_1 H_2(\equiv H_{\text{SM}})$&304& 615 & 113 & 590 & 2.04  \\ \hline% 440 x
%\textcolor{red}{REMOVE}& $H_3 H_2(\equiv H_{\text{SM}})$ & 11 & 1.94 \\  % 1694 x
NMSSM & $H_2 H_1(\equiv H_{\text{SM}})$ &531&45& 205 & 47 & 1.92 \\ \hline % 120 x
\end{tabular}
\caption{Maximum rates at NLO QCD in the $(b\bar{b})(WW)$ final
  state for different mixed Higgs pair final states in the
  investigated models with the non-SM-like
  Higgs decaying into $WW$; the corresponding $K$-factor is given in the
  last column. In case of resonantly enhanced production, we
  give in the third and fourth column, respectively, the mass of the
  resonant Higgs boson and 
  the resonant cross section as defined in the text. The fifth
  column contains the mass of the non-SM-like final state Higgs
  boson. More details on these points can be provided on request. \label{tab:other2b2W}}
\end{center}
\vspace*{-0.6cm}
\end{table}

%%%%%%%%%%%%%%%%%%%%%%%%%%%%%%%%%%%%%%%%%%%%%%%%%%%%%%%%%%%%%%
\subsection{The $(b\bar{b})(t\bar{t})$ Final State}
As the SM-Higgs decay into $t\bar{t}$ is kinematically forbidden, it
is always the non-SM-like Higgs that decays into $t\bar{t}$. We find
the maximum rate for
\beq
\mbox{\underline{N2HDM-I:} } \quad \sigma(pp \to H_2 H_1(\equiv H_{\text{SM}}) \to
(t\bar{t})(b\bar{b}) ) = 88 \mbox{ fb} \;. % 3337
\eeq
The related branching ratios are given by BR$(H_1 \to b\bar{b})$ = 0.595 and
BR($H_2 \to t\bar{t}$) = 0.902.
Information on this benchmark point, together with the rates into other
final states, is given in
Tabs.~\ref{tab:bpn2hdm2b2t} (upper) and (lower). The
maximum rates at NLO QCD into $(b\bar{b})(t\bar{t})$ for all investigated models
in the various
final state Higgs pair combinations are listed in Tab.~\ref{tab:other2b2t}
for the cases that exceed 10~fb at NLO.
\begin{table}[h!]
\begin{center}
\begin{tabular}{|c|c|c|c|c|c|}
\hline
$m_{H_1}$ [GeV] & $m_{H_2}$ [GeV] & $m_{H_3}$ [GeV] & $m_{A}$ [GeV]
  & $m_{H^\pm}$ [GeV] & $\tan\beta$ \\ \hline
125.09 & 443.65 & 633.69 & 445.65 & 584.34 & 1.570 \\ \hline \hline
$\alpha_1$ & $\alpha_2$ & $\alpha_3$ & $v_s$ [GeV] & $\mbox{Re}(m_{12}^2)$
 [GeV$^2$] & \\ \hline
1.027 & -0.046 & -0.832 & 9361 & 52724 &
 \\ \hline
\end{tabular}

\vspace*{0.4cm}
\begin{tabular}{|c|c|c|c|c|c|}
\hline
$\sigma_{H_1(\equiv H_{\text{SM}}) H_2}$ [fb] & $\Gamma^{\text{tot}}_{H_1}$ [GeV] &
$\Gamma^{\text{tot}}_{H_2}$ [GeV] & $\Gamma^{\text{tot}}_{H_3}$ [GeV] & $\Gamma^{\text{tot}}_{A}$ [GeV]  & $\Gamma^{\text{tot}}_{H^\pm}$ [GeV] \\ \hline
 164 & 4.155 $\times\,10^{-3}$ & 1.303 &
16.05 & 7.603 & 14.32 \\ \hline \hline
$(b\bar{b})(\tau\bar{\tau})$ [fb] & $(\tau\bar{\tau})(b\bar{b})$ [fb]
&  $(b\bar{b})(\gamma\gamma)$ [fb] & $(\gamma\gamma)(b\bar{b})$ [fb] & $(b\bar{b})(WW)$ [fb] & $(WW)(b\bar{b})$ [fb] \\ \hline
0.01 & 0.01 & 0.001 & 0 & 4 & 0.02
\\ \hline
\end{tabular}
\caption{\underline{{\tt BP14}} Upper: N2HDM-I input
  parameters. Lower: Further information on this
  point.
\label{tab:bpn2hdm2b2t}}
\end{center}
\vspace*{-0.8cm}
\end{table}
\begin{table}[h!]
\begin{center}
\begin{tabular}{|c|c|c|c|c|c|c|}
\hline
Model & Mixed Higgs State  & $m_\text{res.}$ [GeV] & res. rate [fb] & $m_\Phi$ [GeV] & Rate [fb] & $K$-factor\\ \hline \hline
R2HDM-I
%$H_2 H_1(\equiv H_{\text{SM}})$ & 12.19 \\
& $A H_1(\equiv H_{\text{SM}})$&---&--- & 346 & 11 & 1.94 \\ \hline  % 6807 x
N2HDM-I
& $H_2 H_1(\equiv H_{\text{SM}})$ &634&81& 444 & 88& 1.86 \\ % 3337 x
& $A H_1(\equiv H_{\text{SM}})$ &---&---& 363 &15 & 1.90 \\ \hline % 750 x
%& $A H_2(\equiv H_{\text{SM}})$ & 11.25 \\ \hline
N2HDM-II
& $H_2 H_1(\equiv H_{\text{SM}})$ &813&23&511 & 34 & 1.79 \\ \hline % 3771 x
NMSSM
& $A_1 H_1(\equiv H_{\text{SM}})$ &---&---& 53 & 82 & 1.88\\ % 2251
& $H_2 H_1(\equiv H_{\text{SM}})$ &535&19& 371 & 19 &1.91 \\ \hline % 4150
\end{tabular}
\caption{
Maximum rates at NLO QCD in the $(b\bar{b}) (t\bar{t}) $ final state at
  NLO for different mixed Higgs
  pair final states in the investigated models with the non-SM-like
  Higgs decaying into $t\bar{t}$; the corresponding $K$-factor is given in the
  last column. In case of resonantly enhanced production, we
  give in the third and fourth column, respectively, the mass of the resonant Higgs boson and
  the resonant cross section as defined in the text. The fifth
  column contains the mass of the non-SM-like final state Higgs
  boson. More details on these points can be provided on request. \label{tab:other2b2t}}
\end{center}
\vspace*{-0.8cm}
\end{table}

\subsection{Multi-Higgs Final States}
In non-minimal Higgs models like the C2HDM, N2HDM, and NMSSM we can
have multi-Higgs final states from cascade Higgs-to-Higgs decays. In
the production of a SM-like plus non-SM-like Higgs final state,
$H_{\text{SM}} \Phi$, we found that both the Higgs-to-Higgs decay of
the SM-like Higgs or the non-SM-like one can lead to substantial final
state rates. The largest NLO rates that we found above 10~fb, in the multi-Higgs final
state, are summarised in Tab.~\ref{tab:multihiggs}. In the
C2HDM, we did not find NLO rates above 10~fb. We maintain the ordering of particles with regards to their decay chains, so that it becomes clear which Higgs boson decays into which Higgs pair. We give the rates in the
$(6b)$ final state as they lead to the largest cross sections for all
shown scenarios. In the following, we highlight a few benchmark
scenarios from the table. 

\begin{table}[h!]
\begin{center}
{\small \begin{tabular}{|c|c|c|c|c|c|}
\hline
Model & Mixed Higgs State & $m_{\Phi_1}$ [GeV] & $m_{\Phi_2}$ [GeV]
  & Rate [fb] & $K$-factor \\ \hline \hline
%C2HDM-I & $H_3 H_1 (\equiv H_{\text{SM}}) \to (b\bar{b}) H_1 H_1 \to (b\bar{b})
%  (b\bar{b}) (b\bar{b})$ & 3
%% xxx 1778, sig_LO(H1H3)=4.59/2, BR(H3H1H1)=0.966, BR(H1->bb)=0.845=>3
%\\ \hline
N2HDM-I
& $H_2 H_3(\equiv H_{\text{SM}}) \to H_1 H_1  (b\bar{b}) \to
  (b\bar{b}) (b\bar{b}) (b\bar{b})$ & 98 & 41 & 15& 1.95
  \\ % xxx 86 M_H1=40.52, M_H2 =98.28, BR(H1->bb)=0.8640, BR(H3->bb)=0.55,
     % BR(H2->H1H1)=0.648, sigma(H2H3)=2*27.83 fb
& $H_2 H_1(\equiv H_{\text{SM}}) \to H_1 H_1 (b\bar{b})\to (b\bar{b})
  (b\bar{b}) (b\bar{b})$ & 282 & - & 40 & 1.96 \\ % xxx 5265; M_H2= 282, BR(H2-> H1H1)= 0.369
% BR(H1\to bb)=0.597
& $H_2 H_1(\equiv H_{\text{SM}}) \to A A (b\bar{b}) \to (b\bar{b})
  (b\bar{b}) (b\bar{b})$ & 157 & 73 & 33 & 2.05\\ % xxx 7938 M_H2=
                                % 156.93, M_A = 72.69 
% BR(A\to bb)=0.773, BR(H2->AA)=0.993, BR(A->bb)=0.773; 7938
& $H_1 H_2(\equiv H_{\text{SM}}) \to (b\bar{b}) H_1 H_1 \to (b\bar{b})
  (b\bar{b}) (b\bar{b})$ & 54 & - & 111& 2.09 \\
% xxx 73; M_H1 = 53.77, BR(H_1 \to bb)=0.8567, BR(H_2 \to H_1
% H_1)=0.04011, BR(H2->bb)=0.559; sig(H1 H2) = 2.10443
& $H_3 H_2(\equiv H_{\text{SM}}) \to H_1 H_1 (b\bar{b})\to (b\bar{b})
  (b\bar{b}) (b\bar{b})$ & 212 & 83 & 8 & 1.93  \\% xxx 1684 M_H1=77.04, M_H3=173.98, BR(H3\to
% H1H1)=0.754, BR(H1\to bb)=0.83587, BR(H2->bb)=0.578
%\textcolor{red}{REMOVE}& $H_3 H_2(\equiv H_{\text{SM}}) \to H_1 H_2 (b\bar{b})\to (b\bar{b})  (b\bar{b}) (b\bar{b})$  & 6 & 1.91\\% xxx 559 M_H1=117.55, M_H3=280.84, BR(H3\to
                           % H1H2)=0.93717, BR(H1\to bb)=0.761046138 H2bb = 0.58664929
\hline
N2HDM-II
& $H_2 H_1(\equiv H_{\text{SM}}) \to H_1 H_1 (b\bar{b}) \to (b\bar{b})
  (b\bar{b}) (b\bar{b}) $ & 271 & - & 3 & 1.87\\ % xxx 1316 M_H2=270.81, M_H3=615.04,
                            %BR(H1\tobb)=0.576, BR(H2->H1H1)=0.538,
                            %sig_LO(H1H2)=13.01
\hline
NMSSM
& $H_2 H_1(\equiv H_{\text{SM}}) \to H_1 H_1 (b\bar{b}) \to
  (b\bar{b}) (b\bar{b}) (b\bar{b})$ & 319 & - & 11 & 1.90\\  % xxx 8648 H1->bb: 0.635534596
& $H_2 H_1(\equiv H_{\text{SM}}) \to A_1 A_1 (b\bar{b}) \to
  (b\bar{b}) (b \bar{b}) (b\bar{b})$ & 253 & 116 & 26 & 1.92 \\  % xxx 10059 A->bb:
                                % 0.704142486
\hline
\end{tabular}}

\vspace*{0.4cm}
\begin{tabular}{|c|c|c|c|}
	\hline
	Model & Mixed Higgs State & $m_\text{res.}$ [GeV] & res. rate [fb] 
	 \\ \hline \hline
	%C2HDM-I & $H_3 H_1 (\equiv H_{\text{SM}}) \to (b\bar{b}) H_1 H_1 \to (b\bar{b})
	%  (b\bar{b}) (b\bar{b})$ & 3
	%% xxx 1778, sig_LO(H1H3)=4.59/2, BR(H3H1H1)=0.966, BR(H1->bb)=0.845=>3
	%\\ \hline
	N2HDM-I
	& $H_2 H_3(\equiv H_{\text{SM}}) \to H_1 H_1  (b\bar{b}) \to
	(b\bar{b}) (b\bar{b}) (b\bar{b})$ & --- & ---
	\\ % xxx 86 M_H1=40.52, M_H2 =98.28, BR(H1->bb)=0.8640, BR(H3->bb)=0.55,
	% BR(H2->H1H1)=0.648, sigma(H2H3)=2*27.83 fb
	& $H_2 H_1(\equiv H_{\text{SM}}) \to H_1 H_1 (b\bar{b})\to (b\bar{b})
	(b\bar{b}) (b\bar{b})$ & 441 & 39  \\ % xxx 5265; M_H2= 282, BR(H2-> H1H1)= 0.369
	% BR(H1\to bb)=0.597
	& $H_2 H_1(\equiv H_{\text{SM}}) \to A A (b\bar{b}) \to (b\bar{b})
	(b\bar{b}) (b\bar{b})$ & 294 &  37 \\ % xxx 7938 M_H2=
	% 156.93, M_A = 72.69 
	% BR(A\to bb)=0.773, BR(H2->AA)=0.993, BR(A->bb)=0.773; 7938
	& $H_1 H_2(\equiv H_{\text{SM}}) \to (b\bar{b}) H_1 H_1 \to (b\bar{b})
	(b\bar{b}) (b\bar{b})$ & 229  & 119  \\
	% xxx 73; M_H1 = 53.77, BR(H_1 \to bb)=0.8567, BR(H_2 \to H_1
	% H_1)=0.04011, BR(H2->bb)=0.559; sig(H1 H2) = 2.10443
	& $H_3 H_2(\equiv H_{\text{SM}}) \to H_1 H_1 (b\bar{b})\to (b\bar{b})
	(b\bar{b}) (b\bar{b})$ &---  &---    \\% xxx 1684 M_H1=77.04, M_H3=173.98, BR(H3\to
	% H1H1)=0.754, BR(H1\to bb)=0.83587, BR(H2->bb)=0.578
	%\textcolor{red}{REMOVE}& $H_3 H_2(\equiv H_{\text{SM}}) \to H_1 H_2 (b\bar{b})\to (b\bar{b})  (b\bar{b}) (b\bar{b})$  & 6 & 1.91\\% xxx 559 M_H1=117.55, M_H3=280.84, BR(H3\to
	% H1H2)=0.93717, BR(H1\to bb)=0.761046138 H2bb = 0.58664929
	\hline
	N2HDM-II
	& $H_2 H_1(\equiv H_{\text{SM}}) \to H_1 H_1 (b\bar{b}) \to (b\bar{b})
	(b\bar{b}) (b\bar{b}) $ & 615  & 2 \\ % xxx 1316 M_H2=270.81, M_H3=615.04,
	%BR(H1\tobb)=0.576, BR(H2->H1H1)=0.538,
	%sig_LO(H1H2)=13.01
	\hline
	NMSSM
	& $H_2 H_1(\equiv H_{\text{SM}}) \to H_1 H_1 (b\bar{b}) \to
	(b\bar{b}) (b\bar{b}) (b\bar{b})$ & 560  &  11\\  % xxx 8648 H1->bb: 0.635534596
	& $H_2 H_1(\equiv H_{\text{SM}}) \to A_1 A_1 (b\bar{b}) \to
	(b\bar{b}) (b \bar{b}) (b\bar{b})$ &  518& 26   \\  % xxx 10059 A->bb:
	% 0.704142486
	\hline
\end{tabular}

\caption{Upper: Maximum rates for multi-Higgs final
    states given at NLO QCD. The $K$-factor is given in the last column. In the
  third and fourth column we also give the mass values $m_{\Phi_1}$ and
  $m_{\Phi_2}$ of the non-SM-like Higgs
  bosons involved in the process, in the order of their
  appearance. Lower: In case of resonantly enhanced production the mass of the
  resonantly produced Higgs boson is given together with the NNLO QCD
  production rate. More details on these points can be provided on request. \label{tab:multihiggs}} 
\end{center}
\vspace*{-0.8cm}
\end{table}

\subsubsection{Non-SM-like Higgs Search: Di-Higgs beats Single Higgs}
In the following we present N2HDM-I and NMSSM scenarios with three
SM-like Higgs bosons in the final
states with $H_1$ being SM-like and with NLO rates above 10~fb. These benchmark
points are special in the sense
that the production of the non-SM-like Higgs boson $H_2$ from di-Higgs
states beats, or is at least comparable, to its direct
production.\footnote{For another example where New Physics might first be
  accessible in Higgs pair production in a composite Higgs model, see
  \cite{Grober:2016wmf}.} This
appears in cases where the non-SM-like Higgs is singlet-like and/or
is more down- than up-type like. The latter suppresses direct production from
gluon fusion. The former suppresses all couplings to SM-like
particles. In these cases the heavy non-SM-like Higgs
boson might rather be discovered in 
the di-Higgs channel than in direct single Higgs production. \s

The input parameters for the N2HDM-I point are given in Tab.~\ref{tab:3sm-I}.
\begin{table}[h!]
\begin{center}
\begin{tabular}{|c|c|c|c|c|c|}
\hline
$m_{H_1}$ [GeV] & $m_{H_2}$ [GeV] & $m_{H_3}$ [GeV] & $m_{A}$ [GeV]
  & $m_{H^\pm}$ [GeV] & $\tan\beta$ \\ \hline
125.09 & 281.54 & 441.25 & 386.98 & 421.81 & 1.990
\\ \hline \hline
$\alpha_1$ & $\alpha_2$ & $\alpha_3$ & $v_s$ [GeV] &
$\mbox{Re}(m_{12}^2)$ [GeV$^2$] & \\ \hline
1.153 & 0.159 & 0.989 & 9639 & 29769
& \\ \hline
\end{tabular}
\caption{\underline{{\tt BP15}} N2HDM-I input parameters
\label{tab:3sm-I}}
% Was formerly point 6731. This is excluded by new data. Now it is
% point 5265.
\end{center}
\vspace*{-0.4cm}
\end{table}
With the values for the NLO $H_1 H_2$ cross section and the
branching ratios $\mbox{BR}(H_2\to H_1 H_1)$ and
$\mbox{BR}(H_1 \to b\bar{b})$ we get the following rate in the $6b$
final state,
\beq
\sigma_{H_1 H_2}^{\text{NLO}} \times \mbox{BR}(H_2\to H_1 H_1) \times
\mbox{BR}(H_1 \to b\bar{b})^3 = 509 \cdot 0.37 \cdot
0.60^3 \mbox{ fb} = 40 \mbox{ fb} \;.
\eeq
We can compare this with direct $H_2$ production (we use the NNLO
value calculated with {\tt SusHi}) in either the $4b$
final state from the $H_2 \to H_1 H_1$ decay,
\beq
\sigma^{\text{NNLO}} (H_2) \times \mbox{BR}(H_2 \to H_1 H_1) \times
\mbox{BR}(H_1\to b\bar{b})^2 = 161 \cdot 0.37 \cdot 0.60^2 \mbox{ fb}
= 21 \mbox{ fb} \;, 
\eeq
or direct $H_2$ production in the other dominant decay channel given
by the $WW$ final state,
\beq
\sigma^{\text{NNLO}}  (H_2) \times \mbox{BR}(H_2 \to WW) = 161 \cdot 0.44 \mbox{
  fb} = 71 \mbox{ fb} \;.
\eeq
Note that the $H_2$ branching ratio into $(b\bar{b})$ is tiny.
The second lightest Higgs boson $H_2$
has a significant down-type and large singlet admixture but only a small
up-type admixture so that its production in gluon fusion
is not very large\footnote{The production in association with
  $b$ quarks is very small for the small $\tan\beta$ value of this
  scenario.} and also its decay branching ratios into a lighter Higgs
pair are comparable to the largest decay rates into SM particles. In
this case, the non-SM-like Higgs boson $H_2$ has better chances of being discovered
in di-Higgs when compared to single Higgs channels. Note, that the $W$ bosons
still need to decay into fermionic final
states where additionally the neutrinos are not detectable so that the
$H_2$ mass cannot be reconstructed. \s

The input parameters for the first NMSSM scenario that we discuss here
are given in Tab.~\ref{tab:beat1}. We also specify in Tab.~\ref{tab:beat2} the
parameters required for the
computation of the Higgs pair production cross sections through {\tt
  HPAIR}. \s
\begin{table}[h!]
\begin{center}
\begin{tabular}{|c|c|c|c|c|c|}
\hline
$\lambda$ & $\kappa$ & $A_\lambda$ [GeV] &  $A_\kappa$ [GeV] &
$\mu_{\text{eff}}$ [GeV] & $\tan\beta$
\\ \hline
0.593 & 0.390 & 296 & 5.70 & 200 & 2.815
\\ \hline
$m_{H^\pm}$ [GeV] & $M_1$ [GeV] & $M_2$ [GeV] & $M_3$ [TeV] & $A_t$
[GeV] & $A_b$ [GeV] \\ \hline
505 & 989.204 & 510.544 & 2 &-2064 & -1246\\ \hline
$m_{\tilde{Q}_3}$ [GeV] & $m_{\tilde{t}_R}$ [GeV] & $m_{\tilde{b}_R}$ [GeV] & $A_\tau$
[GeV] & $m_{\tilde{L}_3}$  [GeV] & $m_{\tilde{\tau}_R}$ [GeV] \\ \hline
1377 & 1207 & 3000 & -1575.91 & 3000 & 3000 \\ \hline
\end{tabular}
\caption{\underline{{\tt BP16}} NMSSM input parameters required by
  {\tt NMSSMCALC} for the
  computation of the NMSSM spectrum.  \label{tab:beat1}}
% 10059
\end{center}
\vspace*{-0.8cm}
\end{table}
\begin{table}[h!]
\begin{center}
\begin{tabular}{|c|c|c|c|c|}
\hline
$m_{H_1}$ [GeV] & $m_{H_2}$ [GeV] & $m_{H_3}$ [GeV] & $m_{A_1}$ [GeV] &
$m_{A_2}$ [GeV] \\ \hline
127.78 & 253 & 518 & 116 & 508
\\ \hline
$\Gamma^{\text{tot}}_{H_1}$ [GeV] & $\Gamma^{\text{tot}}_{H_2}$ [GeV] &
$\Gamma^{\text{tot}}_{H_3}$ [GeV] & $\Gamma^{\text{tot}}_{A_1}$ [GeV] &
$\Gamma^{\text{tot}}_{A_2}$ [GeV]
\\ \hline
  4.264 $10^{-3}$ & 0.466 & 3.145 & $9.9 10^{-7}$ & 4.750
\\ \hline
$h_{11}$ & $h_{12}$ & $h_{13}$ & $h_{21}$ & $h_{22}$ \\
\hline
0.325 & 0.939 & -0.112 & 0.234 & 0.034
\\ \hline
$h_{23}$ & $h_{31}$ & $h_{32}$ & $h_{33}$ & $a_{11}$  \\ \hline
0.971 & 0.916 & -0.321 & -0.209 & -0.0063
\\ \hline
$a_{21}$ & $a_{13}$ & $a_{23}$ & & \\
\hline
-0.0022 & 0.999 & 0.0067 & & \\ \hline
\end{tabular}
\caption{\underline{{\tt BP16}} These input parameters and those given in the first line of
  Tab.~\ref{tab:beat1} are required by {\tt HPAIR} for the computation of the
  Higgs pair production cross sections. The total width of the charged
  Higgs boson is not required but given here for completeness
  $\Gamma^{\text{tot}}_{H^\pm}=3.94$ GeV. \label{tab:beat2}}
\end{center}
\vspace*{-0.4cm}
\end{table}

Since $H_2$ is rather singlet-like, its production cross section through
gluon fusion is small and also its decay branching ratios into
SM-final states. The gluon fusion production cross section amounts to
\beq
\sigma^{\text{NNLO}}  (H_2) = 13.54 \mbox{ fb} \;.
\eeq
Its dominant branching ratio is given by the decay into $A_1 A_1$,
reaching
\beq
\mbox{BR} (H_2 \to A_1 A_1) = 0.887 \;.
\eeq
We hence get for direct $H_2$ production in the $A_1 A_1$ final state
the rate
\beq
\sigma^{\text{NNLO}}  (H_2) \times \mbox{BR} (H_2 \to A_1 A_1) = 12.01 \mbox{
  fb} \;.
\eeq
On the other hand, we have for di-Higgs production of $H_1 H_2$ at NLO
QCD where $H_1$ is the SM-like Higgs state,
\beq
\sigma^{\text{NLO}} (H_1 H_2) = 111 \mbox{ fb} \;.
\eeq
With
\beq
\mbox{BR} (H_1 \to b\bar{b}) = 0.539,
\eeq
and the $H_2$ branching ratio into $A_1 A_1$ given above we hence have
\beq
\sigma^{\text{NLO}} (H_1 H_2) \times \mbox{BR} (H_1 \to b\bar{b}) \times \mbox{BR}
(H_2 \to A_1 A_1) = 53 \mbox{ fb} \;.
\eeq
With
\beq
\mbox{BR} (A_1 \to b \bar{b}) = 0.704
\eeq
we then obtain in double Higgs production in the $6b$ final state the rate
\beq
\sigma^{\text{NLO}} (H_1 H_2)_{6b} = 53 \times 0.704^2 \mbox{ fb} = 26 \mbox{ fb} \;.
\eeq
On the other hand, we have in single Higgs production for the $4b$ final state
\beq
\sigma^{\text{NNLO}}  (H_2)_{4b} &=& \sigma^{\text{NNLO}}  (H_2) \times
\mbox{BR}(H_2 \to A_1 A_1) \times \mbox{BR}(A_1 \to b\bar{b})^2
\nonumber \\
&=& 13.54 \times 0.887 \times 0.704^2 \mbox{ fb} = 5.95
\mbox{ fb} \;.
\eeq
Note that direct $H_2$ production with subsequent decay into $W^+W^-$ only reaches a rate of 1~fb. We clearly see that
di-Higgs beats single Higgs production and the 
non-SM-like singlet-dominated state $H_2$ might be first discovered in
di-Higgs production instead directly in single $H_2$ production through gluon fusion. \s

For the second NMSSM benchmark scenario that we present here the input
parameters for {\tt NMSSMCALC} and {\tt 
  HPAIR} are summarized in Tabs.~\ref{tab:beat3} and \ref{tab:beat4}.
\begin{table}[h!]
\begin{center}
\begin{tabular}{|c|c|c|c|c|c|}
\hline
$\lambda$ & $\kappa$ & $A_\lambda$ [GeV] &  $A_\kappa$ [GeV] &
$\mu_{\text{eff}}$ [GeV] & $\tan\beta$
\\ \hline
0.545 & 0.598 & 168 & -739 & 258 & 2.255
\\ \hline
$m_{H^\pm}$ [GeV] & $M_1$ [GeV] & $M_2$ [GeV] & $M_3$ [TeV] & $A_t$
[GeV] & $A_b$ [GeV] \\ \hline
548 & 437.872 &  498.548 &2 & -1028& 1083  \\ \hline
$m_{\tilde{Q}_3}$ [GeV] & $m_{\tilde{t}_R}$ [GeV] & $m_{\tilde{b}_R}$ [GeV] & $A_\tau$
[GeV] & $m_{\tilde{L}_3}$  [GeV] & $m_{\tilde{\tau}_R}$ [GeV] \\ \hline
1729 & 1886 & 3000 & -1679.21 & 3000 & 3000 \\ \hline
\end{tabular}
\caption{\underline{{\tt BP17}} NMSSM input parameters required by
  {\tt NMSSMCALC} for the
  computation of the NMSSM spectrum.  \label{tab:beat3}}
% 8648
\end{center}
\vspace*{-0.4cm}
\end{table}
\begin{table}[h!]
\begin{center}
\begin{tabular}{|c|c|c|c|c|}
\hline
$m_{H_1}$ [GeV] & $m_{H_2}$ [GeV] & $m_{H_3}$ [GeV] & $m_{A_1}$ [GeV] &
$m_{A_2}$ [GeV] \\ \hline
123.20 & 319 & 560 & 545 & 783
\\ \hline
$\Gamma^{\text{tot}}_{H_1}$ [GeV] & $\Gamma^{\text{tot}}_{H_2}$ [GeV] &
$\Gamma^{\text{tot}}_{H_3}$ [GeV] & $\Gamma^{\text{tot}}_{A_1}$ [GeV] &
$\Gamma^{\text{tot}}_{A_2}$ [GeV]
\\ \hline
  3.985 $\times\,10^{-3}$ & 0.010 & 4.207 & 6.399 & 6.913
\\ \hline
$h_{11}$ & $h_{12}$ & $h_{13}$ & $h_{21}$ & $h_{22}$ \\
\hline
0.419 & 0.909 & 0.015 & 0.187 & -0.102
\\ \hline
$h_{23}$ & $h_{31}$ & $h_{32}$ & $h_{33}$ & $a_{11}$  \\ \hline
0.977 & 0.889 & -0.407 & -0.212 & 0.908
\\ \hline
$a_{21}$ & $a_{13}$ & $a_{23}$ & & \\
\hline
-0.104 & 0.114 & 0.994 & & \\ \hline
\end{tabular}
\caption{\underline{{\tt BP17}} These input parameters and those given
  in the first line of
  Tab.~\ref{tab:beat1} are required by {\tt HPAIR} for the computation of the
  Higgs pair production cross sections. The total width of the charged
  Higgs boson is not required but given here for completeness
  $\Gamma^{\text{tot}}_{H^\pm}= 5.503$ GeV. \label{tab:beat4}}
\end{center}
\vspace*{-0.4cm}
\end{table}
The singlet-like $H_2$ dominantly decays into an SM-like pair $H_1
H_1$, and  in the $H_1 H_1$ final state we obtain the rate
\beq
\sigma^{\text{NNLO}} (H_2) \times \mbox{BR}(H_2 \to H_1 H_1) =
134.95 \cdot 0.566 \mbox{ fb} = 76.38 \mbox{ fb} \;.
\eeq
With BR$(H_1 \to b\bar{b})= 0.636$ this results in the $4b$ rate
\beq
 \sigma^{\text{NNLO}} (H_2) \times \mbox{BR}(H_2 \to H_1 H_1) \times
 \mbox{BR}^2 (H_1 \to b\bar{b})= 31.00 \mbox{ fb} \;.
\eeq
On the other hand, with BR$(H_2 \to b\bar{b})= 0.103$, we have the $2b$
final state rate
\beq
\sigma^{\text{NNLO}} (H_2) \times \mbox{BR}(H_2 \to b\bar{b}) =
134.95 \cdot 0.104 \mbox{ fb} = 14.03 \mbox{ fb} \;.
\eeq
The rate for direct $H_2$ production in the $4b$ final state via its
decay into $H_1 H_1$ beats the one of direct $H_2$ production in the $2b$ final state
by more than a factor 2. Note finally that the $6b$ rate for $H_2$
production, through $H_1 H_2$ production and further $H_2$ decay into
Higgs pairs, amounts to
\beq
\sigma^{\text{NLO}} (H_1 H_2) \times \mbox{BR} (H_2 \to H_1 H_1)
\times \mbox{BR}^3 (H_1 \to b\bar{b})= 75 \cdot 0.566 \cdot 0.636^3 \mbox{
fb} = 11 \mbox{ fb} \;,
\eeq
which is not much below the $2b$ final state rate.

%%%%%%%%%%%%%%%%%%%%%%%%%%%%%%%%%%%%%%%%%%%%%%%%%%%%%%%%%%%%%%
\section{Non-SM-Like Higgs Pair Final States \label{sec:nonsm}}
For non-SM-like Higgs pair production, we can have a large plethora of all
possible Higgs pair combinations inducing final states with multiple
Higgs bosons, two or three Higgs bosons in association with one or two gauge bosons,
or also with a top-quark pair, resulting finally in multi-fermion,
multi-photon or multi-fermion plus multi-photon final states. We
present a few selected interesting signatures from non-SM-like Higgs
pair production in Tab.~\ref{tab:selected}. More signatures and benchmark points can be
provided on request. As we can infer from the table, we can have high
rates in non-SM-like Higgs pair production, {\it e.g.}~up to 9~pb in
the $4b$ final state from non-SM-like $H_1H_1$ production in the
N2HDM-I with $H_2 \equiv H_{\text{SM}}$. \s

\begin{table}[h!]
\begin{center}
\begin{tabular}{|c|c|c|c|c|c|}
\hline
Model & SM-like Higgs & Signature  & $m_\Phi$ [GeV] & Rate [fb] & $K$-factor \\ \hline \hline
 N2HDM-I
& $H_3$ & $H_1 H_1 \to (b\bar{b}) (b\bar{b})$ & 41 & 14538 & 2.18 \\
% xxx 86 M_H1=40.52, M_H2 =98.28, BR(H1->bb)=0.8640 gamgam=1.6*10^-3, sig(H1H1)=8.9*2 pb
& $H_3$ & $H_1 H_1 \to (4b) ; (4\gamma)$ & 41 & 4545 ; 700 & 2.24 \\
% xxx 151 M_H1=40.55, M_H2 =93.70, BR(H1->gamgam)=0.252 in bb 0.642,
% sig(H1H1)=4.923*2 pb
& $H_1$ & $A A \to (b\bar{b}) (b\bar{b})$ & 75 & 6117 & 2.11 \\
% xxx 100 mH3=323, mH3=1535, mA=75.43, BR(A->bb)=0.77, sig(AA)=2*4.92 pb,
% sig(AH1)= 872 fb
& $H_1$ & $H_2 H_2 \to (b\bar{b}) (b\bar{b})$ & 146 & 73 & 2.01\\
% xxx 3512 mH2=140.53, mH3=381, mA=230, mHp=259, BR(H2->bb)=0.74,
% sig(H2H2)=2*66 fb
%
%%& $H_1$ & $H_2 H_2 \to (H_1 H_1) (H_1 H_1) \to 4 (b\bar{b})$ & 2.13 \\
%% 2495 m_H2=314.53, BR(H_1->bb)=0.595, BR(H2->H1H1)=0.603
%
& $H_2$ & $A A \to (b\bar{b}) (b\bar{b})$& 80 & 2875 & 2.13 \\
% xxx 1078 mH1=79, mH3=178, mA=80, mHp=172, BR(A->bb)=0.76,
% sig(AA)=2*2.36 pb
& $H_2$ & $A H_1 \to (b\bar{b}) (b\bar{b})$ & $m_A:$ 87& 921 & 2.09\\
&            &                                                    & $m_{H_1}:$ 91  & & \\
% xx 615, mH1=91, mH3=637, mA=87, mHp=331, BR(A->bb)=0.74,
% BR(H1->bb)=0.82, sig(AH1)=2*726 fb, sig(H1H1)=2*376, sig(AA)=2*241
& $H_2$ & $H_1 H_1 \to (b\bar{b}) (b\bar{b})$ & 47 & 8968 & 2.17 \\
% xxx 285 mH1=64.66, mH3=660, mA=432, mHp=401, BR(H1->bb)=0.85,
% sig(H1H1)=2*8.27 pb
\hline
N2HDM-II
& $H_2$ & $H_1 H_1 \to (b\bar{b}) (b\bar{b})$ & 44 & 1146 & 2.18 \\
% xx 10 mH1=43.80, mH3=520, BR(H1->bb)=0.853, sig(H1H1)=2*721 fb
\hline
C2HDM-I
& $H_1$ & $H_2 H_2 \to  (b\bar{b}) (b\bar{b})$ & 128 & 475 & 2.07 \\
% xxx 173, mH2= 128.00, sig(H2H2)=580*2 fb, BR(H2->bb)=0.630
& $H_2$ & $H_1 H_1 \to (b\bar{b}) (b\bar{b})$ & 66 & 814 & 2.16 \\
% xxx 1778, mH1=65.81, BR(H1->bb)=0.845, sig(H1H1)=0.528
%
%
%& $H_2$ & $H_3 H_3 \to H_1 H_1 H_1 H_1 \to 4 (b\bar{b})$ & 5.19 \\
% xx 1778 mH1=65.81, mH3=151.05, BR(H3->H1H1)=0.966, sig(H3H3)=2*5.45 fb,
% BR(H1-> bb)=0.845
& $H_3$ & $H_1 H_1 \to (b\bar{b}) (b\bar{b})$ & 84 & 31 & 2.09 \\
% xxx 77, m_H1=84.39, m_H2=109.39, sig(H1H1)=22.14*2 fb, BR(H1bb)=0.825
\hline
NMSSM
& $H_1$ & $A_1 A_1 \to (b\bar{b}) (b\bar{b})$ & 166 & 359 & 1.95 \\
% xxx 1899, mH1=122.89, mA1=166.45, BR(A1->bb)=0.849, sig(A1A1)=255*2
& $H_1$ & $A_1 A_1 \to (\gamma\gamma) (\gamma\gamma)$ & 179 & 34 & 1.96 \\
% xxx 501, mH1=127.35, mA1=155.64, BR(A1->gamgam)=0.645, bb=0.300,
% sig(A1A1)= 2*98.72
& $H_2$ & $H_1 H_1 \to (b\bar{b}) (b\bar{b})$ & 48 & 3359 & 2.18 \\
% xxx 19337 mH1=48.26, mH3=882, mA1=386, mA2=879, sig(H1H1)=2*1.85 pb,
% BR(H1->bb)=0.91
& $H_2$ & $A_1 A_1 \to (b\bar{b}) (b\bar{b})$ & 54 & 1100 & 2.18\\
% xxx 16817 mH1=56.50, mH2=122.99, mA1=53.65, BR(A1->bb)=0.913,
%sig(A1A1)=2*605 fb
& $H_1$ & $A_1 A_1 \to (t\bar{t}) (t\bar{t})$ & 350 & 20 & 1.82 \\
% xxx 6624 mH1=124.67, mA1=349.6, BR(A1->tt)=0.987
\hline
\end{tabular}

\vspace*{0.4cm}
\begin{tabular}{|c|c|c|c|c|c|}
	\hline
	Model& Signature & $m_\text{res.}$ [GeV] & res. rate [fb]& $m_\text{res.}$ 2 [GeV] & res. rate 2 [fb]  \\ \hline \hline
	N2HDM-I
	& $H_1 H_1 \to (b\bar{b}) (b\bar{b})$& 125.09 & 621 &98& 17137\\
	% xxx 86 M_H1=40.52, M_H2 =98.28, BR(H1->bb)=0.8640 gamgam=1.6*10^-3, sig(H1H1)=8.9*2 pb
	& $H_1 H_1 \to (4b) ; (4\gamma)$ & 125.09&126; 19  & 94& 5445;839\\
	% xxx 151 M_H1=40.55, M_H2 =93.70, BR(H1->gamgam)=0.252 in bb 0.642,
	% sig(H1H1)=4.923*2 pb
	& $A A \to (b\bar{b}) (b\bar{b})$ &1535& $<$0.1 & 323&482\\
	% xxx 100 mH3=323, mH3=1535, mA=75.43, BR(A->bb)=0.77, sig(AA)=2*4.92 pb,
	% sig(AH1)= 872 fb
	 & $H_2 H_2 \to (b\bar{b}) (b\bar{b})$&360&76&---&---\\
	% xxx 3512 mH2=140.53, mH3=381, mA=230, mHp=259, BR(H2->bb)=0.74,
	% sig(H2H2)=2*66 fb
	%
	%%& $H_1$ & $H_2 H_2 \to (H_1 H_1) (H_1 H_1) \to 4 (b\bar{b})$ & 2.13 \\
	%% 2495 m_H2=314.53, BR(H_1->bb)=0.595, BR(H2->H1H1)=0.603
	%
	& $A A \to (b\bar{b}) (b\bar{b})$&178&3191&---&---\\
	% xxx 1078 mH1=79, mH3=178, mA=80, mHp=172, BR(A->bb)=0.76,
	% sig(AA)=2*2.36 pb
	 & $A H_1 \to (b\bar{b}) (b\bar{b})$ &  --- & --- &---&---\\
	% xx 615, mH1=91, mH3=637, mA=87, mHp=331, BR(A->bb)=0.74,
	% BR(H1->bb)=0.82, sig(AH1)=2*726 fb, sig(H1H1)=2*376, sig(AA)=2*241
	 & $H_1 H_1 \to (b\bar{b}) (b\bar{b})$ &588&22&125.09&997 \\
	% xxx 285 mH1=64.66, mH3=660, mA=432, mHp=401, BR(H1->bb)=0.85,
	% sig(H1H1)=2*8.27 pb
	\hline
	N2HDM-II
	 & $H_1 H_1 \to (b\bar{b}) (b\bar{b})$ &520& $<$ 0.1 & 125.09& 1330\\
	% xx 10 mH1=43.80, mH3=520, BR(H1->bb)=0.853, sig(H1H1)=2*721 fb
	\hline
	C2HDM-I
	 & $H_2 H_2 \to  (b\bar{b}) (b\bar{b})$&266& 497  &---&---\\
	% xxx 173, mH2= 128.00, sig(H2H2)=580*2 fb, BR(H2->bb)=0.630
	 & $H_1 H_1 \to (b\bar{b}) (b\bar{b})$&151&598&---& ---\\
	% xxx 1778, mH1=65.81, BR(H1->bb)=0.845, sig(H1H1)=0.528
	%
	%
	%& $H_2$ & $H_3 H_3 \to H_1 H_1 H_1 H_1 \to 4 (b\bar{b})$ & 5.19 \\
	% xx 1778 mH1=65.81, mH3=151.05, BR(H3->H1H1)=0.966, sig(H3H3)=2*5.45 fb,
	% BR(H1-> bb)=0.845
	 & $H_1 H_1 \to (b\bar{b}) (b\bar{b})$&---&---&---&---\\
	% xxx 77, m_H1=84.39, m_H2=109.39, sig(H1H1)=22.14*2 fb, BR(H1bb)=0.825
	\hline
	NMSSM
	 & $A_1 A_1 \to (b\bar{b}) (b\bar{b})$ &552& 31 &453& 332\\
	% xxx 1899, mH1=122.89, mA1=166.45, BR(A1->bb)=0.849, sig(A1A1)=255*2
	 & $A_1 A_1 \to (\gamma\gamma) (\gamma\gamma)$&796&$<0.01$&444& 34\\
	% xxx 501, mH1=127.35, mA1=155.64, BR(A1->gamgam)=0.645, bb=0.300,
	% sig(A1A1)= 2*98.72
	 & $H_1 H_1 \to (b\bar{b}) (b\bar{b})$&882& $<$0.1   &125.59& 4173\\
	% xxx 19337 mH1=48.26, mH3=882, mA1=386, mA2=879, sig(H1H1)=2*1.85 pb,
	% BR(H1->bb)=0.91
	 & $A_1 A_1 \to (b\bar{b}) (b\bar{b})$&676& $<0.1$&122.99&1353\\
	% xxx 16817 mH1=56.50, mH2=122.99, mA1=53.65, BR(A1->bb)=0.913,
	%sig(A1A1)=2*605 fb
	 & $A_1 A_1 \to (t\bar{t}) (t\bar{t})$&741& 7  &705 & 14\\
	% xxx 6624 mH1=124.67, mA1=349.6, BR(A1->tt)=0.987
	\hline
\end{tabular}
\caption{Upper: Selected rates for non-SM-like Higgs pair final states at NLO
  QCD. We specify the model, which of the Higgs bosons is the
  SM-like one, the signature and its rate as well as the
  $K$-factor. In the
  fourth column we also give the mass value $m_{\Phi}$ of the non-SM-like Higgs
  boson involved in the process. 
Lower: In case of resonantly enhanced cross sections, the mass of the
resonantly produced Higgs boson is given together with the NNLO QCD
production rate. Some scenarios contain two heavier Higgs bosons that
can contribute to resonant production. 
All benchmark details can 
  be provided on request. \label{tab:selected}}
\end{center}
\vspace*{-0.4cm}
\end{table}

%%%%%%%%%%%%%%%%%%%%%%%%%%%%%%%%%%%%%%%%%%%%%%%%%%%%%%%%%%%%%%
\paragraph{Cascade Decays with Multiple Higgs Final States \label{sec:cascade}}
As already stated, in non-mimimal Higgs extensions, we can have
Higgs-to-Higgs cascade decays that can lead to multiple Higgs final
states. The largest rate at NLO QCD that we found, for a final state with more than
three Higgs bosons, is given in the N2HDM-I, where we have 
\beq
\sigma(pp \to H_2 H_2 \to H_1 H_1 H_1 H_1 \to 4 (b\bar{b}))=
1.4 \mbox{ fb} \;.
\eeq
The SM-like Higgs is $H_1$ and the $K$-factor for the NLO QCD
production of $H_2 H_2$ is 1.82. Also in the NMSSM and C2HDM we can
have multiple Higgs production but the rates are below 10~fb after the
decays of the Higgs bosons. In the N2HDM, we can even produce up to
eight Higgs bosons in the final states but the rates are too small to be measurable.
%
%\begin{table}[h!]
%\begin{center}
%\begin{tabular}{|c|c|c|c|c|}
%\hline
%Model & SM-like Higgs & Signature & Rate [fb]& K-factor \\ \hline \hline
% N2HDM-I
%& $H_2$ & $H_2 H_1 \to H_1 H_1 H_1  \to (b\bar{b}) (b\bar{b})
%          (b\bar{b})$ & 111 & 2.09 \\
%% xxx 73 sig(H1H2)=2*2.1pb, BR(H2->H1H1)=0.040, BR(H1->bb)=0.857
%%& $H_1$ & $H_1 H_2 \to H_1 H_1 H_1 \to (b\bar{b}) (b\bar{b}) (b\bar{b})$ &40.83 \\
%% xxx 5265 BR(H1->bb)=0.597, BR(H2->H1H1)=0.369, sig(H1H2)=2*260 fb
%& $H_1$ & $H_2 H_2 \to H_1 H_1 H_1 H_1 \to 4 (b\bar{b})$ & 2 & 1.82\\
%% xxx 2495, mH2=315, BR(H2->H1H1)=0.603, sig(H2H2)=2*22fb, BR(H1->bb)=0.596
%%\hline
%\hline
%%NMSSM
%%& $H_1$ & $H_1 H_2 \to H_1 H_1 H_1 \to (b\bar{b})
%%          (b\bar{b})(b\bar{b})$ & 7 \\
%%%  xxx 9339 sig(H1H2)=102, BR(H2-> H1H1)=0.449, BR(H1->bb)=0.532
%\end{tabular}
%\caption{Maximum rates for multi-Higgs final states at NLO
%  QCD. We specify the model, which of the Higgs bosons is the
%  SM-like one, the signature and its rate. All benchmark details can
%  be provided on request.  \label{tab:selectedCas}}
%\end{center}
%\vspace*{-0.4cm}
%\end{table}

\section{Conclusions \label{sec:concl}}
In this paper, we have performed a comprehensive analysis of Higgs
pair production in some archetypical BSM models, namely the R2HDM, the
C2HDM, and the N2HDM as non-SUSY representatives, and the
NMSSM as a SUSY model. After applying the relevant theoretical and
experimental constraints, in particular limits from non-resonant
and resonant di-Higgs searches, we explore the ranges of the parameter spaces of
these models that are still allowed. We find that while the SM-like
Higgs top-Yukawa couplings are constrained to within about 10\% of the
SM model value, there is still some freedom on the trilinear Higgs
self-coupling. In particular, zero values for the SM-like
trilinear Higgs self-coupling are still allowed in all
models. Interestingly, the experimental searches start to constrain
the trilinear couplings of the N2HDM. In general, in order to derive limits on the
couplings both resonant and non-resonant searches will be
required. Overall, the delineation of the parameter space from di-Higgs
production is difficult as in BSM models we have the
Yukawa and trilinear couplings of various Higgs bosons involved and
also their total widths play a role in the size of the cross
section. \s

As for the maximum possible sizes of the resonantly enhanced
cross sections for SM-like Higgs pair production, we find that they
can be quite different across the
models studied, so that the cross section value itself might exclude certain models provided it exceeds a specific limit. We presented benchmark
scenarios for the maximum cross sections. They not only feature cross
sections that can exceed the SM value by up to a factor of 12 but they are also
interesting because they can lead to measurable rates of triple Higgs
production or the production of a Higgs boson pair in association with
a $Z$ boson. In the C2HDM, we presented a scenario where the
simultaneous measurement of Higgs-to-bosons decays would allow for the
test of CP violation. \s

We also investigated to which extent an EFT
approach can reproduce the Higgs pair results in specific UV-complete
models. Since in gluon fusion into Higgs pairs, we integrate across
possible resonances present in extended Higgs sectors due to the
additional non-SM-like Higgs bosons, the EFT approach cannot correctly
reproduce the results. This is particularly true if the resonances
are rather light. Moreover, we found that the value of the resonance
mass from which on the EFT approach starts to approximate the result
in the specific UV-complete model by better than 10\%
depends on the benchmark scenario itself and that the total width of
the intermediate resonance plays an important role. This again emphasizes the
importance of investigating specific models besides a more general EFT
approach in order to get a complete picture of the new physics
landscape. \s

We also presented benchmark points for the production of a SM-like
Higgs together with a non-SM-like one, leading to a plethora of
different final state signatures. The presented benchmark points represent those
with significant rates in certain final state signatures, namely $(b\bar{b})(b\bar{b})$,
$(b\bar{b})(WW)$, and $(b\bar{b})(t\bar{t})$. But also multi-photon
final states can have important rates as shown by our results, in
particular in non-SM-like Higgs pair final states. We
highlighted additionally scenarios that involve further Higgs-to-Higgs decays,
of the mixed SM-like plus non-SM-like final state, leading to
three-Higgs final states with 
significant rates. On top of that, three benchmark scenarios were
presented where new heavy Higgs bosons might rather be discovered
through resonant di-Higgs production than from direct single Higgs
production. This is the case for the singlet extended models N2HDM and
NMSSM, where the couplings of the searched heavy Higgs boson to SM
particles are suppressed due to large singlet admixtures but not its
trilinear coupling to other Higgs bosons. Finally, we gave a short
overview of the maximum possible rates in the production of two
non-SM-like Higgs bosons and also for the production of multiple Higgs
bosons. \s

For all benchmark points, we provided the di-Higgs production cross
sections at NLO QCD in the heavy loop particle limit. We found that
the $K$-factors range betwee 1.79 and 2.24 depending on the model and
the final Higgs pair state. \s

We collected a large amount of viable parameter points in the R2HDM,
C2HDM, N2HDM, and NMSSM with interesting features in the context of
Higgs pair production and necessarily had to restrict ourselves on
specific scenarios to highlight some prominent features. We emphasize,
however, that we can provide benchmarks with specific features on
request and invite the readers to contact us. \s

With this work, we hope to have given a close to complete and
comprehensive overview of possible signatures in di-Higgs or even
multi-Higgs production in representative BSM models and what can be
learnt from them. It can be a starting point for further
investigations in many different final state signatures based on the
data sample that we have generated. Having a guideline at hand of what
can be expected may help us find our way through the vast new
physics landscape and get deeper insights in the mechanism behind
electroweak symmetry breaking. Ultimately helping us to answer some of our
most pressing open questions in the world of elementary particle physics.

%%%%%%%%%%%%%%%%%%%%%%%%%%%%%%%%%%%%%%%%%%%%%%%%%%%%%%%%%%%%%%
\section*{Acknowledgement}
We thank the LHC Higgs Working group in general, and in particular we
are grateful for numerous fruitful discussions in meetings and workshops of the
HH, NMSSM and Extended Higgs sector subgroups. We are also grateful to
their conveners E.~Brost, M.~d'Alfonso, U.~Ellwanger,
R.~Gr\"ober, N.~Lu, J.~Mazzitelli, T.~Robens, N.~Rompotis, N.~Shah,
D.~Winterbottom, L.~Zivkovic. We acknowledge many helpful discussions
with A.~Ferrari, M.~Klute, J.~M\"uller, M.~Spira, J.~Wittbrodt, R.~Wolf. Thanks
go to M.~Gabelmann for providing the NMSSM sample.
M.M. acknowledges support by the BMBF projects 05H18VKCC1 and
05H21VKCCA. D.A. acknowledges support the Deutsche Forschungsgemeinschaft (DFG, German Research Foundation) under grant 396021762 - TRR 257. J. El~F. would like to thank the Abdus Salam International
Centre for Theoretical Physics (ICTP) for hospitality and financial
support where part of this work has been done.  R.S. and P.F. are supported
by CFTC-UL under FCT contracts UIDB/00618/2020, UIDP/00618/2020, and
by the projects CERN/FISPAR/0002/2017 and CERN/FIS-PAR/0014/2019. 
P.F. is supported by the project CERN/FIS-PAR/0004/2019.

%%%%%%%%%%%%%%%%%%%%%%%%%%%%%%%%%%%%%%%%%%%%%%%%%%%%%%%%%%%%%%

%%%%%%%%%%%%%%%%%%%%%%%%%%%%%%%%%%%%%%%%%%%%%%%%%%%%%%%%%%%%%%
\begin{appendix}

\section{Resonant and Non-Resonant Production Cross
  Sections \label{sec:prodsec}}
In this section, we re-present the cross sections of
Figs.~\ref{fig:scattersm1} and \ref{fig:scattersm2} by, first, showing
all the points that have a resonant contribution and, 
second, displaying only points that are considered non-resonant by the
definition given in Sec.~\ref{sec:sep}. Notice that the points displayed for each
part are not complementary in the full sample, {\it i.e.}~some point for
which we show its resonant contribution might be defined as
non-resonant and appear in the non-resonant plots as well. This is
because a point being defined as dominantly non-resonant (by our
definition) is not exempt from being compared to experimental resonant
constraints. On the other hand, it makes no sense to compare points
where resonance production is dominant with experimental non-resonant
limits, hence only points that we define as non-resonant ones can be
compared to non-resonant experimental limits. To be specific, points
with resonant contribution means parameter points where at least one
of the heavier neutral Higgs bosons\footnote{Note that in the C2HDM, N2HDM, and
  NMSSM for the case where the lightest Higgs boson is the SM-like
  one, we can have two possible resonances.} has a mass large enough
to decay into a pair of two SM-like Higgs bosons. And the resonant
production cross sections are 
obtained by calculating with {\tt SusHi} the NNLO QCD production cross section of
the ``resonant'' Higgs boson and subsequently multiplying
it with its branching ratio into the two SM-like Higgs bosons. The
points included in the non-resonant plots on the other hand are all
those scatter points of Figs.~\ref{fig:scattersm1} and
\ref{fig:scattersm2} where resonant production is kinematically not
possible, or where the resonant production cross section accounts for
less than 10\% of the total di-Higgs cross section. \s
%the resonant and 
%non-resonant cross section values corresponding to
%Figs.~\ref{fig:scattersm1} and \ref{fig:scattersm2}. The separation
%between resonant and non-resonant production is done as described in
%Sec.~\ref{sec:sep}. 

\begin{figure}[h!]
\centering
\includegraphics[width=0.45\textwidth]{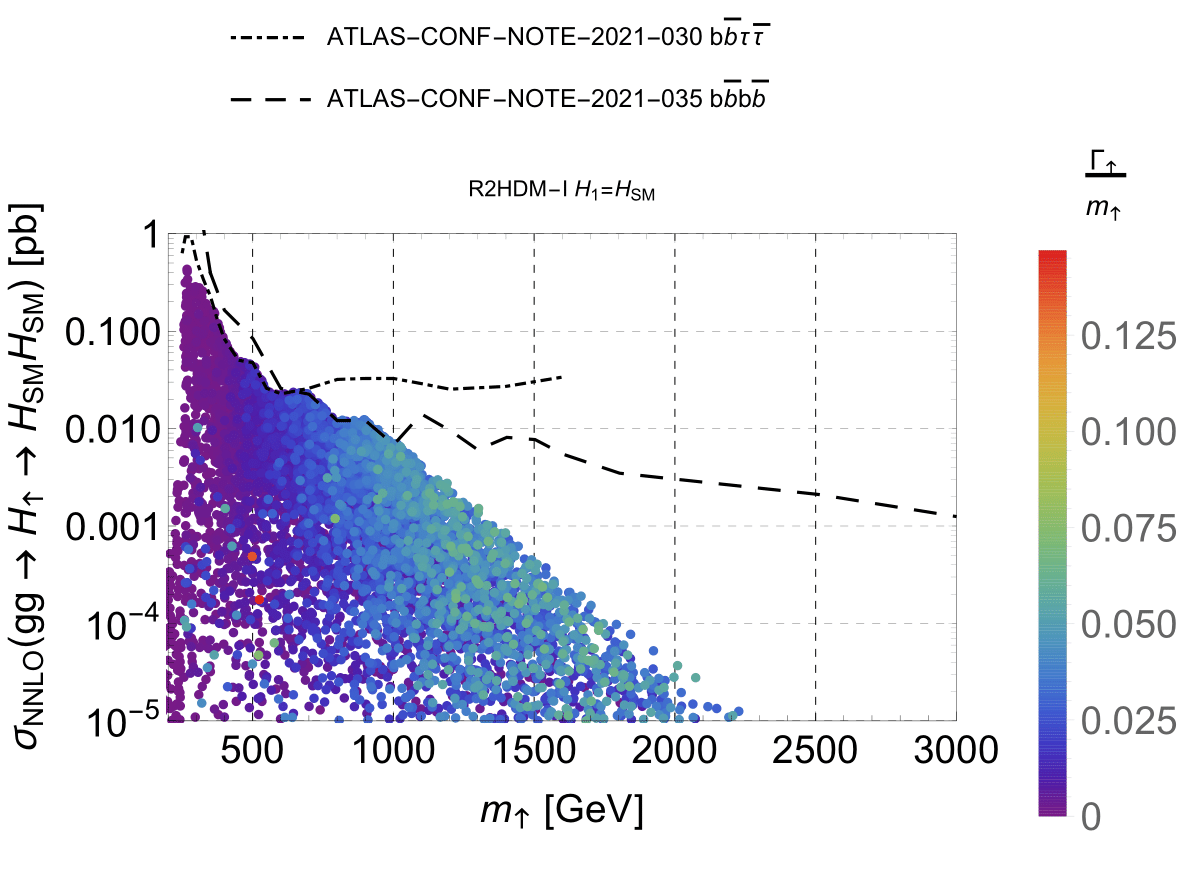}
\hspace*{0.5cm}
\includegraphics[width=0.45\textwidth]{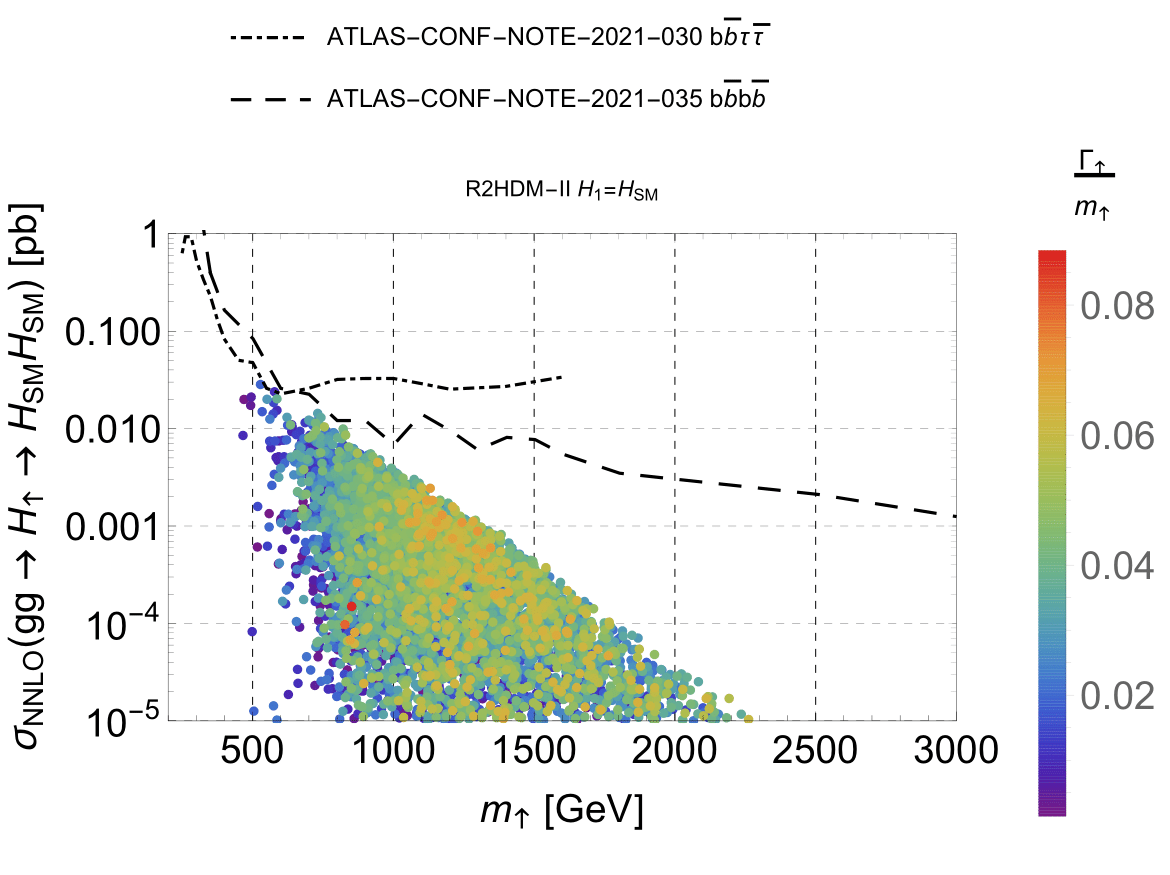}
\caption{Resonant production cross sections for the R2HDM points of
  Fig.~\ref{fig:scattersm1} (upper) for R2HDM-I (left) and R2HDM-II (right) as a
function of the resonantly produced heavier Higgs boson $H_2 \equiv H_\uparrow$. Color
code: ratio of the total width of the resonant Higgs boson and its
mass. \label{fig:resr2hdm}}
\end{figure}

Figure~\ref{fig:resr2hdm} shows the resonant
production cross section values for the R2HDM for the 
points presented in Fig.~\ref{fig:scattersm1} (upper). 
%The cross
%sections are obtained by calculating the production cross section of
%the ``resonant'' Higgs boson, here $H_2$, and subsequently multiplying
%it with its branching ratio into the two SM-like Higgs bosons. 
The color code denotes the ratio of the total width of the resonant Higgs
boson and its mass.  \s

\begin{figure}[h!]
\centering
\includegraphics[width=0.45\textwidth]{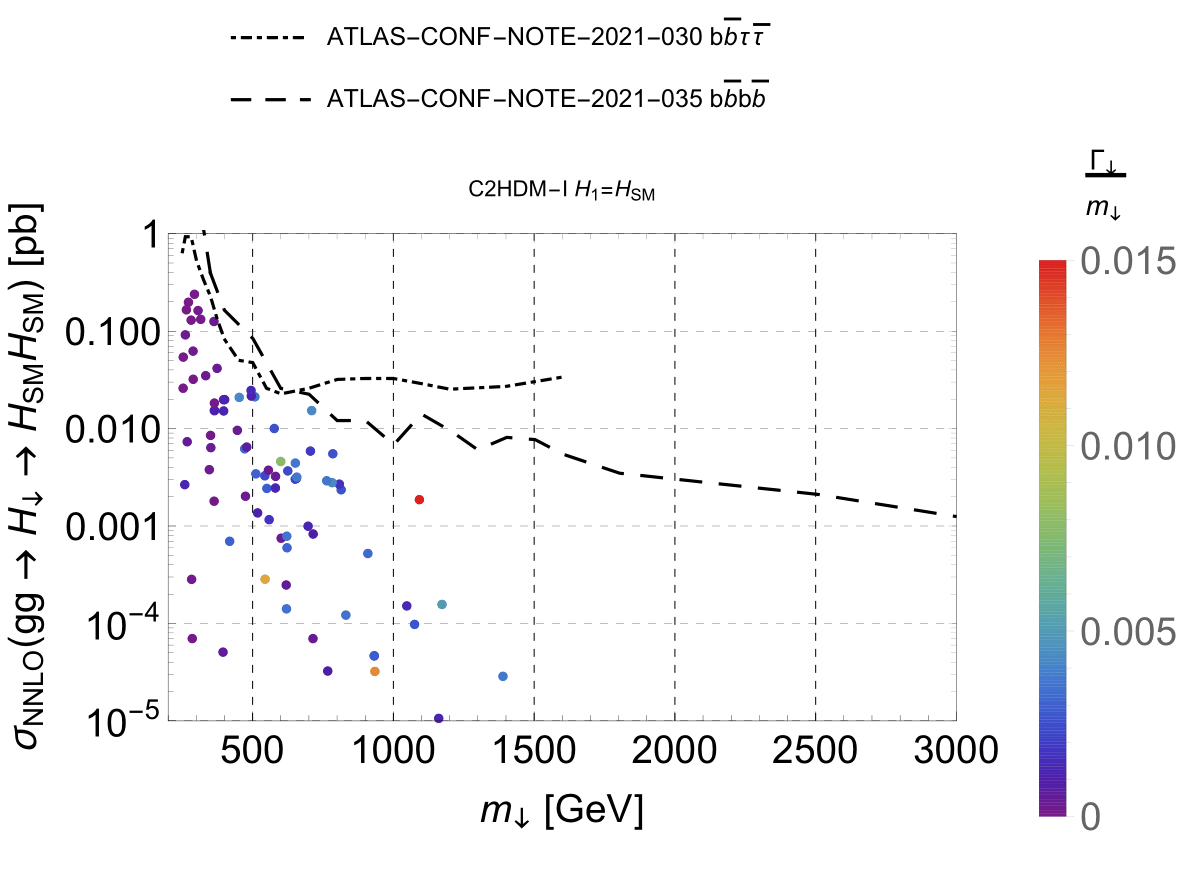}
\hspace*{0.5cm}
\includegraphics[width=0.45\textwidth]{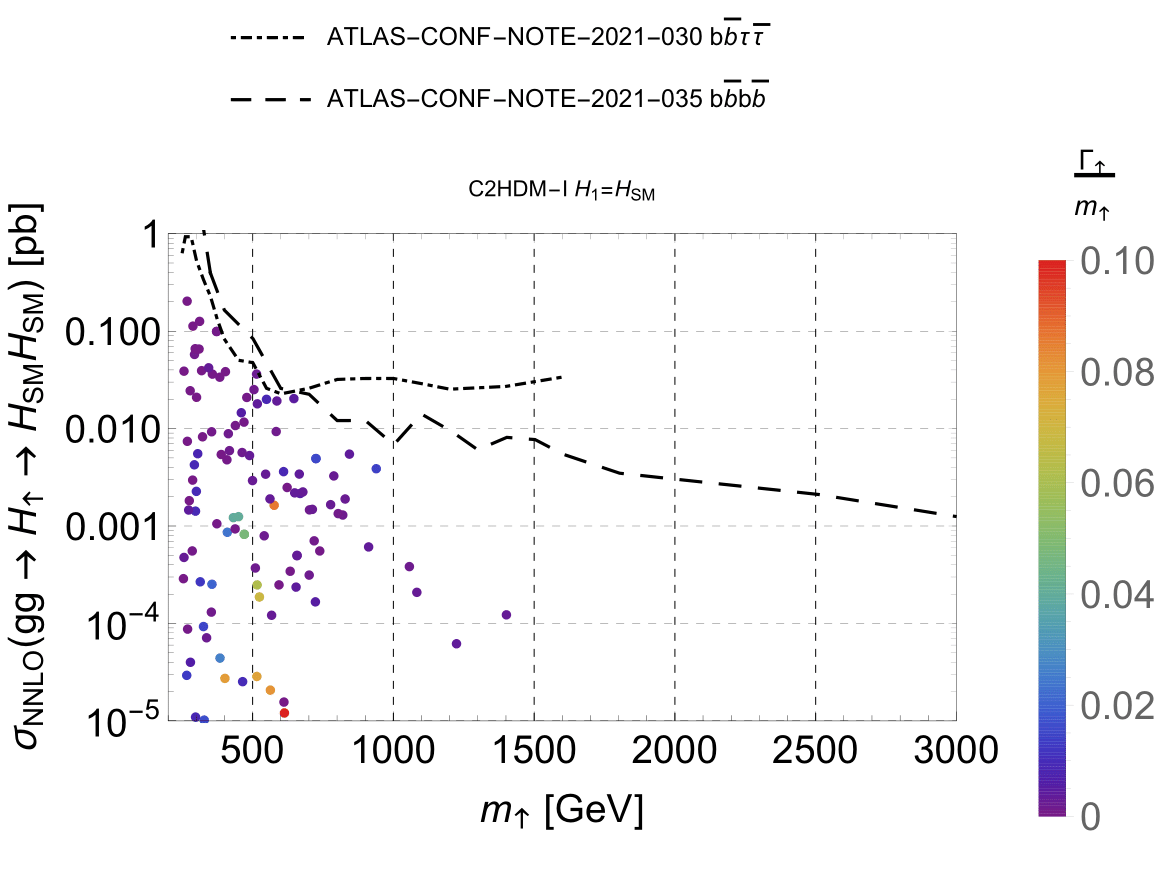} \\
\includegraphics[width=0.45\textwidth]{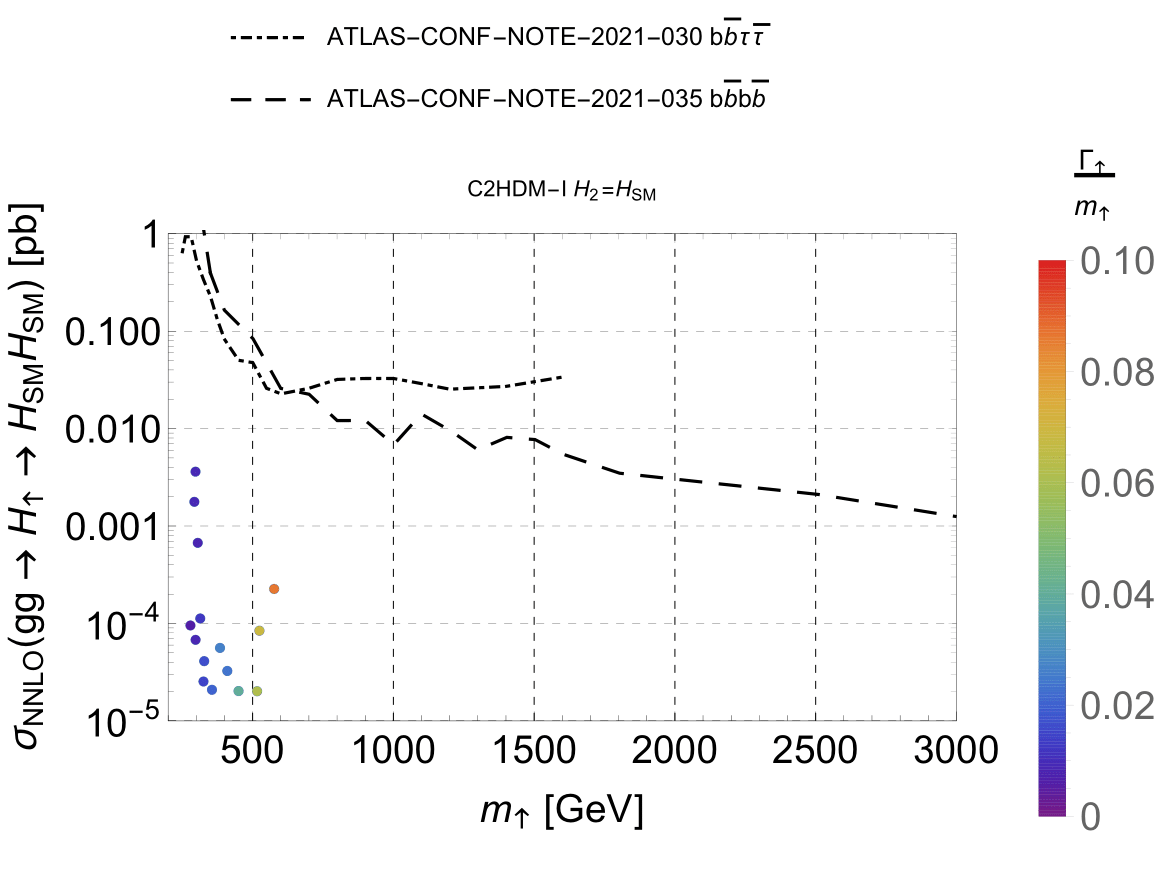} \\
\includegraphics[width=0.45\textwidth]{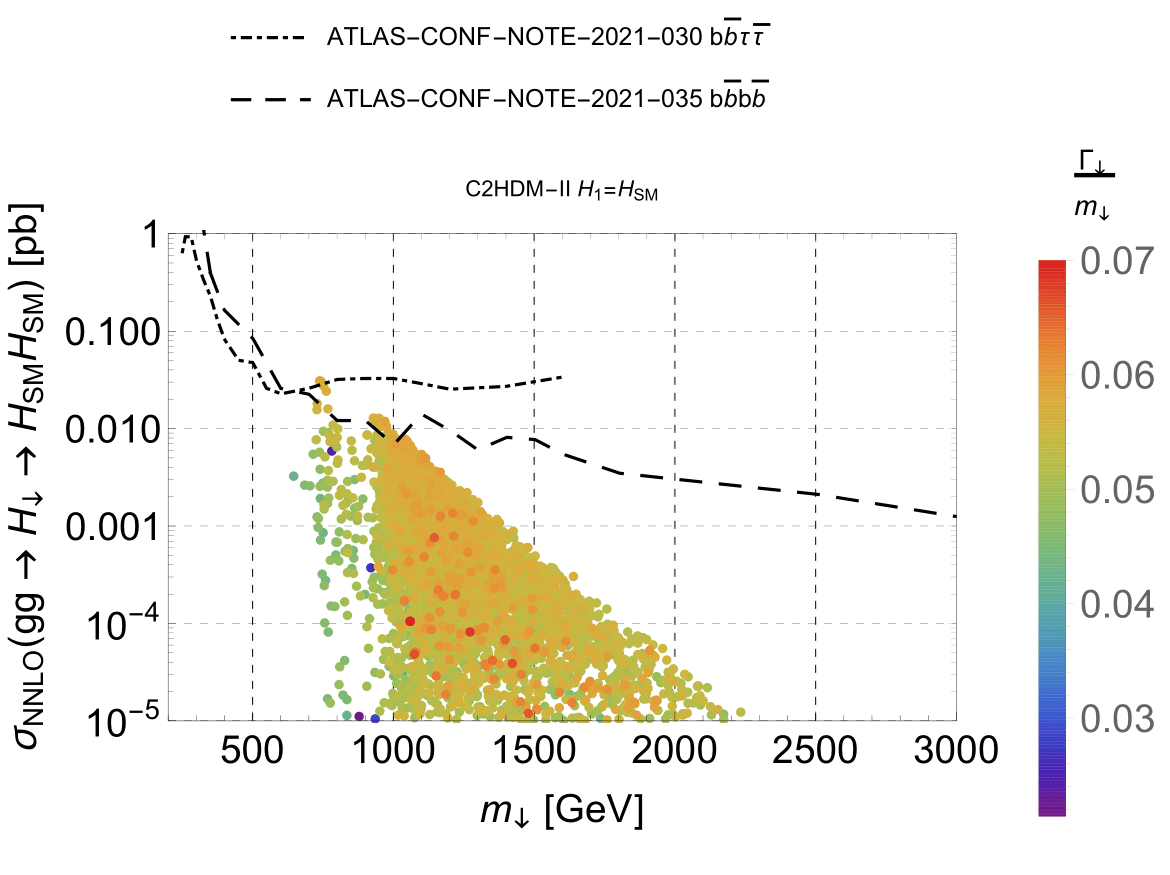} 
\hspace*{0.5cm}
\includegraphics[width=0.45\textwidth]{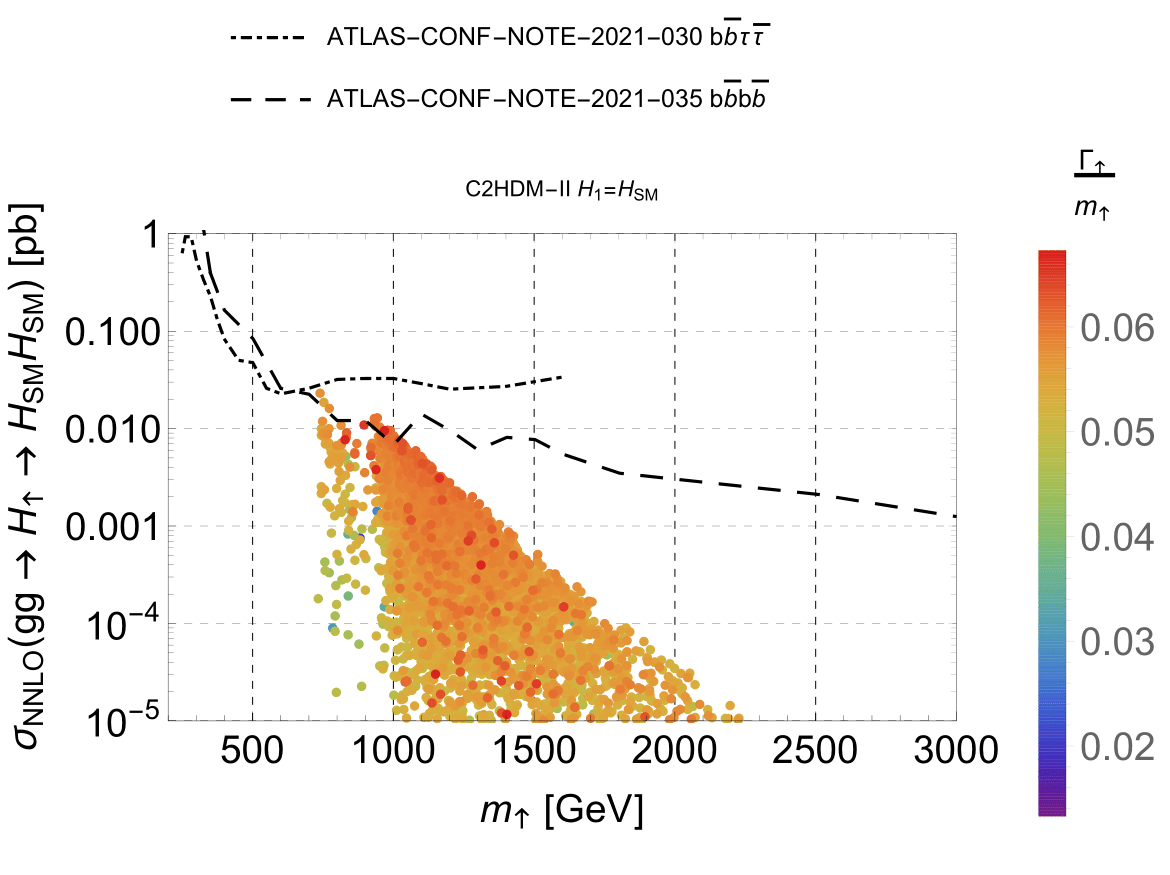} 
\caption{Resonant production cross sections for the C2HDM points of
  Fig.~\ref{fig:scattersm1} (middle) for C2HDM-I (upper two rows) and C2HDM-II
  (lower row) as a
function of the resonantly produced heavier Higgs boson $H_2 \equiv
H_\downarrow$ or $H_3 \equiv H_\uparrow$. Color
code: ratio of the total width of the resonant Higgs boson and its
mass. \label{fig:resc2hdm}}
\end{figure}

In Fig.~\ref{fig:resc2hdm} we display the resonant production cross
sections for the C2HDM for the different cases w.r.t.~to which of the
$H_{1,2}$ is the SM-like Higgs boson. Note that in the C2HDM T2 we do
not have scenarios compatible with all constraints with resonantly
enhanced production in case $H_2$ is the SM-like Higgs boson. In case
$H_1$ is the SM-like Higgs boson, both $H_2$ and $H_3$ can in
principle lead to resonant enhancement. We therefore show in separate
plots their resonant cross sections as a function of their mass. \s

\begin{figure}[h!]
\centering
\includegraphics[width=0.45\textwidth]{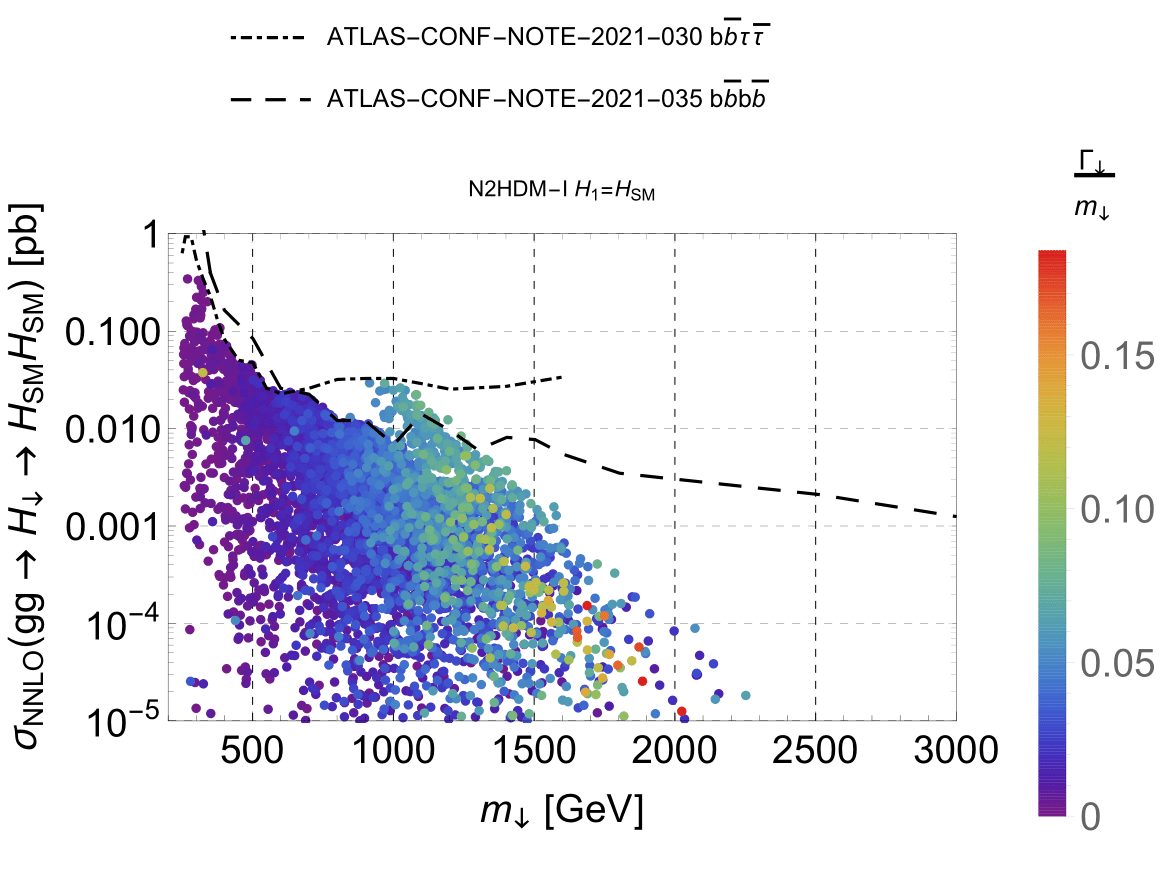}
\hspace*{0.5cm}
\includegraphics[width=0.45\textwidth]{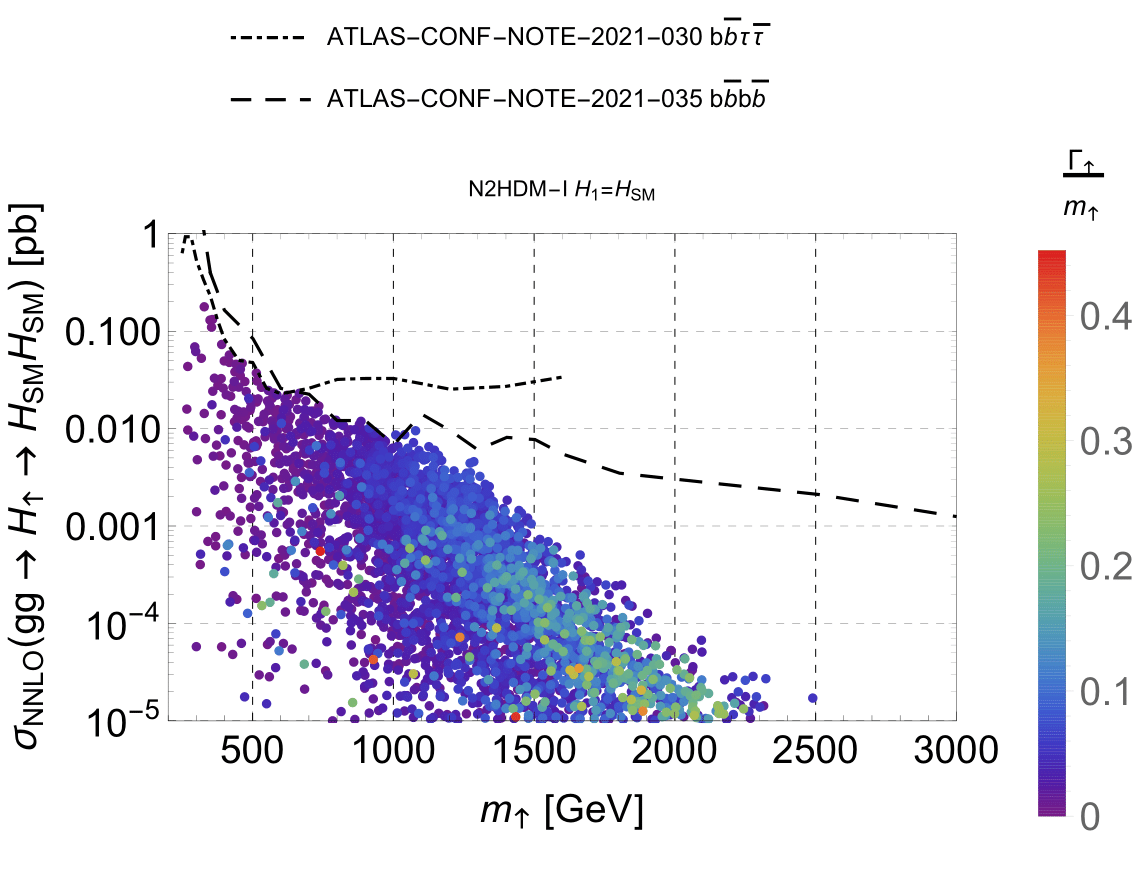} \\
\includegraphics[width=0.45\textwidth]{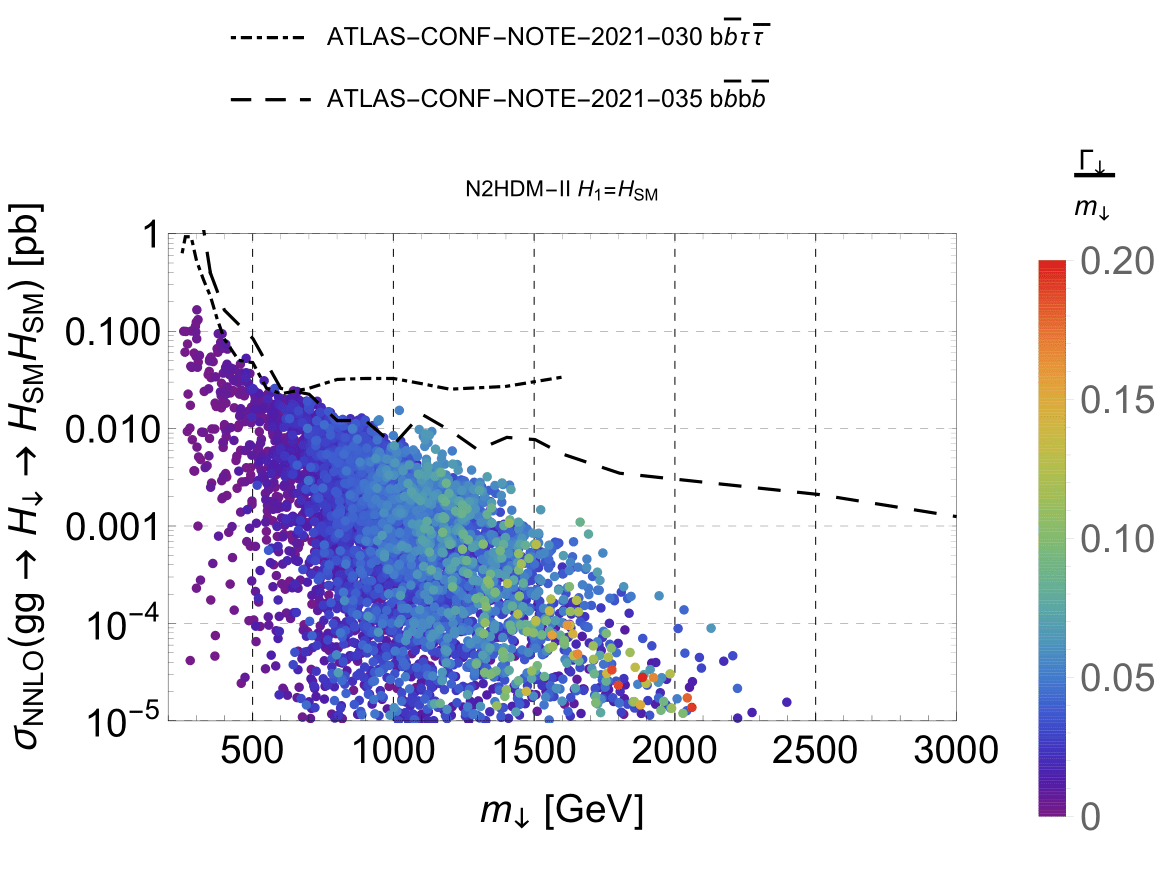} 
\hspace*{0.5cm}
\includegraphics[width=0.45\textwidth]{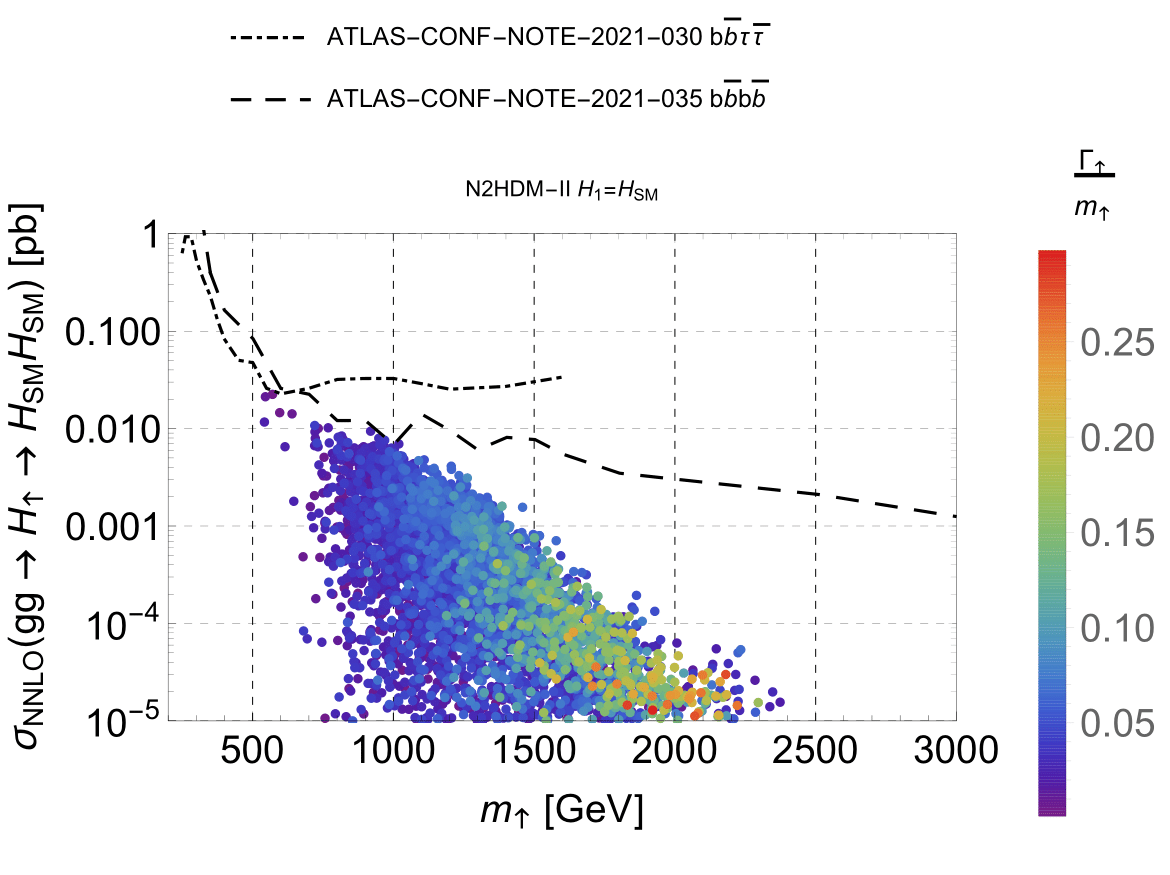} \\
\includegraphics[width=0.45\textwidth]{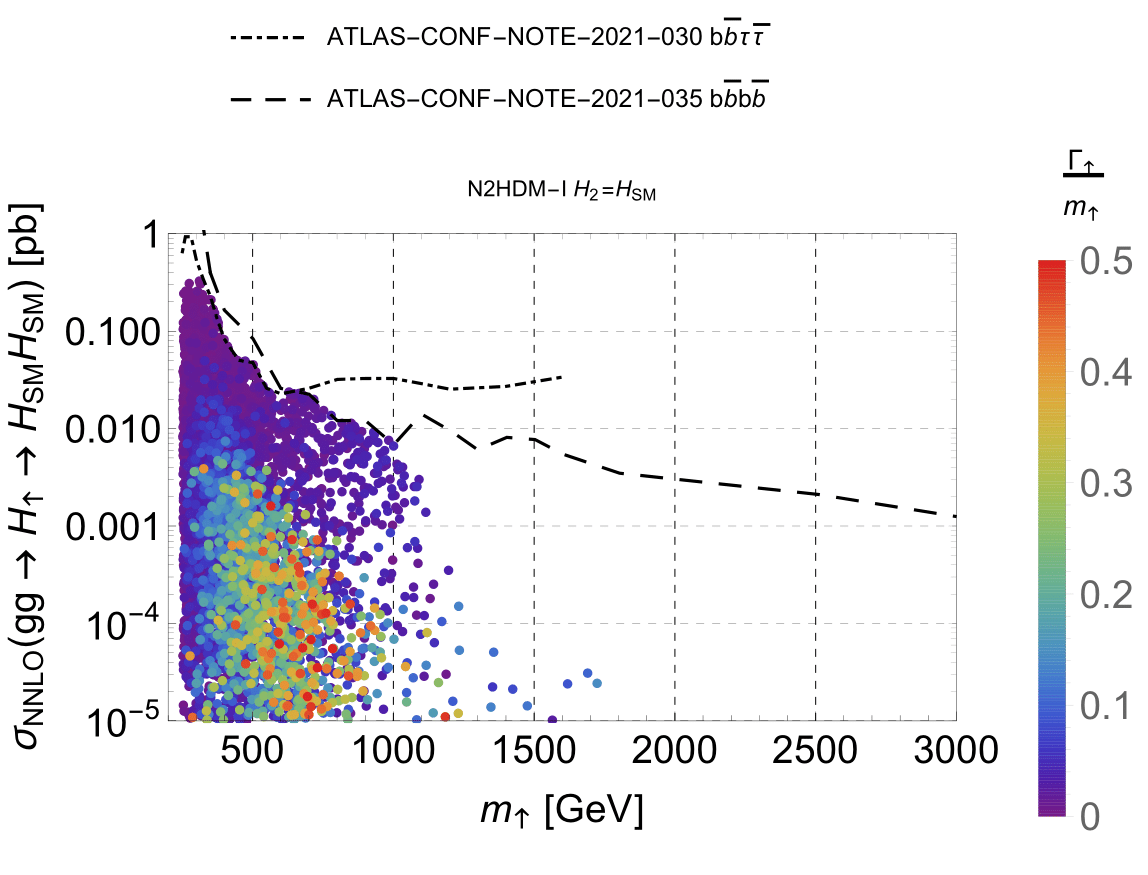} 
\hspace*{0.5cm}
\includegraphics[width=0.45\textwidth]{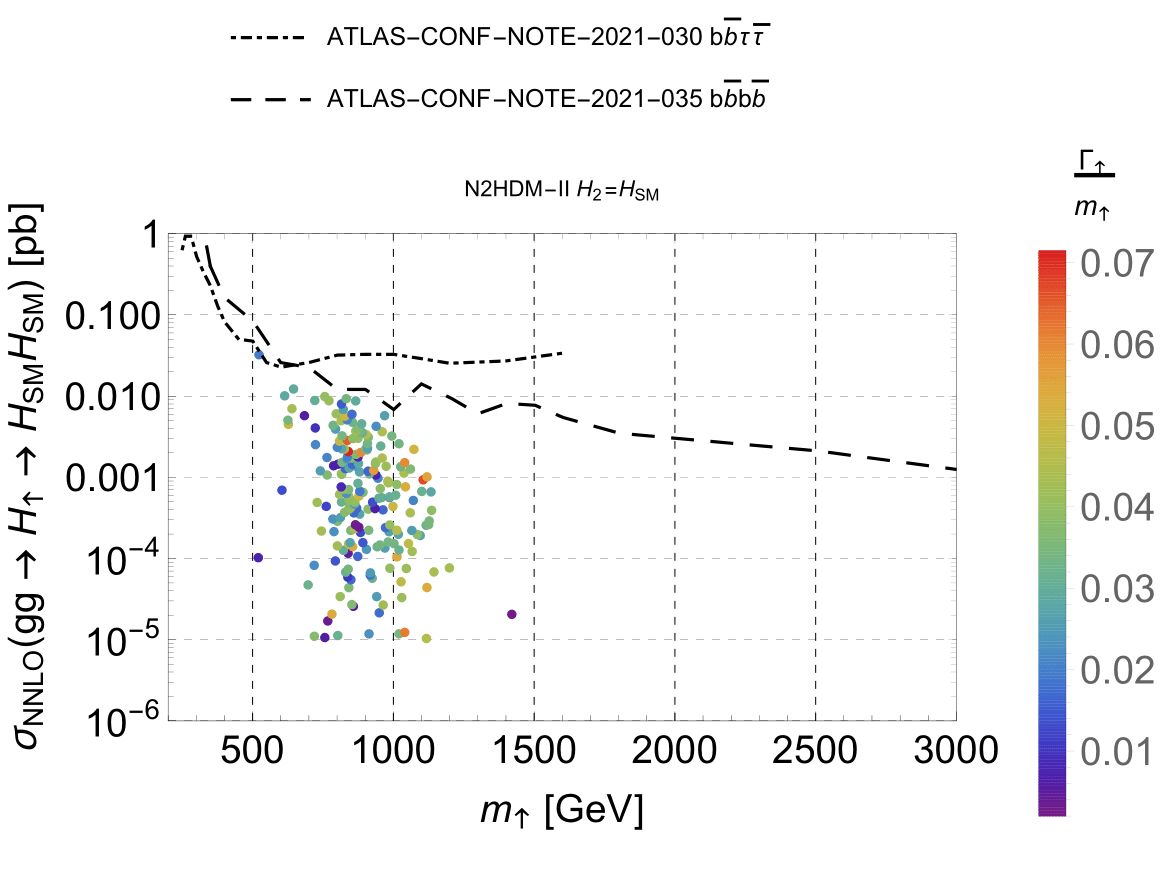} 
\caption{Resonant production cross sections for the N2HDM points of
  Fig.~\ref{fig:scattersm1} (lower) for N2HDM-I (upper) and N2HDM-II (middle)
  with $H_1 \equiv H_{\text{SM}}$. Lower row: $H_2 \equiv
  H_{\text{SM}}$ in case of type 1 (left) and type 2 (right); all plots 
as a
function of the resonantly produced Higgs boson. Color
code: ratio of the total width of the resonant Higgs boson and its
mass. \label{fig:resn2hdm}}
\end{figure}
\begin{figure}[t!]
\centering
\includegraphics[width=0.45\textwidth]{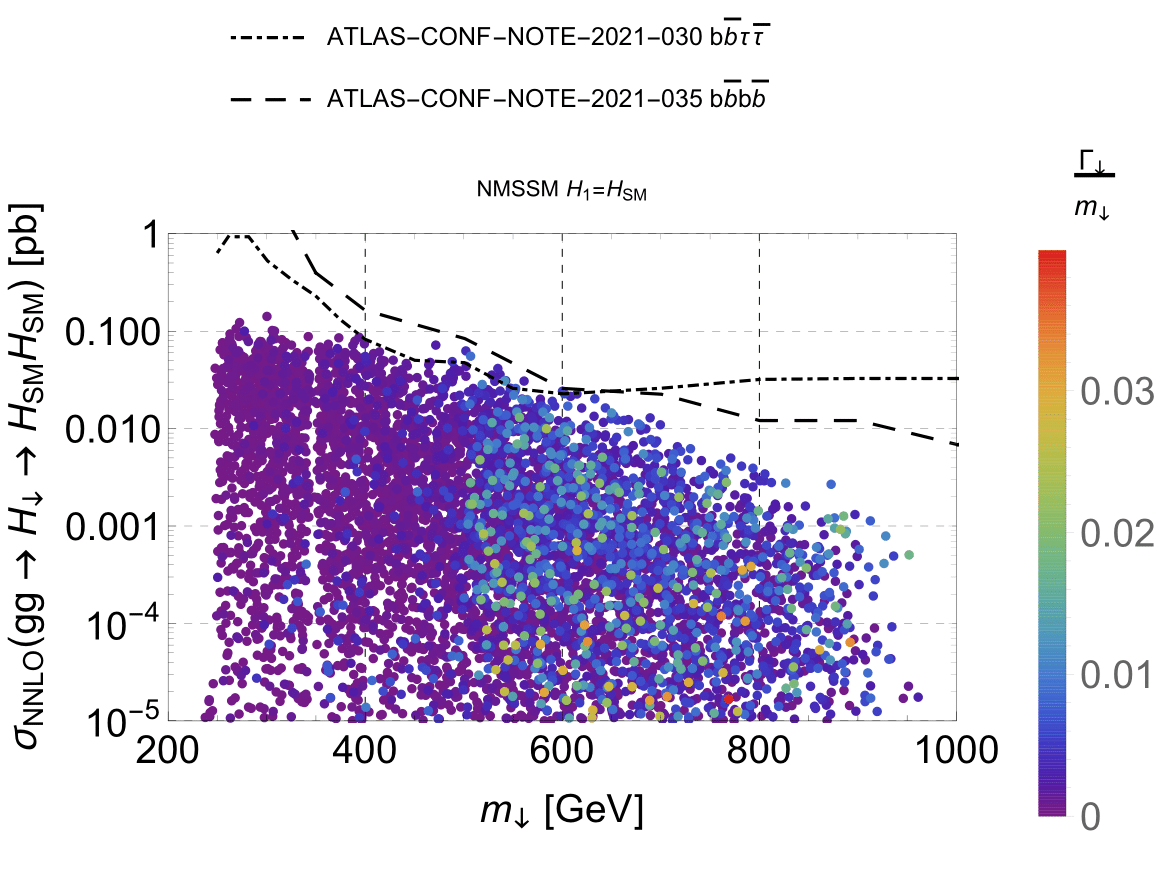}
\hspace*{0.5cm}
\includegraphics[width=0.45\textwidth]{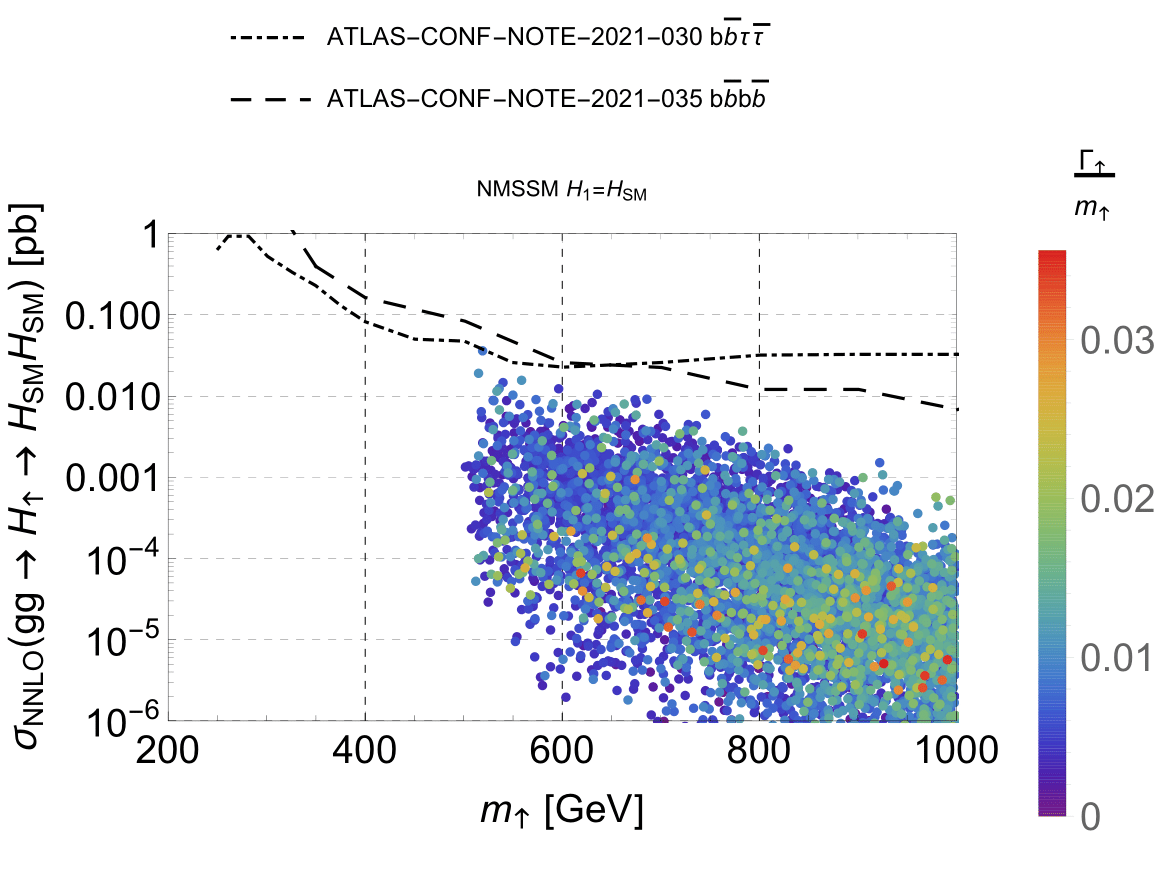} \\
\includegraphics[width=0.45\textwidth]{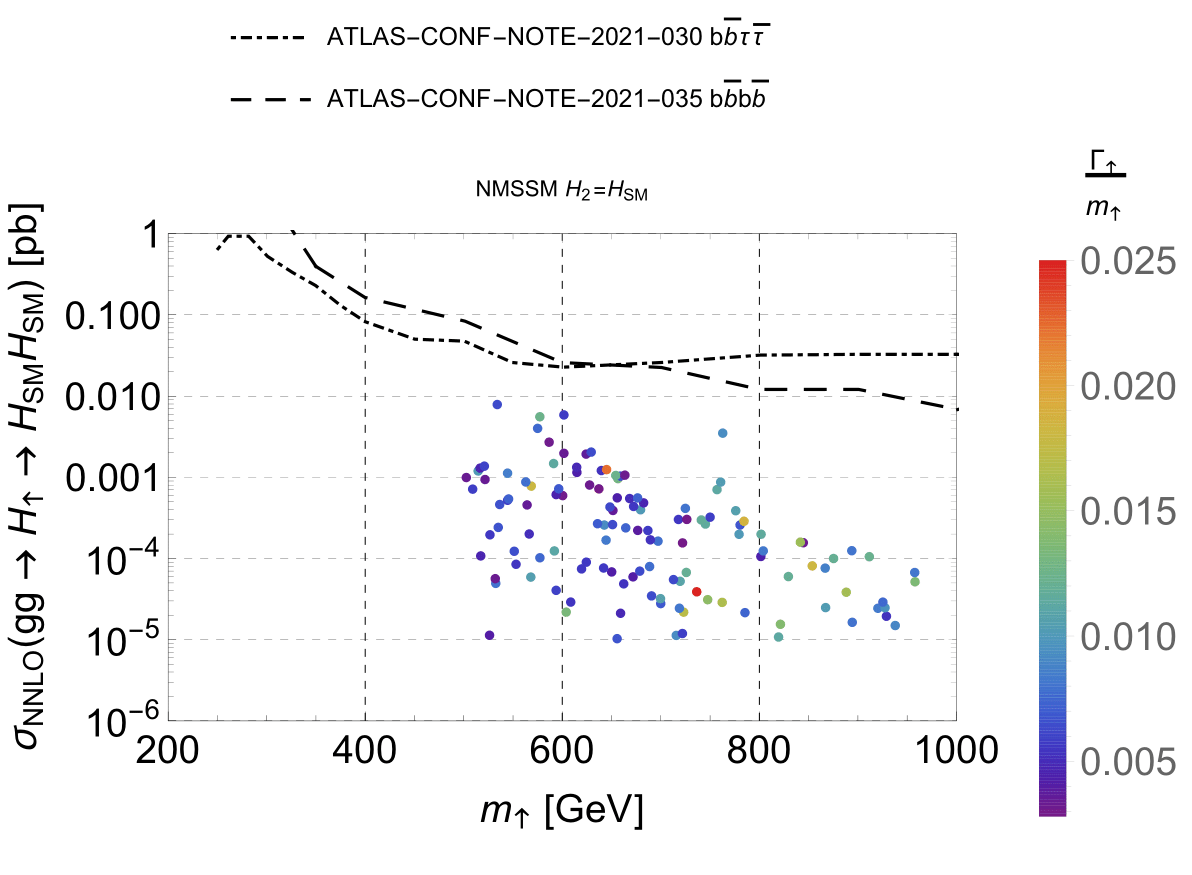}
\caption{Resonant production cross sections for the NMSSM points of
  Fig.~\ref{fig:scattersm2}  for $H_1 \equiv H_{\text{SM}}$  (upper)
  and  $H_2 \equiv H_{\text{SM}}$ (lower) as a
function of the resonantly produced heavier Higgs boson $H_2 \equiv
H_\downarrow$ or $H_3 \equiv H_\uparrow$. Color
code: ratio of the total width of the resonant Higgs boson and its
mass. \label{fig:nmssm}}
\end{figure}

The resonant production cross
sections for the N2HDM for the different cases w.r.t.~to which of the
$H_{1,2}$ is the SM-like Higgs boson are shown in
Fig.~\ref{fig:resn2hdm} and those for the NMSSM in
Fig.~\ref{fig:nmssm}. \s

\begin{figure}[h!]
\centering
\includegraphics[width=0.45\textwidth]{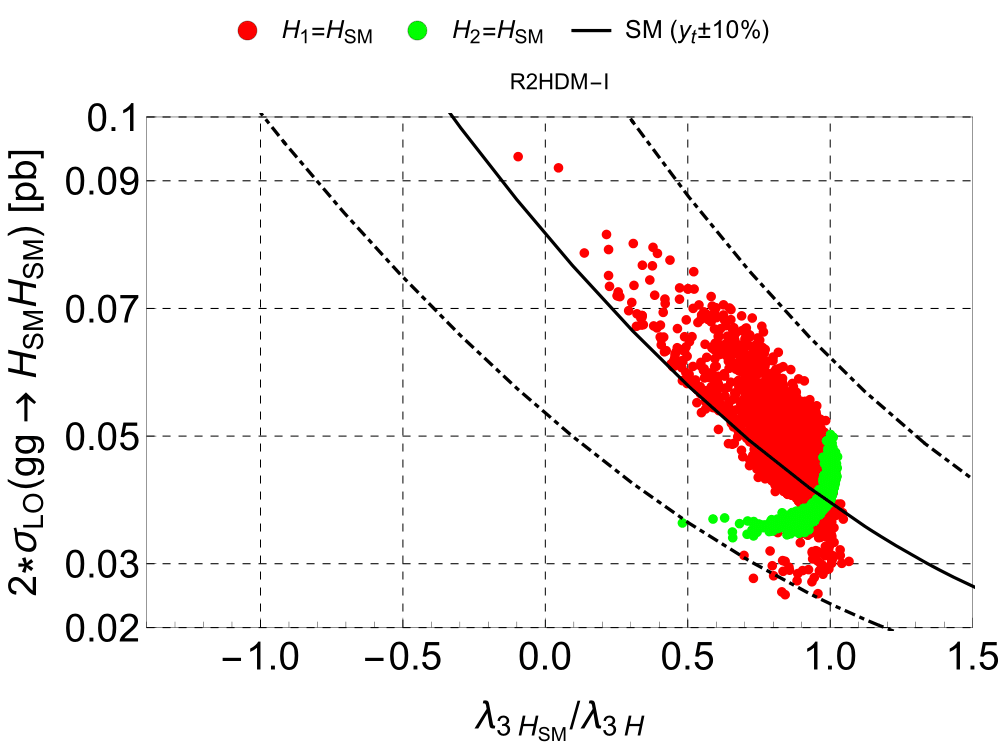}
\hspace*{0.5cm}
\includegraphics[width=0.45\textwidth]{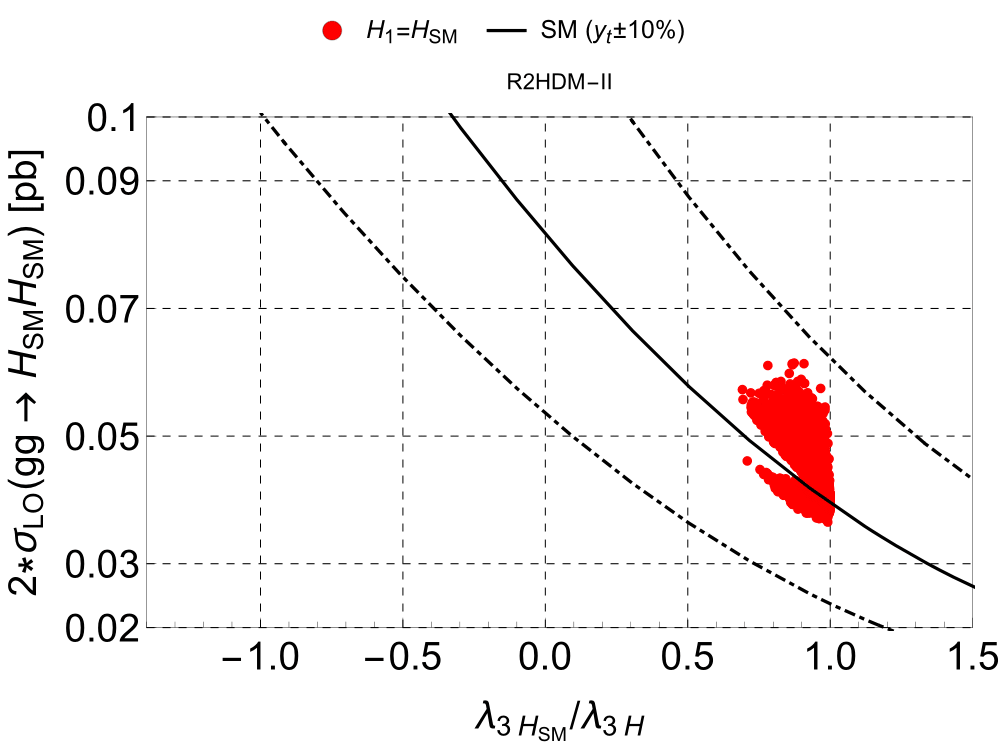} \\
\includegraphics[width=0.45\textwidth]{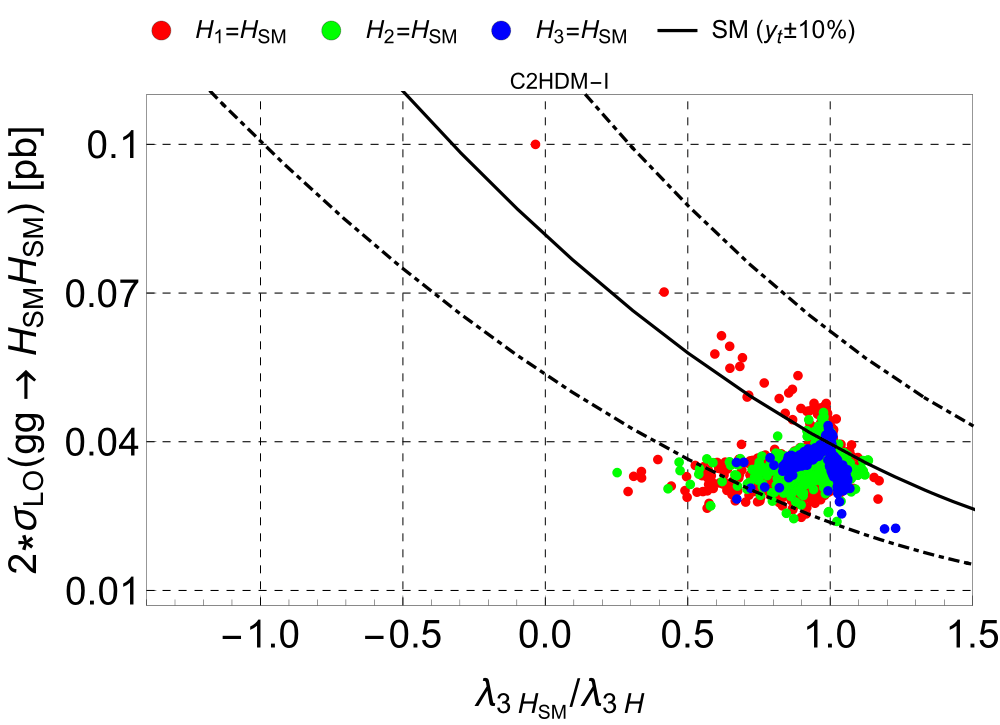}
\hspace*{0.5cm}
\includegraphics[width=0.45\textwidth]{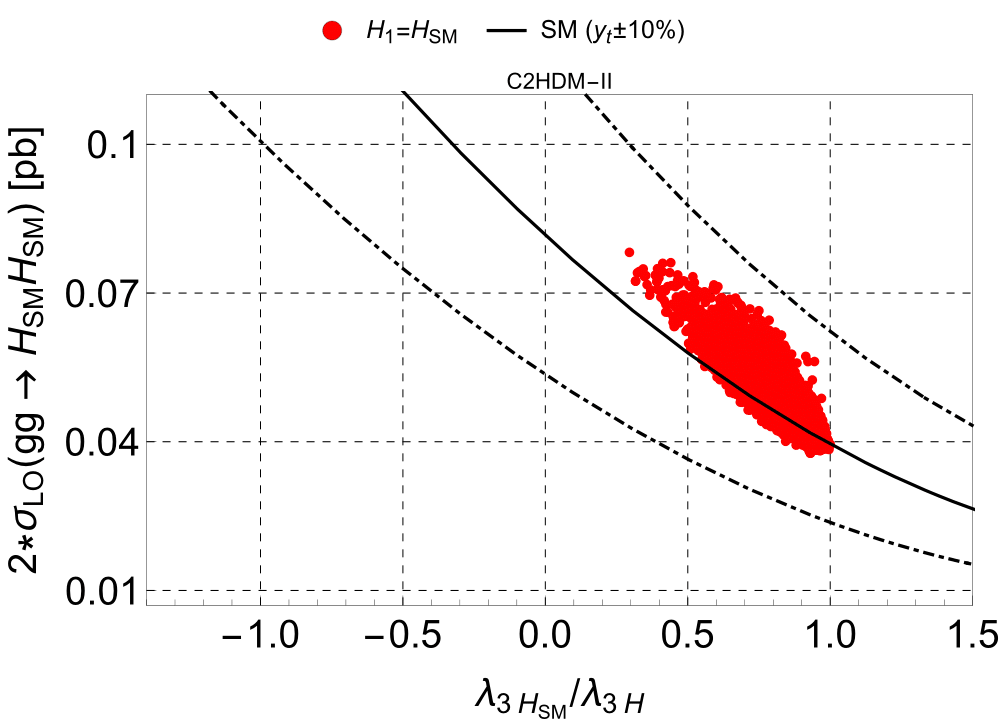} \\
\includegraphics[width=0.45\textwidth]{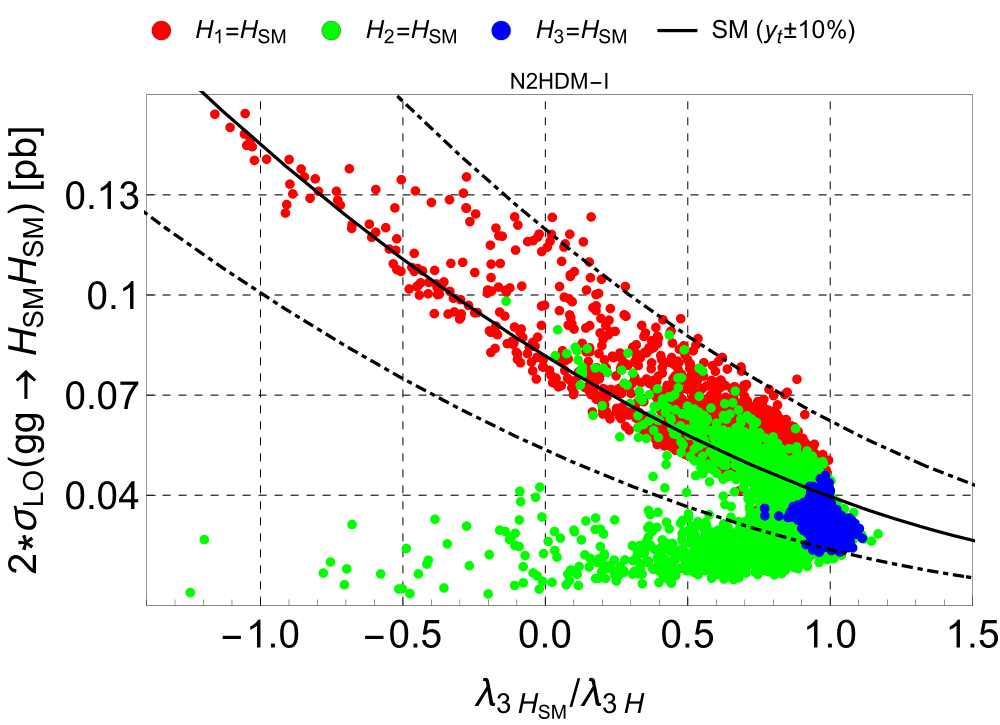}
\hspace*{0.5cm}
\includegraphics[width=0.45\textwidth]{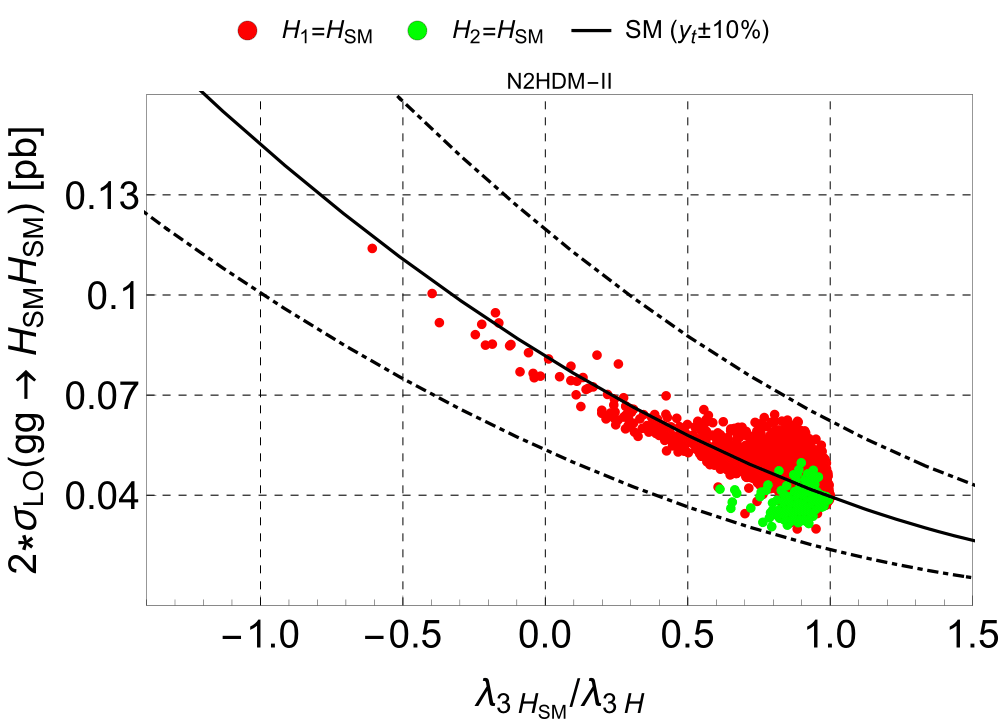} 
\caption{Non-resonant di-Higgs cross sections for the R2HDM (upper),
  C2HDM (middle) and N2HDM (lower) points of
  Fig.~\ref{fig:scattersm2} for type 1 (left) and type 2 (right) as a
  function of the trilinear Higgs self-coupling of the SM-like Higgs
  boson of the respective model 
  normalized to the trilinear coupling of the SM Higgs boson. Note
  the different $y$-axis range in the results for the R2HDM and those
  for the C2HDM and N2HDM. The full line shows the change of the SM
  Higgs pair cross section as a function of the variation of the trilinear Higgs
  self-coupling, the dashed lines show the change, when additionally
  the SM top-Yukawa coupling is varied by $\pm 10$\%. \label{fig:nonrescxn1}}
\end{figure}

\begin{figure}[h!]
\centering
\includegraphics[width=0.45\textwidth]{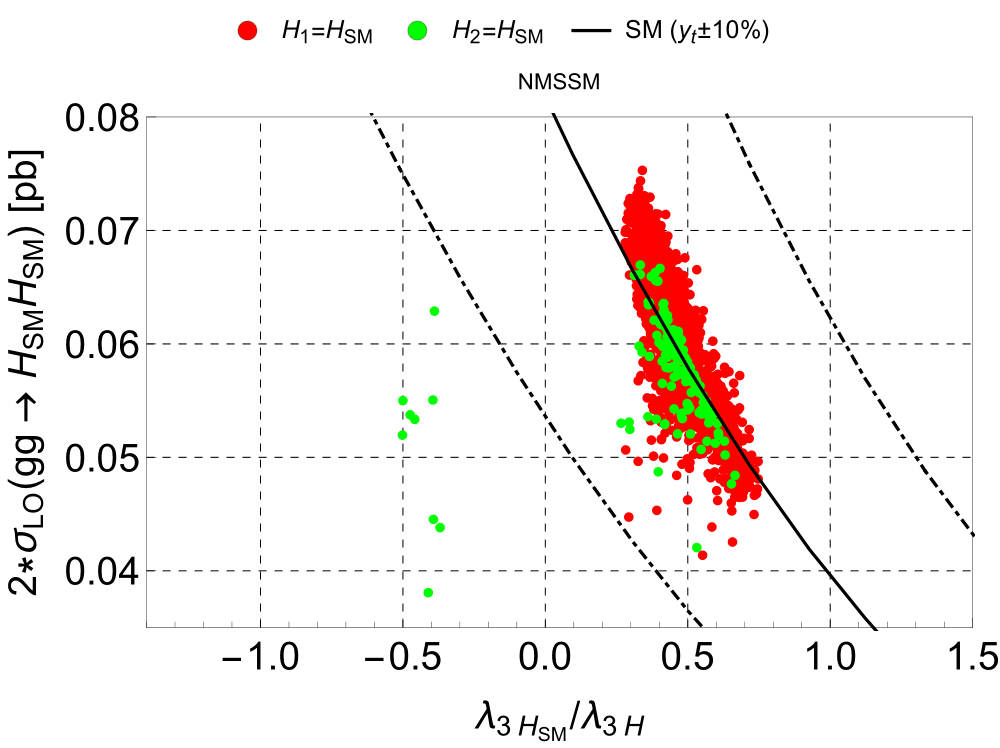}
\caption{Same as Fig.~\ref{fig:nonrescxn1} but for the NMSSM. \label{fig:nonrescxn2}}
\end{figure}

In Fig.~\ref{fig:nonrescxn1} we display the di-Higgs
cross sections for points considered non-resonant by our definition,
for the R2HDM, C2HDM, and N2HDM type 1 and type 2. We show them as a function
  of the trilinear Higgs self-coupling of the SM-like 
  Higgs boson in the respective model normalized to the value of the
  trilinear Higgs self-coupling of the SM, $\lambda_{3H_{\text{SM}}}/\lambda_{3H}$. In
  Fig.~\ref{fig:nonrescxn2} we show the corresponding NMSSM plot. 
We also include in the plots as a full line the change of the SM Higgs
pair production cross section as a function of
$\lambda_{3H_{\text{SM}}}/\lambda_{3H}$. The dashed lines show its
change if additionally the top-Yukawa coupling is varied by 10\% away
from the SM value. From these plots we can infer the importance
of interference effects. We see {\it e.g.}~in
Fig.~\ref{fig:nonrescxn1} (lower left) for the N2HDM-I with $H_2
\equiv H_{\text{SM}}$ green points that are well below the full and dashed
lines. The
suppression of the N2HDM cross section cannot be caused by the variation of
the trilinear or top-Yukawa coupling away from the SM values (indicated
by the full and dashed lines). It is caused by the negative
interference between the triangle diagram contributions of $H_1$ and $H_2$, as we
explicitely verified. 

%%%%%%%%%%%%%%%%%%%%%%%%%%%%%%%%%%%%%%%%%%%%%%%%%%%%%%%%%%%%%%
\section{Alignment Limit in the C2HDM \label{sec:alignc2hdm}}
In the following, we derive the limits that are required to achieve
the alignment limit in the C2HDM. In terms of the matrix elements of
the mixing matrix, defined in Eq.~(\ref{eq:rmixmatrixc2hdm}), the $H_i$ ($i=1,2,3$)
Higgs couplings to massive gauge bosons ($V=Z,W$) and to fermions $f$ in the C2HDM-I and C2HDM-II, read respectively
\beq
g^{\text{C2HDM-I/II}}_{H_iVV} &=& (R_{i1}c_\beta + s_\beta R_{i2}) \, g_{HVV}^{\text{SM}} \;,\\
g^{\text{C2HDM-I}}_{H_iff} &=&
(\frac{R_{i2}}{s_\beta}+i\gamma_5\frac{R_{i3}}{t_\beta}) \, g^{\text{SM}}_{Hff} \;,\\
g^{\text{C2HDM-II}}_{H_iuu} &=& (\frac{R_{i2}}{s_\beta}+i\gamma_5
\frac{R_{i3}}{t_\beta}) \, g^{\text{SM}}_{Huu} \;,
\\
g^{\text{C2HDM-II}}_{H_i,dd(ll)} &=& (\frac{R_{i1}}{c_\beta}+ i \gamma_5
R_{i3} t_\beta) \, g^{\text{SM}}_{Hdd(hll)} \;,
\eeq
where $u$ denotes up-type quarks, $d$ down-type quarks, $l$
leptons and $g_{HXX}^{\text{SM}}$ are the corresponding SM couplings of the
SM Higgs $H$ to $XX$. \s

We define the conditions to get alignment between the C2HDM and the SM
for the SM-like Higgs given by $H_i$ as
\begin{align}
\frac{g^{\text{C2HDM}}_{H_iVV}}{g_{HVV}^{\text{SM}}} &= (R_{i1}c_\beta + s_\beta
                                                       R_{i2}) =1 \;,\\
\frac{g^{\text{C2HDM}}_{H_iff}}{g^{\text{SM}}_{Hff}} & = 1 \quad \quad \Rightarrow \  \frac{R_{i2}}{s_\beta}=\frac{R_{i1}}{c_\beta}=1 \ \mbox{and} \ R_{i3}=0 \;,\\
\frac{g^{\text{C2HDM}}_{H_iH_iH_i}}{g^{\text{SM}}_{3H}} & = 1 \;,
\end{align}
with $m_{H_i}=m_H$. The implications of these limits for the various
possibilities of $H_1$, $H_2$, or $H_3$ being SM-like for the mixing
angles are as follows:

\paragraph*{$H_1$ SM-like scenario:}
To get the condition for alignment in this case, we have to solve
the following equation,
\beq
\frac{g^{\text{C2HDM}}_{H_1VV}}{g_{HVV}^{\text{SM}}}= 1 \quad
  \Rightarrow c_{\alpha_1} c_{\alpha_2} c_\beta + c_{\alpha_2}
    s_{\alpha_1} s_\beta &=&  1 \;, \nonumber \\
\Rightarrow c_{(\alpha_1-\beta)} c_{\alpha_2} &=& 1 \;,
\eeq
from which follows the solution $S_1$,
\begin{equation}
S_1=\{ \alpha_2=0 \quad \mbox{and} \quad \ \beta=\alpha_1 \} \;.
\end{equation}
Using $S_1$ it is easy to verify that
\begin{align}
 \frac{R_{12}}{s_\beta} = 1\ ,\
 \frac{R_{11}}{c_\beta} = 1\, , \
R_{13}=0 \;,
\end{align}
so that also
\beq
 g^{\text{C2HDM}}_{H_1ff} = g^{\text{SM}}_{Hff} \quad \mbox{ and } \quad
g^{\text{C2HDM}}_{H_1H_1H_1}=g^{\text{SM}}_{3H}  \;,
\eeq
as required in the alignment limit.
For the latter, we used the formulae given at the webpage \cite{trilc2hdm}.

\paragraph*{$H_2$ SM-like scenario:}

To get the conditions for alignment in this case, we analogously set
\begin{align}
\frac{g^{\text{C2HDM}}_{H_2VV}}{g_{HVV}^{\text{SM}}}= 1 \quad
  &\Rightarrow -(c_{\alpha_1} s_{\alpha_2} s_{\alpha_3} + s_{\alpha_1}
    c_{\alpha_3})c_\beta + (c_{\alpha_1} c_{\alpha_3} - s_{\alpha_1}
    s_{\alpha_2} s_{\alpha_3}) s_\beta = 1 \nonumber \\
& \Rightarrow
  -c_{\beta-\alpha_1}s_{\alpha_2}s_{\alpha_3}+c_{\alpha_3}s_{\beta-\alpha_1}
  =1 \;,
\end{align}
from which follows
\begin{align}
S_2 =  \big\{&\{ \alpha_3 = 0\ \mbox{and}\ \beta=\pi/2+\alpha_1\}\ \mbox{or} \ \{\alpha_2 = - \alpha_3 = \pm \pi/2\ \mbox{and}\ \beta=\alpha_1  \} \nonumber \\
&\mbox{or} \ \{\alpha_2= \alpha_3 = \pm \pi/2 \ \mbox{and} \ \beta=\alpha_1 + \pi\} \big\}.\;
\end{align}

We will call the first, the second and the third solution  $S_{21}$, $S_{22}$ and $S_{23}$, respectively.
With these conditions, the SM-like Higgs gauge and fermion couplings, as
well as the trilinear Higgs self-coupling approach the corresponding
SM values.

\paragraph*{$H_3$ SM-like scenario:}
For $H_3$ SM-like we get the condition
\begin{align}
\frac{g^{\text{C2HDM}}_{H_3VV}}{g_{HVV}^{\text{SM}}}= 1 \quad
  &\Rightarrow (-c_{\alpha_1} s_{\alpha_2} c_{\alpha_3}+s_{\alpha_1}
    s_{\alpha_3})c_\beta  -(c_{\alpha_1} s_{\alpha_3} +s_{\alpha_1}
    s_{\alpha_2} c_{\alpha_3})s_\beta = 1 \nonumber \\
& \Rightarrow -c_{\beta-\alpha_1}s_{\alpha_2}c_{\alpha_3}-s_{\alpha_3}s_{\beta-\alpha_1}=1 \;,
\end{align}
which is solved by
\begin{align}
S_3= &\big\{ \{ \alpha_3 = 0,\ \alpha_2=\pi/2 \  \mbox{and}\ \beta=\alpha_1  +\pi\} \ \mbox{or} \ \{\alpha_3 = \pm \pi/2 \ \mbox{and} \
\beta=\alpha_1 \mp \pi/2 \} \nonumber \\
 & \mbox{or} \ \{ \alpha_3 = 0, \alpha_2=-\pi/2 \  \mbox{and}\ \beta=\alpha_1\}  \big\} \;.
\end{align}
In the following, we will call the first, the second and the third solution $S_{31}$, $S_{32}$ and $S_{33}$, respectively.
For solution $S_3$ we obtain the SM values of the $H_3$ couplings to
gauge bosons and fermions as well as for the trilinear Higgs
self-coupling.

\paragraph*{Mass dependence in the alignment limit:}
In the C2HDM, the mass of the heaviest neutral scalar is a dependent
parameter, given by
\begin{equation}\hspace*{1cm}
m^2_{H_3}=\frac{m^2_{H_1}R_{13}(R_{12}t_\beta-R_{11})+m_{H_2}^2R_{23}(R_{22}
  t_\beta -R_{21})}{R_{33}(R_{31}-R_{32} t_\beta)} \;. \label{eq:massc2hdm}
\end{equation}
In the limit where $H_1$ is the SM-like Higgs boson, after
applying the solution $S_1$ to Eq.~(\ref{eq:massc2hdm}) we find that the terms that depend on $m_{H_1}$ vanish and we get
\begin{equation}
m_{H_3}=m_{H_2} \;.
\label{m3s1}
\end{equation}
For the case of $H_2$ being SM-like and applying the $S_2$ limits to
Eq.~(\ref{eq:massc2hdm}), we find that the first solution $S_{21}$
gives a similar equality as $S_1$, namely
\beq
m_{H_3}=m_{H_1}  \;,
\eeq
but for the second and third solution $S_{22}$, $S_{23}$, we
first apply the limit $\beta \rightarrow \alpha_1$
and then $\alpha_2 \rightarrow \pm \pi/2$, to obtain
\begin{equation}
m_{H_3}^2 |_{\beta \rightarrow \alpha_1,\alpha_2 \rightarrow \pm
  \pi/2}=\frac{(2m_{H_1}^2-m_{H_2}^2)c_{2 \alpha_1}}{2c_{2(\alpha_1\pm \alpha_3)}c_{\alpha_3}}+\frac{m_{H_2}^2
  }{2c_{\alpha_3}} \;.
\label{m3s221}
\end{equation}
From this limit we can conclude that for $\alpha_2\rightarrow
\pm \frac{\pi}{2}$ the lightest Higgs boson $H_1$ is a CP-odd Higgs boson
\cite{ElKaffas:2007rq} and from Eq.~(\ref{m3s221}) that the mass of
the third Higgs boson $H_3$ goes to infinity. \s

In the scenario where $H_3$ is SM-like, we have to change the
dependent mass from $m_{H_3}$ to $m_{H_1}$ or $m_{H_2}$. By imposing
$m_{H_1}$ as a dependent parameter, we have
\begin{equation}
m_{H_1}^2=\frac{m_{H_3}^2 R_{33} (R_{31}-R_{32} t_\beta)-m_{H_2}^2
  R_{23} (R_{22} t_\beta-R_{21})}{R_{13} (R_{12} t_\beta-R_{11})} \;.
\label{M1rr}
\end{equation}
By applying the solution $S_{31}$ or $S_{33}$, we find that $H_1$ is CP-odd with a mass equal to the mass of the SM-like Higgs boson $m_H$
\beq
m_{H_1}=m_{H_3}=m_{H} \;.
\eeq
For the solution $S_{32}$, we get
\beq
m_{H_1}=m_{H_2} \;.
\eeq
 Note that, if we apply the solution $S_{22}$ or $S_{23}$ to Eq.~(\ref{M1rr}), we obtain
\beq
m_{H_1}=m_{H_2}=m_{H} \;.
\eeq
However, applying the solution $S_{21}$ leads to
\beq
m_{H_1}=m_{H_3}\;.
\eeq

\section{Alignment Limit in the N2HDM \label{sec:alignn2hdm}}
We now turn to the derivation for the alignment limit in the N2HDM.
In terms of the matrix elements of the mixing matrix $R$, defined as
in the C2HDM, {\it cf.}~Eq.~(\ref{eq:rmixmatrixc2hdm}), the $H_i$ ($i=1,2,3$)
Higgs couplings to massive gauge bosons ($V=Z,W$) and to fermions $f$,
respectively, as well as the trilinear Higgs self-couplings, read in the
N2HDM-I and N2HDM-II
\beq
g^{\text{N2HDM-I/II}}_{H_iVV} &=& (R_{i1}c_\beta + s_\beta R_{i2}) \, g_{HVV}^{\text{SM}}\;,\\
g^{\text{N2HDM-I}}_{H_iff} &=& \frac{R_{i2}}{s_\beta} \, g^{\text{SM}}_{Hff}\;,\\
g^{\text{N2HDM-II}}_{H_iuu} &=& \frac{R_{i2}}{s_\beta} \, g^{\text{SM}}_{Huu}
\;,\\
g^{\text{N2HDM-II}}_{H_i,dd(ll)} &=& \frac{R_{i1}}{c_\beta}\,
g^{\text{SM}}_{Hdd(hll)} \;,\\
g^{\text{N2HDM-I/II}}_{H_iH_iH_i} & =&
\frac{3}{v}\Big(-\mu^2 \begin{bmatrix}R_{i2}^2 c_\beta
  (\frac{R_{i2}c_\beta}{s_\beta}-R_{i1})+R^2_{i1}s_\beta
  (\frac{R_{i1}s_\beta}{c_\beta}-R_{i2} )\end{bmatrix} \nonumber \\
&& +\frac{m_{H_i}^2}{v_S}\begin{bmatrix}
R_{i3}^3v + R_{i2}^3\frac{v_S}{s_\beta}+R_{i1}^3\frac{v_S}{c_\beta} \Big)
\\
\end{bmatrix}
\;,
\eeq
where $u$ denotes up-type quarks, $d$ down-type quarks and $l$
leptons, $g_{HXX}^{\text{SM}}$ are the corresponding SM couplings of the
SM Higgs $H$ to $XX$ and we have introduced
\beq
\mu^2=\frac{m_{12}^2}{s_\beta c_\beta} \;.
\eeq
We define the conditions to get alignment between the N2HDM and the SM
for the SM-like Higgs given by $H_i$ as
\begin{align}
\frac{g^{\text{N2HDM}}_{H_iVV}}{g_{HVV}^{\text{SM}}} &= (R_{i1}c_\beta + s_\beta
                                                      R_{i2}) =1 \;,\\
\frac{g^{\text{N2HDM}}_{H_iff}}{g^{\text{SM}}_{Hff}} & = 1 \;,\\
\frac{g^{\text{N2HDM}}_{H_iH_iH_i}}{g^{\text{SM}}_{3H}} & = 1 \;,
\end{align}
with $m_{H_i}=m_H$. The implications of these limits for the various
possibilities of $H_1$, $H_2$, or $H_3$ being SM-like for the mixing
angles are as follows:

\paragraph*{$H_1$ SM-like scenario:}
This is analogous to the C2HDM, where to get the condition of alignment we have to solve the following equation
\beq
\frac{g^{\text{N2HDM}}_{H_1VV}}{g_{HVV}^{\text{SM}}}= 1 \quad
  \Rightarrow c_{\alpha_1} c_{\alpha_2} c_\beta + c_{\alpha_2}
    s_{\alpha_1} s_\beta &=& 1 \;, \nonumber \\
\Rightarrow c_{\alpha_1-\beta} c_{\alpha_2} &=& 1 \;,
\eeq
from which we derive
\begin{equation}
S_1=\{ \alpha_2=0 \quad \mbox{and} \quad \ \beta=\alpha_1 \} \;.
\end{equation}
This leads to
\begin{align}
 \frac{R_{12}}{s_\beta} = 1,\quad
 \frac{R_{11}}{c_\beta} = 1 \; \mbox{and}\ R_{13}=0 \:.
\end{align}
We also get $g_{H_iff}^{\text{N2HDM}}=g_{Hff}^{\text{SM}}$ and $ g^{\text{N2HDM}}_{H_1H_1H_1}=g^{\text{SM}}_{3H} \;$ as required in the alignment limit.

\paragraph*{$H_2$ SM-like scenario:}
To get the condition of the alignment in this case we do the same
calculation as above,
\begin{align}
\frac{g^{\text{N2HDM}}_{H_2VV}}{g_{HVV}^{\text{SM}}}= 1 \quad
  &\Rightarrow -(c_{\alpha_1} s_{\alpha_2} s_{\alpha_3} + s_{\alpha_1}
    c_{\alpha_3})c_\beta + (c_{\alpha_1} c_{\alpha_3} - s_{\alpha_1}
    s_{\alpha_2} s_{\alpha_3}) s_\beta = 1 \nonumber \\
& \Rightarrow
  -c_{\beta-\alpha_1}s_{\alpha_2}s_{\alpha_3}+c_{\alpha_3}s_{\beta-\alpha_1}
  =1 \;,
\end{align}
from which follows
\begin{align}
S_2 =  \big\{&\{ \alpha_3 = 0\ \mbox{and}\ \beta=\pi/2+\alpha_1\}\ \mbox{or} \ \{\alpha_2 = - \alpha_3 = \pm \pi/2\ \mbox{and}\ \beta=\alpha_1  \} \nonumber \\
&\mbox{or} \ \{\alpha_2= \alpha_3 = \pm \pi/2 \ \mbox{and} \ \beta=\alpha_1 + \pi\} \big\}.\;
\end{align}
With these solutions we can find that all couplings of the SM-like Higgs
$H_2$ to massive gauge bosons and fermions, as well as the trilinear
Higgs self-coupling of $H_2$ approaches the alignment limit.

\paragraph*{$H_3$ SM-like scenario :}
For $H_3$ being SM-like, we require
\begin{align}
\frac{g^{\text{N2HDM}}_{H_3VV}}{g_{HVV}^{sm}}= 1 \quad &\Rightarrow (-c_{\alpha_1} s_{\alpha_2} c_{\alpha_3}+s_{\alpha_1} s_{\alpha_3})c_\beta  -(c_{\alpha_1} s_{\alpha_3} +s_{\alpha_1} s_{\alpha_2} c_{\alpha_3})s_\beta = 1 \\
& \Rightarrow -c_{\beta-\alpha_1}s_{\alpha_2}c_{\alpha_3}-s_{\alpha_3}s_{\beta-\alpha_1}=1
\end{align}
leading to
\begin{align}
S_3= &\big\{ \{ \alpha_3 = 0,\ \alpha_2=\pi/2 \  \mbox{and}\ \beta=\alpha_1  +\pi\} \ \mbox{or} \ \{\alpha_3 = \pm \pi/2 \ \mbox{and} \
\beta=\alpha_1 \mp \pi/2 \} \nonumber \\
 & \mbox{or} \ \{ \alpha_3 = 0, \alpha_2=-\pi/2 \  \mbox{and}\ \beta=\alpha_1\}  \big\} \;.
\end{align}
In applying these solutions, we explicitly checked that the $H_3$
couplings approach the alignment values.
\end{appendix}

%%%%%%%%%%%%%%%%%%%%%%%%%%%%%%%%%%%%%%%%%%%%%%%%%%%%%%%%%%%%%%

\clearpage
%\newpage
%\vspace*{1cm}
%\bibliographystyle{h-physrev}
%% \bibliographystyle{h-physrev.bst}
%\bibliography{comparisonshh.bib}

%%%%%%%%%%%%%%%%%%%%%%%%%%%%%%%%%%%%%%%%%%%%%%%%%%%%%%%%%%%%%%

\end{document}